\documentclass[10pt]{article}
\pdfoutput=1
\usepackage{mathStyle}
\usepackage{amsmath,amssymb,amsthm,latexsym,bbm,calc,wasysym,mathtools,empheq, yfonts,upgreek}
\usepackage[english]{babel}
\usepackage{relsize}
\usepackage{graphicx,color}
\usepackage[dvipsnames]{xcolor}
\usepackage[makeroom]{cancel}

\usepackage{tikz-cd}
\usepackage{tikz}
\usetikzlibrary{shapes.geometric}
\usetikzlibrary{arrows,matrix,calc,scopes,decorations.markings,intersections}
\usetikzlibrary{arrows.meta}
\usetikzlibrary{decorations.pathreplacing,angles,quotes,patterns}

\usepackage{xspace}
\usepackage{pifont,dsfont}
\usepackage{marvosym}
\usepackage{slashed}
\usepackage{subfigure}
\usepackage{sidecap}
\usepackage{stmaryrd}
\usepackage{empheq}

\usepackage[export]{adjustbox}
\usepackage{placeins}
\usepackage{enumitem}
\usepackage{verbatim}

\DeclareMathAlphabet{\mathpzc}{OT1}{pzc}{m}{it}

\usepackage{amsthm}
\newtheoremstyle{mydef}%
	{0.9em} %Space above
	{0.7em}% Space below
	{\itshape\hangindent=2em}% Body font 
	{1.8em}% Indent amount
	{\scshape}% ⟨Theorem head font⟩
	{.}% ⟨Punctuation after theorem head ⟩
	{.5em}% Space after theorem head 
	{}%
	
\theoremstyle{mydef}
\newtheorem{definition}{Definition}
\numberwithin{definition}{section}

\theoremstyle{mydef}

\numberwithin{lemma}{section}

\theoremstyle{mydef}

\numberwithin{theorem}{section}

\theoremstyle{mydef}
\newtheorem{example}{Example}
\numberwithin{example}{section}

\theoremstyle{mydef}

\numberwithin{convention}{section}

\theoremstyle{mydef}
\newtheorem{proposition}{Proposition}
\numberwithin{proposition}{section}

\theoremstyle{mydef}

\numberwithin{remark}{section}

\theoremstyle{mydef}

\numberwithin{conjecture}{section}

\newcommand{\rU}{{\rm U}}

\newcommand{\ket}[1]{| #1 \rangle}
\newcommand{\bra}[1]{\langle #1 |}
\newcommand{\Z}{\mc{Z}^{G}_{\omega}}
\newcommand{\Zo}[1]{\mc{Z}^{G}_{\omega}[#1]}
\newcommand{\cob}[3]{{_{#1}#2_{#3}}}
\newcommand{\Hilb}[1]{\mc{H}^{G}_{\omega}[#1]}

\newcommand{\Hambulk}[1]{\mathbb{H}^{G}_{\omega}[#1]^{\rm bulk}}

\newcommand{\gsopen}[1]{\mc{V}^{G}_{\omega}[#1]}
\newcommand{\gsTwo}[1]{\mc{V}^{G,A,B}_{\alpha,\phi,\psi}[#1]}
\newcommand{\gsThree}[1]{\mc{V}^{G,A,B}_{\pi,\lambda,\mu}[#1]}

\newcommand{\I}{\mathbb{I}}
\newcommand{\la}{\langle}
\newcommand{\ra}{\rangle}
\newcommand{\q}{\quad}
\newcommand{\nn}{\nonumber}
\newcommand{\sss}{\scriptstyle}
\newcommand{\ssss}{\scriptscriptstyle}
\newcommand{\bul}{{\sss \bullet}}
\newcommand{\snum}[1]{\text{\small $#1$}}
\newcommand{\mc}[1]{\mathcal{#1}}
\newcommand{\fr}[1]{\mathfrak{#1}}

\newcommand{\biGr}[3]{#1 \! \xrightarrow[{\raisebox{1.2ex-\heightof{$\scriptstyle#3$}}[0pt]{$\scriptstyle \; #3 \;$}}]{\; #2 \;} \!}
\newcommand{\biGrZ}[2]{\xrightarrow[{\raisebox{1.2ex-\heightof{$\scriptstyle#2$}}[0pt]{$\scriptstyle \; #2 \;$}}]{#1}}
\newcommand{\biGrFoot}[3]{#1 {\raisebox{-1pt}[0pt]{$\xrightarrow[{\raisebox{2pt}[0pt]{$\scriptscriptstyle #3$}}]{\scriptscriptstyle #2}$}}}

\newcommand{\grAlg}{\mathbb C[G_{AB}]^\alpha_{\phi \psi}}
\newcommand{\grAlgL}{\mathbb C[\Lambda(G_{AB})]^\alpha_{\phi \psi}}
\newcommand{\grAlgLbis}{\mathbb C[\Lambda(G_{BC})]^\alpha_{\psi \varphi}}
\newcommand{\grAlgLter}{\mathbb C[\Lambda(G_{AC})]^\alpha_{\phi \varphi}}
\newcommand{\grAlgLqua}{\mathbb C[\Lambda(G_{CD})]^\alpha_{\varphi \chi}}
\newcommand{\CC}[6]{\Big[{}^{{\raisebox{1pt}[0pt]{${\scriptstyle #1}$}}}_{#4} {}^{{\raisebox{1pt}[0pt]{${\scriptstyle #2}$}}}_{#5} \Big| {}^{{\raisebox{1pt}[0pt]{${\scriptstyle #3}$}}}_{#6}\Big]} 
\newcommand{\SixJ}[6]{\Big\{{}^{{\raisebox{1pt}[0pt]{${\scriptstyle #1}$}}}_{#4} {}^{{\raisebox{1pt}[0pt]{${\scriptstyle #2}$}}}_{#5} {}^{{\raisebox{1pt}[0pt]{${\scriptstyle #3}$}}}_{#6}\Big\}} 
\newcommand{\exBdry}{\partial \Sigma^{\rm o}{\sss |}_{\rm ex.}}

\newcommand{\sfT}{\textsf{\small T}}
\newcommand{\ssfT}{\textsf{\scriptsize T}}
\newcommand{\qp}{\mathrlap{\hspace{2pt} p}q \, }
\newcommand{\btimes}{\text{\small{\raisebox{-0.4pt}[0pt]{$ \; \boxtimes \; $}}}}
\newcommand{\btimesFt}{\text{\scriptsize{\raisebox{-0.3pt}[0pt]{$ \,  \boxtimes \, $}}}}
\newcommand{\btimesFFt}{\text{\tiny{\raisebox{-0.3pt}[0pt]{$  \boxtimes  $}}}}
\newcommand{\bigbplus}{
	\mathop{
		\vphantom{\bigoplus} 
		\mathchoice
		{\vcenter{\hbox{\resizebox{\widthof{$\displaystyle\bigoplus$}}{!}{$\boxplus$}}}}
		{\vcenter{\hbox{\resizebox{\widthof{$\bigoplus$}}{!}{$\boxplus$}}}}
		{\vcenter{\hbox{\resizebox{\widthof{$\scriptstyle\oplus$}}{!}{$\boxplus$}}}}
		{\vcenter{\hbox{\resizebox{\widthof{$\scriptscriptstyle\oplus$}}{!}{$\boxplus$}}}}
	}\displaylimits 
}
\newcommand{\sfpartial}{\textsf{\reflectbox{6}}}

\tikzset{->1-/.style={decoration={
			markings,
			mark=at position #1 with {\arrow{Triangle[fill=black, width=3.5pt, length=3.5pt]}}},postaction={decorate}}}
\tikzset{->2-/.style={decoration={
					markings,
					mark=at position #1 with {\arrow{Triangle[fill=white, width=5pt, length=5pt]}}},postaction={decorate}}}

\tikzset{->-/.style={decoration={
			markings,
			mark=at position #1 with {\arrow{latex}}},postaction={decorate}}}

\tikzset{-<-/.style={decoration={
			markings,
			mark=at position #1 with {\arrow{latex reversed}}},postaction={decorate}}}

\tikzset{
	every picture/.append style={
		line join=round,
		line cap=round,
		line width = 0.8pt
	}
}

\definecolor{bdry1}{HTML}{4db8ff}
\definecolor{bdry2}{HTML}{ffb366}
\definecolor{bdry3}{HTML}{70db70}

\tikzstyle{dot}=[dotted, line width=0.8pt]
\tikzstyle{op}=[opacity=0.4]
\tikzstyle{op6}=[opacity=0.6]
\tikzstyle{op5}=[opacity=0.5]
\tikzstyle{op4}=[opacity=0.4]
\tikzstyle{op3}=[opacity=0.3]
\tikzstyle{op2}=[opacity=0.2]
\tikzstyle{op1}=[opacity=0.1]
\tikzstyle{sha}=[shorten >= 0.3em]
\tikzstyle{shb}=[shorten <= 0.3em]
\tikzstyle{shab}=[shorten >= 0.3em, shorten <= 0.3em]
\tikzstyle{bulk}=[fill=black,opacity=0.15,postaction={pattern=crosshatch,pattern color=white, opacity=1}]
\tikzstyle{bulkExc}=[fill=black,opacity=0.65,postaction={pattern=crosshatch,pattern color=white, opacity=1}]
\tikzstyle{bulkEdge}=[line width = 0.8pt]
\tikzstyle{bdryEdge1}=[line width = 1.6pt, color=bdry1]
\tikzstyle{bdryEdge2}=[line width = 1.6pt, color=bdry2]
\tikzstyle{bdryEdge3}=[line width = 1.6pt, color=bdry3]
\tikzstyle{bdryPlaq1}=[op4, color=bdry1]
\tikzstyle{bdryPlaq2}=[op4, color=bdry2]
\tikzstyle{bdryPlaq3}=[op4, color=bdry3]

\newcommand{\interL}[5]{
	\path[name path=P#1#2] (#1) -- (#2);
	\path[name path=P#3#4] (#3) -- (#4);
	\path [name intersections={of=P#1#2 and P#3#4,by={sp}}];
	\draw[sha, #5] (#1) -- (sp);
	\draw[shb, #5] (sp) -- (#2);
}

\newcommand{\interLLL}[9]{
	\path[name path=P#1#2] (#1) -- (#2);
	\path[name path=P#3#4] (#3) -- (#4);
	\path[name path=P#5#6] (#5) -- (#6);
	\path[name path=P#7#8] (#7) -- (#8);		
	\path [name intersections={of=P#1#2 and P#3#4,by={sp1}}];
	\path [name intersections={of=P#1#2 and P#5#6,by={sp2}}];
	\path [name intersections={of=P#1#2 and P#7#8,by={sp3}}];		
	\draw[sha, #9] (#1) -- (sp1);
	\draw[shab, #9] (sp1) -- (sp2);
	\draw[shab, #9] (sp2) -- (sp3);
	\draw[shb, #9] (sp3) -- (#2);
}

\def\centerarc[#1](#2)(#3:#4:#5)
{ \draw[#1] ($(#2)+({#5*cos(#3)},{#5*sin(#3)})$) arc (#3:#4:#5); }

\tikzset{
	partial ellipse/.style args={#1:#2:#3}{
		insert path={+ (#1:#3) arc (#1:#2:#3)}
	}
}

\newcommand{\pinchedinterval}[1]{
	\begin{tikzpicture}[scale=0.6,baseline={([yshift=-.5ex]current bounding box.center)}]
	\def\a{1.5}
	\coordinate (1) at (0,0);
	\coordinate (2) at (\a,0);
	\draw[] (1) to [bend right = 60]node[anchor=north]{$\overline{#1}$} (2);
	\draw[] (1) to [bend left =60]node[anchor=south]{$#1$} (2);
	\end{tikzpicture}
}

\newcommand{\interval}[1]{
	\begin{tikzpicture}[scale=0.6,baseline={([yshift=-.5ex]current bounding box.center)}]
	\def\a{1.5}
	\def\b{1.5}
	\coordinate (1) at (0,0);
	\coordinate (2) at (\a,0);
	\coordinate (3) at (0,\b);
	\coordinate (4) at (\a,\b);
	\draw[] (1) to node[anchor=north]{$\overline{#1}$} (2);
	\draw[] (3) to node[anchor=south]{$#1$} (4);
	\draw[] (1) to (3);
	\draw[] (2) to (4);
	\end{tikzpicture}
}

\newcommand{\clst}[1]{
	\begin{tikzpicture}[scale=0.4,baseline=-0.5em, line width=0.8pt]
	\coordinate (0) at (-1,3);
	\coordinate (1) at (0.2,0); 
	\coordinate (2) at (-1,1);
	\coordinate (3) at (2.5,-1);
	\coordinate (4) at (2,1);
	\coordinate (5) at (-2,-1);
	\coordinate (6) at (0,-1.8);
	\coordinate (7) at (-2.5,1);
	\coordinate (8) at (0.3,1.8);
	\coordinate (9) at (-1.3,-2.2);
	\coordinate (10) at (-3,-1.5);
	\ifthenelse{#1 = 1}{
		\draw[] (1) -- (2);
		\draw[] (1) -- (3);
		\draw[] (1) -- (4);
		\draw[] (1) -- (5);
		\draw[] (1) -- (6);
		\draw[] (2) -- (4);	
		\draw[] (4) -- (3);
		\draw[] (3) -- (6);
		\draw[] (6) -- (5);
		\draw[] (5) -- (2);
		\draw[] (2) -- (7);
		\draw[] (7) -- (5);
		\draw[] (2) -- (8);
		\draw[] (8) -- (4);
		\draw[] (5) -- (9);
		\draw[] (9) -- (6);
		\draw[] (7) -- (8);	
		\draw[] (9) -- (10);
		\draw[] (5) -- (10);
		\draw[] (7) -- (10);	
	}{}	
	\ifthenelse{#1 = 2}{
		\draw[] (1) -- (2);
		\draw[] (1) -- (3);
		\draw[] (1) -- (4);
		\draw[] (1) -- (5);
		\draw[] (1) -- (6);
		\draw[] (2) -- (4);	
		\draw[] (4) -- (3);
		\draw[] (3) -- (6);
		\draw[] (6) -- (5);
		\draw[] (5) -- (2);
	}{}
	\ifthenelse{#1 = 3}{
		\draw[] (0) -- (1);
		\draw[] (0) -- (2);
		\draw[] (0) -- (3);
		\draw[] (0) -- (4);
		\draw[] (0) -- (5);
		\draw[] (0) -- (6);		
		\draw[] (1) -- (3);
		\draw[] (1) -- (6);
		\interL{1}{2}{0}{6}{};
		\interL{1}{5}{0}{6}{};
		\interL{1}{4}{0}{3}{};	
		\interLLL{2}{4}{0}{6}{0}{1}{0}{3}{};			
		\draw[] (4) -- (3);
		\draw[] (3) -- (6);
		\draw[] (6) -- (5);
		\draw[] (5) -- (2);
		\node[left] at (0) {${\scriptstyle 0'}$};
	}{}		
	\node[below, xshift=0.3em] at (1) {${\scriptstyle 0}$};
	\end{tikzpicture}	
}

\newcommand{\HamONE}[2]{
	\begin{tikzpicture}[scale=#1,baseline=-0.5em]
	\coordinate (0) at (0,1);
	\coordinate (1) at (1,-0.4);
	\coordinate (2) at (0.55,-1);
	\coordinate (3) at (-1.2,-0.55);
	\coordinate (4) at (0,-0.1);	
	
	\draw[] (0) -- (1);
	\draw[] (1) -- (2);
	\draw[] (2) -- (3);
	\draw[] (0) -- (3);
	\draw[] (0) -- (2);
	\draw[] (3) -- (4);
	\draw[] (0) -- (4);
	\draw[] (2) -- (4);
	\draw[shorten >= 0.3em] (1) -- (intersection of 1--4 and 0--2);
	\draw[shorten <= 0.3em] (intersection of 1--4 and 0--2) -- (4);
	\draw[shorten >= 0.3em] (1) -- (intersection of 1--3 and 0--2);
	\draw[shorten <= 0.3em, shorten >=0.3em] (intersection of 1--3 and 0--2) -- (intersection of 1--3 and 2--4);
	\draw[shorten <= 0.3em] (intersection of 1--3 and 2--4) -- (3);
	
	\ifthenelse{#2 = 1}{	
		\node[above] at (0) {{$\scriptstyle{0}$}};	
		\node[right] at (1) {{$\scriptstyle{1}$}};
		\node[below] at (2) {{$\scriptstyle{2}$}};
		\node[left] at (3) {{$\scriptstyle{3}$}};	
		\node[left, yshift=1.5] at (4) {{$\scriptstyle{4}$}};}{}
	
	\ifthenelse{#2 = 2}{	
		\node[above] at (0) {{$\scriptstyle{0}$}};	
		\node[right] at (1) {{$\scriptstyle{1}$}};
		\node[below] at (2) {{$\scriptstyle{2}$}};
		\node[left] at (3) {{$\scriptstyle{3}$}};	
		\node[left, yshift=1.5] at (4) {{$\scriptstyle{4'}$}};}{}
	\end{tikzpicture}
}

\newcommand{\HamTWO}[2]{
	\begin{tikzpicture}[scale=#1,baseline=-0.5em]
	\coordinate (0) at (0,1);
	\coordinate (1) at (1,-0.4);
	\coordinate (2) at (0.55,-1);
	\coordinate (3) at (-1.2,-0.55);
	\coordinate (4) at (0,-0.1);
	\coordinate (5) at (-1.5,1.1);	
	
	\draw[] (0) -- (1);
	\draw[] (1) -- (2);
	\draw[] (2) -- (3);
	\draw[shorten >= 0.3em] (0) -- (intersection of 0--3 and 2--5);
	\draw[shorten <= 0.3em] (intersection of 0--3 and 2--5) -- (3);
	\draw[] (0) -- (2);
	\draw[shorten >= 0.3em] (3) -- (intersection of 3--4 and 2--5);
	\draw[shorten <= 0.3em] (intersection of 3--4 and 2--5) -- (4);
	\draw[] (0) -- (4);
	\draw[] (2) -- (4);
	\draw[] (0) -- (5);
	\draw[] (3) -- (5);
	\draw[shorten >= 0.3em] (1) -- (intersection of 1--4 and 0--2);
	\draw[shorten <= 0.3em] (intersection of 1--4 and 0--2) -- (4);
	\draw[shorten >= 0.3em] (1) -- (intersection of 1--3 and 0--2);
	\draw[shorten <= 0.3em, shorten >=0.3em] (intersection of 1--3 and 0--2) -- (intersection of 1--3 and 2--4);
	\draw[shorten <= 0.3em, shorten >= 0.3em] (intersection of 1--3 and 2--4) -- (intersection of 1--3 and 2--5);
	\draw[shorten <= 0.3em] (intersection of 1--3 and 2--5) -- (3);
	\draw[shorten >= 0.3em] (4) -- (intersection of 0--3 and 4--5);
	\draw[shorten <= 0.3em] (intersection of 0--3 and 4--5) -- (5);
	\draw[] (2) -- (5);
	\draw[shorten >= 0.3em] (1) -- (intersection of 1--5 and 0--2);
	\draw[shorten >= 0.3em, shorten <= 0.3em] (intersection of 1--5 and 0--2) -- (intersection of 1--5 and 0--4);
	\draw[shorten >= 0.3em, shorten <= 0.3em] (intersection of 1--5 and 0--4) -- (intersection of 1--5 and 0--3);
	\draw[shorten <= 0.3em] (intersection of 1--5 and 0--3) -- (5);

	\ifthenelse{#2 = 1}{	
		\node[above] at (0) {{$\scriptstyle{0}$}};	
		\node[right] at (1) {{$\scriptstyle{1}$}};
		\node[below] at (2) {{$\scriptstyle{2}$}};
		\node[left] at (3) {{$\scriptstyle{3}$}};
		\node[above] at (5) {{$\scriptstyle{4'}$}};		
		\node[left, yshift=1.2, xshift=-0.05em] at (4) {{$\scriptstyle{4}$}};}{}
	\end{tikzpicture}
}

\newcommand{\snippetBulk}[2]{
	\begin{tikzpicture}[scale=#1, baseline=0.1em]
	\coordinate (1) at (0,0);
	\coordinate (2) at (1,0.5);

	\ifthenelse{#2=1}{
		\fill[bulk] (1) rectangle (2);
	}{}	
	\ifthenelse{#2=2}{
		\fill[bulkExc] (1) rectangle (2);
	}{}
	\end{tikzpicture}
}

\newcommand{\exampleTwoD}[2]{
	\begin{tikzpicture}[scale=#1, baseline=1.8em]
	\coordinate (1) at (0,1);
	\coordinate (2) at (3,1);
	\coordinate (3) at (3,-0.1);
	\coordinate (4) at (0,-0.1);
	\coordinate (5) at (0.5,1);
	\coordinate (6) at (1.5,0.2);
	\coordinate (7) at (2.5,1);
	\coordinate (8) at (1.5,1);
	\coordinate (9) at (1.5,2);
	\coordinate (10) at (2.5,0.4);

	\ifthenelse{#2 = 1}{
		\fill[bulk] (1) -- (2) -- (3) -- (4) -- (1);
		\draw[] (5) -- (6) -- (7);
		\draw[] (6) -- (8);
		\draw[bdryEdge2] (1) -- (2);
		\node[above] at (5) {${\sss 0}$};
		\node[above] at (8) {${\sss 1}$};
		\node[above] at (7) {${\sss 2}$};
		\node[below] at (6) {${\sss 3}$};
	}{}
	\ifthenelse{#2 = 2}{
		\fill[bulk] (5) -- (6) -- (7) -- (5);
		\draw[] (5) -- (6) -- (7);
		\draw[] (6) -- (8);
		\draw[bdryEdge2] (5) -- (7);
		\node[above] at (5) {${\sss 0}$};
		\node[above] at (8) {${\sss 1}$};
		\node[above] at (7) {${\sss 2}$};
		\node[below] at (6) {${\sss 3}$};
	}{}
	\ifthenelse{#2 = 3}{
		\fill[bulk] (5) -- (6) -- (7) -- (5);
		\draw[] (5) --node[left,xshift=-0.1em]{${\sss ag}$} (6) --node[right]{${\sss a'^{-1}g}$} (7);
		\draw[] (6) --node[right]{${\sss g}$} (8);
		\draw[bdryEdge2] (5) --node[above, text=black, opacity=1]{${\sss a}$} (8);
		\draw[bdryEdge2] (8) --node[above, text=black, opacity=1]{${\sss a'}$} (7);
		\node[above] at (5) {${\sss 0}$};
		\node[above] at (8) {${\sss 1}$};
		\node[above] at (7) {${\sss 2}$};
		\node[below] at (6) {${\sss 3}$};
	}{}
	\ifthenelse{#2 = 4}{
		\fill[bulk] (5) -- (7) -- (10) -- (5);
		\draw[bdryEdge2] (5) -- (7);
		\fill[bulk] (9) -- (10) -- (7) -- (9);
		\draw[bdryEdge2] (8) -- (9);
		\draw[] (10) -- (8);
		\fill[bulk] (9) -- (10) -- (5) -- (9);
		\draw[] (5) -- (10) -- (7);
		\draw[] (10) -- (9);
		\draw[bdryEdge2] (5) -- (9) -- (7);
		\node[left] at (5) {${\sss 0}$};
		\node[above, xshift=-0.5em] at (8) {${\sss 1}$};
		\node[right] at (7) {${\sss 2}$};
		\node[below] at (10) {${\sss 3}$};
		\node[above] at (9) {${\sss \tilde 1}$};
	}{}
	\ifthenelse{#2 = 5}{
		\fill[bulk] (5) -- (6) -- (7) -- (5);
		\draw[] (5) --node[left,xshift=-0.1em]{${\sss ag}$} (6) --node[right, pos=0.4]{${\sss a'^{-1}g}$} (7);
		\draw[] (6) --node[right, pos=0.7,xshift=-0.2em]{${\sss \tilde a^{-1} \!g}$} (8);
		\draw[bdryEdge2] (5) --node[above, text=black, opacity=1]{${\sss a \tilde a}$} (8);
		\draw[bdryEdge2] (8) --node[above, text=black, opacity=1]{${\sss \tilde a^{-1}a'}$} (7);
		\node[above] at (5) {${\sss 0}$};
		\node[above] at (8) {${\sss 1}$};
		\node[above] at (7) {${\sss 2}$};
		\node[below] at (6) {${\sss 3}$};
	}{}
	\ifthenelse{#2 = 6}{
		\fill[bulk] (5) -- (6) -- (7) -- (5);
		\draw[] (5) --node[left,xshift=-0.1em]{${\sss ag}$} (6) --node[right]{${\sss a'^{-1}g}$} (7);
		\draw[bdryEdge2] (5) --node[above, text=black, opacity=1]{${\sss aa'}$} (7);
		\node[above] at (5) {${\sss 0}$};
		\node[above] at (7) {${\sss 2}$};
		\node[below] at (6) {${\sss 3}$};
	}{}
	\ifthenelse{#2 = 7}{
		\fill[bulk] (0.5,-0.75) -- (1,0) -- (2,-0.6);
		\draw[bdryEdge2] (0.5,-0.75) -- (1,0);
		\fill[bulk] (0,0) -- (1,0) -- (2,-0.6) -- (0,0);
		\fill[bulk] (0,0) -- (2,-0.6) -- (0.5,-0.75) -- (0,0);
		\draw[] (1,0) -- (2,-0.6) -- (0,0);
		\draw[] (2,-0.6) -- (0.5,-0.75);
		\draw[bdryEdge2] (0,0) -- (1,0);
		\draw[bdryEdge2] (0,0) -- (0.5,-0.75);
		\node[above] at (0,0) {${\sss 0}$};
		\node[below] at (0.5,-0.75) {${\sss 1}$};
		\node[above] at (1,0) {${\sss 2}$};
		\node[above] at (2,-0.6) {${\sss 3}$};
	}{}
	\end{tikzpicture}
}

\newcommand{\bdryInterface}[1]{
	\begin{tikzpicture}[scale=#1, baseline=1.3em]
	\coordinate (1) at (0,1);
	\coordinate (2) at (4,1);
	\coordinate (3) at (0,0);
	\coordinate (4) at (4,0);
	\coordinate (5) at (2,1);

	\fill[bulk] (1) -- (2) -- (4) -- (3) -- (1);
	\draw[bdryEdge2] (1) -- (5);
	\draw[bdryEdge1] (5) -- (2);
	\node[circle, fill = black, inner sep=1.6pt] at (5) {};
	\node[above] at (1,1) {${\sss A_\phi}$};
	\node[above] at (3,1) {${\sss B_\psi}$};
	\end{tikzpicture}
}

\newcommand{\bdryInterfaceBis}[1]{
	\begin{tikzpicture}[scale=#1, baseline=1.3em]
	\coordinate (1) at (0,1);
	\coordinate (2) at (4,1);
	\coordinate (3) at (0,0);
	\coordinate (4) at (4,0);
	\coordinate (5) at (2,1);
	\coordinate (6) at (6,0);
	\coordinate (7) at (6,1);
	
	\fill[bulk] (3) rectangle (7);
	\draw[bdryEdge2] (1) -- (5);
	\draw[bdryEdge1] (5) -- (2);
	\draw[bdryEdge3] (2) -- (7);
	\node[circle, fill = black, inner sep=1.6pt] at (5) {};
	\node[circle, fill = black, inner sep=1.6pt] at (2) {};
	\node[above] at (1,1) {${\sss A_\phi}$};
	\node[above] at (3,1) {${\sss B_\psi}$};
	\node[above] at (5,1) {${\sss C_\varphi}$};
	\end{tikzpicture}
}

\newcommand{\bdryInterfaceExc}[2]{
	\begin{tikzpicture}[scale=#1, baseline=1.3em]
	\coordinate (1) at (0,1);
	\coordinate (2) at (4,1);
	\coordinate (3) at (0,0);
	\coordinate (4) at (4,0);
	\coordinate (5) at (2,1);
	\coordinate (6) at (1.6,1);
	\coordinate (7) at (2.4,1);
	
	\ifthenelse{#2=1}{
		\begin{scope}
		\clip (1) -- (6) to[out=-90,in=-90,looseness=1.5] (7) -- (2) -- (4) -- (3) -- (1);
		\fill[bulk] (1) -- (2) -- (4) -- (3) -- (1);
		\end{scope}
		\fill[bulkExc] (6) to[out=-90,in=-90,looseness=1.5] (7);	
		\draw[bdryEdge2] (1) -- (5);
		\draw[bdryEdge1] (5) -- (2);
		\node[circle, fill = black, inner sep=1.6pt] at (5) {};
	}{}
	\ifthenelse{#2=2}{
		\begin{scope}
		\clip (1) -- (6) to[out=-90,in=-90,looseness=1.5] (7) -- (2) -- (4) -- (3) -- (1);
		\fill[bulk] (1) -- (2) -- (4) -- (3) -- (1);
		\end{scope}
		\draw[bulkEdge] (6) to[out=-90,in=-90,looseness=1.5] (7);
		\draw[bdryEdge2] (1) -- (6);
		\draw[bdryEdge1] (7) -- (2);
	}{}
	\end{tikzpicture}
}

\newcommand{\twoDiskExc}[2]{
	\begin{tikzpicture}[scale=#1, baseline=2.5em]
	\coordinate (1) at (0,0);
	\coordinate (2) at (0,2);
	\coordinate (3) at (-0.4,2);
	\coordinate (4) at (0.4,2);
	\coordinate (5) at (-0.4,0);
	\coordinate (6) at (0.4,0);
	\coordinate (7) at (-0.4,1.5);
	\coordinate (8) at (0.4,1.5);
	\coordinate (9) at (-0.4,0.5);
	\coordinate (10) at (0.4,0.5);
	
	\ifthenelse{#2=1}{
		\begin{scope}
		\clip (3) to[out=-90,in=-90,looseness=1.5] (4);
		\fill[bulkExc] (1) arc (-90:270:1);
		\end{scope}
		\begin{scope}
		\clip (5) to[out=90,in=90,looseness=1.5] (6);
		\fill[bulkExc] (1) arc (-90:270:1);
		\end{scope}
		\begin{scope}
		\clip (-1,2) -- (3) to[out=-90,in=-90,looseness=1.5] (4) -- (1,2) -- (1,0) -- (6) to[out=90,in=90,looseness=1.5] (5) -- (-1,0) -- (-1,2);
		\fill[bulk] (1) arc (-90:270:1);
		\end{scope}
		\draw[bdryEdge1] (1) arc[start angle=-90, end angle=90, radius=1];
		\draw[bdryEdge2] (1) arc[start angle=270, end angle=90, radius=1];
		\node[circle, fill = black, inner sep=1.6pt] at (1) {};
		\node[circle, fill = black, inner sep=1.6pt] at (2) {};	
	}{}
	\ifthenelse{#2=2}{		
		\begin{scope}
		\clip (-1,2) -- (3) to[out=-90,in=-90,looseness=1.5] (4) -- (1,2) -- (1,0) -- (6) to[out=90,in=90,looseness=1.5] (5) -- (-1,0) -- (-1,2);
		\fill[bulk] (1) arc (-90:270:1);
		\end{scope}
		\begin{scope}
		\clip (1) arc[start angle=-90, end angle=270, radius=1];
		\draw[bulkEdge] (3) to[out=-90,in=-90,looseness=1.5] (4);
		\end{scope}
		\begin{scope}
		\clip (1) arc[start angle=-90, end angle=270, radius=1];
		\draw[bulkEdge] (5) to[out=90,in=90,looseness=1.5] (6);
		\end{scope}
		\centerarc[bdryEdge1](0,1)(-67:67:1)
		\centerarc[bdryEdge2](0,1)(113:247:1)
	}{}
	\ifthenelse{#2=3}{	
		\fill[bulk] (7) rectangle (10);	
		\draw[bulkEdge] (7) -- (8);
		\draw[bulkEdge] (9) -- (10);
		\draw[bdryEdge2] (7) -- (9);
		\draw[bdryEdge1] (8) -- (10);
	}{}
	\end{tikzpicture}
}

\newcommand{\symmetryMap}[2]{
	\begin{tikzpicture}[scale=#1, baseline=1.3em]
	\coordinate (1) at (0,1);
	\coordinate (2) at (4,1);
	\coordinate (3) at (0,0);
	\coordinate (4) at (4,0);
	\coordinate (5) at (2,1);
	\coordinate (6) at (1.6,1);
	\coordinate (7) at (2.4,1);
	
	\coordinate (8) at (1.6,1.5);
	\coordinate (9) at (2.4,1.5);
	\coordinate (10) at (1.6,2.5);
	\coordinate (11) at (2.4,2.5);
	
	\coordinate (12) at (1.6,2);
	\coordinate (13) at (2.4,2);
	
	\coordinate (14) at (1.6,2.5);
	\coordinate (15) at (2.4,2.5);
	\coordinate (16) at (1.6,3.5);
	\coordinate (17) at (2.4,3.5);
	
	\ifthenelse{#2=1}{
		\begin{scope}
		\clip (1) -- (6) to[out=-90,in=-90,looseness=1.5] (7) -- (2) -- (4) -- (3) -- (1);
		\fill[bulk] (1) -- (2) -- (4) -- (3) -- (1);
		\end{scope}
		\draw[bulkEdge] (6) to[out=-90,in=-90,looseness=1.5] (7);
		\draw[bdryEdge2] (1) -- (6);
		\draw[bdryEdge1] (7) -- (2);
		
		\fill[bulk] (8) rectangle (11);	
		\draw[bulkEdge] (8) -- (9);
		\draw[bulkEdge] (10) -- (11);
		\draw[bdryEdge2] (8) -- (10);
		\draw[bdryEdge1] (9) -- (11);
		\node[] at (2,1.1) {${\cup}$};
	}{}
	
	\ifthenelse{#2=2}{
		
		\fill[bulk] (1) -- (6) -- (12) -- (13) -- (7) -- (2) -- (4) -- (3) -- (1);
		\draw[bdryEdge2] (1) -- (6);
		\draw[bdryEdge1] (7) -- (2);		
		
		\draw[bulkEdge] (12) -- (13);
		\draw[bdryEdge2] (6) -- (12);
		\draw[bdryEdge1] (7) -- (13);
	}{}
	
	\ifthenelse{#2=3}{
		\begin{scope}
		\clip (1) -- (6) to[out=-90,in=-90,looseness=1.5] (7) -- (2) -- (4) -- (3) -- (1);
		\fill[bulk] (1) -- (2) -- (4) -- (3) -- (1);
		\end{scope}
		\draw[bulkEdge] (6) to[out=-90,in=-90,looseness=1.5] (7);
		\draw[bdryEdge2] (1) -- (6);
		\draw[bdryEdge1] (7) -- (2);
	}{}
	
	\ifthenelse{#2=4}{
		\fill[bulk] (1) -- (6) -- (12) -- (13) -- (7) -- (2) -- (4) -- (3) -- (1);
		\draw[bdryEdge2] (1) -- (6);
		\draw[bdryEdge1] (7) -- (2);		
		
		\draw[bulkEdge] (12) -- (13);
		\draw[bdryEdge2] (6) -- (12);
		\draw[bdryEdge1] (7) -- (13);

		\fill[bulk] (14) rectangle (17);	
		\draw[bulkEdge] (14) -- (15);
		\draw[bulkEdge] (16) -- (17);
		\draw[bdryEdge2] (14) -- (16);
		\draw[bdryEdge1] (15) -- (17);
		\node[] at (2,2.24) {${\cup}$};
	}{}
	\end{tikzpicture}
}

\newcommand{\gluingCyl}[2]{
	\begin{tikzpicture}[scale=#1, baseline=1.3em]
	\coordinate (1) at (0,0);
	\coordinate (2) at (0.8,0);
	\coordinate (3) at (0,1);
	\coordinate (4) at (0.8,1);
	
	\coordinate (5) at (0,1.5);
	\coordinate (6) at (0.8,1.5);
	\coordinate (7) at (0,2.5);
	\coordinate (8) at (0.8,2.5);
	
	\coordinate (9) at (0,2);
	\coordinate (10) at (0.8,2);

	\ifthenelse{#2=1}{		
		\fill[bulk] (1) rectangle (4);	
		\draw[bulkEdge] (1) -- (2);
		\draw[bulkEdge] (3) -- (4);
		\draw[bdryEdge2] (1) -- (3);
		\draw[bdryEdge1] (2) -- (4);
		\fill[bulk] (5) rectangle (8);	
		\draw[bulkEdge] (5) -- (6);
		\draw[bulkEdge] (7) -- (8);
		\draw[bdryEdge2] (5) -- (7);
		\draw[bdryEdge1] (6) -- (8);
		\node[] at (0.4,1.25) {${\cup}$};
	}{}
	\ifthenelse{#2=2}{		
		\fill[bulk] (9) rectangle (2);	
		\draw[bulkEdge] (9) -- (10);
		\draw[bulkEdge] (1) -- (2);
		\draw[bdryEdge2] (1) -- (9);
		\draw[bdryEdge1] (2) -- (10);
	}{}
	\ifthenelse{#2=3}{		
		\fill[bulk] (1) rectangle (4);	
		\draw[bulkEdge] (1) -- (2);
		\draw[bulkEdge] (3) -- (4);
		\draw[bdryEdge2] (1) -- (3);
		\draw[bdryEdge1] (2) -- (4);
	}{}
	\end{tikzpicture}
}

\newcommand{\tubeTwoD}[2]{
	\begin{tikzpicture}[scale=#1, baseline=1.1em]
	\coordinate (1) at (0,0);
	\coordinate (2) at (1,0);
	\coordinate (3) at (0,0.8);
	\coordinate (4) at (1,0.8);
	\coordinate (5) at (2,0);
	\coordinate (6) at (2,0.8);

	\ifthenelse{#2=1}{		
		\fill[bulk] (1) rectangle (4);	
		\draw[bulkEdge] (1) -- (3);
		\draw[bulkEdge] (2) -- (4);
		\draw[bulkEdge] (3) -- (2);
		\draw[bdryEdge1] (1) -- (2);
		\draw[bdryEdge2] (3) -- (4);
		\node[above] at (3) {${\sss 0}$};
		\node[above] at (4) {${\sss 1}$};
		\node[below] at (1) {${\sss 0'}$};
		\node[below] at (2) {${\sss 1'}$};
	}{}
	
	\ifthenelse{#2=2}{		
		\fill[bulk] (1) rectangle (4);	
		\draw[bulkEdge] (1) --node[left]{${\sss g}$} (3);
		\draw[bulkEdge] (2) --node[right]{${\sss a^{-1}gb}$} (4);
		\draw[bulkEdge] (3) -- (2);
		\draw[bdryEdge1] (1) --node[below, text=black, opacity=1]{${\sss b}$} (2);
		\draw[bdryEdge2] (3) --node[above, text=black, opacity=1]{${\sss a}$} (4);
		\node[above] at (3) {${\sss 0}$};
		\node[above] at (4) {${\sss 1}$};
		\node[below] at (1) {${\sss 0'}$};
		\node[below] at (2) {${\sss 1'}$};
	}{}
	
	\ifthenelse{#2=3}{		
		\fill[bulk] (1) rectangle (4);	
		\draw[bulkEdge] (1) --node[left]{${\sss g'}$} (3);
		\draw[bulkEdge] (2) --node[right]{${\sss a'^{-1}g'b'}$} (4);
		\draw[bdryEdge1] (1) --node[below, text=black, opacity=1]{${\sss b'}$} (2);
		\draw[bdryEdge2] (3) --node[above, text=black, opacity=1]{${\sss a'}$} (4);
		\draw[bulkEdge] (3) -- (2);
		\node[above] at (3) {${\sss 1}$};
		\node[above] at (4) {${\sss 2}$};
		\node[below] at (1) {${\sss 1'}$};
		\node[below] at (2) {${\sss 2'}$};
	}{}
	
	\ifthenelse{#2=4}{		
		\fill[bulk] (1) rectangle (6);	
		\draw[bulkEdge] (1) --node[left]{${\sss g}$} (3);
		\draw[bulkEdge] (2) -- (4);
		\draw[bulkEdge] (5) --node[right]{${\sss (aa')^{-1}gbb'}$} (6);
		\draw[bulkEdge] (3) -- (2);
		\draw[bulkEdge] (4) -- (5);		
		\draw[bdryEdge1] (1) --node[below, text=black, opacity=1]{${\sss b}$} (2);
		\draw[bdryEdge1] (2) --node[below, text=black, opacity=1]{${\sss b'}$} (5);
		\draw[bdryEdge2] (3) --node[above, text=black, opacity=1]{${\sss a}$} (4);
		\draw[bdryEdge2] (4) --node[above, text=black, opacity=1]{${\sss a'}$} (6);
		\node[above] at (3) {${\sss 0}$};
		\node[above] at (4) {${\sss 1}$};
		\node[below] at (1) {${\sss 0'}$};
		\node[below] at (2) {${\sss 1'}$};
		\node[below] at (5) {${\sss 2'}$};
		\node[above] at (6) {${\sss 2}$};
	}{}
	
	\ifthenelse{#2=5}{		
		\fill[bulk] (1) rectangle (6);	
		\draw[bulkEdge] (1) --node[left]{${\sss g}$} (3);
		\draw[bulkEdge] (2) -- (4);
		\draw[bulkEdge] (5) --node[right]{${\sss (aa')^{-1}g'bb'}$} (6);
		\draw[bulkEdge] (3) -- (2);
		\draw[bulkEdge] (4) -- (5);
		\draw[bdryEdge1] (1) --node[below, text=black, opacity=1]{${\sss b \tilde b}$} (2);
		\draw[bdryEdge1] (2) --node[below, text=black, opacity=1]{${\sss \tilde b^{-1}b'}$} (5);
		\draw[bdryEdge2] (3) --node[above, text=black, opacity=1]{${\sss a \tilde a}$} (4);
		\draw[bdryEdge2] (4) --node[above, text=black, opacity=1]{${\sss \tilde a^{-1}a'}$} (6);
		\node[above] at (3) {${\sss 0}$};
		\node[above] at (4) {${\sss 1}$};
		\node[below] at (1) {${\sss 0'}$};
		\node[below] at (2) {${\sss 1'}$};
		\node[below] at (5) {${\sss 2'}$};
		\node[above] at (6) {${\sss 2}$};
	}{}
	
	\ifthenelse{#2=6}{		
		\fill[bulk] (1) rectangle (4);	
		\draw[bulkEdge] (1) --node[left]{${\sss g}$} (3);
		\draw[bulkEdge] (3) -- (2);
		\draw[bulkEdge] (2) --node[right]{${\sss (aa')^{-1}gbb'}$} (4);
		\draw[bdryEdge1] (1) --node[below, text=black, opacity=1]{${\sss bb'}$} (2);
		\draw[bdryEdge2] (3) --node[above, text=black, opacity=1]{${\sss aa'}$} (4);
		\node[above] at (3) {${\sss 1}$};
		\node[above] at (4) {${\sss 2}$};
		\node[below] at (1) {${\sss 1'}$};
		\node[below] at (2) {${\sss 2'}$};
	}{}
	\end{tikzpicture}
}

\newcommand{\tubeTwoDPinched}[4]{
	\begin{tikzpicture}[scale=#1, baseline=-1em]
	\coordinate (1) at (0,0);
	\coordinate (2) at (2,0);
	\coordinate (3) at (1,-1.5);
	\coordinate (4) at (-2,1);
	\coordinate (5) at (0,1);
	\coordinate (6) at (-1,-0.5);
	
	\fill[bulk] (2) -- (5) -- (6) -- (3) -- (2);
	\draw[bdryEdge2] (5) -- (6);
	\draw[bulkEdge] (2) -- (6);
	
	\fill[bulk] (1) -- (4) -- (6) -- (3) -- (1);
	\draw[bdryEdge2] (4) -- (6);
	\fill[bulk] (1) -- (2) -- (5) -- (4) -- (1);
	\draw[bulkEdge] (2) -- (5);
	\draw[bulkEdge] (1) -- (4);
	\draw[bulkEdge] (2) -- (4); 
	\draw[bulkEdge] (3) -- (4);
	\draw[bulkEdge] (3) -- (6);
	\draw[bdryEdge1] (1) -- (3);
	\draw[bdryEdge1] (2) -- (3);	
	\draw[bdryEdge1] (1) -- (2);
	\draw[bdryEdge2] (4) -- (5);
	
	\node[above] at (1) {{${\sss #3'}$}};
	\node[above] at (2) {{${\sss #2'}$}};
	\node[below] at (3) {{${\sss #4'}$}};
	\node[above] at (4) {{${\sss #3}$}};
	\node[above] at (5) {{${\sss #2}$}};
	\node[below] at (6) {{${\sss #4}$}};
	\end{tikzpicture}
}

\newcommand{\tubeTwoDPinchedBis}[4]{
	\begin{tikzpicture}[scale=#1, baseline=2em]
	\coordinate (1) at (0,0);
	\coordinate (2) at (2,0);
	\coordinate (3) at (-2,1);
	\coordinate (4) at (0,1);
	\coordinate (5) at (4,0);
	\coordinate (6) at (2,1);
	\coordinate (7) at (2,2);
	\coordinate (8) at (0,3);
	
	\fill[bulk] (1) -- (2) -- (5) -- (6) -- (4) -- (3) -- (1);
	\draw[bulkEdge] (2) -- (4);
	\draw[bulkEdge] (2) -- (3);
	\draw[bulkEdge] (4) -- (5);
	\draw[bdryEdge1] (1) -- (5);
	
	\draw[bdryEdge2] (3) -- (6) -- (8) -- (3);
	\draw[bdryEdge2] (4) -- (8);
	\fill[bulk] (5) -- (6) -- (8) -- (7) -- (5);
	\draw[bulkEdge] (5) -- (6);
	\draw[bulkEdge] (8) -- (5);
	\draw[bulkEdge] (7) -- (4);
	\fill[bulk] (1) -- (7) -- (8) -- (3) -- (1);
	\draw[bulkEdge] (3) -- (7);
	\draw[bulkEdge] (7) -- (8);
	\draw[bulkEdge] (1) -- (3);
	
	\draw[bdryEdge1] (2) -- (7);
	\draw[bdryEdge1] (1) -- (7) -- (5) -- (1);
	
	\node[below] at (1) {{${\sss #2'}$}};
	\node[below] at (2) {{${\sss #3'}$}};
	\node[below] at (5) {{${\sss #4'}$}};
	\node[above] at (7) {{${\sss \tilde #3'}$}};
	\node[below] at (3) {{${\sss #2}$}};
	\node[left, yshift=4pt] at (4) {{${\sss #3}$}};
	\node[right, yshift=2pt, xshift=-0.5pt] at (6) {{${\sss #4}$}};
	\node[above] at (8) {{${\sss \tilde #3}$}};
	\end{tikzpicture}
}

\newcommand{\annulus}[1]{
	\begin{tikzpicture}[scale=#1, baseline=0em]
	\coordinate (0) at (0,0);
	\coordinate (1) at (0.8,0);
	
	\fill[bulk] (1) arc (0:360:1.3) -- (0) arc (0:-360:0.5) -- (1);
	\draw[bdryEdge1] (0) arc (0:360:0.5);
	\draw[bdryEdge2] (1) arc (0:360:1.3);

	\end{tikzpicture}
}

\newcommand{\trinion}[2]{
	\begin{tikzpicture}[scale=#1, baseline=0em]
	\coordinate (0) at (0,0);
	\coordinate (1) at (0,-1);
	\coordinate (2) at (0,1);
	\coordinate (3) at (-0.4,1);
	\coordinate (4) at (0.4,1);
	\coordinate (5) at (-0.4,-1);
	\coordinate (6) at (0.4,-1);
	\coordinate (7) at (-0.4,0.5);
	\coordinate (8) at (0.4,0.5);
	\coordinate (9) at (-0.4,-0.5);
	\coordinate (10) at (0.4,-0.5);
	
	\coordinate (20) at ({-0.4*cos(120)+sin(120)},{-0.4*sin(120)-cos(120)});
	\coordinate (21) at ({0.4*cos(120)+sin(120)},{0.4*sin(120)-cos(120)});
	\coordinate (22) at ({-0.4*cos(240)+sin(240)},{-0.4*sin(240)-cos(240)});
	\coordinate (23) at ({0.4*cos(240)+sin(240)},{0.4*sin(240)-cos(240)});
	\coordinate (24) at (-0.4,-1);
	\coordinate (25) at (0.4,-1);
	\coordinate (26) at ({-0.4*cos(240)+sin(240)},1);
	\coordinate (27) at ({0.4*cos(120)+sin(120)},1);
	
	\ifthenelse{#2=1}{
		\begin{scope}
		\clip (24) to[out=90,in=90,looseness=1.5] (25);
		\fill[bulkExc] (0,-1) arc (-90:270:1);
		\end{scope}
		\begin{scope}
		\clip (20) to[out=-150,in=-150,looseness=1.5] (21);
		\fill[bulkExc] (0,-1) arc (-90:270:1);
		\end{scope}
		\begin{scope}
		\clip (22) to[out=-30,in=-30,looseness=1.5] (23);
		\fill[bulkExc] (0,-1) arc (-90:270:1);
		\end{scope}
		\begin{scope}
		\clip (23) to[out=-30,in=-30,looseness=1.5] (22) -- (26) -- (27) -- (21) to[out=-150,in=-150,looseness=1.5] (20) -- (1,-1) -- (25) to[out=90,in=90,looseness=1.5] (24) -- (-1,-1) -- (23);
		\fill[bulk] (1) arc (-90:270:1);
		\end{scope}
		\centerarc[bdryEdge3](0)(-90:30:1)
		\centerarc[bdryEdge1](0)(30:150:1)
		\centerarc[bdryEdge2](0)(150:270:1)
		\node[circle, fill = black, inner sep=1.6pt] at (0,-1) {};
		\node[circle, fill = black, inner sep=1.6pt] at ({cos(30)},{sin(30)}) {};
		\node[circle, fill = black, inner sep=1.6pt] at ({-cos(30)},{sin(30)}) {};
		\node[above] at (0,1) {${\sss A_\phi}$};	
		\node[below] at (-0.85,-0.6) {${\sss B_\psi}$};	
		\node[below] at (0.85,-0.6) {${\sss C_\varphi}$};		
	}{}
	\ifthenelse{#2=2}{	
		\begin{scope}
		\clip (23) to[out=-30,in=-30,looseness=1.5] (22) -- (26) -- (27) -- (21) to[out=-150,in=-150,looseness=1.5] (20) -- (1,-1) -- (25) to[out=90,in=90,looseness=1.5] (24) -- (-1,-1) -- (23);
		\fill[bulk] (1) arc (-90:270:1);
		\end{scope}
		\begin{scope}
		\clip (1) arc[start angle=-90, end angle=270, radius=1];
		\draw[bulkEdge] (24) to[out=90,in=90,looseness=1.5] (25);
		\end{scope}
		\begin{scope}
		\clip (1) arc[start angle=-90, end angle=270, radius=1];
		\draw[bulkEdge] (20) to[out=-150,in=-150,looseness=1.5] (21);
		\end{scope}
		\begin{scope}
		\clip (1) arc[start angle=-90, end angle=270, radius=1];
		\draw[bulkEdge] (22) to[out=-30,in=-30,looseness=1.5] (23);
		\end{scope}
		\centerarc[bdryEdge3](0)(-67:7:1)
		\centerarc[bdryEdge1](0)(53:127:1)
		\centerarc[bdryEdge2](0)(173:247:1)
	}{}
	\end{tikzpicture}
}

\newcommand{\trinionTr}[2]{
	\begin{tikzpicture}[scale=#1, baseline=0em]
	\coordinate (1) at (0,0);
	\coordinate (2) at (0.4,0);
	\coordinate (3) at (-0.4,0);
	\coordinate (4) at (0,{0.8*sin(60)});
	\coordinate (5) at (-0.4,-1);
	\coordinate (6) at (0.4,-1);
	
	\coordinate (7) at ({-cos(30)},{0.8*sin(60)+sin(30)});
	\coordinate (8) at ({-0.4-cos(30)},{sin(30)});
	\coordinate (9) at ({cos(30)},{0.8*sin(60)+sin(30)});
	\coordinate (10) at ({0.4+cos(30)},{sin(30)});

	\ifthenelse{#2=1}{		
		\fill[bulk] (5) -- (3) -- (8) -- (7) -- (4) -- (9) -- (10) -- (2) -- (6) -- (5);	
		\draw[bulkEdge] (5) -- (6);
		\draw[bulkEdge] (7) -- (8);
		\draw[bulkEdge] (9) -- (10);
		\draw[bdryEdge2] (8) -- (3) -- (5);
		\draw[bdryEdge3] (6) -- (2) -- (10);
		\draw[bdryEdge1] (7) -- (4) -- (9);
	}{}
	\ifthenelse{#2=2}{		
		\fill[bulk] (5) -- (3) -- (8) -- (7) -- (4) -- (9) -- (10) -- (2) -- (6) -- (5);	
		\draw[bulkEdge] (2) -- (3) -- (4) -- (2);
		\draw[bulkEdge] (5) --node[below]{${\sss a'g_1g_2c'^{-1}}$} (6);
		\draw[bulkEdge] (7) --node[above, text = black, sloped]{${\sss a^{-1}g_1b}$} (8);
		\draw[bulkEdge] (9) --node[above, text = black, sloped]{${\sss b'^{-1}g_2c}$} (10);
		\draw[bulkEdge] (3) -- (7);
		\draw[bulkEdge] (4) -- (10);
		\draw[bulkEdge] (2) -- (5);
		\draw[bdryEdge2] (8) --node[below, text = black]{${\sss a}$} (3) --node[left, text = black, xshift=1pt]{${\sss a'}$} (5);
		\draw[bdryEdge3] (6) --node[right, text = black]{${\sss c'}$} (2) --node[below, text = black]{${\sss c}$} (10);
		\draw[bdryEdge1] (7) --node[above, text = black]{${\sss b}$} (4) --node[above, text = black]{${\sss b'}$} (9);
		
		\node[below] at (8) {{${\sss 2}$}};
		\node[above] at (7) {{${\sss 2'}$}};
		\node[above] at (4) {{${\sss 1'}$}};
		\node[above] at (9) {{${\sss 3'}$}};
		\node[below] at (10) {{${\sss 3''}$}};
		\node[left] at (5) {{${\sss 0}$}};
		\node[right] at (6) {{${\sss 0''}$}};
		\node[left, yshift=-5pt] at (3) {{${\sss 1}$}};
		\node[right, yshift=-5pt] at (2) {{${\sss 1''}$}};
	}{}
	\ifthenelse{#2=3}{		
		\fill[bulk] (5) -- (3) -- (8) -- (7) -- (4) -- (9) -- (10) -- (2) -- (6) -- (5);	
		\draw[bulkEdge] (2) -- (3) -- (4) -- (2);
		\draw[bulkEdge] (5) -- (6);
		\draw[bulkEdge] (7) -- (8);
		\draw[bulkEdge] (9) -- (10);
		\draw[bulkEdge] (3) -- (7);
		\draw[bulkEdge] (4) -- (10);
		\draw[bulkEdge] (2) -- (5);
		\draw[bdryEdge2] (8) -- (3) -- (5);
		\draw[bdryEdge3] (6) -- (2) -- (10);
		\draw[bdryEdge1] (7) -- (4) -- (9);
		
		\node[below] at (8) {{${\sss 2}$}};
		\node[above] at (7) {{${\sss 2'}$}};
		\node[above] at (4) {{${\sss 1'}$}};
		\node[above] at (9) {{${\sss 3'}$}};
		\node[below] at (10) {{${\sss 3''}$}};
		\node[left] at (5) {{${\sss 0}$}};
		\node[right] at (6) {{${\sss 0''}$}};
		\node[left, yshift=-5pt] at (3) {{${\sss 1}$}};
		\node[right, yshift=-5pt] at (2) {{${\sss 1''}$}};
	}{}
	\end{tikzpicture}
}

\newcommand{\trinionPinched}[1]{
	\begin{tikzpicture}[scale=#1, baseline=1.5em]
	\coordinate (0) at (0,0);
	\coordinate (1) at (2,0);
	\coordinate (2) at (0.5,0.8);
	\coordinate (6) at (1.5,-1);
	\coordinate (7) at (3.5,-1);
	\coordinate (8) at (-2.4,0.4);
	\coordinate (9) at (-1.9,1.2);
	\coordinate (10) at (4,0.6);
	\coordinate (11) at (2.5,1.4);
	\coordinate (3) at (0,2.3);
	\coordinate (4) at (2,2.3);
	\coordinate (5) at (0.5,3.1);
	
	\fill[bulk] (6) -- (0) -- (8) -- (9) -- (2) -- (11) -- (10) -- (1) -- (7) -- (6);
	\draw[bulkEdge] (0) -- (9);
	\draw[bulkEdge] (2) -- (10);
	\draw[bulkEdge] (1) -- (6);
	\fill[bulk] (4) -- (5) -- (11) -- (10) -- (4);
	\draw[bulkEdge] (5) -- (10);
	\draw[bdryEdge1] (11) -- (5);
	\draw[bdryEdge1] (9) -- (2) -- (11);
	\draw[bulkEdge] (0) -- (2) -- (1);
	\draw[bdryEdge1] (2) -- (5);
	\fill[bulk] (8) -- (9) -- (5) -- (3) -- (8);
	\draw[bulkEdge] (3) -- (9);
	\draw[bulkEdge] (0) -- (5);
	\draw[bulkEdge] (1) -- (5);
	\draw[bulkEdge] (0) -- (1); 
	\draw[bulkEdge] (0) -- (4);
	\draw[bdryEdge3] (4) -- (1) -- (10);
	\draw[bdryEdge3] (1) -- (7);
	\fill[bulk] (3) -- (4) -- (7) -- (6) -- (3);
	\draw[bulkEdge] (4) -- (6);
	
	\fill[bulk] (3) -- (4) -- (5) -- (3);
	\draw[bulkEdge] (10) -- (11);
	\draw[bulkEdge] (8) -- (9);
	\draw[bulkEdge] (6) -- (7);
	\draw[bulkEdge] (3) -- (4) -- (5) -- (3);
	\draw[bdryEdge3] (4) -- (10);
	\draw[bdryEdge3] (4) -- (7);
	\draw[bdryEdge1] (5) -- (9);
	\draw[bdryEdge2] (3) -- (8);
	\draw[bdryEdge2] (0) -- (3);
	\draw[bdryEdge2] (3) -- (6);
	\draw[bdryEdge2] (8) -- (0) -- (6);

	\node[below] at (6) {{${\sss 0}$}};
	\node[below] at (7) {{${\sss 0''}$}};
	\node[below] at (0) {{${\sss 1}$}};
	\node[below] at (8) {{${\sss 2}$}};
	\node[left] at (9) {{${\sss 2'}$}};
	\node[right] at (10) {{${\sss 3''}$}};
	\node[above] at (5) {{${\sss \tilde 1'}$}};
	\node[above] at (4) {{${\sss \tilde 1''}$}};
	\node[right] at (1) {{${\sss 1''}$}};
	\node[left, yshift=5pt, xshift=4pt] at (2) {{${\sss 1'}$}};
	\node[yshift=3pt,xshift=-2pt] at (3) {{${\sss \tilde 1}$}};
	\node[left, yshift=1pt] at (11) {{${\sss 3'}$}};
	
	\end{tikzpicture}
}

\newcommand{\bdryInterfaceThreeExc}[2]{
	\begin{tikzpicture}[scale=#1, baseline=3em]
	\coordinate (1) at (0,1.5);
	\coordinate (2) at (4,1.5);
	\coordinate (3) at (0,0);
	\coordinate (4) at (4,0);
	\coordinate (5) at (2,1.5);
	\coordinate (6) at (-1.5,2.5);
	\coordinate (7) at (2.5,2.5);
	\coordinate (8) at (-1.5,1);
	\coordinate (10) at (0.5,2.5);
	\coordinate (20) at (-0.15,2);
	\coordinate (21) at (0.65,2);
	\coordinate (22) at (1.85,2);
	\coordinate (23) at (2.65,2);
	
	\ifthenelse{#2=1}{
		\fill[black, opacity=0.65] (21) to[out=-90, in=-90, looseness=1.7] (22) to[out=130, in=130, looseness=0.6] (23) to[out=-90, in=-90, looseness=1.7] (20) to[out=130, in=130, looseness=0.6] (21);	
		\fill[bulk] (1) -- (2) -- (4) -- (3) -- (1);
		\fill[bulk] (1) -- (6) -- (8) -- (3) -- (1);
		\fill[bdryPlaq2] (1) -- (6) -- (10) -- (5) -- (1);
		\fill[bdryPlaq1] (2) -- (5) -- (10) -- (7) -- (2);
		\draw[op2] (1) -- (3);
		\draw[op2] (6) -- (8);
		\draw[op2] (2) -- (4);
		\draw[bdryEdge2] (5) -- (1) -- (6) -- (10);
		\draw[bdryEdge1] (5) -- (2) -- (7) -- (10);
		\draw[line width = 1.6pt] (5) -- (10);
		\fill[black,opacity=0.65] (20) to[out=130, in=130, looseness=0.6] (21) to[out=-50, in=-50, looseness=0.6] (20);
		\fill[black, opacity=0.65] (22) to[out=130, in=130, looseness=0.6] (23) to[out=-50, in=-50, looseness=0.6] (22);
		\node[below] at (-0.8,2.5) {${\sss A_\lambda}$};	
		\node[below] at (2.5,2.5) {${\sss B_\mu}$};	
	}{}
	\ifthenelse{#2=2}{
		\draw[bulkEdge, dotted] (21) to[out=-90, in=-90, looseness=1.7] (22);	
		\draw[bulkEdge, dotted] (20) to[out=-90, in=-90, looseness=1.7] (23);
		\fill[bulk] (1) -- (2) -- (4) -- (3) -- (1);
		\fill[bulk] (1) -- (6) -- (8) -- (3) -- (1);
		\fill[bdryPlaq2] (1) -- (6) -- (10) -- (5) -- (1);
		\fill[bdryPlaq1] (2) -- (5) -- (10) -- (7) -- (2);
		\draw[op2] (1) -- (3);
		\draw[op2] (6) -- (8);
		\draw[op2] (2) -- (4);
		\draw[bdryEdge2] (5) -- (1) -- (6) -- (10);
		\draw[bdryEdge1] (5) -- (2) -- (7) -- (10);
		\draw[line width = 1.6pt] (5) -- (10);
		\fill[bulk] (20) to[out=130, in=130, looseness=0.6] (21) to[out=-50, in=-50, looseness=0.6] (20);
		\fill[bulk] (22) to[out=130, in=130, looseness=0.6] (23) to[out=-50, in=-50, looseness=0.6] (22);
		\draw[bdryEdge1] (22) to[out=130, in=130, looseness=0.6] (23) to[out=-50, in=-50, looseness=0.6] (22);
		\draw[bdryEdge2] (20) to[out=130, in=130, looseness=0.6] (21) to[out=-50, in=-50, looseness=0.6] (20);
	}{}
	\end{tikzpicture}
}

\newcommand{\bdryInterfaceThreeExcFusion}[1]{
	\begin{tikzpicture}[scale=#1, baseline=3em]
	\coordinate (1) at (0,1.5);
	\coordinate (2) at (4,1.5);
	\coordinate (3) at (0,0);
	\coordinate (4) at (4,0);
	\coordinate (5) at (2,1.5);
	\coordinate (6) at (-1.5,2.5);
	\coordinate (7) at (2.5,2.5);
	\coordinate (8) at (-1.5,1);
	\coordinate (10) at (0.5,2.5);
	\coordinate (11) at (6,0);
	\coordinate (12) at (6,1.5);
	\coordinate (13) at (4.5,2.5);
	\coordinate (20) at (-0.15,2);
	\coordinate (21) at (0.65,2);
	\coordinate (22) at (1.85,2);
	\coordinate (23) at (2.65,2);
	\coordinate (24) at (3.85,2);
	\coordinate (25) at (4.65,2);

	\fill[black, opacity=0.65] (21) to[out=-90, in=-90, looseness=1.7] (22) to[out=130, in=130, looseness=0.6] (23) to[out=-90, in=-90, looseness=1.7] (20) to[out=130, in=130, looseness=0.6] (21);
	\fill[black, opacity=0.65] (23) to[out=-90, in=-90, looseness=1.7] (24) to[out=130, in=130, looseness=0.6] (25) to[out=-90, in=-90, looseness=1.7] (22) to[out=130, in=130, looseness=0.6] (23);	
	\fill[bulk] (1) -- (12) -- (11) -- (3) -- (1);
	\fill[bulk] (1) -- (6) -- (8) -- (3) -- (1);
	\fill[bdryPlaq2] (1) -- (6) -- (10) -- (5) -- (1);
	\fill[bdryPlaq1] (2) -- (5) -- (10) -- (7) -- (2);
	\fill[bdryPlaq3] (12) -- (2) -- (7) -- (13) -- (12);
	\draw[op2] (1) -- (3);
	\draw[op2] (6) -- (8);
	\draw[op2] (11) -- (12);
	\draw[bdryEdge2] (5) -- (1) -- (6) -- (10);
	\draw[bdryEdge1] (5) -- (2);
	\draw[bdryEdge1] (7) -- (10);
	\draw[bdryEdge3] (2) -- (12) -- (13) -- (7);
	\draw[line width = 1.6pt] (5) -- (10);
	\draw[line width = 1.6pt] (2) -- (7);
	\fill[black,opacity=0.65] (20) to[out=130, in=130, looseness=0.6] (21) to[out=-50, in=-50, looseness=0.6] (20);
	\fill[black, opacity=0.65] (22) to[out=130, in=130, looseness=0.6] (23) to[out=-50, in=-50, looseness=0.6] (22);
	\fill[black, opacity=0.65] (24) to[out=130, in=130, looseness=0.6] (25) to[out=-50, in=-50, looseness=0.6] (24);
	\node[below] at (-0.8,2.55) {${\sss A_\lambda}$};	
	\node[below] at (1.2,2.55) {${\sss B_\mu}$};
	\node[below] at (3.2,2.55) {${\sss C_\nu}$};	

	\end{tikzpicture}
}

\newcommand{\threeDiskExc}[2]{
	\begin{tikzpicture}[scale=#1, baseline=-0.5em]
	\coordinate (0) at (0,0);
	\coordinate (1) at (0,2);
	\coordinate (2) at (0,-2);
	\coordinate (3) at (-1.6,-0.4);
	\coordinate (4) at (-1.6,0.4);
	\coordinate (5) at (1.6,-0.4);
	\coordinate (6) at (1.6,0.4);
	\coordinate (7) at (-0.4,2);
	\coordinate (8) at (0.4,2);
	\coordinate (9) at (-0.4,-2);
	\coordinate (10) at (0.4,-2);
	
	\ifthenelse{#2=1}{
		\path[name path = Parc] (2) arc[start angle=-90, end angle=270, radius=2];
		\path[name path = PL] (7) to[out=-50, in=50, looseness=0.8] (9);
		\path[name path = PR] (8) to[out=-50, in=50, looseness=0.8] (10);
		\path [name intersections={of=Parc and PL,by={sp1,sp2}}];
		\path [name intersections={of=Parc and PR,by={sp3,sp4}}];
		
		\draw[line width = 1.6pt, op5] (1) to[out=-130, in =130, looseness=0.8] (2);
		\begin{scope}		
		\clip (2) arc[start angle=-90, end angle=270, radius=2];
		\fill[black,opacity=0.3] (sp1) to[out=-135, in=135, looseness=0.8] (sp2) -- (sp3) to[out=135, in=-135, looseness=0.8] (sp4) -- (sp1);
		\fill[black, opacity=0.4] (sp1) to[out=-55, in=-125, looseness =1] (sp4);
		\fill[black, opacity=0.4] (sp2) to[out=55, in=125, looseness =1] (sp3);
		\end{scope}
		\fill[black,opacity=0.65] (3) to[out=0, in=0, looseness=1.2] (4) to[out=15, in=165, looseness =1] (6) to[out=-180, in=180, looseness=1.2] (5) to[out=165, in=15, looseness=1] (3);
		\begin{scope}		
		\clip (2) arc[start angle=-90, end angle=270, radius=2];
		\fill[black,opacity=0.8] (7) to[out=-50, in=50, looseness=0.8] (9) -- (10) to[out=50, in=-50, looseness=0.8] (8) -- (7);
		\end{scope}
		\fill[bdryPlaq1] (2) arc (-90:90:2) to[out=-50, in =50, looseness=0.8] (2);
		\fill[bdryPlaq2] (1) arc (90:270:2) to[out=50, in =-50, looseness=0.8] (1);
		\centerarc[bdryEdge1](0)(-90:90:2)
		\centerarc[bdryEdge2](0)(90:270:2)
		\draw[line width = 1.6pt] (1) to[out=-50, in =50, looseness=0.8] (2);
		\fill[black,opacity=0.65] (3) to[out=0, in=0, looseness=1.2] (4) to[out=-180, in=180, looseness=1] (3);
		\fill[black,opacity=0.65] (5) to[out=0, in=0, looseness=1] (6) to[out=-180, in=180, looseness=1.2] (5);
	}{}
	\ifthenelse{#2=2}{
		\path[name path = Parc] (2) arc[start angle=-90, end angle=270, radius=2];
		\path[name path = PL] (7) to[out=-50, in=50, looseness=0.8] (9);
		\path[name path = PR] (8) to[out=-50, in=50, looseness=0.8] (10);
		\path [name intersections={of=Parc and PL,by={sp1,sp2}}];
		\path [name intersections={of=Parc and PR,by={sp3,sp4}}];

		\fill[bulk] (sp1) to[out=-55, in=-125, looseness =1] (sp4) to[out=-135, in=135, looseness=0.8] (sp3) to[out=125, in=55, looseness =1] (sp2) to[out=135, in=-135, looseness=0.8] (sp1);	
		
		\draw[bdryEdge2] (sp1) to[out=-135, in=135, looseness=0.8] (sp2);
		\draw[bdryEdge1] (sp3) to[out=135, in=-135, looseness=0.8] (sp4);
		
		\draw[bulkEdge, dotted]  (4) to[out=15, in=165, looseness =1] (6);
		\draw[bulkEdge, dotted] (5) to[out=165, in=15, looseness=1] (3);
		
		\fill[bulk] (sp1) to[out=-55, in=-125, looseness =1] (sp4) to[out=-50, in=50, looseness=0.8] (sp3) to[out=125, in=55, looseness =1] (sp2) to[out=50, in=-50, looseness=0.8] (sp1);
		
		\fill[bdryPlaq1] (sp3) arc (-78:78:2) to[out=-135, in =135, looseness=0.8] (sp3);
		\fill[bdryPlaq2] (sp1) arc (100:260:2) to[out=55, in =-55, looseness=0.8] (sp1);
		
		\draw[bulkEdge] (sp1) to[out=-55, in=-125, looseness =1] (sp4);
		\draw[bulkEdge] (sp2) to[out=55, in=125, looseness =1] (sp3);
		\draw[bdryEdge2] (sp1) to[out=-50, in=50, looseness=0.8] (sp2);
		\draw[bdryEdge1] (sp3) to[out=50, in=-50, looseness=0.8] (sp4);
		
		\fill[bdryPlaq1] (sp3) arc (-78:78:2) to[out=-50, in =50, looseness=0.8] (sp3);
		\fill[bdryPlaq2] (sp1) arc (100:260:2) to[out=50, in =-50, looseness=0.8] (sp1);
		\centerarc[bdryEdge1](0)(-77:77:2)
		\centerarc[bdryEdge2](0)(101:259:2)
		
		\fill[bulk] (3) to[out=0, in=0, looseness=1.2] (4) to[out=-180, in=180, looseness=1] (3);
		\draw[bdryEdge2] (3) to[out=0, in=0, looseness=1.2] (4) to[out=-180, in=180, looseness=1] (3);
		\fill[bulk] (5) to[out=0, in=0, looseness=1] (6) to[out=-180, in=180, looseness=1.2] (5);
		\draw[bdryEdge1] (5) to[out=0, in=0, looseness=1] (6) to[out=-180, in=180, looseness=1.2] (5);
	}{}
	\end{tikzpicture}
}

\newcommand{\tubeThreeDBeauty}[2]{
	\begin{tikzpicture}[scale=#1, baseline=3em]
	\coordinate (0) at (0,0);
	\coordinate (1) at (0,0.8);
	\coordinate (2) at (0,1.6);
	\coordinate (3) at (0,2.4);
	
	\coordinate (4) at (3,0);
	\coordinate (5) at (3,0.8);
	\coordinate (6) at (3,1.6);
	\coordinate (7) at (3,2.4);
	
	\ifthenelse{#2 = 1}{
		\fill[bdryPlaq1] (4) to[out=180, in=180, looseness=1] (7) -- (6) to[out=180, in=180, looseness=1] (5) to[out=0, in=0, looseness=1] (6) -- (7) to[out=0, in=0, looseness=1] (4);
		
		\draw[bulkEdge, dotted] (1) -- (5);
		\draw[bulkEdge, dotted] (2) -- (6);
		
		\draw[bdryEdge1, op5] (5) to[out=180, in=180, looseness=1] (6);
		\draw[bdryEdge1, op5] (5) to[out=0, in=0, looseness=1] (6);
		\draw[bdryEdge1, op5] (4) to[out=180, in=180, looseness=1] (7);
		
		\fill[bulk] (0) to[out=0, in=0, looseness=1] (3) -- (7) to[out=0, in=0, looseness=1] (4) -- (0);
		\draw[bulkEdge] (0) -- (4);
		\draw[bulkEdge] (3) -- (7);
		
		\fill[bdryPlaq2] (0) to[out=180, in=180, looseness=1] (3) -- (2) to[out=180, in=180, looseness=1] (1) to[out=0, in=0, looseness=1] (2) -- (3) to[out=0, in=0, looseness=1] (0);
		\fill[bulk] (1) to[out=180, in=180, looseness=1] (2) to[out=0, in=0, looseness=1] (1);
		\draw[bdryEdge2] (0) to[out=180, in=180, looseness=1] (3);
		\draw[bdryEdge2] (0) to[out=0, in=0, looseness=1] (3);
		\draw[bdryEdge2] (1) to[out=180, in=180, looseness=1] (2);
		\draw[bdryEdge2] (1) to[out=0, in=0, looseness=1] (2);
		
		\draw[bdryEdge1] (4) to[out=0, in=0, looseness=1] (7);
	}{}
	\ifthenelse{#2 = 2}{
		\fill[black, opacity=0.15] (4) to[out=180, in=180, looseness=1] (7) -- (6) to[out=180, in=180, looseness=1] (5) to[out=0, in=0, looseness=1] (6) -- (7) to[out=0, in=0, looseness=1] (4);
		
		\draw[bulkEdge, dotted, ->1-=0.5] (1) -- (5);
		\draw[bulkEdge, dotted] (2) -- (6);
		
		\draw[op5] (5) to[out=180, in=180, looseness=1] (6);
		\draw[op5, ->1- =0.5] (5) to[out=0, in=0, looseness=1, edge node={coordinate[pos=0.5] (sp1)}] (6);
		\draw[op5] (4) to[out=180, in=180, looseness=1] (7);
		\draw[->1-=0.6] (4) -- (5);
		
		\fill[bulk] (0) to[out=0, in=0, looseness=1] (3) -- (7) to[out=0, in=0, looseness=1] (4) -- (0);
		\draw[->1-=0.5] (0) -- (4) node[pos=0.5, below] {${\sss g}$};
		\draw[] (3) -- (7);
		
		\fill[white, opacity=0.5] (0) to[out=180, in=180, looseness=1] (3) -- (2) to[out=180, in=180, looseness=1] (1) to[out=0, in=0, looseness=1] (2) -- (3) to[out=0, in=0, looseness=1] (0);
		\fill[black, opacity=0.15] (0) to[out=180, in=180, looseness=1] (3) -- (2) to[out=180, in=180, looseness=1] (1) to[out=0, in=0, looseness=1] (2) -- (3) to[out=0, in=0, looseness=1] (0);
		\fill[bulk] (1) to[out=180, in=180, looseness=1] (2) to[out=0, in=0, looseness=1] (1);
		\draw[] (0) to[out=180, in=180, looseness=1] (3);
		\draw[->1-=0.5] (0) to[out=0, in=0, looseness=1, edge node={coordinate[pos=0.5] (sp3)}] (3);
		\draw[] (1) to[out=180, in=180, looseness=1] (2);
		\draw[->1-=0.5] (1) to[out=0, in=0, looseness=1, edge node={coordinate[pos=0.5] (sp4)}] (2);
		\draw[->1-=0.5] (4) to[out=0, in=0, looseness=1, edge node={coordinate[pos=0.5] (sp2)}] (7);
		\draw[->1-=0.6] (0) -- (1) node[pos=0.5,right] {${\sss x_1}$};
		
		\path[] (1) -- (5) node[pos=0.5, below] {${\sss x_1^{-1}gy_1}$};
		\path[] (4) -- (5) node[pos=0.5, right] {${\sss y_1}$};
		\node[left] at (sp1) {${\sss y_2^{y_1} \!\!\!}$};
		\node[left] at (sp4) {${\sss x_2^{x_1} \!\!\!}$};
		\node[right] at (sp2) {${\sss y_2}$};
		\node[right] at (sp3) {${\sss x_2}$};
	}{}
	\end{tikzpicture}
}

\newcommand{\tubeThreeD}[7]{
	\begin{tikzpicture}[scale=#1, baseline=2.5em]
	\coordinate (0) at (0,0);
	\coordinate (1) at (2,0);
	\coordinate (2) at (1.2,0.5);
	\coordinate (3) at (-0.8,0.5);
	\coordinate (4) at (0,2);
	\coordinate (5) at (2,2);
	\coordinate (6) at (1.2,2.5);
	\coordinate (7) at (-0.8,2.5);
	
	\ifthenelse{#7 = 1}{
		\fill[bulk] (1) -- (2) -- (6) -- (5) -- (1);
		\fill[bulk] (0) -- (1) -- (2) -- (3) -- (0);
		\draw[] (1) -- (2);
		\draw[] (1) -- (3);
		\draw[] (2) -- (5);	
		\draw[bdryEdge2] (2) -- (6);
		\draw[bdryEdge2] (3) -- (6);
		\draw[bdryEdge2] (2) -- (3);	
		\fill[bulk] (4) -- (5) -- (6) -- (7) -- (4);
		\draw[] (5) -- (7);
		\draw[] (5) -- (6);
		\draw[bdryEdge2] (6) --node[above, color=black]{${\sss #5_1}$} (7);
		\draw[] (3) -- (5);	
		\fill[bulk] (0) -- (3) -- (7) -- (4) -- (0);
		\draw[] (3) -- (4);
		\draw[] (0) --node[left, pos=0.3, yshift=-0.2em, color=black]{${\sss #4}$} (3);
		\draw[] (4) -- (7);
		\draw[bdryEdge2] (3) --node[left, text=black, xshift=0.1em]{${\sss #5_2}$} (7);
		\draw[bdryEdge1] (4) -- (5);
		\draw[bdryEdge1] (0) --node[right, text=black, pos=0.4, xshift=-0.1em]{${\sss #6_2}$} (4);
		\draw[bdryEdge1] (1) -- (5);
		\draw[bdryEdge1] (0) -- (5);
		\draw[bdryEdge1] (0) --node[below, text=black]{${\sss #6_1}$} (1);	
	}{}	
	\ifthenelse{#7 = 2}{
		\fill[bulk] (1) -- (2) -- (6) -- (5) -- (1);
		\fill[bulk] (0) -- (1) -- (2) -- (3) -- (0);
		\draw[] (1) -- (2);
		\draw[] (1) -- (3);
		\draw[] (2) -- (5);	
		\draw[bdryEdge2] (2) -- (6);
		\draw[bdryEdge2] (3) -- (6);
		\draw[bdryEdge2] (2) -- (3);	
		\fill[bulk] (4) -- (5) -- (6) -- (7) -- (4);
		\draw[] (5) -- (7);
		\draw[] (5) -- (6);
		\draw[bdryEdge2] (6) --node[above, color=black]{${\sss #5_1 #5_1'}$} (7);
		\draw[] (3) -- (5);	
		\fill[bulk] (0) -- (3) -- (7) -- (4) -- (0);
		\draw[] (3) -- (4);
		\draw[] (0) --node[left, pos=0.3, yshift=-0.2em, color=black]{${\sss #4}$} (3);
		\draw[] (4) -- (7);
		\draw[bdryEdge2] (3) --node[left, text=black, xshift=0.1em]{${\sss #5_2}$} (7);
		\draw[bdryEdge1] (4) -- (5);
		\draw[bdryEdge1] (0) --node[right, text=black, pos=0.4, xshift=-0.1em]{${\sss #6_2}$} (4);
		\draw[bdryEdge1] (1) -- (5);
		\draw[bdryEdge1] (0) -- (5);
		\draw[bdryEdge1] (0) --node[below, text=black]{${\sss #6_1 #6_1'}$} (1);	
	}{}	
	\ifthenelse{#7 = 3}{
		\fill[bulk] (1) -- (2) -- (6) -- (5) -- (1);
		\fill[bulk] (0) -- (1) -- (2) -- (3) -- (0);
		\draw[] (1) -- (2);
		\draw[] (1) -- (3);
		\draw[] (2) -- (5);	
		\draw[] (2) -- (6);
		\draw[] (3) -- (6);
		\draw[] (2) -- (3);	
		\fill[bulk] (4) -- (5) -- (6) -- (7) -- (4);
		\draw[] (5) -- (7);
		\draw[] (5) -- (6);
		\draw[] (6) --node[above, color=black]{${\sss #5_1}$} (7);
		\draw[] (3) -- (5);	
		\fill[bulk] (0) -- (3) -- (7) -- (4) -- (0);
		\draw[] (3) -- (4);
		\draw[] (0) --node[left, pos=0.3, yshift=-0.2em, color=black]{${\sss #4}$} (3);
		\draw[] (4) -- (7);
		\draw[] (3) --node[left, text=black, xshift=0.1em]{${\sss #5_2}$} (7);
		\draw[] (4) -- (5);
		\draw[] (0) --node[right, text=black, pos=0.4, xshift=-0.1em]{${\sss #6_2}$} (4);
		\draw[] (1) -- (5);
		\draw[] (0) -- (5);
		\draw[] (0) --node[below, text=black]{${\sss #6_1}$} (1);	
	}{}	
	
	\node[below] at (0) {{$\scriptstyle{#2}'$}}; 
	\node[below] at (1) {{$\scriptstyle{#3'}$}}; 
	\node[right, yshift=0.1em] at (2) {{$\scriptstyle{#3}$}}; 
	\node[left] at (3) {{$\scriptstyle{#2}$}}; 
	\node[below, xshift=0.5em, yshift=0.2em] at (4) {{$\scriptstyle{\tilde{#2}'}$}};	
	\node[above] at (5) {{$\scriptstyle{\tilde{#3}'}$}}; 
	\node[above] at (6) {{$\scriptstyle{\tilde{#3}}$}}; 
	\node[above] at (7) {{$\scriptstyle{\tilde{#2}}$}}; 	
	\end{tikzpicture}
}

\newcommand{\tubeThreeDDouble}[8]{
	\begin{tikzpicture}[scale=#1, baseline=2.5em]
	\coordinate (0) at (0,0);
	\coordinate (1) at (2,0);
	\coordinate (2) at (1.2,0.5);
	\coordinate (3) at (-0.8,0.5);
	\coordinate (4) at (0,2);
	\coordinate (5) at (2,2);
	\coordinate (6) at (1.2,2.5);
	\coordinate (7) at (-0.8,2.5);
	\coordinate (8) at (4,0);	
	\coordinate (9) at (3.2,0.5);
	\coordinate (10) at (4,2);
	\coordinate (11) at (3.2,2.5);
	
	\ifthenelse{#8 = 1}{
		\fill[bulk] (8) -- (9) -- (11) -- (10) -- (8);
		\fill[bulk] (1) -- (2) -- (9) -- (8) -- (1);
		\draw[] (8) -- (9);
		\draw[] (2) -- (8);
		\draw[] (9) -- (10);
		\draw[bdryEdge2] (9) -- (11);
		\draw[bdryEdge2] (2) -- (9);
		\draw[bdryEdge2] (6) -- (9);
		\fill[bulk] (5) -- (6) -- (11) -- (10) -- (5);
		\draw[] (6) -- (10);
		\draw[] (10) -- (11);
		\draw[bdryEdge2] (6) --node[above, color=black]{${\sss #6_1'}$} (11);
		\draw[] (2) -- (10);
		\draw[bdryEdge1] (5) -- (10);
		\draw[bdryEdge1] (8) -- (10);
		\draw[bdryEdge1] (1) -- (10);
		\draw[bdryEdge1] (1) --node[below, text=black]{${\sss #7_1'}$} (8);
		
		\fill[bulk] (1) -- (2) -- (6) -- (5) -- (1);
		\fill[bulk] (0) -- (1) -- (2) -- (3) -- (0);
		\draw[] (1) -- (2);
		\draw[] (1) -- (3);
		\draw[] (2) -- (5);	
		\draw[bdryEdge2] (2) -- (6);
		\draw[bdryEdge2] (3) -- (6);
		\draw[bdryEdge2] (2) -- (3);	
		\fill[bulk] (4) -- (5) -- (6) -- (7) -- (4);
		\draw[] (5) -- (7);
		\draw[] (5) -- (6);
		\draw[bdryEdge2] (6) --node[above, color=black]{${\sss #6_1}$} (7);
		\draw[] (3) -- (5);	
		\fill[bulk] (0) -- (3) -- (7) -- (4) -- (0);
		\draw[] (3) -- (4);
		\draw[] (0) --node[left, pos=0.3, yshift=-0.2em, color=black]{${\sss #5}$} (3);
		\draw[] (4) -- (7);
		\draw[bdryEdge2] (3) --node[left, text=black, xshift=0.1em]{${\sss #6_2}$} (7);
		\draw[bdryEdge1] (4) -- (5);
		\draw[bdryEdge1] (0) --node[right, text=black, pos=0.4, xshift=-0.1em]{${\sss #7_2}$} (4);
		\draw[bdryEdge1] (1) -- (5);
		\draw[bdryEdge1] (0) -- (5);
		\draw[bdryEdge1] (0) --node[below, text=black]{${\sss #7_1}$} (1);
	}{}
	\ifthenelse{#8 = 2}{
		\fill[bulk] (8) -- (9) -- (11) -- (10) -- (8);
		\fill[bulk] (1) -- (2) -- (9) -- (8) -- (1);
		\draw[] (8) -- (9);
		\draw[] (2) -- (8);
		\draw[] (9) -- (10);
		\draw[bdryEdge2] (9) -- (11);
		\draw[bdryEdge2] (2) -- (9);
		\draw[bdryEdge2] (6) -- (9);
		\fill[bulk] (5) -- (6) -- (11) -- (10) -- (5);
		\draw[] (6) -- (10);
		\draw[] (10) -- (11);
		\draw[bdryEdge2] (6) --node[above, color=black]{${\sss \tilde #6^{-1} #6_1'}$} (11);
		\draw[] (2) -- (10);
		\draw[bdryEdge1] (5) -- (10);
		\draw[bdryEdge1] (8) -- (10);
		\draw[bdryEdge1] (1) -- (10);
		\draw[bdryEdge1] (1) --node[below, text=black]{${\sss \tilde #7^{-1} #7_1'}$} (8);
		
		\fill[bulk] (1) -- (2) -- (6) -- (5) -- (1);
		\fill[bulk] (0) -- (1) -- (2) -- (3) -- (0);
		\draw[] (1) -- (2);
		\draw[] (1) -- (3);
		\draw[] (2) -- (5);	
		\draw[bdryEdge2] (2) -- (6);
		\draw[bdryEdge2] (3) -- (6);
		\draw[bdryEdge2] (2) -- (3);	
		\fill[bulk] (4) -- (5) -- (6) -- (7) -- (4);
		\draw[] (5) -- (7);
		\draw[] (5) -- (6);
		\draw[bdryEdge2] (6) --node[above, color=black]{${\sss #6_1 \tilde #6}$} (7);
		\draw[] (3) -- (5);	
		\fill[bulk] (0) -- (3) -- (7) -- (4) -- (0);
		\draw[] (3) -- (4);
		\draw[] (0) --node[left, pos=0.3, yshift=-0.2em, color=black]{${\sss #5}$} (3);
		\draw[] (4) -- (7);
		\draw[bdryEdge2] (3) --node[left, text=black, xshift=0.1em]{${\sss #6_2}$} (7);
		\draw[bdryEdge1] (4) -- (5);
		\draw[bdryEdge1] (0) --node[right, text=black, pos=0.4, xshift=-0.1em]{${\sss #7_2}$} (4);
		\draw[bdryEdge1] (1) -- (5);
		\draw[bdryEdge1] (0) -- (5);
		\draw[bdryEdge1] (0) --node[below, text=black]{${\sss #7_1 \tilde #7}$} (1);
	}{}
	
	\node[below] at (0) {{$\scriptstyle{#2}'$}}; 
	\node[below] at (1) {{$\scriptstyle{#3'}$}}; 
	\node[left, yshift=0.5em, xshift=0.1em] at (2) {{$\scriptstyle{#3}$}}; 
	\node[left] at (3) {{$\scriptstyle{#2}$}}; 
	\node[below, xshift=0.5em, yshift=0.2em] at (4) {{$\scriptstyle{\tilde{#2}'}$}};	
	\node[below, xshift=0.6em, yshift=0.1em] at (5) {{$\scriptstyle{\tilde{#3}'}$}}; 
	\node[above] at (6) {{$\scriptstyle{\tilde{#3}}$}}; 
	\node[above] at (7) {{$\scriptstyle{\tilde{#2}}$}}; 
	\node[below] at (8) {{$\scriptstyle{#4}'$}}; 
	\node[right, yshift=0.1em] at (9) {{$\scriptstyle{#4}$}};	
	\node[above] at (10) {{$\scriptstyle{\tilde{#4}'}$}}; 
	\node[above] at (11) {{$\scriptstyle{\tilde{#4}}$}}; 
	\end{tikzpicture}
}

\makeatletter
\newcommand{\xRrightarrow}[2][]{\ext@arrow 0359\Rrightarrowfill@{#1}{#2}}
\newcommand{\Rrightarrowfill@}{\arrowfill@\equiv\equiv\Rrightarrow}
\newcommand{\xLleftarrow}[2][]{\ext@arrow 3095\Lleftarrowfill@{#1}{#2}}
\newcommand{\Lleftarrowfill@}{\arrowfill@\Lleftarrow\equiv\equiv}
\makeatother

\usepackage{soul}

\hypersetup{%
	bookmarks=true,         % show bookmarks bar first?
	bookmarksopen=true,
	unicode=true,           % to show unicode characters in Acrobat’s bookmarks
	pdftoolbar=true,        % show Acrobat’s toolbar?
	pdfmenubar=true,        % show Acrobat’s menu?
	pdffitwindow=false,     % window fit to page when opened
	pdfstartview={FitH},    % fits the width of the page to the window
	pdftitle={Gapped boundaries and string-like excitations in (3+1)d gauge models of topological phases},
	pdfauthor={},
	pdfsubject={},          % subject of the document
	pdfcreator={},          % creator of the document
	pdfproducer={pdfLaTeX}, % producer of the document
	pdfkeywords={topological phases} {category theory},
	pdfnewwindow=true,      % links in new window
	colorlinks,
	citecolor=BrickRed,
	filecolor=BrickRed,
	linkcolor=BlueViolet,
	urlcolor=BrickRed,
	linktocpage=true
}

\title{\boldmath Gapped boundaries and string-like excitations in (3+1)d gauge models of topological phases}

\author[\Square]{Alex Bullivant,}
\author[\pentagon, \hexagon]{Clement Delcamp}

\affiliation[\Square]{Department of Pure Mathematics, University of Leeds, Leeds, LS2 9JT, UK}

\affiliation[\pentagon]{Max-Planck-Institut f{\"u}r Quantenoptik \\ Hans-Kopfermann-Str. 1, 85748 Garching, Germany}
\affiliation[\hexagon]{Munich Center for Quantum Science and Technology (MCQST)\\ Schellingstr. 4, D-80799 M{\"u}nchen}

\emailAdd{bullivant.alex@gmail.com}
\emailAdd{clement.delcamp@mpq.mpg.de}

\abstract{\\~\\ We study lattice Hamiltonian realisations of (3+1)d Dijkgraaf-Witten theory with gapped boundaries. In addition to the bulk loop-like excitations, the Hamiltonian yields bulk dyonic string-like excitations that terminate at gapped boundaries. Using a tube algebra approach, we classify such excitations and derive the corresponding representation theory. Via a dimensional reduction argument, we relate this tube algebra to that describing (2+1)d boundary point-like excitations at interfaces between two gapped boundaries. Such point-like excitations are well known to be encoded into a bicategory of module categories over the input fusion category. Exploiting this correspondence, we define a bicategory that encodes the string-like excitations ending at gapped boundaries, showing that it is a sub-bicategory of the centre of the input bicategory of group-graded 2-vector spaces. In the process, we explain how gapped boundaries in (3+1)d can be labelled by so-called pseudo-algebra objects over this input bicategory.}

\begin{document} 
	\vspace*{-2em}
	\maketitle
	\flushbottom
	\newpage
	
\section{Introduction}
A prominent class of gapped quantum phases of matter are given by so-called \emph{topological phases of matter}. Such phases can be defined as equivalence classes of gapped quantum models whose low-energy effective descriptions realise \emph{topological quantum field theories} (TQFTs) \cite{Atiyah:1989vu}. In (2+1)d, \emph{spherical fusion categories} can be used to define a state-sum TQFT known as the \emph{Turaev-Viro-Barrett-Westbury} TQFT \cite{Turaev:1992hq,Barrett:1993ab}. Given such data, one can define an exactly solvable Hamiltonian model on a closed manifold, in a canonical manner, that describes non-chiral topological phases in (2+1)d \cite{Levin:2004mi, 2010AnPhy.325.2707K, alex2011stringnet}.
Such models support topological excitations referred to as \emph{anyons}, which display exotic braiding and fusion statistics. Topological excitations are typically described via the so-called \emph{Drinfel'd center} of the input spherical fusion category \cite{majid2000foundations}. For any spherical fusion category, the center construction defines a \emph{modular} tensor category, which is widely accepted as being the right classification tool for anyons in (2+1)d \cite{Kitaev1997, Kitaev:2006lla}. 

Given an open manifold, it is often possible to extend the lattice Hamiltonian to the boundary, while preserving the gap. Equivalence classes of such extensions define the notion of \emph{gapped boundaries}, which realise anomalous TQFTs. These are found to be described by indecomposable \emph{module categories} over the input spherical category. Furthermore, boundary Hamiltonians yield point-like excitations that can be classified through the language of \emph{module category functors} \cite{kitaev2012models}. Domain walls between distinct topological phases can be considered in a similar fashion. By iterating the procedure, it is possible to further extend such models to interfaces between different gapped boundaries. The corresponding zero-dimensional Hamiltonians yield point-like excitations in their own right. These different settings have received a lot of attention in recent years within the topological order community \cite{Bravyi:1998sy, 2011CMaPh.306..663B, kitaev2012models, Barkeshli:2014cna, Bullivant:2017qrv, wang2018gapped, cong2016topological, Yoshida:2017xqa, PhysRevB.78.115421,  PhysRevLett.105.030403, PhysRevX.8.031048}, partly due to their application to the field of \emph{topological quantum computation} \cite{PhysRevB.87.045130, cong2016topological}. Mathematically, these fit in the wider topic of \emph{defect} TQFTs \cite{morrison2011higher, Carqueville:2016nqk, Carqueville:2017ono, Carqueville:2016kdq, Carqueville:2017aoe, Carqueville:2018sld, Fuchs:2014ema, Fuchs:2013gha}. 

Despite tremendous progress in our understanding of (2+1)d topological models, a lot of questions remain open regarding generalizations to higher dimensions. It is expected that topological models in (3+1)d should take as input a spherical fusion bicategory. Although the precise definition of such notion remains partly elusive, a compelling definition has been recently put forward by Douglas et al. in \cite{douglas2018fusion}. In this manuscript, the authors show that their definition encompasses a large class of four-dimensional state-sum invariants. Ultimately, we would like to derive properties of (3+1)d topological models within this general higher category theoretical framework, which is admittedly tantalizing but difficult. In order to make progress in this direction, we decide to focus on so-called \emph{gauge models of topological phases}, i.e. models that have a lattice gauge theory interpretation \cite{Kitaev1997,Hu:2012wx,Wan:2014woa,Bullivant:2019fmk}. These models are interesting for diverse reasons. Technically, they are particularly manageable allowing to carry out computations in full detail, and they are easily definable in any dimensions. Physically, they happen to be extremely relevant in (3+1)d as they seem to encapsulate a large class of Bosonic models displaying topological order \cite{lan2017classification,zhu2018topological,Thorngren:2020aph, johnsonfreyd2020classification}. 

In (2+1)d, topological gauge models are obtained by choosing as input the category of $G$-graded vector spaces, with $G$ a finite group and monoidal structure twisted by a  cohomology class in $H^{3}(G,\rU(1))$. The corresponding state-sum invariant is referred to as the Dijkgraaf-Witten invariant \cite{dijkgraaf1990topological}. In this context, (bulk) anyonic excitations are described in terms of the so-called \emph{twisted quantum double} of the group, whose irreducible representations provide the simple objects of the Drinfel'd centre of the category of $G$-graded vector spaces \cite{Drinfeld:1989st,Dijkgraaf1991}. Gapped boundaries are found to be labelled by a simple set of data, namely a subgroup of the input group and a 2-cochain that is compatible with the input 3-cocycle \cite{Beigi_2011, Bullivant:2017qrv}, and their excitations have been considered for instance in \cite{Lan:2013wia, cong2016topological, Bridgeman:2019wyu, Fuchs:2013gha, PhysRevB.78.115421}.

More generally, given a closed ($d$+1)-manifold, the input data of Dijkgraaf-Witten theory is a finite group $G$ and a cohomology class $[\omega]\in H^{d+1}(G,\rU(1))$. It is always possible to define a lattice Hamiltonian realization of the theory on a $d$-dimensional hypersurface $\Sigma$, such that the ground state subspace of the model is provided by the image of the partition function assigned to the cobordism $\Sigma \times [0,1]$. In (3+1)d, the resulting gauge models are known to yield loop-like excitations, i.e. excitations with the topology of a circle. In general such a loop-like excitation corresponds to a loop-like magnetic flux to which a point-like charge is attached, while being threaded by an auxiliary flux. This threading flux plays a crucial role as it can constrain the quantum numbers associated with the other flux and the charge. In \cite{Delcamp:2017pcw, Bullivant:2019fmk}, their classification and statistics were found to be described in terms of the so-called \emph{twisted quantum triple} of the group, which is a natural extension of the twisted quantum double. Although a general theory of gapped boundaries in (3+1)d is still lacking, examples have already been proposed in the case of topological gauge models \cite{wang2018gapped,PhysRevX.8.031048}. These are labelled by a set of data akin to (2+1)d, namely a subgroup of the input group and a  3-cochain compatible with the input 4-cocycle. The main objective of our manuscript is to study excitations for such gapped boundaries in (3+1)d.

In order to reveal the algebraic structure underlying the bulk excitations in arbitrary spatial dimension, several strategies exist. Our focus is on the so-called \emph{tube algebra approach} \cite{2010AnPhy.325.2707K, Lan:2013wia, bultinck2017anyons, Aasen:2017ubm,DDR1, Delcamp:2017pcw,  Bullivant:2019fmk, Bullivant:2019tbp}, which is a generalization of Ocneanu's tube algebra \cite{ocneanu1994chirality, ocneanu2001operator}. In general, the `tube' refers to the manifold $\partial \Sigma \times [0,1]$, where $\partial \Sigma$ is the boundary left by removing a regular neighbourhood of the excitation in question, and the `algebra' to an algebraic extension of the gluing operation  $(\partial \Sigma \times [0,1]) \cup_{\partial \Sigma} (\partial \Sigma \times [0,1])  \simeq (\partial \Sigma \times [0,1])$ to the Hilbert space of states on the tube. For instance, the twisted quantum double and the twisted quantum triple are found to be isomorphic to the tube algebras associated with the manifolds $\mathbb S^1 \times [0,1]$ and $\mathbb T^2 \times [0,1]$, respectively. This approach relies on the fact that properties of a given excitation are encoded into the boundary conditions that the model assigns to the boundary $\partial \Sigma$ \cite{Bullivant:2019fmk}. This strategy has been extensively applied to general two-dimensional models, and more recently to gauge and higher gauge models in three dimensions \cite{Bullivant:2019fmk, Bullivant:2019tbp}.

The tube algebra approach can be adapted in order to study excitations on defects and gapped boundaries, and has been employed in some specific cases in \cite{kitaev2012models, Lan:2013wia, Bridgeman:2019wyu, Williamson:2017uzx}. In this context, the tube possesses two kinds of boundary: a physical gapped boundary that corresponds to the one of the spatial manifold, and a boundary obtained by removing a local neighbourhood of an excitation incident on the boundary of the spatial manifold. Although, the method is very general and could be used to study any pattern of excitations in (3+1)d, we shall focus on a specific configuration, namely bulk string-like excitations that terminate at gapped boundaries. There are several motivations to consider these specific excitations. The first one is that, due to the topology of the problem, we can relate the corresponding tube algebra to the one relevant to the study of point-like excitations at the zero-dimensional interface of two gapped boundaries in (2+1)d. This is a generalization of what happens in the bulk, where upon dimensional reduction, bulk loop-like excitations can be treated as point-like anyons \cite{Wang:2014xba,Wang:2014oya}. In \cite{Bullivant:2019fmk}, this mechanism was made precise in terms of so-called \emph{lifted} models, where we showed that higher-dimensional tube algebras could be recast in terms of lower-dimensional analogues using the language of loop groupoids. We generalize these techniques in this manuscript by introducing the notion  of \emph{relative groupoid algebras}, which we use to unify both the (2+1)d and (3+1)d tube algebras.

Although this correspondence between two seemingly very different types of excitations is interesting per se, it turns out to be a precious technical tool. Indeed, since it allows us to recast the (3+1)d tube algebra as a (2+1)d one, we can use the (2+1)d scenario, which is easier to visualise and intuit, as a guideline for the more complex (3+1)d case. Using this framework, we derive the irreducible representations of the (3+1)d tube algebra, which classify the elementary string-like excitations whose endpoints lie on gapped boundaries. We further define a notion of tensor product that encodes the concatenation of these excitations, and compute the Clebsch-Gordan series compatible with this tensor product. Moreover, we find the 6j-symbols  that ensure the \emph{quasi-coassociativity} of this tensor product. All these mathematical notions can then be put to use in order to define canonical bases of ground states or excited states in the presence of gapped boundaries.

The second reason we decide to focus on such open string-like excitations pertains to category theory. The same way the relevant category theoretical data to describe gauge models in (2+1)d is the category of $G$-graded vector spaces, the one relevant to describe (3+1)d gauge models is the \emph{bicategory of $G$-graded 2-vector spaces}. In a recent work \cite{Kong:2019brm},  Kong et al. applied the generalised \emph{centre} construction to this bicategory and demonstrated that the result was given by the bicategory of module categories over the multi-fusion category of loop-groupoid-graded vector spaces. This is a categorification of the well-know result that the centre of the category of group-graded vector spaces can be described as the category of modules for the loop-groupoid algebra \cite{willerton2008twisted}.  The latter relation can be appreciated from the point of view of the tube algebra approach, which we use to argue that the centre of the bicategory of $G$-graded 2-vector spaces describes string-like excitations together with boundary conditions for the string endpoints.

In order to prove this statement, we construct explicitly the bicategory of module categories over the multi-fusion category of groupoid-graded vector spaces. To do so, we rely on the familiar correspondence between indecomposable module categories and category of module over \emph{algebra objects} \cite{2001math.....11139O, etingof2016tensor, ostrik2002module}. When applied to the group treated as a one-object groupoid, this provides a description for (2+1)d point-like excitations at the interface between two gapped boundaries. When applied to the loop-groupoid of the group, we demonstrate that it describes the string-like excitations and their endpoints boundary conditions, which string-like excitations ending at gapped boundaries is a subclass of.

\begin{center}
	\textbf{Organisation of the paper}
\end{center}
\noindent
In sec.~\ref{sec:DW} we review the construction of the lattice Hamiltonian realization of Dijkgraf-Witten theory in any spatial dimension. We then describe an extension of the Hamiltonian model to introduce gapped boundary conditions. In the subsequent discussion, we apply the tube algebra approach to point-like excitations at the interface of two one-dimensional gapped boundaries in sec.~\ref{sec:tube2D}. In sec.~\ref{sec:tube3D}, we consider string-like bulk excitations that terminate at gapped boundaries and apply the tube algebra approach to this scenario.
We also introduce in this section the notion of relative groupoid algebra that unifies the (2+1)d and (3+1)d computations. The representation theory of the tube algebras is presented in full detail in sec.~\ref{sec:rep}. Finally, the category theoretical structures capturing the properties of boundary excitations in (2+1)d and (3+1)d are developed in sec.~\ref{sec:Highercat}. The correspondence with the centre construction of the bicategory of group-graded 2-vector spaces is also established in this section.

\newpage
\section{Dijkgraaf-Witten Hamiltonian Model\label{sec:DW}}
\emph{In this section, we first review the definition of the Dijkgraaf-Witten theory and the construction of its Hamiltonian realisation. We then generalise the construction to include gapped boundaries.}
\subsection{Partition function}\label{sec:DWpartitionfunction}

The input for the ($d$+1)-dimensional \emph{Dijkgraaf-Witten} theory is given by a pair $(G,[\omega])$ where $G$ is a finite group and $[\omega]\in H^{d+1}(G,\rU(1))$ is a ($d$+1)-cohomology class.\footnote{Here $\rU(1)$ denotes the circle group as a $G$-module with action $\triangleright:G\times \rU(1)\rightarrow \rU(1)$ given by $g\triangleright u=u$ for all $g\in G$ and $u\in \rU(1)$.} Given a closed manifold, this theory can be conveniently expressed as a sigma model with target space the classifying space $BG$ of the group $G$. In order to extend the definition of the partition function to open manifolds, it is necessary to endow the manifold with a triangulation, in which case the partition function is obtained by summing over $G$-labellings of the 1-simplices  that satisfy compatibility constraints. Ultimately, we are interested in lattice Hamiltonian realisations of such theory, for which we need the expression of the partition function that the Dijkgraaf-Witten theory assigns to a special class of open manifolds referred to as \emph{pinched interval cobordisms}. We shall directly define the partition function for this special class of manifolds. Details regarding more basic aspects of this theory can be found in \cite{dijkgraaf1990topological, Bullivant:2019fmk}.

Let $\Xi$ be a compact, oriented $d$-manifold with a  possibly non-empty boundary. We define the \emph{pinched interval cobordism} $\Xi\times_{\rm p}\I$ over $\Xi$ as the quotient manifold
\begin{align}
	\Xi\times_{\rm p}\I\equiv\Xi\times \I \; /\sim \; ,
\end{align}
where $\I\equiv [0,1]$ denotes the unit interval, and the equivalence relation $\sim$ is such that $ (x,i)\sim (x,i')$, for all $(x,i),(x,i')\in \partial\Xi\times \I $. By definition, we have $\partial (\Xi\times_{\rm p}\I)=\overline{\Xi}\cup_{\partial\Xi} \Xi$ and $\overline{\Xi}\cap \Xi=\partial\Xi$, where $\overline{\Xi}$ is the manifold $\Xi$ with reversed orientation. In contrast, the boundary of the \emph{interval cobordism} $\Xi \times \I$ over $\Xi$ reads $\partial (\Xi\times \I)=\overline{\Xi}\cup \Xi\cup (\partial\Xi\times \I)$. To illustrate this distinction, we can consider the following simple examples:
\begin{equation*}
	[0,1]\times_{\rm p}[0,1]=
	\pinchedinterval{}
	\q , \q
	[0,1]\times [0,1]=
	\interval{} \; .
\end{equation*}
Naturally, if $\partial \Xi=\varnothing$, then we have the identification $\Xi\times_{\rm p}\I=\Xi\times \I$.

In order to define the Dijkgraaf-Witten partition function, we shall further require our pinched interval (spacetime) manifold be equipped with a choice of \emph{triangulation}, i.e. a $\Delta$-complex whose geometric realisation is homeomorphic to the manifold. We shall further assume that every triangulation has a chosen total ordering of its $0$-simplices (vertices), referred to as a \emph{branching structure}. A choice of branching structure for a triangulation naturally encodes the structure of a directed graph on the corresponding one-skeleton. By convention, we choose the 1-simplices (edges) to be directed from the lowest ordered vertex to the highest ordered vertex. Given a compact, oriented $d$-manifold $\Xi$, we notate a triangulation of the pinched interval cobordism $\Xi\times_{\rm p}\mathbb{I}$ by ${_{\triangle'}\Xi_{\triangle}}$, such that $\partial({_{\triangle'}\Xi_{\triangle}})=\overline{\Xi_{\triangle}}\cup_{\partial \Xi_{\triangle'}} \Xi_{\triangle'}$, where $\Xi_{\triangle}$ and $\Xi_{\triangle'}$ denote two possibly different triangulations of $\Xi$. Let us remark that by definition, we have $\partial (\Xi_{\triangle})=\partial( \Xi_{\triangle'})$. 

Let $\Xi\times_{\rm p}\mathbb{I}$ be a ($d$+1)-dimensional pinched interval cobordism endowed with a triangulation ${_{\triangle'}\Xi_{\triangle}}$. We define a $G$-colouring of ${_{\triangle'}\Xi_{\triangle}}$ as an assignment of group elements $g_{v_i v_j} \in G$ to every oriented 1-simplex $(v_{i}v_{j})\subset {_{\triangle'}\Xi_{\triangle}}$, with $v_{i}<v_{j}$, such that for every 2-simplex $(v_{i}v_{j}v_{k})\subset {_{\triangle'}\Xi_{\triangle}}$, with $v_{i}<v_{j}<v_{k}$, the condition $g_{v_i v_j}g_{v_j v_k}=g_{v_i v_k}$ is satisfied. The set of $G$-colourings on ${_{\triangle'}\Xi_{\triangle}}$ is notated by ${\rm Col}({_{\triangle'}\Xi_{\triangle}},G)$. Given a $G$-colouring $g\in {\rm Col}({_{\triangle'}\Xi_{\triangle}},G)$ and an $n$-simplex $\triangle^{(n)}=(v_{0}v_{1}\ldots v_{n})\subset {_{\triangle'}\Xi_{\triangle}}$, we denote by $g[v_{0}v_{1}\ldots v_{n}]\equiv(g_{v_0v_1},\ldots,g_{v_{n-1}v_n})\in G^{n}$, the $n$ group elements specifying the restriction of $g$ to a $G$-colouring of $(v_{0}v_{1}\ldots v_{n})$. Using this notation, we further define the evaluation of a ($d$+1)-cocycle $\omega \in Z^{d+1}(G,\rU(1))$ on a $G$-colouring $g\in {\rm Col}({_{\triangle'}\Xi_{\triangle}},G)$ restricted to a ($d$+1)-simplex $(v_{0}\ldots v_{d+1})\subset {_{\triangle'}\Xi_{\triangle}}$ as
\begin{align*}
	\omega(g[v_{0}\ldots v_{d+1}])
	\equiv
	\omega(g_{v_0v_1},\ldots,g_{v_dv_{d+1}})\;.
\end{align*}
Equipped with the above, let us now define the partition function that the ($d$+1)-dimensional Dijkgraaf-Witten theory assigns to a given pinched interval cobordism. Letting $\Xi$ be a compact, oriented $d$-manifold and ${_{\triangle'}\Xi_{\triangle}}$ a triangulation of $\Xi\times_{\rm p}\mathbb{I}$, the partition function defines a linear operator
\begin{align*}
	\Z[{_{\triangle'}\Xi_{\triangle}}]
	:
	\Hilb{\Xi_{\triangle}}
	\rightarrow
	\Hilb{\Xi_{\triangle'}} \; ,
\end{align*}
where the Hilbert spaces $\Hilb{\Xi_\triangle}$ and $\Hilb{\Xi_{\triangle'}}$ are defined according to
\begin{align}
	\label{eq:microhilbpartitionfunction}
	\Hilb{\Xi_{*}}\equiv \bigotimes_{\triangle^{(1)}\subset \Xi_{*}}
	\mathbb{C}[G] \; .
\end{align}
In the equation above, the tensor product is over all 1-simplices $\triangle^{(1)}$ in the corresponding triangulation, and $\mathbb{C}[G]$ denotes the Hilbert space spanned by $\{\ket{g} \}_{\forall \, g\in G}$ with inner product $\langle g|h\rangle=\delta_{g,h}$, $\forall \, g,h\in G$. Explicitly, the linear operator $\Z[{_{\triangle'}\Xi_{\triangle}}]$ reads
\begin{align*}
	\Z[{{_{\triangle'}\Xi_{\triangle}}}]
	\equiv
	\frac{1}{|G|^{\#({_{\triangle'}\Xi_{\triangle}})}}
	\!\!
	\sum_{g\in {\rm Col}({{_{\triangle'}\Xi_{\triangle}}},G)}
	\prod_{\triangle^{(d+1)}\subset {{_{\triangle'}\Xi_{\triangle}}}}
	\!\!\!\!
	\omega(g[\triangle^{(d+1)}])^{\epsilon(\triangle^{(d+1)})}
	\!\!\!
	\bigotimes_{\triangle^{(1)}\subset \Xi_{\triangle'}} 
	\!\!\!\!
	\ket{g[\triangle^{(1)}]}
	\!\!
	\bigotimes_{\triangle^{(1)}\subset \Xi_{\triangle}} 
	\!\!\!\!
	\bra{g[\triangle^{(1)}]} \; ,
\end{align*}
where $\#({_{\triangle'}\Xi_{\triangle}}) := |{_{\triangle'}\Xi_{\triangle}}^{(0)}|-\frac{1}{2}|\partial{_{\triangle'}\Xi_{\triangle}}^{(0)}|-\frac{1}{2}|\partial\Xi_{\triangle}^{(0)}|$ and $\epsilon(\triangle^{(d+1)})\in\pm 1$ denotes the orientation of the ($d$+1)-simplex $\triangle^{(d+1)}\subset {_{\triangle'}\Xi_{\triangle}}$.

Before concluding this section, let us describe some of the salient features of the partition function above. Firstly, given a pinched interval cobordism $\Xi\times_{\rm p}\mathbb{I}$ and two choices of triangulation ${_{\triangle'}\Xi_{\triangle}}$ and ${_{\triangle'}\tilde{\Xi}_{\triangle}}$ such that $\partial({_{\triangle'}\Xi_{\triangle}}) = \partial({_{\triangle'}\tilde{\Xi}_{\triangle}})$, we find the operators $\Z[{_{\triangle'}\Xi_{\triangle}}]=\Z[{{_{\triangle'}\tilde{\Xi}_{\triangle}}}]$ to be equal. This property follows directly from the ($d$+1)-cocycle condition satisfied by $\omega$, i.e. $d^{(d+1)}\omega=1$. This implies that the operator $\Z$ is \emph{boundary relative triangulation independent}, i.e. it remains invariant under retriangulation of the interior ${\rm int}({_{\triangle'}\Xi_{\triangle}}) := {_{\triangle'}\Xi_{\triangle}} \backslash \partial({_{\triangle'}\Xi_{\triangle}})$ of ${_{\triangle'}\Xi_{\triangle}}$ but does depend on a choice of boundary triangulation. Using this boundary relative triangulaton independence, we find the crucial relation
\begin{align*}
	\Z[{{_{\triangle''}\Xi_{\triangle'}}}] \, 
	\Z[{{_{\triangle'}\Xi_{\triangle}}}]
	=
	\Z[{{_{\triangle''}\Xi_{\triangle}}}] \;.
\end{align*}
Secondly, given a $d$-manifold $\Sigma$ equipped with a triangulation $\Sigma_\triangle$ and $\Xi_{\triangle}$ a subcomplex of ${\rm int}(\Sigma_{\triangle})$, there is a natural action of $\Z[{{_{\triangle'}\Xi_{\triangle}}}]$ on $\Hilb{\Sigma_{\triangle}}$ such that
\begin{align*}
	\Z[{{_{\triangle'}\Xi_{\triangle}}}]
	:
	\mathcal{H}^G_\omega[\Sigma_{\triangle}]
	\rightarrow
	\mathcal{H}^G_\omega[\Sigma_{\triangle'}]
\end{align*}
where $\Sigma_{\triangle'}$ is a triangulation of $\Sigma$ induced from $\Sigma_{\triangle}$ by replacing the subcomplex $\Xi_{\triangle}\subset {\rm int}(\Sigma_{\triangle})$ with $\Xi_{\triangle'}$, while keeping the remaining triangulation the same. On the subspace
\begin{align}
	\label{eq:physSpace}
	\gsopen{\Sigma_{\triangle}}:= {\rm Im} \, \Zo{{_{\triangle'}\Xi_{\triangle}}}
	\subset
	\Hilb{\Xi_{\triangle}} \; ,
\end{align}
the operator $\Z[{{_{\triangle'}\Xi_{\triangle}}}]$ further defines a unitary isomorphism
\begin{align}
	\label{eq:isoTriangulations}
	\Z[{{_{\triangle'}\Xi_{\triangle}}}]
	:
	\gsopen{\Sigma_{\triangle}}
	\xrightarrow{\sim}
	\gsopen{\Sigma_{\triangle'}} \; .
\end{align}
This follows directly from the boundary relative triangulation independence of $\Z$ as well as the Hermicity condition
\begin{align}\label{eq:Hermicity}
	\Zo{{{_{\triangle'}\Xi_{\triangle}}}}^{\dagger}
	=
	\Zo{{{_{\triangle}\Xi_{\triangle'}}}}
	\;.
\end{align}

\subsection{Hamiltonian realisation of Dijkgraaf-Witten theory}\label{sec:DWHam}

Let us now construct an exactly solvable model that is the lattice Hamiltonian realisation of Dijkgraaf-Witten theory in $d$ spatial dimensions \cite{Hu:2012wx,Wan:2014woa,Bullivant:2019fmk}. The input of the model is a pair $(G,\omega)$ where $G$ is a finite group and $\omega$ a normalised representative of a cohomology class in $H^{d+1}(G,\rU(1))$. Given an oriented (possibly open) $d$-manifold $\Sigma$ representing the spatial manifold of the theory, and a choice of triangulation $\Sigma_{\triangle}$, the microscopic Hilbert space of the model is given by
\begin{align*}
	\Hilb{\Sigma_{\triangle}}\equiv \bigotimes_{\triangle^{(1)}\subset \Sigma_{\triangle}}\mathbb{C}[G] \; ,
\end{align*}
as in \eqref{eq:microhilbpartitionfunction}.
A natural choice of basis for $\Hilb{\Sigma_{\triangle}}$ is given by an assignment of $g_{v_{i}v_{j}}\in G$ for each oriented edge $(v_{i}v_{j})\subset\Sigma_{\triangle}$ defined by the vertices $v_{i}<v_{j}$. Henceforth, we shall refer to such states as \emph{graph-states}.

The bulk Hamiltonian is obtained as a sum of mutually commuting projectors that come in two families. Firstly, to every $2$-simplex $(v_0v_1v_2) \subset {\rm int}(\Sigma_{\triangle})$ of the interior of $\Sigma_\triangle$, we assign an operator $\mathbb{B}_{(v_0v_1v_2)}$ that is defined via the following action on a graph-state $\ket{g}\in\Hilb{\Sigma_{\triangle}}$:
\begin{equation*}
	\mathbb B_{(v_0v_1v_2)} : | g \rangle \mapsto \delta_{g_{v_0v_1} g_{v_1v_2}\, , \,g_{v_0v_2}} | g \rangle \; .
\end{equation*}
This definition can be extended linearly to an operator on any state $\ket{\uppsi}\in\Hilb{\Sigma_{\triangle}}$. Secondly, to every 0-simplex $(v_0) \subset {\rm int}(\Sigma_\triangle)$, we assign an operator $\mathbb A_{(v_0)}$ which acts on a local neighbourhood of $(v_0)$ defined as the subcomplex $\Xi_{v_{0}}:={\rm cl} \circ {\rm st}(v_0)\subset \Sigma_{\triangle}$. Here ${\rm st}(-)$ and ${\rm cl}(-)$ are the \emph{star} and the \emph{closure} operations, respectively, so that $\Xi_{v_0}$ corresponds to the smallest subcomplex of $\Sigma_\triangle$ that include all the simplices of which $(v_0)$ is a subsimplex.  The definition of $\mathbb A_{(v_0)}$ requires the triangulated pinched interval cobordism $\cob{\Xi_{v_{0}}\!}{\Xi}{\, \Xi_{v_{0}}}$ defined as
\begin{align*}
	\cob{\Xi_{v_{0}}\!}{\Xi}{\, \Xi_{v_{0}}}:=
	(\snum{v_0'}) \sqcup_{\rm j} {\rm cl} \circ {\rm st}(v_0)\; ,
\end{align*}
where $ {-}\sqcup_{\rm j} {-}$ denotes the \emph{join} operation.  Given two simplices $ \triangle^{(n)} \equiv (v_0 v_1 \ldots v_n) $ and $\triangle^{(n')} \equiv (v_{n+1}v_{n+2} \ldots v_{n+n'+1})$, the join operation creates the new simplex $\triangle^{(n)} \sqcup_{\rm j} \triangle^{(n')} \equiv (v_0v_1 \ldots v_{n+n'+1})$. In the definition above, $(v_0')$ refers to an  auxiliary vertex such that $v_{0}<v_{0}'<v_1$, and which follows the ordering of $(v_{0})$ with respect to the other vertices in $\Sigma_{\triangle}$. For the sake of concreteness, we illustrate these various definitions with the following two-dimensional example:
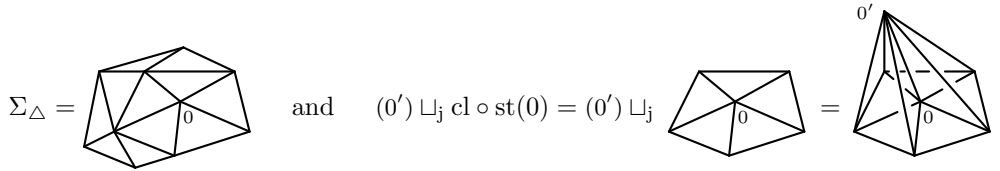
\begin{equation*}
	\Sigma_{ \triangle} = \clst{1} \q\;\; \text{and} \q\;\;
	\snum{(0')} \sqcup_{\rm j} {\rm cl} \circ {\rm st} (0) = 
	\snum{(0')} \sqcup_{\rm j} \; \clst{2} \; =  	\clst{3} \; .
\end{equation*}
Finally, given a state $\ket{\uppsi}\in\Hilb{\Sigma_{\triangle}}$, the action of the operator $\mathbb{A}_{(v_{0})}$ is defined via
\begin{align}
	\label{eq:opA}
	\mathbb{A}_{(v_{0})}: \ket{\uppsi}
	\mapsto
	\Zo{(\snum{v_0'}) \sqcup_{\rm j} {\rm cl} \circ {\rm st}(v_0)} \, \ket{\uppsi} \; .
\end{align}
For instance, in (3+1)d the action of the operator $\mathbb{A}_{(4)}$ on a vertex $\snum{(4)}$ shared by four 3-simplices explicitly reads
\begin{align*}
	\mathbb{A}_{(4)}  \Bigg| \! \HamONE{0.9}{1} \!\! \Bigg\rangle 
	&= \mathcal{Z}_\pi^G\Bigg[ \! \HamTWO{0.9}{1} \! \Bigg]  \, \Bigg| \! \HamONE{0.9}{1} \!\! \Bigg\rangle \\[-1.6em] \nn
	& = \frac{1}{|G|}\sum_{k \in G} 
	\frac{\pi(g_{01}g_{12},g_{23},g_{34},g_{44'}) \, \pi(g_{01},g_{12},g_{23}g_{34},g_{44'})}{\pi(g_{12},g_{23},g_{34},g_{44'}) \, \pi(g_{01},g_{12}g_{23},g_{34},g_{44'})}
	\Bigg| \! \HamONE{0.9}{2} \!\! \Bigg\rangle  ,
\end{align*}
where $\pi \in Z^4(G,\rU(1))$. 
The lattice Hamiltonian is finally obtained as
\begin{align}
	\label{eq:Ham}
	\Hambulk{\Sigma_{\triangle}}
	=-\sum_{\triangle^{(2)}\subset {\rm int}(\Sigma_{\triangle})} \!\! \mathbb{B}_{\triangle^{(2)}}
	\, -\sum_{\triangle^{(0)}\subset{\rm int}(\Sigma_{\triangle})} \!\!\mathbb{A}_{\triangle^{(0)}}\;,
\end{align}
where the sums run over all the 2-simplices and 0-simplices in the interior of $\Sigma_{\triangle}$, respectively. It follows from the definitions and the boundary relative triangulation independence that the operators $\{\mathbb{A}_{\triangle^{(0)}},\mathbb{B}_{\triangle^{(2)}}\}_{\forall \,  \triangle^{(0)}\! ,\triangle^{(2)}\subset {\rm int}(\Sigma_{\triangle})}$
satisfy the algebra
\begin{gather*}
	\mathbb{A}_{(v_{i})}\mathbb{A}_{(v_{i})}=\mathbb{A}_{(v_{i})}
	\; , \q 
	\mathbb{A}_{(v_{i})}\mathbb{A}_{(v_{j})}=\mathbb{A}_{(v_{j})}\mathbb{A}_{(v_{i})} \; ,
	\\
	\mathbb{B}_{(v_{j}v_{k}v_{l})}\mathbb{B}_{(v_{j}v_{k}v_{j})}=\mathbb{B}_{(v_{j}v_{k}v_{l})}
	\; , \q
	\mathbb{B}_{(v_{j}v_{k}v_{l})}\mathbb{B}_{(v_{j}'v_{k}'v_{l}')} = \mathbb{B}_{(v_{j'}v_{k'}v_{l'})}\mathbb{B}_{(v_{j}v_{k}v_{l})} \; ,
	\\
	\mathbb{A}_{(v_{i})}\mathbb{B}_{(v_{j}v_{k}v_{l})}
	=
	\mathbb{B}_{(v_{j}v_{k}v_{l})}\mathbb{A}_{(v_{i})} \; ,
\end{gather*}
for all $(v_{i}), (v_{i'}),(v_{j}v_{k}v_{l}),(v_{j}'v_{k}'v_{l}')\subset \Sigma_{\triangle}$. All the operators are mutually commuting projectors and the Hamiltonian is exactly solvable. It follows that the \emph{ground state projector} $\mathbb{P}^{\rm bulk}_{\Sigma_{\triangle}}$ simply reads
\begin{align}
	\label{eq:projGS}
	\mathbb{P}^{\rm bulk}_{\Sigma_{\triangle}}:= \prod_{\triangle^{(0)}\subset {\rm int}(\Sigma_{\triangle})}\mathbb A_{\triangle^{(0)}} \prod_{\triangle^{(2)}\subset {\rm int}(\Sigma_{\triangle})}\mathbb B_{\triangle^{(2)}} \;.
\end{align}
Notice that the ordering in the product is superfluous by the commutativity of the operators. Furthermore it follows from inspection that
\begin{align}
\mathbb{P}^{\rm bulk}_{\Sigma_{\triangle}}=\mc{Z}^{G}_{\omega}[{_{\triangle}}{\Sigma}_{\triangle}]\,,
\end{align}
such that the ground state subspace of $\Hambulk{\Sigma_{\triangle}}$ is given by
\begin{align}
	\label{eq:idSub}
	{\rm Im} \, \mathbb{P}^{\rm bulk}_{\Sigma_{\triangle}}
	= {\rm Im} \, \mc{Z}^{G}_{\omega}[{_{\triangle}}{\Sigma}_{\triangle}] \,
	\equiv \gsopen{\Sigma_{\triangle}} \; ,
\end{align}
with the last equality following from \eqref{eq:physSpace}. This is the space spanned by linear superpositions $| \uppsi \ra$ of graph-states fulfilling the stabiliser constraints $\mathbb A_{\triangle^{(0)}} | \uppsi \ra = | \uppsi \ra$ and $\mathbb B_{\triangle^{(2)}}| \uppsi \ra = | \uppsi \ra$ at every $\triangle^{(0)}, \triangle^{(2)} \subset {\rm int}(\Sigma_{\triangle})$.

Let us conclude this construction by making two observations. The first one is that we showed in \eqref{eq:isoTriangulations} how given two triangulations $\Sigma_\triangle$ and $\Sigma_{\triangle'}$ of $\Sigma$ such that $\partial \Sigma_\triangle = \partial \Sigma_\triangle'$, the subspaces $\gsopen{\Sigma_{\triangle}}$ and $\gsopen{\Sigma_{\triangle'}}$ were unitarily isomorphic. This signifies that it is always possible to perform local changes of the triangulation in the interior of $\Sigma$ while remaining in the same gapped phase. This will turns out to be very useful when performing explicit computations. In particular, we shall often apply unitary isomorphisms obtained from pinched interval cobordisms describing so-called \emph{Pachner moves}. The second observation is that the Hamiltonian operators do not mix ground states with differing boundary $G$-colourings, so that there exists a natural decomposition of the Hilbert space as 
\begin{align}\label{eq:boundarydecomp}
	\gsopen{\Sigma_{\triangle}}=\bigoplus_{a\in {\rm Col}(\partial\Sigma_{\triangle},G)}
	\!\! 
	\gsopen{\Sigma_{\triangle}}_{a}
\end{align}
where $\gsopen{\Sigma_{\triangle}}_{a}\subseteq \gsopen{\Sigma_{\triangle}}$ denotes the subspace of states identified by the boundary colouring $a\in {\rm Col}(\partial\Sigma_{\triangle},G)$. More details regarding the construction up to that point can be found in \cite{Bullivant:2019fmk}.

\subsection{Gapped boundary partition function}
Given an open $d$-dimensional surface $\Sigma$ endowed with a triangulation $\Sigma_\triangle$, we reviewed above how to define an exactly solvable model as the Hamiltonian realisation of Dijkgraaf-Witten theory whose input data is a finite group $G$ and normalised ($d$+1)-cocycle in $H^{d+1}(G,\rU(1))$. The lattice Hamiltonian $\mathbb H^G_\omega[\Sigma_\triangle]^{\rm bulk}$ was obtained as a sum of mutually commuting projectors that act on the interior of $\Sigma_\triangle$. We would like to extend this Hamiltonian to $\partial \Sigma_\triangle$ while preserving the gap of the system, giving rise to the notion of gapped boundaries. In order to do so, we shall first  define a generalisation  of the partition function introduced in sec.~\ref{sec:DWpartitionfunction} for spacetime ($d$+1)-manifolds presenting two types of boundaries.

Let us begin by introducing the notion of \emph{relative pinched interval cobordisms}. Let $\Xi$ be a compact, oriented, $d$-manifold with non-empty boundary and $\Omega \subseteq \partial \Xi$ a choice of ($d$$-$1)-dimensional submanifold of the boundary. The relative pinched interval cobordism $\Xi \times^{\Omega}_{\rm p}\mathbb{I}$ over $\Xi$ with respect to $\Omega$ is defined as the quotient manifold
\begin{align}
	\Xi\times^{\Omega}_{\rm p}\mathbb{I} \equiv \Xi \times \mathbb{I}/\sim_{\Omega},
\end{align}
where $\sim_{\Omega}$ is defined such that $(x,i)\sim_{\Omega}(x,i')$, for all $(x,i),(x,i')\in (\partial \Xi\backslash {\rm int}(\Omega))\times \mathbb{I}$. By definition, we have $\partial(\Xi \times^{\Omega}_{\rm p}\mathbb{I})=\overline{\Xi\cup_{\Omega}(\Omega \times_{\rm p}\mathbb{I})}\cup_{\partial\Xi}\Xi$ and $\overline{\Xi}\cap\Xi=\partial\Xi\backslash{\rm int}(\Omega)$. To illustrate this definition we consider the following simple examples:
\begin{equation*}
	[0,1]\times_{\rm p}[0,1]=
	\pinchedinterval{}
	\q , \q
	[0,1]\times^{\Omega}_{\rm p} [0,1]=
	\begin{tikzpicture}[scale=0.6,baseline={([yshift=-.5ex]current bounding box.center)}]
	\def\a{1.5}
	\def\b{1.5}
	\coordinate (1) at (0,0);
	\coordinate (2) at (\a,0);
	\coordinate (3) at (0,\b);
	\coordinate (4) at (\a,\b);
	\draw[] (1) to node[anchor=north]{} (2);
	\draw[] (1) to (3);
	\draw[] (3) to (2);
	%\draw[] (2) to (4);
	\end{tikzpicture} \; ,
\end{equation*}
with $\Omega \equiv 0\subset \{0,1\}=\partial\mathbb{I}$. Henceforth, we shall utilise the convention that $\Xi\times^{\Omega}_{\rm p}\mathbb{I}$ defines a cobordism
\begin{align}
	\Xi\times^{\Omega}_{\rm p}\mathbb{I}:\Xi\rightarrow \Xi\,,
\end{align}
and refer to ${\Omega \times_{\rm p}\mathbb I}\subset \partial(\Xi\times^{\Omega}_{\rm p}\mathbb{I})$ as a \emph{time-like boundary}. A triangulation of $\Xi\times^{\Omega}_{\rm p}\mathbb I$ can be constructed as follows: Let $\Xi_{\triangle}$, $\Xi_{\triangle'}$ be a pair of triangulations of $\Xi$ such that $\Omega_{\triangle}\subset\partial \Xi_{\triangle}$ and $\Omega_{\triangle'}\subset\partial \Xi_{\triangle'}$ define two possibly different triangulations of $\Omega$ satisfying
\begin{align}
	\partial\Xi_{\triangle}\backslash{\rm int}(\Omega_{\triangle})
	=
	\partial\Xi_{\triangle'}\backslash{\rm int}(\Omega_{\triangle'}) \; .
\end{align}
Considering a triangulation ${_{\triangle'}\Omega_{\triangle}}$ of the time-like boundary $\Omega\times_{\rm p} \mathbb I$, we define ${_{\triangle'}{\Xi^{\Omega}}_{\triangle}}$ as the triangulation of the relative pinched interval cobordism $\Xi\times^{\Omega}_{\rm p} \mathbb I$ whose boundary reads
$\overline{
	\Xi_{\triangle}\cup_{\Omega_{\triangle}} {_{\triangle'}\Omega_{\triangle}}
}
\cup_{\partial \Xi_{\triangle'}}\Xi_{\triangle'}$.

Given a triangulation ${_{\triangle'}{\Xi^{\Omega}}_{\triangle}}$ of $\Xi\times^{\Omega}_{\rm p} \mathbb I$, let us now define a generalisation of the ($d$+1)-dimensional Dijkgraaf-Witten theory with input data $(G,\omega)$ such that the corresponding partition function evaluated on ${_{\triangle'}{\Xi^{\Omega}}_{\triangle}}$ remains invariant under triangulation changes of both the interior of
${_{\triangle'}{\Xi^{\Omega}}_{\triangle}}$ and the interior of the time-like boundary ${_{\triangle'}{\Omega}_{\triangle}}$. 
Let $\Omega = \sqcup_{i} \Omega_i$ be a decomposition of $\Omega$ into connected components $\Omega_{i}$, each with triangulations $\Omega_{\triangle,i}\subset\partial \Xi_{\triangle}$ and $\Omega_{\triangle',i}\subset\partial \Xi_{\triangle'}$. The generalised theory associates to each connected component $\Omega_i$ a pair $(A_{i},\phi_{i})$, where $A_{i}\subset G$ is a subgroup and $\phi_{i}\in C^{d}(A_{i},\rU(1))$ a normalised group $d$-cochain such that $d^{(d)}\phi_{i}=\omega^{-1}{\sss |}_{A_{i}}$. We refer to the data $(A_{i},\phi_{i})$ as a choice of \emph{gapped boundary condition}.\footnote{In sec.~\ref{sec:Highercat}, we shall revisit gapped boundary conditions from a category theoretical point of view.} We define a $(G,\{A_i\})$-colouring $g$ of ${_{\triangle'}{\Xi^{\Omega}}_{\triangle}}$ as a $G$-colouring such that $g[{_{\triangle'}{\Omega}_{\triangle,i}}]\in {\rm Col}({_{\triangle'}{\Omega}_{\triangle,i}},A_{i})$. The set of $(G,\{A_i\})$-colourings on ${_{\triangle'}{\Xi^{\Omega}}_{\triangle}}$ is denoted by ${\rm Col}({_{\triangle'}{\Xi^{\Omega}}_{\triangle}},G,\{A_i\})$. Equipped with such choices, we define the generalised partition function as follows:
\begin{align}
	\label{eq:relativepartitionfunction}
	\mc{Z}^{G,\{A_{i}\}}_{\omega,\{\phi_{i}\}}[{_{\triangle'}{\Xi^{\Omega}}_{\triangle}}]
	&  =
	\frac{1}{|G|^{\#({_{\triangle'}{\Xi^{\Omega}}_{\triangle}}) } \prod_i |A_i|^{\#({_{\triangle'}{\Omega}_{\triangle,i}})} }
	\\
	\nn
	&\hspace{-2.2em} \sum_{g\in{\rm Col}({_{\triangle'}{\Xi^{\Omega}}_{\triangle}},G,\{A_{i}\})}
	\;\prod_{i}\bigg(
	\prod_{\triangle^{(d)}\subset {_{\triangle'}{\Omega}_{\triangle,i}} }
	\!\!\! \phi_{i}(g[\triangle^{(d)}])^{\epsilon(\triangle^{(d)})}
	\bigg)
	\prod_{\triangle^{(d+1)}\subset {_{\triangle'}{\Xi^{\Omega}}_{\triangle}}}
	\!\!\!\!
	\omega(g[\triangle^{(d+1)}])^{\epsilon(\triangle^{(d+1)})}
	\\
	\nn
	&\hspace{5.3em}\bigotimes_{\triangle^{(1)}\subset \Xi_{\triangle'}} 
	\!\!\!\!
	\ket{g[\triangle^{(1)}]}
	\!\!
	\bigotimes_{\triangle^{(1)}\subset \Xi_{\triangle}} 
	\!\!\!\!
	\bra{g[\triangle^{(1)}]} \; ,
\end{align}
where $\#({_{\triangle'}{\Xi^{\Omega}}_{\triangle}}):=|{\rm int}({_{\triangle'}{\Xi}_{\triangle}})^{(0)}|+\frac{1}{2}|{\rm int}(\Xi_{\triangle'})^{(0)}|+\frac{1}{2}|{\rm int}(\Xi_{\triangle})^{(0)}|$ and $\#({_{\triangle'}{\Omega}_{\triangle,i}}):=|{\rm int}({_{\triangle'}{\Omega}_{\triangle,i}})^{(0)}|+\frac{1}{2}|{\rm int}(\Omega_{\triangle',i})^{(0)}|+\frac{1}{2}|{\rm int}(\Omega_{\triangle,i})^{(0)}|$.

As stated previously, the partition function remains invariant under retriangulation of the interior of ${_{\triangle'}\Omega_{\triangle}}$ as well as the interior of ${_{\triangle'}{\Xi^{\Omega}}_{\triangle}}$. In this manner, the partition function  $\mc{Z}^{G,\{A_{i}\}}_{\omega,\{\phi_{i}\}}[{_{\triangle}{\Xi^{\Omega}}_{\triangle}}]$ defines a projection operator and we associate to the triangulation $\Xi_\triangle$ the following Hilbert space:
\begin{align}
	\mc{V}^{G,\{A_{i}\}}_{\omega,\{\phi_{i}\}}[\Xi_{\triangle}]
	:=
	{\rm Im} \; \mc{Z}^{G,\{A_{i}\}}_{\omega,\{\phi_{i}\}}[{_{\triangle}{\Xi^{\Omega}}_{\triangle}}]\; .
\end{align}
Furthermore, akin to equations \eqref{eq:isoTriangulations} and \eqref{eq:Hermicity}, the triangulation invariance properties of the partition function together with the Hermicitiy condition
	\begin{align}
	\mc{Z}^{G,\{A_{i}\}}_{\omega,\{\phi_{i}\}}[{_{\triangle'}{\Xi^{\Omega}}_{\triangle}}]^{\dagger}
	=
	\mc{Z}^{G,\{A_{i}\}}_{\omega,\{\phi_{i}\}}[{_{\triangle}{\Xi^{\Omega}}_{\triangle'}}] 
\end{align}
demonstrate that the operator
\begin{align}
	\mc{Z}^{G,\{A_{i}\}}_{\omega,\{\phi_{i}\}}[{_{\triangle'}{\Xi^{\Omega}}_{\triangle}}]
	:
	\mc{V}^{G,\{A_{i}\}}_{\omega,\{\phi_{i}\}}[\Xi_{\triangle}]
	\xrightarrow{\sim}
	\mc{V}^{G,\{A_{i}\}}_{\omega,\{\phi_{i}\}}[\Xi_{\triangle'}]
\end{align}
defines a unitary isomorphism of Hilbert spaces.

\subsection{Hamiltonian model in the presence of gapped boundaries}\label{sec:bdryHam}
 
In sec. \ref{sec:DWHam}, we described the Hamiltonian realisation $\mathbb H^G_\omega[\Sigma_\triangle]^{\rm bulk}$ of the Dijkgraaf-Witten theory in $d$ spatial dimensions in the presence of open boundary conditions. Utilising the partition function \eqref{eq:relativepartitionfunction} introduced in the previous section, we shall now define an extension of the Hamiltonian model to include gapped boundary conditions \cite{Bullivant:2017qrv,wang2018gapped}.

Let us consider an oriented $d$-manifold $\Sigma$ with non-empty boundary and a choice of triangulation $\Sigma_\triangle$. The input of the model is a pair $(G,\omega)$ and a choice of gapped boundary conditions $\{(A_i ,\phi_i)\}$ for each connected component $\partial \Sigma_{\triangle,i} \subset \Sigma_\triangle$, where $A_i \subset G$ is a subgroup and $\phi_{i}\in C^{d}(A_{i},\rU(1))$ is a normalised group $d$-cochain satisfying the condition $d^{(d)}\phi_{i}=\omega^{-1}{\sss |}_{A_{i}}$. In the interior of $\Sigma_\triangle$, the (bulk) Hamiltonian was defined in eq.~\ref{eq:Ham}. Given such a choice of gapped boundary conditions, let us now define an operator that acts on a local neighbourhood of a boundary vertex $(v_{0})\subset \partial\Sigma_{\triangle,i}$. Mimicking the definition of the bulk vertex operator, we consider the subcomplex $\Xi_{v_{0}}:={\rm cl}\circ {\rm st}(v_{0})$, which corresponds to the smallest subcomplex that includes all the simplices of which $(v_0)$ is a subsimplex. We next define the triangulated relative pinched interval cobordism over $\Xi_{v_0}$ with respect to $\Omega := {\rm cl}\circ {\rm st}(v_{0})\cap \partial \Sigma_{\triangle,i}$
\begin{align}
	\label{eq:pinchedBdryOp}
	\cob{\Xi_{v_{0}}\!}{\Xi}{\, \Xi_{v_{0}}}:=(v'_{0})\sqcup_{\rm j}  {\rm cl}\circ {\rm st}(v_{0}) \;,
\end{align}
whose boundary is given by
\begin{align}
	\partial (\cob{\Xi_{v_{0}}\!}{\Xi}{\, \Xi_{v_{0}}})
	=
	\overline{
	\Xi_{v_{0}}\cup_{\Omega}(\cob{\Omega_{v_{0}}\!}{\Omega}{\, \Omega_{v_{0}}}) 
	}
	\cup_{\partial\Xi_{v_{0}}}\Xi_{v_{0}}
\end{align}
where  $\Omega_{v_0} := (v'_{0})\sqcup_{\rm j}\Omega$. Given this relative pinched interval cobordism, we define the action of the operator $\mathbb{A}^{A_{i},\phi_{i}}_{(v_{0})}$ on a state $| \uppsi \ra \in \mathcal H^{G,A_i}_{\omega,\phi_i}[\Sigma_\triangle]$ via
\begin{align}
	\mathbb{A}^{A_{i},\phi_{i}}_{(v_{0})}:\ket{\uppsi}
	\mapsto
	\mc{Z}^{G,A_{i}}_{\omega,\phi_{i}}[\cob{\Xi_{v_{0}}\!}{\Xi}{\, \Xi_{v_{0}}}] 
	\ket{\uppsi} \; .
\end{align}
The gapped boundary Hamiltonian is finally defined as
\begin{align}
	\mathbb{H}^{G,\{A_{i}\}}_{\omega,\{\phi_{i}\}}[\Sigma_{\triangle}]
	=
	\Hambulk{\Sigma_{\triangle}}
	+
	\sum_{\partial\Sigma_{\triangle,i}\subset\partial \Sigma_{\triangle}}
	\mathbb{H}^{G,A_{i}}_{\omega,\phi_{i}}[\partial \Sigma_{\triangle,i}]^{\rm bdry} \; ,
\end{align}
where
\begin{align}
	\mathbb{H}^{G,A_{i}}_{\omega,\phi_{i}}[\partial \Sigma_{\triangle,i}]^{\rm bdry}
	:=
	-\sum_{\triangle^{(0)}\subset \partial\Sigma_{\triangle,i}}\mathbb{A}^{A_{i},\phi_{i}}_{\triangle^{(0)}}\; .
\end{align}
From the triangulation invariance properties of the partition function $\mc{Z}^{G,\{A_{i}\}}_{\omega,\{\phi_{i}\}}$ follows that the Hamiltonian is a sum of mutually commuting projection operators, and as such it is still exactly solvable. Furthermore, analogously to  the bulk Hamiltonian, we can identify the ground-state subspace $\mc{V}^{{G,\{A_{i}\}}}_{\omega,\{\phi_{i}\}}[\Sigma_{\triangle}]$ with
\begin{align}
	{\rm Im}\; \mc{Z}^{{G,\{A_{i}\}}}_{\omega,\{\phi_{i}\}}[{_{\triangle}{\Sigma^{\partial \Sigma}}_{\triangle}}]
	\equiv
	\mc{V}^{{G,\{A_{i}\}}}_{\omega,\{\phi_{i}\}}[\Sigma_{\triangle}] \; ,
\end{align} 
and verify that the unitary isomorphism
\begin{align}
	\mc{Z}^{{G,\{A_{i}\}}}_{\omega,\{\phi_{i}\}}[{_{\triangle'}{\Sigma^{\partial \Sigma}}_{\triangle}}]:
	\mc{V}^{{G,\{A_{i}\}}}_{\omega,\{\phi_{i}\}}[\Sigma_{\triangle}]
	\xrightarrow{\sim}
	\mc{V}^{{G,\{A_{i}\}}}_{\omega,\{\phi_{i}\}}[\Sigma_{\triangle'}]
\end{align}
commutes with the Hamiltonian. This last statement implies that we can always replace a given triangulated subcomplex $\Omega_\triangle \subset \partial \Sigma_\triangle$ by $\Omega_{\triangle'}$ while remaining in the ground state sector.

Note finally that in the subsequent discussion, we shall also refer to \emph{gapped interfaces} between several gapped boundaries. However, we will not require an explicit form of the Hamiltonian for such interfaces, and as such we omit here the explicit definition. Despite such an omission, the corresponding Hamiltonian can be explicitly defined in close analogy with the construction of the gapped boundary Hamiltonian presented in this section.

\bigskip \noindent
In order to illustrate the definition and some properties of the gapped boundary Hamiltonian, let us now specialize to two dimensions (see also \cite{Bullivant:2017qrv}). We consider a two-dimensional surface $\Sigma$ endowed with a triangulation $\Sigma_\triangle$ and a single connected boundary component $\partial\Sigma_{\triangle}$. The input data for the bulk Hamiltonian is a finite group $G$ and a normalised group $3$-cocycle $\alpha$. Furthermore, we define on $\partial \Sigma_\triangle$ a gapped boundary whose input data is a pair $(A, \phi)$, where $A \subset G$ is a subgroup and $\phi$ a group 2-cochain satisfying $d^{(2)}\phi=\alpha^{-1}{\sss |}_{A}$ which is explicitly expressed via
\begin{equation}
\alpha^{-1}(a,a',a'') \stackrel{!}{=} d^{(2)}\phi(a,a',a'') 
=
\frac{\phi(a',a'') \, \phi(a, a'a'')}{\phi(aa',a'') \, \phi(a,a')} \; ,
\end{equation}
for every $a,a',a'' \in A \subset G$. We consider the following situation:
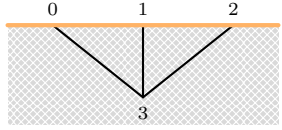
\begin{equation*}
\exampleTwoD{1.2}{1}
\end{equation*}
where the dashed area $\snippetBulk{0.6}{1}$ represents the bulk of the manifold, whereas the coloured line stands for the gapped boundary. The black lines represent the 1-simplices on the interior $\Sigma_\triangle$ that are included in ${\rm cl}\circ {\rm st}\snum{(1)}$. We first want to write down the action of the boundary operator at the vertex $\snum{(1)}$ on graph-states of the form
\begin{equation}
\label{eq:exampleHilb}
{\rm Span}_{\mathbb C}\bigg\{\bigg| \mathfrak{g} \bigg[ \exampleTwoD{1.2}{2}  \bigg] \bigg\ra \bigg\}_{\substack{\forall \, \mathfrak{g} \in {\rm Col}({\rm cl} \circ {\rm st}(1), G,A) }} 
\!\!\! \equiv
{\rm Span}_{\mathbb C}\bigg\{\bigg|  \exampleTwoD{1.2}{3} \! \bigg\ra \bigg\}_{\substack{\hspace{-0.9em} \forall \, g \in G \\ \forall \, a,a' \in A }} \!\!\!\!  .
\end{equation}
The boundary vertex operator $\mathbb{A}^{A,\phi}_{(1)}$ boils down to evaluating the partition function \eqref{eq:relativepartitionfunction} on the relative pinched interval cobordism $\snum{(023)}\times_{\rm p}^{(02)} \mathbb I$ defined by
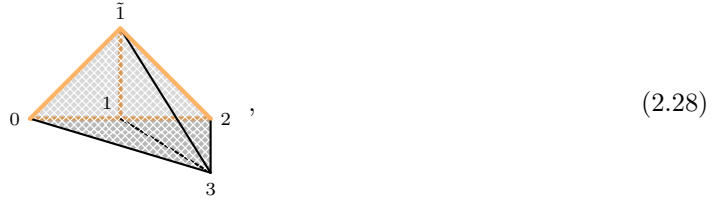
\begin{equation}
\raisebox{-1.9em}{\exampleTwoD{1.2}{4}} ,
\end{equation}
such that $\snum{0} < \snum{1} < \snum{\tilde 1} < \snum{2} < \snum{3}$ and the orange edges represent the time-like boundary.
Explicitly, the action of this boundary vertex operator reads
\begin{align}
	 \mathbb{A}^{A,\phi}_{(1)} \bigg | \exampleTwoD{1.2}{3}  \! \bigg\ra  = \frac{1}{|A|}\sum_{\tilde a \in A} \frac{\alpha(a, \tilde a, \tilde a^{-1}g) \, \phi(\tilde a, \tilde a^{-1}a')}{\alpha(\tilde a ,\tilde a^{-1}a',a'^{-1}g) \, \phi(a,\tilde a)} 
	\bigg|  \exampleTwoD{1.2}{5} \! \bigg\ra \; .
\end{align}
Let us now compute a triangulation changing boundary operator on a graph state \eqref{eq:exampleHilb}. More specifically, let us construct the isomorphism that replaces the boundary subcomplex $\snum{(01)} \cup \snum{(12)}$ by a single 1-simplex $\snum{(02)}$. The corresponding operator is conveniently obtained by evaluating the partition function \eqref{eq:relativepartitionfunction} on the relative pinched interval cobordism
\begin{equation}
\raisebox{3.7em}{\exampleTwoD{1.2}{7}} ,
\end{equation}
with time-like boundary $\snum{(012)}$, implementing the isomorphism
\begin{equation}
\bigg| \exampleTwoD{1.2}{3} \! \bigg\ra \simeq 
\frac{1}{|A|^\frac{1}{2}}  \frac{\alpha(a,a',a'^{-1}g)}{\phi(a,a')}
\bigg| \exampleTwoD{1.2}{6} \! \bigg\ra \; .
\end{equation}
We can now confirm that this triangulation changing operator does commute with the Hamiltonian operator. This follows from the cocycle relations $d^{(2)}\phi(a, \tilde a, \tilde a^{-1}a') = \alpha^{-1}(a,\tilde a, \tilde a^{-1}a')$ and $d^{(3)}\alpha(a,\tilde a, \tilde a^{-1}a',a'^{-1}g)$ = 1.

\section{Tube algebra for gapped boundary excitations in (2+1)d\label{sec:tube2D}}
\emph{In this section, we apply the tube algebra approach in order to derive the algebraic structure underlying the boundary point-like excitations in two spatial dimensions.}

\subsection{Definition}

Let us consider an open two-dimensional surface $\Sigma$. Its boundary $\partial \Sigma$ is referred to as the \emph{physical boundary} of the system. In the previous section, we explained how to construct the lattice Hamiltonian realisation of Dijkgraaf-Witten theory on a triangulation of $\Sigma$. We further detailed how this model could be extended to the physical boundary of $\Sigma$ in such way as to remain gapped. Bulk excitations of this model were studied in detail in general dimensions in \cite{Bullivant:2019fmk}. In addition to bulk excitations, the lattice Hamiltonian yields point-like boundary excitations that are excitations obtained by violating some of the stabiliser constraints on the boundary. We are interested in the classification and the statistics of such gapped boundary excitations.  More specifically, we consider the situation where two different one-dimensional gapped boundaries meet at a zero-dimensional interface, and are interested in the point-like excitations living at such interface. This situation can be locally depicted as follows:
\begin{equation}
	\bdryInterface{0.9} \; .
\end{equation}
Given that the input data for the bulk theory is a pair $(G,\alpha)$, where $\alpha$ is a normalized representative of a cohomology class in $H^3(G,\rU(1))$, the thick coloured lines stand for two gapped boundaries characterized by the boundary conditions $A_\phi \equiv(A ,\phi)$ and $B_{\psi} \equiv (B,\psi)$, respectively, while the black dot illustrates the binary interface between them. The boundary conditions $A_\phi$ and $B_\psi$, which were defined in the previous section, are such that $A,B \subset G$, $d^{(2)}\phi = \alpha^{-1}{\sss |}_A$ and $d^{(2)}\psi = \alpha^{-1}{\sss |}_B$. We denote the lattice Hamiltonian for this specific choice of boundary conditions by $\mathbb H^{G,A, B}_{\alpha, \phi,\psi}[\Sigma]$, and its associated ground state subspace by $\gsTwo{\Sigma}$. In the following discussion, we will suppose that the Hamiltonian is further extended to the interface, but we do not require the explicit form of the corresponding operator. 
Note that although we restrict our attention to gapped boundaries, our exposition could be easily generalised to accommodate \emph{domain walls}, which can be thought of as shared gapped boundaries between two (possibly different) topological phases.

By definition, given a point-like excitation at the interface of two one-dimensional gapped boundaries, there is a local neighbourhood of $\Sigma$ for which the energy density is higher than the one of the ground state. Removing such a local neighbourhood leaves a new boundary component, referred to as the \emph{excitation boundary}, that is incident on the physical boundary $\partial \Sigma$ of the manifold. We denote the resulting manifold by $\Sigma^{\rm o}$ and the excitation boundary by $\exBdry$. 
We illustrate this configuration as follows:
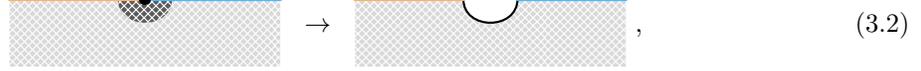
\begin{equation}
	\bdryInterfaceExc{0.9}{1} \;\; \to \;\; \bdryInterfaceExc{0.9}{2} \; ,
\end{equation}
where the dashed area $\snippetBulk{0.6}{2}$ represents the region whose energy density is higher than the one of the ground state. The black line represents the excitation boundary, whose topology is the one of the unit interval $\mathbb I \equiv [0,1]$. Endowing $\Sigma^{\rm o}$ with a triangulation, we are interested in the lattice Hamiltonian $\mathbb H^{G,A, B}_{\alpha, \phi,\psi}[\Sigma^{\rm o}_\triangle  \backslash \partial \Sigma^{\rm o}_\triangle {\sss |}_{\rm ex.}]$ obtained by removing all the operators whose supports are on $\partial \Sigma^{\rm o}_\triangle {\sss |}_{\rm ex.}$. In a way reminiscent to the bulk Hamiltonian in sec.~\ref{sec:DWHam}, this Hamiltonian displays \emph{open boundary conditions} such that the corresponding ground state subspace can be decomposed over them. Properties of the point-like excitations can then be encoded into the boundary conditions, so that a classification of the boundary conditions induces a classification of the corresponding point-like excitations.
In other words, ground states in $\gsTwo{\Sigma^{\rm o}_\triangle}$, which are characterised by a given excitation boundary colouring, define specific excitations with respect to ground states in the Hilbert space $\gsTwo{\Sigma_\triangle}$.
In general, any such excitation is a superposition of \emph{elementary point-like excitations}. In order to find these point-like elementary boundary excitations, we apply the \emph{tube algebra} approach, whose general construction can be found in \cite{Bullivant:2019fmk}.

Let us consider the manifold $\partial \Sigma^{\rm o}{\sss |}_{\rm ex.} \times \mathbb I$. Naturally, it has the topology of a 2-cell but we would like to emphasize the fact that it has two kinds of boundary components, namely a pair of physical boundary components and a pair of excitation boundary components.
More precisely, it is the system obtained by removing from the two-disk $\mathbb D^2$ local neighbourhoods at the interface of two different physical boundaries:
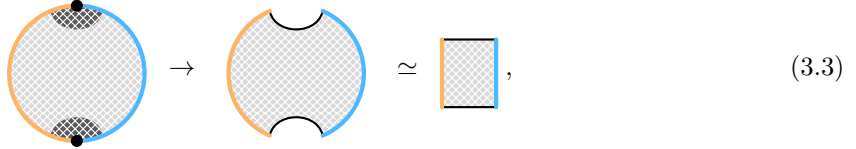
\begin{equation}
	\label{eq:cyl}
	\twoDiskExc{0.9}{1} \;\;  \to \;\; \twoDiskExc{0.9}{2} 
	\;\; \simeq  \;\; \twoDiskExc{0.9}{3} \; ,
\end{equation}
where the nomenclature is the same as before.
A crucial, yet trivial, fact is that we can always glue a copy of $\partial \Sigma^{\rm o}{\sss |}_{\rm ex.} \times \mathbb I$ to $\Sigma^{\rm o}$ along $\partial \Sigma^{\rm o}{\sss |}_{\rm ex.}$ without modifying its topology, i.e.
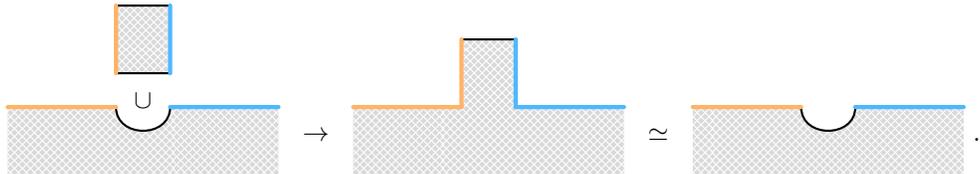
\begin{equation*}
	\symmetryMap{0.9}{1} \;\; \to \;\; \symmetryMap{0.9}{2} \;\; \simeq \;\; \symmetryMap{0.9}{3} \; .
\end{equation*} 
As explained in more detail in \cite{Bullivant:2019fmk}, given a triangulation of $\Sigma^{\rm o} $ and making use of the triangulation changing unitary isomorphisms, this simple gluing operation  induces a \emph{symmetry} map on the ground state subspace, whose \emph{simple modules} classify the boundary conditions on $\exBdry$ and as such the corresponding point-like boundary excitations. In order to compute these simple modules, we further remark that it is always possible to apply a diffeomorphism so that a local neighbourhood of $\exBdry$ is of the form $\exBdry \times \mathbb I$ so that the corresponding ground state subspaces are isomorphic. The effect of such diffeomorphism is to localise the action of the symmetry map so that it only involves degrees of freedom living within $\exBdry \times \mathbb I$. Consequently, it is enough to consider the symmetry map that corresponds to the gluing of two copies of the manifold $\exBdry \times \mathbb I$, i.e.
\begin{equation} 
	(\exBdry \times \mathbb I) \cup_{\exBdry} (\exBdry \times \mathbb I) \simeq \exBdry \times \mathbb I \; .
\end{equation}
We pictorially summarize these operations below:
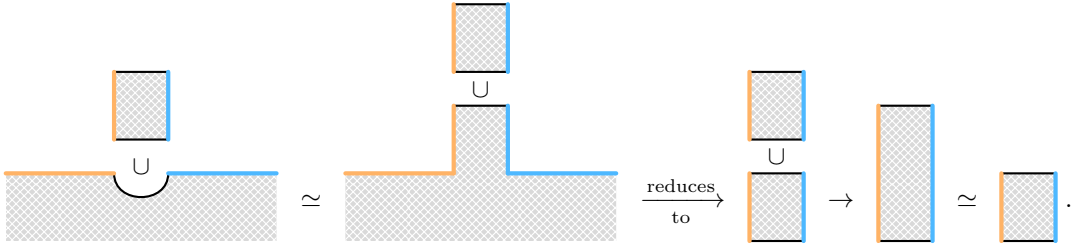
\begin{equation*}
	\symmetryMap{0.9}{1} \;\; \simeq \;\;  \symmetryMap{0.9}{4} \;\; \xrightarrow[\text{to}]{\text{reduces}} \;\; \gluingCyl{0.9}{1} \;\; \to \;\; \gluingCyl{0.9}{2}  \;\; \simeq \;\; \gluingCyl{0.9}{3} \; .
\end{equation*}
Given a triangulation of $\exBdry \times \mathbb I$, this symmetry map in turn endows the associated ground state subspace with a finite-dimensional algebraic structure referred to as the \emph{tube algebra}. Irreducible representations of the tube algebra label the simple modules of the original symmetry map, classifying boundary conditions on $\exBdry$, and thus the corresponding point-like boundary excitations.

\subsection{Computation of the tube algebra}

Let us now derive the tube algebra for the configuration described above so as to determine the elementary boundary excitations at the interface of two one-dimensional gapped boundaries. First, we need to specify the ground state subspace on $\exBdry \times \mathbb I$ by picking a triangulation. Crucially, the choice of triangulation does not matter. Indeed, given a triangulation of the excitation boundary, changing the discretisation of the physical boundary or the bulk of $\exBdry \times \mathbb I$ yields an isomorphic ground state subspace, which would in turn induce an isomorphic tube algebra. Furthermore, a different choice of triangulation for the excitation boundary would yield a \emph{Morita equivalent} tube algebra, which by definition has the same simple modules as the original algebra. As such, we should make the simplest choice of triangulation possible. We choose to discretise the excitation boundary by a single 1-simplex and $\exBdry \times \mathbb I$ as a triangulated 2-cell. The resulting triangulated manifold is denoted by $\mathfrak{T}[\mathbb I_\triangle]$ and the corresponding ground state subspace explicitly reads\footnote{Note that we rotated the drawings by $90^\circ$ for convenience.}
\begin{align}
	\nn
	\gsTwo{\mathfrak{T}[\mathbb I]}
	:= 
	{\rm Span}_{\mathbb C}\bigg\{\bigg| \mathfrak{g} \bigg[ \tubeTwoD{1.2}{1}  \bigg] \bigg\ra \bigg\}_{\substack{\forall \, \mathfrak{g} \in {\rm Col}(\mathfrak{T}[\mathbb I], G,A,B) }}  \!\!\!\!
	&\equiv
	{\rm Span}_{\mathbb C}\bigg\{\bigg| \! \tubeTwoD{1.2}{2} \! \bigg\ra \bigg\}_{\substack{\hspace{-2.5em} \forall \, g \in G \\ \forall \, (a,b) \in A \times  B}}
	\\ 
	\label{eq:basisTube}
	&\equiv
	{\rm Span}_{\mathbb C}\big\{\big| \biGr{g}{a}{b} \big\ra \big\}_{\substack{\hspace{-2.5em} \forall \, g \in G \\ \forall \, (a,b) \in A \times  B}} \; ,
\end{align}
where some labellings are left implicit since they can be deduced from the flatness constraints, i.e. the stabiliser constraints with respect to the $\mathbb B_{\triangle^{(2)}}$-operators.
The tube algebra can be computed using the following algorithm:\footnote{We refer the reader to \cite{Bullivant:2019fmk} for a general and more detailed definition of the tube algebra.} Recall that the tube algebra is an extension of the gluing operation $\mathfrak{T}[\mathbb I] \cup_\mathbb I \mathfrak{T}[\mathbb I] \simeq \mathfrak{T}[\mathbb I]$ to the ground state subspace $\gsTwo{\mathfrak{T}[\mathbb I]}$. Using the relation \eqref{eq:boundarydecomp}, we obtain the following decomposition of the Hilbert space $\gsTwo{\mathfrak{T}[\mathbb I]}$:
\begin{align*}
	\gsTwo{\mathfrak{T}[\mathbb I]}
	=
	\bigoplus_{\substack{
			g_1\in {\rm Col}(\mathbb I \times \{0\},G)
			\\
			g_2\in {\rm Col}(\mathbb I \times \{1\},G)
	}}
	\!\!
	\gsTwo{\mathfrak{T}[\mathbb I]}_{g_1,g_2} \; .
\end{align*}
The gluing itself is then performed via an injective map $\textsf{\small GLU}$ defined according to
\begin{align*}
	\textsf{\small GLU}:
	\gsTwo{\mathfrak{T}[\mathbb I]}
	\otimes
	\gsTwo{\mathfrak{T}[\mathbb I]}
	\rightarrow
	\!\!\!\!
	\bigoplus_{\substack{
			g_1,g_1'\in {\rm Col}(\mathbb I \times \{0\},G)
			\\
			g_2,g_2'\in {\rm Col}(\mathbb I \times \{1\},G)
	}}
	\!\!
	\gsTwo{\mathfrak{T}[\mathbb I]}_{g_1,g_2}
	\otimes
	\gsTwo{\mathfrak{T}[\mathbb I]}_{g_1',g_2'}
	\; ,
\end{align*}
which acts on states $\ket{\uppsi_{g_1,g_2}}\in \gsTwo{\mathfrak{T}[\mathbb I]}_{g_1,g_2}$ and $\ket{\uppsi'_{g_1',g_2'}}\in \gsTwo{\mathfrak{T}[\mathbb I]}_{g_1',g_2'}$ via identification of the boundary conditions along the gluing interface, i.e.
\begin{align*}
	\textsf{\small GLU}:
	\ket{\uppsi_{g_1,g_2}}\otimes \ket{\uppsi'_{g_1',g_2'}}
	\mapsto 
	\delta_{g_2,g_1'} \, 
	\ket{\uppsi_{g_1,g_2}}\otimes \ket{\uppsi'_{g_2,g_2'}}\;.
\end{align*}
This map can be linearly extended to states displaying mixed grading. Importantly, the image of this map typically differs from the ground state subspace $\gsTwo{\mathfrak{T}[\mathbb I] \cup_\mathbb I \mathfrak{T}[\mathbb I]}$ since all the stabiliser constraints might not be satisfied along the gluing interface. This can be resolved by applying the Hamiltonian projection operator $	\mathbb P_{\mathfrak{T}[\mathbb I] \cup_{\mathbb I} \mathfrak{T}[\mathbb I]}$ with respect to the full Hamiltonian $\mathbb H^{G,A,B}_{\alpha,\phi,\psi}[\mathfrak{T}[\mathbb I] \cup_{\mathbb I} \mathfrak{T}[\mathbb I]]$, which was defined in sec.~\ref{sec:bdryHam}. Finally, we can apply a triangulation changing isomorphism in order to obtain a  final state in $\gsTwo{\mathfrak{T}[\mathbb I]}$. Putting everything together, this defines a $\star$-product, which together with $\gsTwo{\mathfrak{T}[\mathbb I]}$ defines the tube algebra:
\begin{equation*}
	\star: \gsTwo{\mathfrak{T}[\mathbb I]} \otimes \gsTwo{\mathfrak{T}[\mathbb I]}
	\xrightarrow{\textsf{\scriptsize GLU}} 
	\mc{H}^{G,A,B}_{\alpha,\phi,\psi}[\mathfrak{T}[\mathbb I] \cup_\mathbb I \mathfrak{T}[\mathbb I]]
	\xrightarrow{\mathbb P_{\mathfrak{T}[\mathbb I] \cup_\mathbb I \mathfrak{T}[\mathbb I]}}
	\gsTwo{\mathfrak{T}[\mathbb I] \cup_\mathbb I \mathfrak{T}[\mathbb I]} \xrightarrow{\sim}
	\gsTwo{\mathfrak{T}[\mathbb I]} \; .
\end{equation*}
Given two basis states of $\gsTwo{\mathfrak{T}[\mathbb I]}$ as defined in \eqref{eq:basisTube}, let us now compute explicitly this $\star$-product. Firstly, the $G$-colourings along the gluing interface are identified via the map $\textsf{\small GLU}$, i.e. 
\begin{align*}
	\textsf{\small GLU}
	\bigg( \bigg| \! \tubeTwoD{1.2}{2} \! \bigg\ra \otimes \bigg| \! \tubeTwoD{1.2}{3} \!\! \bigg\ra \! \bigg)
	=
	\delta_{g',a^{-1}gb} 
	\bigg| \! \tubeTwoD{1.2}{4}  \!\! \bigg\ra \; .
\end{align*}
Secondly, we apply the Hamiltonian projector $\mathbb P_{\mathfrak{T}[\mathbb I] \cup_{\mathbb I} \mathfrak{T}[\mathbb I]}$ in order to enforce the gauge invariance at the physical boundary vertices that are along the gluing interface. This operator is obtained by evaluating the partition function \eqref{eq:relativepartitionfunction} on the relative pinched interval cobordism
\begin{equation}
	\tubeTwoDPinchedBis{0.65}{0}{1}{2} \; ,
\end{equation}
and its action explicitly reads
\begin{align*}
	&\mathbb P_{\mathfrak{T}[\mathbb I] \cup_{\mathbb I} \mathfrak{T}[\mathbb I]}
	\bigg( \bigg| \! \tubeTwoD{1.2}{4} \!\! \bigg\ra \bigg) 
	\\[-0.5em]
	& \q = 
	\frac{1}{|A||B|}\sum_{\substack{(\tilde a ,\tilde b) \in A \times B}}
	\frac{\vartheta^{AB}_g(a,\tilde a|b,\tilde b)}{\vartheta^{AB}_{a^{-1}gb}(\tilde a,\tilde a^{-1}a'|\tilde b,\tilde b^{-1}b')}
	\bigg| \!  \tubeTwoD{1.2}{5} \!\! \bigg\ra \; ,
\end{align*}
where we introduced the cocycle data
\begin{equation}
	\label{eq:defCoc2Gr}
	\vartheta^{AB}_{g}(a,a'|b,b') := 	\frac{\psi(b,b')}{\phi(a,a')}\frac{\alpha(a,a',a'^{-1}a^{-1}gbb')\, \alpha(g,b,b')}{\alpha(a,a^{-1}gb,b')} \; .
\end{equation}
It follows from $\alpha^{-1}{\sss |}_{A} = d^{(2)} \phi$ and $\alpha^{-1}{\sss |}_{B} = d^{(2)}\psi$, as well as the cocycle conditions
\begin{align*}
	d^{(3)}\alpha(a,a',a'',a''^{-1}a'^{-1}a^{-1}gbb'b'') &= 1
	&
	d^{(3)}\alpha(a,a^{-1}gb,b',b'') &= 1
	\\
	d^{(3)}\alpha(a,a',a'^{-1}a^{-1}gbb',b'') &= 1
	&
	d^{(3)}\alpha(g,b,b',b'') &= 1
\end{align*}
that $\vartheta^{AB}$ satisfies
\begin{align}
	\label{eq:coc2Gr}
	d^{(2)}\vartheta^{AB}_g(a,a',a''|b,b',b''):=
	\frac{	\vartheta^{AB}_{a^{-1}gb}(a',a''|b',b'') \, \vartheta^{AB}_{g}(a,a'a''|b,b'b'') }{\vartheta^{AB}_{g}(aa',a''|bb',b'') \, \vartheta^{AB}_{g}(a,a'|b,b')}
	= 1
	\; ,
\end{align}
which in particular implies the following property
\begin{equation}
\vartheta^{AB}_{a^{-1}gb}(a^{-1},a|b^{-1},b) = \vartheta^{AB}_{g}(a,a^{-1}|b,b^{-1}) \; .
\end{equation}
Furthermore, given that $\alpha$, $\phi$ and $\psi$ are normalized cocycles, we have the normalisation conditions:
\begin{equation}
	\label{eq:normCoc}
	\vartheta^{AB}_{g}(\mathbbm 1_A,a'|\mathbbm 1_B,b') = \vartheta^{AB}_g(a,\mathbbm 1_A|b , \mathbbm 1_B ) = 1 = 
	\vartheta^{AB}_{g}(\mathbbm 1_A,a'|b, \mathbbm 1_B) = \vartheta^{AB}_{g}(a, \mathbbm 1_A|\mathbbm 1_B,b') \; .
\end{equation}
Going back to the tube algebra, it remains to apply a triangulation changing isomorphism in order to recover the initial triangulation. This can be done by evaluating the partition function for the pinched interval cobordism $(012)^+ \times \mathbb I$ endowed with the triangulation depicted below:
\begin{equation}
	(012)^+ \times \mathbb I := \tubeTwoDPinched{0.65}{2}{0}{1}
	\equiv (00'1'2')^+ \cup (011'2')^- \cup (0122')^+ \; .
\end{equation}
The corresponding operator implements the isomorphism
\begin{equation*}
	\bigg| \!  \tubeTwoD{1.2}{5} \!\! \bigg\ra  \simeq\frac{1}{|A|^\frac{1}{2}|B|^\frac{1}{2}} \, 
	\vartheta^{AB}_g(a \tilde a,\tilde a^{-1}a'|b \tilde b,\tilde b^{-1}b') \, 
	\bigg| \! \tubeTwoD{1.2}{6} \! \bigg\ra \; .
\end{equation*}
Putting everything together, we obtain
\begin{equation*}
	\bigg| \! \tubeTwoD{1.2}{2} \! \bigg\ra \star \bigg| \! \tubeTwoD{1.2}{3} \!\! \bigg\ra
	=
	\frac{\delta_{g',a^{-1}gb}}{|A|^\frac{1}{2}|B|^\frac{1}{2}} \, \vartheta^{AB}_g(a,a'|b,b') \, 
	\bigg| \! \tubeTwoD{1.2}{6} \! \bigg\ra \; ,
\end{equation*}
where we used the cocycle relation $d^{(2)}\vartheta^{AB}_g(a,\tilde a,\tilde a^{-1}a'|b,\tilde b,\tilde b^{-1}b') = 1$. Using the more symbolic notation introduced in \eqref{eq:basisTube}, the $\star$-product reads
\begin{align}
	\big| \biGr{g}{a}{b} \big\ra
	\star
	\big| \biGr{g'}{a'}{b'} \big\ra
	=
	\frac{\delta_{g',a^{-1}gb}}{|A|^\frac{1}{2}|B|^\frac{1}{2}} \,
	\vartheta^{AB}_g(a,a'|b,b') \,
	\, \big| \biGr{g}{aa'}{bb'} \big\ra \; .
\end{align}

\subsection{Groupoid algebra\label{sec:groupoids}}

Before concluding this section about boundary point-like excitations in (2+1)d, we are going to show that the tube algebra derived above can be recast as a \emph{twisted groupoid algebra}\cite{willerton2008twisted}. Although this might seem a little bit artificial at the moment, this will turn out to be very useful in the subsequent sections. Indeed, we will show that in the language of groupoid algebras, both the tube algebras in (2+1)d and in (3+1)d can be unified allowing for a simultaneous study of the corresponding representation theories.

Let us first review some basic category theoretical definitions. More details can be found for example in \cite{ etingof2016tensor, mac1971category}. Given a category $\mc C$, the set of objects and the set of morphisms between objects are denoted by ${\rm Ob}({\mc C})$ and ${\rm Hom}({\mc C})$, respectively. Given two objects $X,Y \in {\rm Ob}({\mc C})$, the set of morphisms from $X$ to $Y$ is written ${\rm Hom}_{{\mc C}}(X,Y) \ni f:X \to Y$, such that $X= {\rm s}(f)$ and $Y = {\rm t}(f)$ are the \emph{source} and \emph{target} objects of $f$, respectively. Composition rule of morphisms is defined according to
\begin{equation*}
	X \xrightarrow{\;\; f \;\;}Y \xrightarrow{\;\; f' \;\;}Z = X \xrightarrow{\;\; ff' \;\;} Z \; .
\end{equation*}
Furthermore, for every object $X\in {\rm Ob}({\mc C})$, the corresponding identity morphisms is denoted by ${\rm id}_X \in {\rm Hom}_{{\mc C}}(X,X)$. Finally, we notate the set of $n$ composable morphims in $\mc{C}$ by $\mc{C}_{\rm comp}^n:=\{(f_{1},\ldots,f_{n})\in {\rm Hom}(\mc{C})^{n}\,|\, {\rm t}(f_{i})={ \rm s}(f_{i+1}), \; \forall \, i\in 1,\ldots,n-1\}$. Let us now specialize to \emph{groupoids}:

\begin{definition}[Groupoids]
	A (finite) groupoid $\mc{G}$ is a category whose object and morphism sets are finite and all morphisms are invertible, i.e. for each morphism $\fr g \in {\rm Hom}_{\mc{G}}(X,Y)$, there exists a morphism $\fr g^{-1}\in {\rm Hom}_{\mc{G}}(Y,X)$ such that $\fr g \fr g^{-1}={\rm id}_{X}$ and $\fr g^{-1}\fr g={\rm id}_{Y}$.
\end{definition}
\noindent
Every finite group provides a finite one-object groupoid refers to as the \emph{delooping} of the group:
\begin{example}[Delooping of a group]
	Let $G$ be a finite group. The delooping of $G$ is the one-object groupoid $\overline{G}$ with ${\rm Ob}(\overline{G})=\{ \bul \}$ and morphism set ${\rm Hom}_{\mc{G}}(\bul,\bul)=G$ with the composition rule being provided by the group multiplication in $G$.
\end{example}
\noindent
Henceforth, we shall identify any group $G$ and its delooping $\overline G$, denoting both by $G$. 
Generalizing the notion of group cohomology in an obvious way, we obtain the notion of groupoid cohomology:\footnote{Analogously to group cohomology, groupoid cohomology of a groupoid is implicitly defined as the simplicial cohomology of its classifying space.}
\begin{definition}[Groupoid cohomology]
	Let $\mc{G}$ be a finite groupoid and $\mc M$ a $\mc{G}$-module. Given the set of $n$ composable morphisms $\mc{G}^{n}_{\rm comp}$ in $\mc{G}$, we define an $n$-cochain on $\mc{G}$ as a map $\omega_n: \mc{G}_{\rm comp}^n \to \mc M$. On the space $C^n(\mc{G}, \mc M)$ of $n$-cochains, the coboundary operator $d^{(n)}: C^{n}(\mc{G},\mc M) \to C^{n+1}(\mc{G},\mc M)$ is defined via
	\begin{align}
		\label{eq:coBdry}
		&d^{(n)}\omega_n({\fr g}_1, \ldots, {\fr g}_{n+1}) 
		\\[-0.1em] \nn
		& \q := {\fr g}_1 \triangleright \omega_n({\fr g}_2, \ldots, {\fr g}_{n+1})\, \omega_n({\fr g}_1, \ldots {\fr g}_n)^{(-1)^{n+1}} \prod_{i=1}^n \omega_n({\fr g}_1, \ldots,{\fr g}_{i-1},{\fr g}_{i}{\fr g}_{i+1}, {\fr g}_{i+2}, \ldots, {\fr g}_{n+1})^{(-1)^i} \; .
	\end{align}
	The $n$-th cohomology group of groupoid cocycles is then defined as usual by
	\begin{equation}
	H^n(\mc{G},\mc M) :=  \frac{{\rm Ker}\, d^{(n)}}{{\rm Im}\, d^{(n-1)}} \equiv \frac{Z^n(\mc{G},\mc M)}{B^n(\mc{G},\mc M)} \; .
	\end{equation}
\end{definition}
\noindent
Throughout this manuscript, we shall always consider cohomology groups of the form $H^{n}(\mc{G},\rU(1))$, where $\rU(1)$ is taken to be the $\mc{G}$-module with the trivial groupoid action. Naturally, the cohomology of a group coincides with the groupoid cohomology of its delooping. Furthermore, we shall often require, without loss of generality, that cocycles are \emph{normalised}:
\begin{definition}[Normalised cocycles]
	Given a groupoid $n$-cocycle $[\omega_{n}]\in H^{n}(\mc{G},\rU(1))$, we call $\omega_{n}\in[\omega_{n}]$ a normalised representative if $\omega_{n}({\fr g}_{1},\ldots,{\fr g}_n)=1$, whenever any of the arguments is an identity morphism. In particular there always exists a normalised representative of each n-cocycle equivalence class $[\omega_{n}]\in H^{n}(\mc{G},\rU(1)).$
\end{definition}
\noindent
Utilising the technology of groupoid cohomology, we can now introduce \emph{twisted groupoid algebras}, generalising the theory of twisted group algebras \cite{willerton2008twisted}:
\begin{definition}[Twisted groupoid algebra]
	Given a finite groupoid $\mc{G}$ and a normalised 2-cocycle $\vartheta\in Z^{2}(\mc{G},\rU(1))$, the twisted groupoid algebra $\mathbb{C}[\mc{G}]^{\vartheta}$ is the algebra defined over the vector space
	\begin{align}
		{\rm Span}_{\mathbb{C}}\{ \ket{\fr g}\,|\, \forall \,  \fr g\in {\rm Hom}(\mc{G}) \}
	\end{align}
	with algebra product
	\begin{align}
		\ket{\fr g}\star \ket{\fr g'} := \delta_{{\rm t}(\fr g),{\rm s}(\fr g')} \, \vartheta(\fr g,\fr g') \, \ket{\fr g \fr g'} \; .
	\end{align}
	The requirement that $\vartheta$ is a 2-cocycle ensures that $\mathbb C[\mathcal G]^\vartheta$ is an associative algebra.
\end{definition}
\noindent
Putting everything together, let us now recast the (2+1)d tube algebra as a twisted groupoid algebra. Let $G_{AB}$ be the (finite) groupoid whose objects are given by group elements in $G$, and whose morphisms read $\biGr{g}{a}{b}a^{-1}gb \equiv \biGr{g}{a}{b}\,$, where $(a,b)\in A \times B$ with the composition given by the multiplication in $G$:
\begin{align}
	\biGr{g}{a}{b} \biGr{a^{-1}gb}{a'}{b'}a'^{-1}a^{-1}gbb' 
	=
	\biGr{g}{aa'}{bb'}a'^{-1}a^{-1}gbb'\; .
\end{align}
Utilising this definition, we can conveniently redefine $\vartheta^{AB}$ as a normalised groupoid 2-cocycle in $H^2(G_{AB},{\rm U}(1))$, in such a way that the tube algebra defined earlier  is isomorphic to the groupoid algebra $\mathbb C[G_{AB}]^{\vartheta^{AB}} \equiv \mathbb C[G_{AB}]^\alpha_{{\phi \psi}}$ of $G_{AB}$ twisted by $\vartheta^{AB}$.\footnote{Notice that the normalization conditions \eqref{eq:normCoc} do not state that the cocycle is equal to one whenever any of the entry is one, but instead whenever any of the morphism in the corresponding groupoid is the identity. It is therefore compatible with the definition given earlier.}

\section{Tube algebra for gapped boundary excitations in (3+1)d\label{sec:tube3D}}
\emph{In this section, we apply the tube algebra approach to study excitations in the presence of gapped boundaries in (3+1)d. Although the excitation content of the model is rich in (3+1)d, we focus on a special configuration, which turns out to be related to that considered in the previous section via a dimensional reduction argument.}

\subsection{Definition}
The strategy we presented in sec.~\ref{sec:tube2D} applies identically in three dimensions. Given a pattern of two-dimensional gapped boundaries, excitations can be classified by considering boundary conditions of the manifold obtained by removing local neighbourhoods of these excitations. Given that the input data for the bulk theory is a pair $(G,\pi)$, where is $\pi$ a normalized representative of a cohomology class in $H^4(G,\rU(1))$, we are interested in the situation where two two-dimensional gapped boundaries characterized by the boundary conditions $A_\lambda \equiv (A, \lambda)$ and $B_\mu \equiv (B,\mu)$ meet at a one-dimensional interface. The boundary conditions are such that $A,B \subset G$, $d^{(3)}\lambda = \pi^{-1}{\sss |}_{A}$ and $d^{(3)}\mu = \pi^{-1}{\sss |}_{B}$. We denote the Hamiltonian defined according to \eqref{eq:Ham} for these boundary conditions as $\mathbb H^{G,A, B}_{\pi, \lambda,\mu}[\Sigma]$.

Given this situation, several types of excitations could be studied. For instance, we could investigate point-like boundary excitations at the one-dimensional interface. Instead, we consider a bulk string-like excitation that terminates at two (possibly different) gapped boundaries. This situation can be depicted as follows:
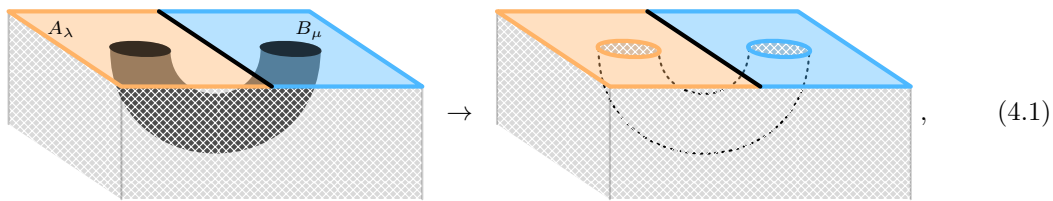
\begin{equation}
	\bdryInterfaceThreeExc{1}{1} \;\; \to \;\;
	\bdryInterfaceThreeExc{1}{2} \; ,
\end{equation}
where the dark volume represents a local neighbourhood of the string-like excitation, and thus the region whose energy density is higher than that of the ground state. Removing this local neighbourhood leaves an excitation boundary $\exBdry$ that has the topology of cylinder. Classifiying boundary conditions on such cylinder corresponds to classifying the string-like excitations.

Let us consider the manifold $\exBdry \times \mathbb I$. This manifold has the topology of a hollow cylinder, which has two kinds of boundary components, namely a pair of physical boundary components and a pair of excitation boundary components. Given the 3-ball endowed with two gapped boundaries, the same manifold can be obtained by removing local neighbourhoods of the interface and of a string terminating at the two gapped boundaries:
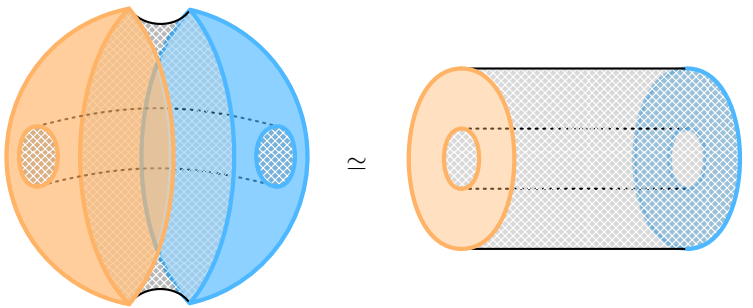
\begin{equation*}
	\threeDiskExc{1}{2} \;\; \simeq \;\; \tubeThreeDBeauty{1}{1} \! .
\end{equation*}
By construction, this manifold can be glued to the original system along the excitation boundary $\exBdry$ without affecting its topology. It follows from the discussion in sec.~\ref{sec:tube2D} that there is a tube algebra associated with the gluing of two copies of this tube-like manifold, whose irreducible representations classify this special type of string-like excitations. 

\subsection{Computation of the tube algebra}

Let us derive the tube algebra for the special configuration described above. As before, we first need to specify  the ground state subspace on $\exBdry \times \mathbb I$ by choosing a discretisation. We choose to discretise $\exBdry \times \mathbb I$ as a triangulated cube with two opposite faces identified. The resulting triangulated manifold is denoted by $\mathfrak{T}[\mathbb S^1 \times \mathbb I]$ and the corresponding ground state subspace explicitly reads 
\begin{align}
	\label{eq:basisTube3D}
	\gsThree{\mathfrak{T}[\mathbb S^1 \times \mathbb I]}
	& :=  {\rm Span}_{\mathbb C}\Bigg\{\Bigg| \! \tubeThreeD{0.9}{0}{1}{g}{a}{b}{1} \Bigg\ra  \Bigg\}_{\substack{\hspace{0em}\forall \, g \in G \, | \, g= a_2^{-1}gb_2 \\\hspace{-2.6em} \forall \, a_1,a_2 \in A \\ \hspace{-2.7em} \forall \, b_1,b_2 \in B }}
	\\
	&\equiv {\rm Span}_{\mathbb C}\big\{ \biGr{\big| (g,a_2,b_2)}{a_1}{b_1} \big\ra \big\}_{\substack{\hspace{0em}\forall \, g \in G \, | \, g= a_2^{-1}gb_2 \\\hspace{-2.6em} \forall \, a_1,a_2 \in A \\ \hspace{-2.7em} \forall \, b_1,b_2 \in B }} \; ,
\end{align}
where we make the identifications $\snum{(0)} \equiv \snum{(\tilde 0)}$, $\snum{(0')} \equiv \snum{(\tilde 0')}$, $\snum{(1)} \equiv \snum{(\tilde 1)}$, $\snum{(1')} \equiv \snum{(\tilde 1')}$, $\snum{(00')} \equiv \snum{(\tilde 0 \tilde 0')}$, $\snum{(01)} \equiv \snum{(\tilde 0 \tilde 1)}$, $\snum{(0'1')} \equiv \snum{(\tilde 0' \tilde 1')}$ and $\snum{(11')} \equiv \snum{(\tilde 1 \tilde 1')}$. As before, some labellings are left implicit since they can be deduced from the flatness constraints. Let us now compute the $\star$-product for two such states adapting in the obvious way the definition of the previous section. Firstly, colourings along the gluing interface are identified via the map $\textsf{\small GLU}$, i.e.
\begin{align*}
	&\textsf{\small GLU}
	\Bigg( \Bigg| \! \tubeThreeD{0.9}{0}{1}{g}{a}{b}{1} \! \Bigg\ra 
	\otimes 
	\Bigg| \! \tubeThreeD{0.9}{1}{2}{g'}{a'}{b'}{1} \!\! \Bigg\ra \! \Bigg)
	\\[-2em]
	& \hspace{11em} =
	\delta_{g',a_1^{-1}gb_1}
	\, \delta_{a_2',a_2^{a_1}} \, \delta_{b_2',b_2^{b_1}}
	 \Bigg| \! \tubeThreeDDouble{0.9}{0}{1}{2}{g}{a}{b}{1} \! \Bigg\ra \; ,
\end{align*}
where we introduced the notation $x^y := y^{-1}xy$.
Secondly, we apply the Hamiltonian projector $\mathbb P_{\mathfrak{T}[\mathbb S^1 \times \mathbb I] \cup_{\mathbb S^1 \times \mathbb I}\mathfrak{T}[\mathbb S^1 \times \mathbb I]}$ in order to enforce the twisted gauge invariance at the physical boundary vertices along the gluing interface. This operator can be expressed by evaluating the partition function \eqref{eq:relativepartitionfunction} on the relevant pinched cobordism. The result reads
\begin{align}
	&\mathbb P_{\mathfrak{T}[\mathbb S^1 \times \mathbb I] \cup_{\mathbb S^2 \times \mathbb I}\mathfrak{T}[\mathbb S^1 \times \mathbb I]}
	\Bigg( \Bigg| \! \tubeThreeDDouble{0.9}{0}{1}{2}{g}{a}{b}{1} \! \Bigg\ra \Bigg)
	\\[-2em]
	\nn
	& \q = 
	\frac{1}{|A||B|}\sum_{\substack{(\tilde a ,\tilde b) \in A \times B}}
	\frac{\varrho^{AB}_{g,a_2,b_2}(a_1,\tilde a|b_1,\tilde b)}{\varrho^{AB}_{a_1^{-1}gb_1,a_2^{a_1},b_2^{b_1}}(\tilde a,\tilde a^{-1}a_1'|\tilde b,\tilde b^{-1}b_1')} \,
	\Bigg| \!  \tubeThreeDDouble{0.9}{0}{1}{2}{g}{a}{b}{2} \! \Bigg\ra \; ,
\end{align}
where we introduced the cocycle data
\begin{equation}
	\label{eq:defCoc2Gr3D}
	\varrho^{AB}_{g,a_2,b_2}(a_1,a_1'|b_1,b_1') := 	\frac{{\sfT}_{b_2}(\mu)(b_1,b_1')}{{\sfT}_{a_2}(\lambda)(a_1,a_1')}\frac{{\sfT}_{a_2}(\pi)(a_1,a_1',a_1'^{-1}a_1^{-1}gb_1b_1') \, {\sfT}_{a_2}(\pi)(g,b_1,b_1')}{{\sfT}_{a_2}(\pi)(a_1,a_1^{-1}gb_1,b_1')} 
\end{equation}
in terms of the cocycle data ${\sfT}(\lambda)$, ${\sfT}(\mu)$ and ${\sfT}(\pi)$ that are itself defined according to
\begin{align*}
	{\sfT}_{x}(\alpha)(y_1,y_2) &:= \frac{\alpha(x,y_1,y_2) \, \alpha(y_1,y_2,x^{y_1y_2})}{\alpha(y_1,x^{y_1},y_2)} \; ,
	\\
	{\sfT}_x(\pi)(y_1,y_2,y_3) &:=
	\frac{\pi(y_1,x^{y_1},y_2,y_3) \, \pi(y_1,y_2,y_3,x^{y_1y_2y_3})}{\pi(x,y_1,y_2,y_3) \, \pi(y_1,y_2,x^{y_1y_2},y_3)} \; ,
\end{align*}
for any group elements $x,y_1,y_2,y_3 \in H$ in a finite group $H$ and group cochains $\alpha \in C^3(H, \rU(1))$, $\pi \in C^4(H, \rU(1))$. Defining
\begin{align}
	d^{(2)}{\sfT}_{x}(\alpha)(y_1,y_2,y_3)&:= 
	\frac{{\sfT}_{x^{y_1}}(\alpha)(y_2,y_3) \, {\sfT}_x(\alpha)(y_1,y_2y_3)}{{\sfT}_x(\alpha)(y_1y_2,y_3) \, {\sfT}_x(\alpha)(y_1,y_2)} \; ,
	\\
	d^{(3)}{\sfT}_{x}(\pi)(y_1,y_2,y_3,y_4)&:= 
	\frac{{\sfT}_{x^{y_1}}(\pi)(y_2,y_3,y_4) \, {\sfT}_x(\alpha)(y_1,y_2y_3,y_4) \, {\sfT}_x(\pi)(y_1,y_2,y_3)}{{\sfT}_x(\pi)(y_1y_2,y_3,y_4) \, {\sfT}_x(\alpha)(y_1,y_2,y_3y_4)} \; ,
\end{align}
it follows from the cocycle conditions $d^{(4)}\pi = 1$, $d^{(3)}\lambda = \pi^{-1}{\sss |}_{A}$ and $d^{(3)}\mu = \pi^{-1}{\sss |}_{B}$ that $d^{(3)}{\sfT}(\pi) = 1$, $d^{(2)} {\sfT}(\lambda) = {\sfT}(\pi)^{-1}{\sss |}_{A} $ and $ d^{(2)}{\sfT}(\mu) =  {\sfT}(\pi)^{-1}{\sss |}_{B}$. Utilising the cocycle conditions
\begin{align*}
	d^{(3)}{\sfT}_{a_2}(\pi)(a_1,a_1',a_1'',a_1''^{-1}a_1'^{-1}a_1^{-1}gb_1b_1'b_1'') &= 1
	&
	d^{(3)}{\sfT}_{a_2}(\pi)(a_1,a_1^{-1}gb_1,b_1',b_1'') &= 1
	\\
	d^{(3)}{\sfT}_{a_2}(\pi)(a_1,a_1',a_1'^{-1}a_1^{-1}gb_1b_1',b_1'') &= 1
	&
	d^{(3)}{\sfT}_{a_2}(\pi)(g,b_1,b_1',b_1'') &= 1 \; ,
\end{align*}
we finally obtain that  $\varrho^{AB}$ satisfies
\begin{align}
	d^{(2)}\varrho^{AB}_{g,a_2,b_2}(a_1,a_1',a_1''|b_1,b_1',b_1''):=
	\frac{	\varrho^{AB}_{a_1^{-1}gb_1,a_2^{a_1},b_2^{b_1}}(a_1',a_1''|b_1',b_1'') \, \varrho^{AB}_{g,a_2,b_2}(a_1,a_1'a_1''|b_1,b_1'b_1'') }{\varrho^{AB}_{g,a_2,b_2}(a_1a_1',a_1''|b_1b_1',b_1'') \, \varrho^{AB}_{g,a_2,b_2}(a_1,a_1'|b_1,b_1')}
	= 1
	\; .
\end{align}
Going back to the tube algebra, it remains to apply a triangulation changing isomorphism in order to recover the initial triangulation, and thus a state in $\gsThree{\mathfrak{T}[\mathbb S^1 \times \mathbb I]}$. This is done by evaluating the partition function for the pinched interval cobordism $(012)^+ \times \mathbb S^1 \times \mathbb I$ endowed with the triangulation defined as
\begin{alignat}{5}
	&\snum{(012)}^{+} \times \mathbb S^1 \times \mathbb I
	& \,=  \,& \,
	\snum{(0122'\tilde{2}')}^{+} &\, \cup \,& \snum{(012\tilde{2}\tilde{2}')}^{-} &\, \cup \,& \snum{(01\tilde{1}\tilde{2}\tilde{2}')}^{+} &\, \cup \,& \snum{(0\tilde{0}\tilde{1}\tilde{2}\tilde{2}')}^{-}
	\nn\\
	&{} &\, \cup \,& \,
	\snum{(011'2'\tilde{2}')}^{-} &\, \cup \,& \snum{(011'\tilde{1}'\tilde{2}')}^{+}&\, \cup \,& \snum{(01\tilde{1}\tilde{1}'\tilde{2}')}^{-}&\, \cup \,& \snum{(0\tilde{0}\tilde{1}\tilde{1}'\tilde{2}')}^{+}
	\nn\\
	& {} &\, \cup \,& \,
	\snum{(00'1'2'\tilde{2}')}^{+}&\, \cup \,& \snum{(00'1'\tilde{1}'\tilde{2}')}^{-}&\, \cup \,& \snum{(00'\tilde{0}'\tilde{1}'\tilde{2}')}^{+}&\, \cup \,& \snum{(0\tilde{0}\tilde{0}'\tilde{1}'\tilde{2}')}^{-} \; .
\end{alignat}
The corresponding operator implements the isomorphism
\begin{equation*}
	\Bigg| \!  \tubeThreeDDouble{0.9}{0}{1}{2}{g}{a}{b}{2} \! \Bigg\ra  \simeq
	\frac{1}{|A|^\frac{1}{2}|B|^\frac{1}{2}} \varrho^{AB}_{g,a_2,b_2}(a_1 \tilde a , \tilde a^{-1}a_1' | b_1 \tilde b , \tilde b^{-1} b_1')
	\Bigg| \! \tubeThreeD{0.9}{0}{2}{g}{a}{b}{2} \! \Bigg\ra . 
\end{equation*}
Putting everything together, we obtain
\begin{align}
	&\Bigg| \! \tubeThreeD{0.9}{0}{1}{g}{a}{b}{1} \! \Bigg\ra 
	\star
	\Bigg| \! \tubeThreeD{0.9}{1}{2}{g'}{a'}{b'}{1} \!\! \Bigg\ra 
	\\[-2em]
	\nn
	& \q =
	\frac{\delta_{g',a_1^{-1}gb_1}
	\, \delta_{a_2',a_2^{a_1}} \, \delta_{b_2',b_2^{b_1}}}{|A|^\frac{1}{2}|B|^\frac{1}{2}} \,
	\varrho^{AB}_{g,a_2,b_2}(a_1,a_1'|b_1,b_1') \,
	\Bigg| \! \tubeThreeD{0.9}{0}{2}{g}{a}{b}{2} \! \Bigg\ra ,
\end{align}
where we used the cocycle relation $d^{(2)}\varrho^{AB}_{g,a_2,b_2}(a_1,\tilde a, \tilde a^{-1}a_1'|b_1, \tilde b, \tilde b^{-1}b_1')$.
Using the more symbolic notation introduced in \eqref{eq:basisTube3D}, we obtained
\begin{align*}
	\big| \biGr{(g,a_2,b_2)}{a_1}{b_1} \big\ra
	\star
	\big| \biGr{(g',a_2',b_2')}{a_1'}{b_1'} \big\ra
	=
	\frac{\delta_{g',a_1^{-1}gb_1}
	\, \delta_{a_2',a_2^{a_1}} \, \delta_{b_2',b_2^{b_1}}}{|A|^\frac{1}{2}|B|^\frac{1}{2}} \,
	\varrho^{AB}_{g,a_2,b_2}(a_1,a_1'|b_1,b_1') \,
	\, \big| \biGr{(g,a_2,b_2)}{a_1a_1'}{b_1b_1'} \big\ra \; .
\end{align*}

\subsection{Relative groupoid algebra\label{sec:relatGr}}

Similarly to its (2+1)d analogue, the tube algebra found above can be recast as a twisted groupoid algebra.  Interestingly, due to the topology of the problem, we shall notice how in this language the (3+1)d tube algebra can be recast in terms of the (2+1)d one, unifying both computations. This is reminiscent of the notion of lifted models and lifted tube algebras developed in \cite{Bullivant:2019fmk} in the context of bulk excitations.

An important ingredient of our construction is the notion of \emph{loop groupoid}:
\begin{definition}[Loop groupoid]
	Given a finite groupoid $\mc{G}$, the loop groupoid $\Lambda \mc{G}$ is the groupoid with object set $\{\fr g \in {\rm End}_{\mc{G}}(X) \, | \, \forall \, X \in {\rm Ob}(\mc{G})\}$ and morphisms of the form $\fr h : \fr g \to \fr h^{-1}\fr g \fr h$, for every $\fr g \in {\rm End}_{\mc{G}}(X)$ and $\fr h \in {\rm Hom}_{\mc{G}}(X,Y)$. Composition in $\Lambda \mc{G}$ is inherited from the one in $\mc{G}$.
\end{definition}
\noindent
Specialising to the case where the finite groupoid is taken to be the delooping of a finite group $G$, we obtain that $\Lambda  G$ is the groupoid with object set ${\rm Ob}(\Lambda  G)=G$ and morphism set ${\rm Hom}(\Lambda  G)=\{ g\xrightarrow{a}a^{-1}ga \, | \, \forall \,  g,a\in G\}$. Composition is given by multiplication in $G$ such that
\begin{equation*}
	\biGr{g}{a}{} \biGr{a^{-1}ga}{a'}{} (aa')^{-1}gaa'=g\xrightarrow{aa'} (aa')^{-1}gaa' \; ,
\end{equation*}  
for all $g,a,a'\in G$. Using this terminology, we can check that the cocycle data ${\sfT}(\pi)$, ${\sfT}(\lambda)$ and ${\sfT}(\mu)$ defined in \eqref{eq:defCoc2Gr3D} actually correspond to loop groupoid cocycles in $Z^3(\Lambda  G, \rU(1))$, $Z^2(\Lambda  A, \rU(1))$ and $Z^2(\Lambda  B, \rU(1))$, respectively. More generally, for any group $G$, we have a map ${\sfT}: Z^{\bullet}(G,\rU(1)) \to Z^{\bullet -1}(\Lambda  G, \rU(1))$ referred to as the $\mathbb S^1$-transgression map. More details regarding this map can be found in \cite{willerton2008twisted,bartlett2009unitary, Bullivant:2019fmk}. We further require the notion of \emph{relative groupoid}:

\begin{definition}[Relative groupoid]\label{def:relativegroupoid}
	Given a groupoid $\mc{G}$, and a pair of subgroupoids $\mc{A}, \mc{B} \subseteq \mc{G}$, the \emph{relative groupoid} $\mc{G}_{\mc{A} \mc{B}}$ is the groupoid with object set ${\rm Ob}(\mc{G}_{\mc{A} \mc{B}}) := \{ \fr g \in {\rm Hom}(\mc{G})\,|\, {\rm s}(\fr g) \in {\rm Ob}(\mc{A}) ,\, {\rm t}(\fr g )\in{\rm Ob}(\mc{B})\}$ and morphism set provided by
	\begin{align}
		\label{eq:relativeGrNot}
		\biGr{\fr g}{\fr a}{\fr  b}\fr a^{-1}\fr g\fr b \equiv \biGr{\fr g}{\fr a}{\fr b} \;\; ,
	\end{align}
	for all $\fr g\in {\rm Ob}(\mc{G}_{\mc{A}\mc{B}})$, $\fr a\in {\rm Hom}_{\mc A}({\rm s}(\fr g),-)$ and $\fr b\in {\rm Hom}_{\mc B}({\rm t}(\fr g),-)$. Composition is defined by 
	\begin{align}
		\biGr{\fr g}{\fr a}{\fr b} \biGr{\fr a^{-1} \fr g \fr b}{\fr a'}{\fr b'} \fr a^{'-1} \fr a^{-1}\fr g \fr b \fr  b'
		=
		\biGr{\fr g}{\fr a \fr a'}{\fr b \fr b'} \fr a^{'-1}\fr a^{-1}\fr g \fr b \fr b' \; ,
	\end{align}
	for all composable pairs $(\fr a, \fr a')\in {\mc A}^{2}_{\rm comp}$ and $(\fr b,\fr b')\in {\mc B}^{2}_{\rm comp}$.
\end{definition}
\noindent
It follows immediately from the definition above that the groupoid $G_{AB}$, whose twisted groupoid algebra is isomorphic to the (2+1)d tube algebra, actually corresponds to the relative groupoid defined for the delooping of the groups. We are almost ready to define the (3+1)d tube algebra in this language. The last item we require is a notion of normalised cocycle for relative groupoid. 
To this end we introduce $(\mc{G},\alpha)$-subgroupoids:
\begin{definition}\label{def:twistedsubgroupoid}
	Given a finite groupoid $\mc{G}$ and a normalised 3-cocycle $\alpha\in Z^{3}(\mc{G},\rU(1))$, we call a pair $(\mc A,\phi)$ a $(\mc{G},\alpha)$-subgroupoid when $\mc{A}\subseteq \mc{G}$ is a subgroupoid of $\mc{G}$ and $\phi\in C^{2}(\mc A,\rU(1))$ is a 2-cochain satisfying the condition
	$d^{(2)}\phi(\fr a, \fr a', \fr a'')=\alpha^{-1}(\fr a, \fr a' , \fr a''){\sss |}_{\mc A}$
	for all composable $(\fr a,\fr a',\fr a'')\in \mc{A}^{3}_{\rm comp}$.
\end{definition}
\noindent
For any pair of $(\mc{G},\alpha)$-subgroupoids $(\mc A,\phi)$ and $(\mc B,\psi)$, we construct a normalised 2-cocycle $\vartheta^{\mc A \mc B}\in Z^{2}(\mc{G}_{\mc A \mc B},\rU(1))$ for the relative groupoid $\mc{G}_{\mc A \mc B}$ via:
\begin{align}
	\label{eq:defvartheta}
	\vartheta^{\mc A \mc B}(
	\biGr{\fr g}{\fr a}{\fr b} \, ,
	\biGr{\fr a^{-1}\fr g \fr b}{\fr a'}{\fr b'} \, )
	&:=
	\frac{\psi(\fr b, \fr b')}{\phi(\fr a,\fr a')}
	\frac{\alpha(\fr a,\fr a',\fr a'^{-1}\fr a^{-1}\fr g \fr b\fr b')\, \alpha(\fr g,\fr b,\fr b')}{\alpha(\fr a,\fr a^{-1}\fr g \fr b,\fr b')}
	\\
	&\equiv
	\vartheta^{\mc A \mc B}_{\fr g}(\fr a,\fr a'|\fr b,\fr b')
\end{align}
for all composable morphisms 
\begin{equation}
	\biGr{\fr g}{\fr a}{\fr b}\, , \, \biGr{\fr a^{-1}\fr g \fr b}{\fr a'}{\fr b'} \;\, \in \mc{G}_{\mc A \mc B} \; ,
\end{equation}
where we are using the shorthand notation introduced in \eqref{eq:relativeGrNot}.
It follows from $\alpha^{-1}{\sss |}_{\mc A} = d^{(2)} \psi$ and $\alpha^{-1}{\sss |}_{\mc B} = d^{(2)}\phi$, as well as the cocycle conditions
\begin{align*}
	d^{(3)}\alpha(\fr a,\fr a',\fr a'',\fr a''^{-1}\fr a'^{-1}\fr a^{-1}\fr g \fr b \fr b' \fr b'') &= 1
	&
	d^{(3)}\alpha(\fr a,\fr a^{-1}\fr g \fr b,\fr b',\fr b'') &= 1
	\\
	d^{(3)}\alpha(\fr a,\fr a',\fr a'^{-1}\fr a^{-1}\fr g \fr b \fr b',\fr b'') &= 1
	&
	d^{(3)}\alpha(\fr g,\fr b,\fr b',\fr b'') &= 1
\end{align*}
that $\vartheta^{\mc A \mc B}$ satisfies the 2-cocycle relation
\begin{align}
	d^{(2)}\vartheta^{\mc A \mc B}_{\fr g}(\fr a,\fr a',\fr a''|\fr b,\fr b',\fr b''):=
	\frac{	\vartheta^{\mc A\mc B}_{\fr a^{-1}\fr g\fr b}(\fr a',\fr a''|\fr b',\fr b'') \, \vartheta^{\mc A \mc B}_{\fr g}(\fr a,\fr a'\fr a''|\fr b,\fr b'\fr b'') }{\vartheta^{\mc A \mc B}_{\fr g}(\fr a \fr a',\fr a''|\fr b \fr b',\fr b'') \, \vartheta^{\mc A \mc B}_{\fr g}(\fr a ,\fr a'|\fr b,\fr b')}
	= 1
	\; .
\end{align}
Unsurprisingly, this equation mimics \eqref{eq:coc2Gr}.
Furthermore, given that $\alpha$ is a normalized cocycle, we have the normalisation conditions:
\begin{align*}
	\vartheta^{\mc A \mc B}_{\fr g}({\rm id}_{{\rm s}(\fr a')},\fr a'|{\rm id}_{{\rm s}(\fr b')},\fr b') = \vartheta^{\mc A \mc B}_{\fr g}(\fr a,{\rm id}_{{\rm t}(\fr a)}|\fr b , {\rm id}_{{\rm t}( \fr b)} ) & = 1 
	\\
	\vartheta^{\mc A \mc B}_{\fr g}({\rm id}_{{\rm s}(\fr a')},\fr a'|\fr b, {\rm id}_{{\rm t}(\fr b)}) = \vartheta^{\mc A \mc B}_{\fr g}(\fr a, {\rm id}_{{\rm t}(\fr a)}|{\rm id}_{{\rm s}(\fr b')},\fr b') & =1
	\; ,
\end{align*}
which further imply
\begin{equation}
	\vartheta^{\mc A \mc B}_{\fr a^{-1}\fr g \fr b}(\fr a^{-1},\fr a|\fr b^{-1},\fr b) = \vartheta^{\mc A \mc B}_{\fr g}(\fr a,\fr a^{-1}|\fr b,\fr b^{-1}) \; .
\end{equation}

\bigskip \noindent
Let $G$ be a finite group and $\pi \in Z^4(G,\rU(1))$. We consider two subgroups $A,B \subset G$ and $\lambda \in C^3(A,\rU(1))$, $\mu \in C^3(B,\rU(1))$ such that $d^{(3)}\lambda = \pi^{-1}{\sss |}_A$ and $d^{(3)}\mu = \pi^{-1}{\sss |}_B$. It follows from the computations in sec.~\ref{sec:tube3D} that $(\Lambda A, \sfT(\lambda))$ and $(\Lambda B, \sfT(\mu))$ are $(\Lambda G, \sfT(\pi))$-subgroupoids. We define $\vartheta^{\Lambda A \Lambda B}$ by applying the formula \eqref{eq:defvartheta} for $\alpha \equiv \sfT(\pi)$, $\phi \equiv \sfT(\lambda)$ and $\psi \equiv \sfT(\mu)$. Putting everything together, we obtain the \emph{twisted relative groupoid algebra} $\mathbb C[\Lambda  G_{\Lambda  A \Lambda  B}]^{\vartheta^{\Lambda  A \Lambda  B}}$. We can show that this twisted relative groupoid algebra is isomorphic to the (3+1)d tube algebra by identifying 
\begin{align}
	\label{eq:identGrTube}
	\biGr{(g,a_2,b_2)}{a_1}{b_1} \;\,
	 \equiv \biGr{\fr g}{\fr a_1}{\fr b_1} \; ,
\end{align}
such that $a_{2}\xrightarrow{g}b_{2} \equiv \fr g \in {\rm Ob}(\Lambda  G_{\Lambda A \Lambda B})$, $a_{2}\xrightarrow{a_{1}}a_{2}^{a_{1}} \equiv \fr a_1\in {\rm Hom}_{\Lambda  A}({\rm s}(\fr g),-)$ and $b_{2}\xrightarrow{b_{1}}b_{2}^{b_{1}} \equiv \fr b_1 \in {\rm Hom}_{\Lambda  B}({\rm t}(\fr g),-)$, as well as 
$\vartheta^{\Lambda  A \Lambda  B} \equiv \varrho^{AB}$, which was defined in \eqref{eq:defCoc2Gr3D}. 

Thereafter, we make use of the shorthand notations $\Lambda(G_{AB}) \equiv \Lambda G_{\Lambda A \Lambda B}$ and $\mathbb C[\Lambda(G_{AB})]^{\vartheta ^{\Lambda(AB)}} \equiv \grAlgL \equiv  \mathbb C[\Lambda  G_{\Lambda  A \Lambda  B}]^{\vartheta^{\Lambda  A \Lambda  B}}  $ to refer to this relative groupoid algebra. We purposefully choose a notation very similar to describe the (2+1)d and (3+1)d tube algebras in order to emphasize the fact that the framework presented in this section unifies both. As a matter of fact, we can obtain the (2+1)d algebra from  the (3+1)d one by restricting the loop groupoid $\Lambda  G$ to morphisms whose source and target objects are the identity in $G$ and by replacing the loop groupoid 3-cocycle $\alpha \equiv \sfT(\pi) \in Z^3(\Lambda  G, \rU(1))$, where $\pi \in Z^4(G,\rU(1))$, by a group 3-cocycle $\alpha \in Z^3(G,\rU(1))$. In virtue of this last remark, we may now focus on the algebra relevant to the (3+1)d scenario, namely $\grAlgL$, and deduce the results for the (2+1)d gapped boundary excitations as a limiting case. 

We conclude this section with a remark regarding the notation. Since the morphisms $\fr a_1 \in {\rm Hom}_{\Lambda  A}({\rm s}(\fr g),-)$ and $\fr b_1 \in {\rm Hom}_{\Lambda  B}({\rm t}(\fr g),-)$ in \eqref{eq:identGrTube} are specified by a choice of group variables in the finite groups $A$ and $B$, respectively, we shall often loosely identify both in the following for notational convenience.

\newpage
\section{Representation theory and elementary gapped boundary excitations\label{sec:rep}}

\emph{In this section, we derive the irreducible representations of the algebra $\grAlgL$, and elucidate their physical interpretation as a classifier for the elementary string-like excitations in (3+1)d. As mentioned earlier, due to the topology of the problem, and the common description as relative groupoid algebras, this study can be straightforwardly applied to describe elementary boundary excitations in (2+1)d.}

\subsection{Simple modules}

Given a finite group $G$, two subgroups $A,B \subset G$ and cocycle data $\pi \in Z^4(G, \rU(1))$, $\lambda \in C^3(A, \rU(1))$, $\mu \in C^3(B, \rU(1))$ satisfying $d^{(4)}\pi = 1$, $d^{(3)}\lambda = \pi^{-1}{\sss |}_A$, $d^{(3)}\mu = \pi^{-1}{\sss |}_B$, respectively, we define $\alpha \equiv \sfT(\pi) \in Z^3(\Lambda  G, \rU(1))$, $\phi \equiv \sfT(\lambda) \in C^2(\Lambda A, \rU(1))$ and $\psi \equiv \sfT(\mu) \in C^2(\Lambda B, \rU(1))$.
We explained above that the simple modules of the groupoid algebra $\grAlgL \equiv \mathbb C[\Lambda(G_{AB})]^{\vartheta^{\Lambda (AB)}}$ classify elementary string-like excitations terminating at gapped boundaries. Let us now derive these simple modules.  We shall find that they are labelled by a pair $(\mathcal{O},R)$, where $\mathcal{O}$ is an equivalence class of boundary colourings with respect to the action of the tube algebra, and $R$ is a \emph{projective} group representation that decomposes the symmetry action of the tube algebra on a given boundary colouring.

\bigskip \noindent
We begin by first decomposing the algebra $\grAlgL$ into a direct sum of subalgebras. To this end, we notice that the tube algebra defines an action on the set of boundary colourings yielding an equivalence relation on ${\rm Ob}(\Lambda(G_{AB}))$ given by 
\begin{equation*}
	\fr g \sim \fr g' \; , \q  {\rm if } \; \exists \; \biGr{\fr g}{\fr a}{\fr b} \;\,  \in {\rm Hom}(\Lambda(G_{AB})) \;\; \text{such that} \;\;  \fr g' = {\rm t}\big(\biGr{\fr g}{\fr a}{\fr b}\big) \; .
\end{equation*}
The subsets of ${\rm Ob}(\Lambda(G_{AB}))$, i.e. boundary colourings of the tube, that are in the same equivalence class form a partition of ${\rm Ob}(\Lambda(G_{AB}))$ into disjoint sets. Let us denote by $\mathcal{O}_{AB},\mathcal{O}_{AB}'\subseteq {\rm Ob}(\Lambda(G_{AB}))$ two such equivalence classes. Considering two basis elements of the form
\begin{equation}
	\big| \biGr{\fr g}{\fr a}{\fr b} \big\ra \, , \, \big| \biGr{\fr g'}{\fr a'}{\fr b'} \big\ra
\end{equation}
such that $\fr g \in \mathcal{O}_{AB}$ and $\fr g' \in \mathcal{O}_{AB}'$, it follows from the definition of the algebra that the product of these two states necessarily vanishes. Consequently, each equivalence class of ${\rm Ob}(\Lambda(G_{AB}))$ defines a subalgebra $(\grAlgL)_{\mathcal{O}_{AB}} \subset \grAlgL$ whose defining vector space is
\begin{equation}
	{\rm Span}_{\mathbb C}\big\{\big|\biGr{\fr g}{\fr a}{\fr b} \big\ra \big\}_{\substack{
	   \forall \, \fr g\xrightarrow[\fr b]{\fr a} \in {\rm Hom}(\Lambda(G_{AB}))
	\\ \hspace{-3.6em} {\rm s.t. }\; \fr g\in \mathcal{O}_{AB}
	}} \; .
\end{equation}
Since orbits $\mathcal{O}_{AB}$ form a partition of ${\rm Ob}(\Lambda(G_{AB}))$, we have the following decomposition
\begin{equation}
	\grAlgL = \bigoplus_{\mathcal{O}_{AB} \subset G} (\grAlgL)_{\mathcal{O}_{AB}} \; .
\end{equation}
Given an equivalence class $\mathcal{O}_{AB}$, we notate its elements by $\{\mathfrak{o}_i\}_{i = 1, \ldots, |\mathcal{O}_{AB}|}$ and call $\mathfrak{o}_1$ the \emph{representative} element of $\mathcal{O}_{AB}$. We further consider the set $\{\fr p_i,\fr q_i\}_{i=1, \ldots, |\mathcal{O}_{AB}|}\subseteq {\rm Hom}(\Lambda A)\times {\rm Hom}(\Lambda B)$ defined by a choice of morphism
\begin{equation*}
	\mathfrak{o}_{i}\xrightarrow[\fr q_{i}]{\fr p_{i}}\mathfrak{o}_{1}\in {\rm Hom}(\Lambda(G_{AB})) \; , \q \forall \, \mathfrak{o}_{i}\in\mathcal{O}_{AB}
\end{equation*}  
and the requirement $(\fr p_1,\fr q_1) = ({\rm id}_{{\rm s}(\mathfrak{o}_{1})}, {\rm id}_{{\rm t}(\mathfrak{o}_{1})})$. The stabiliser group of $\mathcal{O}_{AB}$ is then defined as
\begin{equation}
	Z_{\mathcal{O}_{AB}} := \{(\fr a,\fr b) \in {\rm Hom}(\Lambda A) \times {\rm Hom}(\Lambda B) \; | \; \mathfrak{o}_1 = \fr a^{-1}\mathfrak{o}_1 \fr b \} \; .
\end{equation}
Remark that the orbit-stabiliser theorem implies $|Z_{\mathcal{O}_{AB}}| \cdot |\mathcal{O}_{AB}| = |A||B|$. Finally, we construct the twisted group algebra $\mathbb C[Z_{\mathcal{O}_{AB}}]$ as the algebra with defining vector space
\begin{equation}
	{\rm Span}_\mathbb{C}\big\{ \big|\biGrZ{\fr a}{\fr b}\big\ra \big\}_{\forall \, (\fr a,\fr b) \in Z_{\mathcal{O}_{AB}}}
\end{equation}
and product rule
\begin{equation}
	\big|\biGrZ{\fr a}{\fr b}\big\ra \star \big|\biGrZ{\fr a'}{\fr b'}\big\ra = \vartheta^{\Lambda(AB)}_{\mathfrak{o}_1}(\fr a,\fr a'|\fr b,\fr b') \big|\biGrZ{\fr a \fr a'}{\fr b \fr b'}\big\ra \; .
\end{equation}
Given that $\alpha$ is normalized, it follows from definition \eqref{eq:defvartheta} that $\vartheta^{\Lambda(AB)}_{\mathfrak{o}_1}$ is a representative normalised group 2-cocycle in $H^2(Z_{\mathcal{O}_{AB}}, {\rm U}(1))$. For each simple unitary $\vartheta^{\Lambda(AB)}_{\mathfrak{o}_1}$-projective representation $(\mc{D}^R,V_R)$ of $Z_{\mathcal{O}_{AB}}$, we can define a simple representation of the relative groupoid algebra $\grAlgL$ via a homomorphism $\mc{D}^{\mathcal{O}_{AB},R}:  \grAlgL \to {\rm End}(V_{\mathcal{O}_{AB},R})$ where
\begin{equation}
	V_{\mathcal{O}_{AB},R} := {\rm Span}_{\mathbb C}\{|\mathfrak{o}_i,v_m \ra \}_{\substack{\hspace{-1.2em}\forall \, i =1,\ldots,|\mathcal{O}_{AB}| \\ \forall \, m=1,\ldots,{\rm dim}(V_R)}} \; .
\end{equation}
For $i,j \in \{1,\ldots,|\mathcal{O}_{AB}|\}$, $m,n \in \{1,\ldots,{\rm dim}(V_R)\}$ the matrix elements are defined to be
\begin{equation}
	\label{eq:defD}
	\mc{D}^{\mathcal{O}_{AB},R}_{[im][jn]}\big( \big| \biGr{\fr g}{\fr a}{\fr b} \big\ra \big) = 
	\delta_{\fr g,\mathfrak{o}_i} \, \delta_{\fr a^{-1} \fr g \fr b,\mathfrak{o}_{j}} \, 
	\frac{\vartheta^{\Lambda(AB)}_{\mathfrak{o}_1}(\fr p_i^{-1},\fr a|\fr q_i^{-1},\fr b)}{\vartheta^{\Lambda(AB)}_{\mathfrak{o}_1}(\fr p_i^{-1}\fr a \fr p_j,\fr p_j^{-1}|\fr q_i^{-1}\fr b \fr q_j,\fr q_j^{-1})} \, 
	\mc{D}^R_{mn}\big (\big| \biGrZ{\fr p_i^{-1} \fr a \fr p_j}{\fr q_i^{-1} \fr b \fr q_j}\big\ra \big)
\end{equation}
such that 
\begin{equation}
	\ket{\mathfrak{o}_{i},v_{m}}\triangleright
	\mc{D}^{\mathcal{O}_{AB},R}\big( \big| \biGr{\fr g}{\fr a}{\fr b} \big\ra \big) 
	= \sum_{i,j=1}^{|\mathcal{O}_{AB}|}\sum_{m,n=1}^{{\rm dim}(V_R)}
	\mc{D}^{\mathcal{O}_{AB},R}_{[im][jn]}\big( \big| \biGr{\fr g}{\fr a}{\fr b} \big\ra \big)|\mathfrak{o}_j,v_n \ra  \; .
\end{equation}
Henceforth, we make use of the shorthand notation $\rho_{AB} \equiv (\mathcal{O}_{AB},R)$, $I \equiv [im]$, $J \equiv [jn]$ and $d_{\rho_{AB}} \equiv d_{\mathcal{O}_{AB},R} = |\mathcal{O}_{AB}| \cdot {\rm dim}(V_R)$.
It follows immediately from the definition and the linearity of the $\vartheta^{\Lambda(AB)}_{\mathfrak{o}_1}$-projective representations of $Z_{\mathcal{O}_{AB}}$ that these matrices define an algebra homomorphism, i.e.
\begin{equation}
	\sum_{K}	
	\mc{D}^{\rho_{AB}}_{IK}\big( \big| \biGr{\fr g}{\fr a}{\fr b} \big\ra \big)
	\mc{D}^{\rho_{AB}}_{KJ}\big( \big| \biGr{\fr g'}{\fr a'}{\fr b'} \big\ra \big)
	= \delta_{\fr{g',\fr{a^{-1}gb}}} \, \vartheta^{\Lambda(AB)}_{\fr g}(\fr{a,a'|b,b'}) \, 
	\mc{D}^{\rho_{AB}}_{IJ}\big( \big| \biGr{\fr g}{\fr a \fr a'}{\fr b \fr b'} \big\ra \big) \; .
\end{equation}
Furthermore, the matrix elements satisfy the conjugation relation
\begin{equation}
	\label{eq:complexConj}
	\overline{\mc{D}^{\mc \rho_{AB}}_{IJ}\big( \big| \biGr{\fr g}{\fr a}{\fr b} \big\ra \big)} = 
	\frac{1}{\vartheta^{\Lambda(AB)}_{\fr g}(\fr{a,a^{-1}|b,b^{-1}})}
	\mc{D}^{\mc \rho_{AB}}_{JI}\big( \big| \biGr{\fr{a^{-1}gb}}{\fr a^{-1}}{\fr b^{-1}} \big\ra \big) \; ,
\end{equation}
which follows from the unitarity of the projective representation $\mc{D}^R$ of the stabilizer subgroup $Z_{\mathcal{O}_{AB}}$, inducing a unitary representation of $\grAlgL$. This endows $\grAlgL$ with the structure of a *-algebra which in turn implies its \emph{semi-simplicity} due to finiteness. Finally, the representations matrices satisfy the following \emph{orthogonality} and \emph{completeness} conditions
\begin{gather}
	\label{eq:ortho}
	\frac{1}{|A||B|}
	\! \sum_{ \biGrFoot{\fr g}{\fr a}{\fr b} \in \Lambda(G_{AB})}
	\!\!\!\!
	\overline{\mc{D}^{\rho_{AB}}_{IJ}\big( \big| \biGr{\fr g}{\fr a}{\fr b} \big\ra \big)} 
	\mc{D}^{\rho_{AB}'}_{I'J'}\big( \big| \biGr{\fr g}{\fr a}{\fr b} \big\ra \big) 
	= \frac{\delta_{\rho_{AB},\rho_{AB}'}}{d_{\rho_{AB}}}\, \delta_{I,I'} \, \delta_{J,J'} 
	\\
	\frac{1}{|A||B|}\sum_{\rho_{AB}}\sum_{I,J}d_{\rho_{AB}} 
	\mc{D}^{\rho_{AB}}_{IJ}\big( \big| \biGr{\fr g}{\fr a}{\fr b} \big\ra \big)
	\overline{\mc{D}^{\rho_{AB}}_{IJ}\big( \big| \biGr{\fr g'}{\fr a'}{\fr b'} \big\ra \big)} = \delta_{\fr g, \fr g'} \, \delta_{\fr a,\fr a'} \, \delta_{\fr b,\fr b'} \; .
\end{gather}
A proof of the orthogonality relation can be found in app.~\ref{sec:app_ortho}, the completeness following from similar arguments.

\newpage
\subsection{Comultiplication map and concatenation of string-like excitations \label{sec:comult}}
The simple modules of the relative groupoid algebra $\grAlgL$  classify string-like bulk excitations terminating at gapped boundaries labelled by $A_{\lambda}$ and $B_{\mu}$, such that $\phi \equiv \sfT(\lambda)$ and $\psi \equiv \sfT(\mu)$. Let us now delve deeper into the exploration of the properties of this algebra, in relation to the concatenation of the corresponding excitations. We consider the following system of three gapped boundaries and string-like excitations terminating at these gapped boundaries:
\begin{equation}
	\bdryInterfaceThreeExcFusion{1}
	\; .
\end{equation} 
The two string-like excitations depicted above are characterized by the relative groupoid algebras $\grAlgL$ and $\grAlgLbis$, respectively, where $\varphi \equiv \sfT(\nu)$. We will show that these string-like excitations can be concatenated, and the result of this concatenation is a string-like  excitation terminating at the gapped boundaries labelled by $A_\lambda$ and $C_\nu$.\footnote{Because of the geometry of the operation under consideration, we refrain from referring to this process as the `fusion' of the corresponding string-like excitations. That being said, in (2+1)d, the same map defines the usual fusion of point-like excitations.} More specifically, we will demonstrate that a pair of modules for the relative groupoid algebras $\grAlgL$ and $\grAlgLbis$ can be composed to form a module for the relative groupoid algebra $\grAlgLter$.

Let us consider a pair of elementary string-like excitations with internal Hilbert spaces $V_{\rho_{AB}}$ and $V_{\rho_{BC}}$, respectively. In the absence of external constraints, the corresponding join Hilbert space is provided by the tensor product $V_{\rho_{AB}} \otimes V_{\rho_{BC}}$. It remains to understand how the tube algebra acts on this join Hilbert space.  We introduce an algebra homomorphism $\Delta_B :\grAlgLter \rightarrow \grAlgL \otimes \grAlgLbis$ defined by
\begin{align}
	\Delta_B\big(\big|\biGr{\fr g}{\fr a}{\fr c}\big\ra \big)
	&:=
	\frac{1}{|B|} \!\!\!\!
	\sum_{\substack{\fr g_1 \in {\rm Ob}(\Lambda (G_{AB})) \\ \fr g_2 \in {\rm Ob}(\Lambda (G_{BC})) \\ \fr g_{1}\fr g_{2}= \fr g
	\\
	\fr b\in {\rm Hom}_{\Lambda B}({\rm t}(\fr g_{1}),{\rm s}(\fr g_{2}))
	}} \!\!\!\!\!
	\zeta^{\Lambda(ABC)}_{\fr{a,b,c}}(\fr g_1,\fr g_2)
	\, \big|\biGr{\fr g_1}{\fr a}{\fr b} \big\ra \otimes \big| \biGr{\fr g_2}{\fr b}{\fr c} \big\ra
\end{align}
where
\begin{equation}
	\zeta^{\Lambda(ABC)}_{\fr{a,b,c}}(\fr g_1, \fr g_2):=\frac{\alpha(\fr g_{1},\fr g_{2},\fr c) \, \alpha(\fr a, \fr a^{-1}\fr g_{1}\fr b,\fr b^{-1}\fr g_{2}\fr c)}{\alpha(\fr g_{1},\fr b,\fr b^{-1}\fr g_{2}\fr c)} \; .
\end{equation}
As mentioned earlier, when no confusion is possible, we shall loosely identify $\fr b\in {\rm Hom}_{\Lambda B}({\rm t}(\fr g_{1}),{\rm s}(\fr g_{2}))$ and the group variable $\fr b \in B$ it evaluates to in order to make the notation lighter.
By analogy with the theory of Hopf algebras, we refer in the following to $\Delta_B$ as the $B$-\emph{comultiplication map} of the twisted groupoid algebra $\grAlgLter$.
It follows from the cocycle conditions
\begin{align*}
	d^{(3)}\alpha(\fr a,\fr a',\fr a'^{-1}\fr a^{-1}\fr g_1 \fr b \fr b',\fr b'^{-1}\fr b^{-1}\fr g_2 \fr c \fr c') &= 1
	&
	d^{(3)}\alpha(\fr a,\fr a^{-1} \fr g_1 \fr b,\fr b^{-1}\fr g_2 \fr c, \fr c') & = 1
	\\
	d^{(3)}\alpha(\fr a,\fr a^{-1}\fr b,\fr b',\fr b'^{-1}\fr b^{-1}\fr g_2 \fr c \fr c') &= 1
	&
	d^{(3)}\alpha(\fr g_1,\fr b,\fr b^{-1}\fr g_2\fr c,\fr c') &= 1
	\\
	d^{(3)}\alpha(\fr g_1,\fr b,\fr b',\fr b'^{-1}\fr b^{-1}\fr g_2 \fr c \fr c') &= 1
	&
	d^{(3)}\alpha(\fr g_1,\fr g_2,\fr c,\fr c') &= 1
\end{align*}
that $\zeta^{\Lambda(ABC)}_{\fr{a,b,c}}$ satisfies the relation
\begin{equation}
	\frac{\vartheta^{\Lambda(AB)}_{\fr g_1}(\fr{a,a'|b,b'}) \, \vartheta^{\Lambda(BC)}_{\fr g_2}(\fr{b,b'|c,c'})}{\vartheta^{\Lambda(AC)}_{\fr g_1 \fr g_2}(\fr{a,a'|c,c'})}
	=
	\frac{\zeta^{\Lambda(ABC)}_{\fr{aa',bb',cc'}}(\fr g_1, \fr g_2)}{\zeta^{\Lambda(ABC)}_{\fr{a,b,c}}(\fr g_1, \fr g_2) \, \zeta^{\Lambda(ABC)}_{\fr{a',b',c'}}(\fr{a^{-1}g_1b,b^{-1}g_2c})}
\end{equation}
ensuring that the map $\Delta_B$ is an algebra homomorphism, i.e.
\begin{align}
	\Delta_B\big(\big|\biGr{\fr g}{\fr a}{\fr c}\big\ra \big)
	\circ
	\Delta_B\big(\big|\biGr{\fr g'}{\fr a'}{\fr c'}\big\ra \big)
	=
	\Delta_B\big(\big|\biGr{\fr g}{\fr a}{\fr c}\big\ra
	\star
	\big|\biGr{\fr g'}{\fr a'}{\fr c'}\big\ra \big) \; .
\end{align}
Putting everything together, given the relative groupoid algebras $\grAlgL$, $\grAlgLbis$ and a pair of representations $(\mc{D}^{\rho_{AB}},V_{\rho_{AB}})$ and $(\mc{D}^{\rho_{BC}},V_{\rho_{BC}})$, the comultiplication $\Delta_B$ allows us to define the tensor product representation $ ((\mc{D}^{\rho_{AB}}\otimes \mc{D}^{\rho_{BC}}) \circ \Delta_B, V_{\rho_{AB}} \otimes V_{\rho_{BC}})$, where
\begin{equation}
	(\mc{D}^{\rho_{AB}}\otimes \mc{D}^{\rho_{BC}}) \circ \Delta_B : \grAlgL \otimes \grAlgLbis \to {\rm End}(V_{\rho_{AB}} \otimes V_{\rho_{BC}})
\end{equation}
such that
\begin{equation*}
	(\mc{D}^{\rho_{AB}} \otimes \mc{D}^{\rho_{BC}}) \big(\Delta_B\big(\big|\biGr{\fr g}{\fr a}{\fr c}\big\ra \big)\big)
	= \frac{1}{|B|} 
	\!\!
	\sum_{\substack{\fr g_1 \in {\rm Ob}(\Lambda (G_{AB})) \\ \fr g_2 \in {\rm Ob}(\Lambda (G_{BC})) \\ \fr g_{1}\fr g_{2}= \fr g
			\\
			\fr b\in B
	}} \!\!\!
	\zeta_{\fr{a,b,c}}^{\Lambda(ABC)}(\fr g_1, \fr g_2)
	\mc{D}^{\rho_{AB}}\big(\big|\biGr{\fr g_1}{\fr a}{\fr b}\big\ra \big)
	\otimes
	\mc{D}^{\rho_{BC}}\big(\big|\biGr{\fr g_2}{\fr b}{\fr c}\big\ra \big) \, , 
\end{equation*}
where we loosely identified $\fr b\in {\rm Hom}_{\Lambda B}({\rm t}(\fr g_{1}),{\rm s}(\fr g_{2}))$ and the corresponding group variable for notational convenience.
In the following, it will be often useful to write the so-called \emph{truncated tensor product} $\otimes_B$ of representation matrices defined as
\begin{equation}
	\mc{D}^{\rho_{AB}} \otimes_{B}  \mc{D}^{\rho_{BC}} := (\mc{D}^{\rho_{AB}} \otimes  \mc{D}^{\rho_{BC}}) \circ \Delta_B \; .
\end{equation}
Using the semisimplicity of relative groupoid algebras, the tensor product representations defined above are generically not simple and as such admit a decomposition into direct sum of simple representations, i.e.
\begin{equation}
	\mc{D}^{\rho_{AB}}\otimes_{B} \mc{D}^{\rho_{BC}} \cong \bigoplus_{\rho_{AC}}N^{\rho_{AC}}_{\rho_{AB},\rho_{BC}} \mc{D}^{\rho_{AC}} \; ,
\end{equation}
where the number $N^{\rho_{AC}}_{\rho_{AB},\rho_{BC}} \in \mathbb Z^+_0$ is referred to as the \emph{multiplicity} of the simple $\grAlgLter$ representation $(\mc{D}^{\rho_{AC}},V_{\rho_{AC}})$ appearing in the tensor product of the representations $(\mc{D}^{\rho_{AB}},V_{\rho_{AB}})$ and $(\mc{D}^{\rho_{BC}},V_{\rho_{BC}})$. 
Henceforth, we assume \emph{multiplicity-freeness} of the multifusion category of representations, i.e. $N^{\rho_{AC}}_{\rho_{AB},\rho_{BC}} \in \{0,1\}$ in order to simplify the notations. Note however that it is straightforward to lift this assumption. Using the orthogonality relations of the irreducible representations, we find a useful expression to compute explicitly this number, namely 
\begin{equation}
	N^{\rho_{AC}}_{\rho_{AB},\rho_{BC}}
	=
	\frac{1}{|A||C|}
	\!
	\sum_{\biGrFoot{\fr g}{\fr a}{\fr c} \in \Lambda(G_{AC})}
	\!\!\!
	{\rm tr}\Big[\, (\mc{D}^{\rho_{AB}} \otimes_B \mc{D}^{\rho_{BC}}) \big(\big|\biGr{\fr g}{\fr a}{\fr c}\big\ra \big)
	\overline{\mc{D}^{\rho_{AC}}\big( \big|\biGr{\fr g}{\fr a}{\fr c}\big\ra \big)}\,\Big] \; .
\end{equation}
Note finally that given the algebras $\mathbb{C}[\Lambda(G_{AA})]^\alpha_{\phi \phi}$ and $\mathbb{C}[\Lambda(G_{BB})]^\alpha_{\psi \psi}$, the regular modules\footnote{The regular module of an algebra is defined as the algebra viewed as a module over itself.} $\mathbb{C}[\Lambda(G_{AA})]^\alpha_{\phi \phi}$ $\mathbb{C}[\Lambda(G_{BB})]^\alpha_{\psi \psi}$ satisfy the \emph{unit module} properties 
\begin{align}
	\mathbb{C}[\Lambda(G_{AA})]^\alpha_{\phi \phi}\otimes_{A}\rho_{AB}\cong \rho_{AB}\cong \rho_{AB}\otimes_{B}\mathbb{C}[\Lambda(G_{BB})]^\alpha_{\psi \psi}
\end{align}
as $\grAlgL$ modules.

As explained above, thanks to our formulation in terms of relative groupoid algebras, we can easily extract all the relevant structures for the (2+1)d algebra as a limiting case. This is done in the next section, where we define a canonical basis of excited states. In this scenario, the comultiplication map yields the fusion of the corresponding point-like excitations.  

\subsection{Clebsch-Gordan series}
In preparation for the later discussion, let us study further the properties of the comultiplication map introduced earlier. Since the comultiplication map $\Delta_B$ is an algebra homomorphism, there exist intertwining \emph{unitary} maps
\begin{equation}
	\mathcal{U}^{\rho_{AB}, \rho_{BC}}:
	\bigoplus_{\rho_{AC}}V_{\rho_{AC}}\to V_{\rho_{AB}}\otimes_{B} V_{\rho_{BC}} \; ,
\end{equation}
where the sum is over labels $\rho_{AC}$ such that $\mc{D}^{\rho_{AC}} \in \mc{D}^{\rho_{AB}} \otimes_{B} \mc{D}^{\rho_{BC}}$, that satisfy the defining relation
\begin{equation*}
	(\mc{D}^{\rho_{AB}}_{I_{AB}J_{AB}} \otimes_{B} \mc{D}^{\rho_{BC}}_{I_{BC}J_{BC}}) \big(\big|\biGr{\fr g}{\fr a}{\fr c}\big\ra \big)
	= \!\!\! \sum_{\substack{\rho_{AC} \\ I_{AC}, J_{AC}}} \!\!\!
	\mathcal{U}^{\rho_{AB}, \rho_{BC}}_{[I_{AB}I_{BC}][\rho_{AC}I_{AC}]}
	\mc{D}^{\rho_{AC}}_{I_{AC}J_{AC}}\big( \big|\biGr{\fr g}{\fr a}{\fr c}\big\ra \big)
	\overline{\mathcal{U}^{\rho_{AB}, \rho_{BC}}_{[J_{AB}J_{BC}][\rho_{AC}J_{AC}]}}  \, .
\end{equation*}
Henceforth, we will denote the matrix elements of this unitary map as
\begin{equation*}
	\CC{\rho_{AB}}{\rho_{BC}}{\rho_{AC}}{I_{AB}}{I_{BC}}{I_{AC}} := 	\mathcal{U}^{\rho_{AB}, \rho_{BC}}_{[I_{AB}I_{BC}][\rho_{AC}I_{AC}]} \; ,
\end{equation*}
and refer to them as \emph{Clebsch-Gordan} coefficients.
Using the orthogonality of the representation matrices, we obtain the equivalent defining relation
\begin{equation*}
	\frac{d_{\rho_{AC}}}{|A||C|}
	\sum_{\biGrFoot{\fr g}{\fr a}{\fr c} \in \Lambda(G_{AC}) } \!\!\!
	(\mc{D}^{\rho_{AB}}_{I_{AB}J_{AB}} \otimes_B \mc{D}^{\rho_{BC}}_{I_{BC}J_{BC}}) \big(\big|\biGr{\fr g}{\fr a}{\fr c}\big\ra \big)
	\overline{\mc{D}^{\rho_{AC}}_{I_{AC}J_{AC}}\big( \big|\biGr{\fr g}{\fr a}{\fr c}\big\ra \big)}
	= 
	\CC{\rho_{AB}}{\rho_{BC}}{\rho_{AC}}{I_{AB}}{I_{BC}}{I_{AC}} 
	\overline{\CC{\rho_{AB}}{\rho_{BC}}{\rho_{AC}}{J_{AB}}{J_{BC}}{J_{AC}} }  .
\end{equation*}
The unitarity of $\mathcal{U}^{\rho_{AB},\rho_{BC}}$ imposes the following \emph{orthogonality} and \emph{completeness} relations:
\begin{align}
	\sum_{I_{AB},I_{BC}}
	\CC{\rho_{AB}}{\rho_{BC}}{\rho_{AC}}{I_{AB}}{I_{BC}}{I_{AC}}
	\overline{	\CC{\rho_{AB}}{\rho_{BC}}{\rho'_{AC}}{I_{AB}}{I_{BC}}{I'_{AC}}}
	=
	\delta_{I_{AC},I'_{AC}}
	\delta_{\rho_{AC},\rho'_{AC}}
	\\
	\sum_{\rho_{AC},I_{AC}}
	\overline{	\CC{\rho_{AB}}{\rho_{BC}}{\rho_{AC}}{I_{AB}}{I_{BC}}{I_{AC}}}
	\CC{\rho_{AB}}{\rho_{BC}}{\rho_{AC}}{I'_{AB}}{I'_{BC}}{I_{AC}}	
	=
	\delta_{I_{AB},I'_{AB}}
	\delta_{I_{BC},I'_{BC}} \; .
\end{align}
Furthermore, the Clebsch-Gordan coefficients satisfy the following crucial property
\begin{equation}
	\label{eq:gauge}
	\sum_{\fr g \in {\rm Hom}({\rm s}(\fr a), {\rm s}(\fr c))} \!
	\sum_{\{J\}}	
	(\mc{D}^{\rho_{AB}}_{I_{AB}J_{AB}} \otimes_B \mc{D}^{\rho_{BC}}_{I_{BC}J_{BC}}) \big(\big|\biGr{\fr g}{\fr a}{\fr c}\big\ra \big)
	\overline{	\mc{D}^{\rho_{AC}}_{I_{AC}J_{AC}}\big( \big|\biGr{\fr g}{\fr a}{\fr c}\big\ra \big)}
	\CC{\rho_{AB}}{\rho_{BC}}{\rho_{AC}}{J_{AB}}{J_{BC}}{J_{AC}}
	= 
	\CC{\rho_{AB}}{\rho_{BC}}{\rho_{AC}}{I_{AB}}{I_{BC}}{I_{AC}}
\end{equation}
referred to as the \emph{gauge invariance} of the coefficients. This property can be checked as follows: Firstly, utilise the unitarity of the intertwining maps to rewrite the defining equation as the \emph{intertwining property}
\begin{equation*}
	\sum_{J_{AB},J_{BC}}
	(\mc{D}^{\rho_{AB}}_{I_{AB}J_{AB}} \otimes_B \mc{D}^{\rho_{BC}}_{I_{BC}J_{BC}}) \big(\big|\biGr{\fr g}{\fr a}{\fr c}\big\ra \big)
	\CC{\rho_{AB}}{\rho_{BC}}{\rho_{AC}}{J_{AB}}{J_{BC}}{J_{AC}}
	= \sum_{I_{AC}}
	\CC{\rho_{AB}}{\rho_{BC}}{\rho_{AC}}{I_{AB}}{I_{BC}}{I_{AC}}
	\mc{D}^{\rho_{AC}}_{I_{AC}J_{AC}}\big( \big|\biGr{\fr g}{\fr a}{\fr c}\big\ra \big) \; .
\end{equation*}
Secondly, multiply this equation on both side by $\overline{\mc{D}^{\rho_{AC}}_{K_{AC}J_{AC}}\big( \big| \biGr{\fr g}{\fr a}{\fr c} \big\ra \big)}$ and use the identity
\begin{align*}
	& 
	\sum_{\fr g \in {\rm Hom}({\rm s}(\fr a), {\rm s}(\fr c))} \!
	\sum_{J_{AC}}	
	\mc{D}^{\rho_{AC}}_{I_{AC}J_{AC}}\big( \big| \biGr{\fr g}{\fr a}{\fr c} \big\ra \big)
	\overline{\mc{D}^{\rho_{AC}}_{K_{AC}J_{AC}}\big( \big| \biGr{\fr g}{\fr a}{\fr c} \big\ra \big)}	
	\\
	& \q =
		\sum_{\fr g \in {\rm Hom}({\rm s}(\fr a), {\rm s}(\fr c))}\!
	\sum_{J_{AC}}
	\frac{1}{\vartheta^{\Lambda(AC)}_{\fr g}(\fr{a,a^{-1}|c,c^{-1}})}	
	\mc{D}^{\rho_{AC}}_{I_{AC}J_{AC}}\big( \big| \biGr{\fr g}{\fr a}{\fr c} \big\ra \big)
	\mc{D}^{\rho_{AC}}_{J_{AC}K_{AC}}\big( \big| \biGr{\fr{a^{-1}gc}}{\fr a^{-1}}{\fr c^{-1}} \big\ra \big)
	\\
	& \q =
		\sum_{\fr g \in {\rm Hom}({\rm s}(\fr a), {\rm s}(\fr c))} \!
	\mc{D}^{\rho_{AC}}_{I_{AC}K_{AC}}\big( \big| \biGr{\fr g}{\mathbbm 1_A}{\mathbbm 1_C} \big\ra \big)
	= \delta_{I_{AC},K_{AC}} \; ,
\end{align*}
where we used \eqref{eq:complexConj}. Note that we use the notation $\mathbbm 1_A$ to refer to the morphism in ${\rm Hom}({\rm s}(\fr a),-)$ characterized by the group variable $\mathbbm 1_A \in A$, and similarly for $\mathbbm 1_C$.
Summing over $J_{AC} = 1 , \ldots, d_{\rho_{AC}}$ finally yields the gauge invariance. This invariance of the Clebsch-Gordan coefficients further implies
\begin{align}
	\label{eq:gaugeBIS}
	&\sum_{\{J\}}
	\mc{D}^{\rho_{AB}}_{I_{AB}J_{AB}}\big( \big|\biGr{\fr g_1}{\fr a}{\fr b}\big\ra \big)
	\mc{D}^{\rho_{BC}}_{I_{BC}J_{BC}} \big( \big|\biGr{\fr g_2}{\fr b'}{\fr c}\big\ra\big)
	\CC{\rho_{AB}}{\rho_{BC}}{\rho_{AC}}{J_{AB}}{J_{BC}}{J_{AC}}
	\mc{D}^{\rho_{AC}}_{J_{AC}I_{AC}} \big( \big|\biGr{\fr g_3}{\fr a'}{\fr c'}\big\ra\big)
	\\
	\nn
	& \q = 
	\frac{1}{|B|}\sum_{\fr{\tilde b} \in B}
	\frac{\vartheta^{\Lambda(AB)}_{\fr g_1}(\fr{a,\tilde a|b,\tilde b}) \, \vartheta^{\Lambda(BC)}_{\fr g_2}(\fr{b',\tilde b|c,\tilde c})  \, \zeta^{\Lambda(ABC)}_{\fr{\tilde a,\tilde b,\tilde c}}(\fr a^{-1} \fr g_1 \fr b,\fr b'^{-1}\fr g_2\fr c)}{\vartheta^{\Lambda(AC)}_{\fr g_3}(\fr{\tilde a, \tilde a^{-1}a'|\tilde c, \tilde c^{-1}c'})} 
	\delta_{\fr g_3,\fr a^{-1}\fr g_1 \fr b \fr b'^{-1}\fr g_2 \fr c}
	\\[-0.5em]
	\nn
	& \hspace{2.9em} \times 	
	\sum_{\{K\}}
	\mc{D}^{\rho_{AB}}_{I_{AB}K_{AB}}\big(\big|\biGr{\fr g_1}{\fr {a\tilde a}}{\fr {b\tilde b}}\big\ra \big)
	\mc{D}^{\rho_{BC}}_{I_{BC}K_{BC}}\big(\big|\biGr{\fr g_2}{\fr {b'\tilde b}}{\fr {c\tilde c}}\big\ra \big)
	\CC{\rho_{AB}}{\rho_{BC}}{\rho_{AC}}{K_{AB}}{K_{BC}}{K_{AC}}
	\mc{D}^{\rho_{AC}}_{K_{AC}I_{AC}}\big( \big|\biGr{\fr{\tilde a}^{-1}\fr g_3\fr{\tilde c}}{\fr{\tilde a^{-1}a'}}{\fr{\tilde c^{-1}c'}}\big\ra \big) \, ,
\end{align}
which is true for all composable morphisms $\fr a,\fr{\tilde{a}}$ in $\Lambda A$ and $\fr{c,\tilde{c}}$ in $\Lambda C$. A proof of this identity can be found in app.~\ref{sec:app_gaugeBIS}. It is straightforward to check that this last relation induces another one, namely
\begin{align}
	\label{eq:gaugeConj}
	&\sum_{\{J\}}
	\overline{\mc{D}^{\rho_{AB}}_{J_{AB}I_{AB}}\big( \big|\biGr{\fr g_1}{\fr a}{\fr b}\big\ra \big)} \,
	\overline{\mc{D}^{\rho_{BC}}_{J_{BC}I_{BC}} \big( \big|\biGr{\fr g_2}{\fr b'}{\fr c}\big\ra\big)}
	\CC{\rho_{AB}}{\rho_{BC}}{\rho_{AC}}{J_{AB}}{J_{BC}}{J_{AC}}
	\overline{\mc{D}^{\rho_{AC}}_{I_{AC}J_{AC}} \big( \big|\biGr{\fr g_3}{\fr a'}{\fr c'}\big\ra\big)}
	\\
	\nn
	& \q = 
	\frac{1}{|B|}\sum_{\fr{\tilde b} \in B}
	\vartheta^{\Lambda(AB)}_{\fr g_1}(\fr{\tilde a,\tilde a^{-1} a|\tilde b,\tilde b^{-1} b}) \, \vartheta^{\Lambda(BC)}_{\fr g_2}(\fr{\tilde b, \tilde b^{-1} b'|\tilde c,\tilde c^{-1} c})  \, \zeta^{\Lambda(ABC)}_{\fr{\tilde a,\tilde b,\tilde c}}(\fr g_1, \fr g_2) \, \vartheta^{\Lambda(AC)}_{\fr g_3}(\fr{a', \tilde a|c',\tilde c})^{-1} 
	\\[-0.5em]
	\nn
	& \hspace{2.9em} \times 	
	\sum_{\{K\}}
	\delta_{\fr a'^{-1}\fr g_3 \fr c',\fr g_1 \fr g_2} \, 
	\overline{\mc{D}^{\rho_{AB}}_{K_{AB}I_{AB}}\big(\big|\biGr{\fr{\tilde a}^{-1} \fr g_1 \fr{\tilde b}}{\fr{\tilde a}^{-1} \fr a }{\fr{\tilde b}^{-1}\fr  b}\big\ra \big)} \, 
	\overline{\mc{D}^{\rho_{BC}}_{K_{BC}I_{BC}}\big(\big|\biGr{\fr{\tilde b}^{-1}\fr  g_2 \fr{\tilde c}}{\fr{\tilde b}^{-1} \fr b'}{\fr{\tilde c}^{-1}\fr c'}\big\ra \big)}
	\\[-0.2em]
	\nn
	& \hspace{4.5em} \times
	\CC{\rho_{AB}}{\rho_{BC}}{\rho_{AC}}{K_{AB}}{K_{BC}}{K_{AC}}
	\overline{\mc{D}^{\rho_{AC}}_{I_{AC}K_{AC}}\big( \big|\biGr{\fr g_3}{\fr{a' \tilde a}}{\fr{c' \tilde c}}\big\ra \big)} \, .
\end{align}

\subsection{Associativity and $6j$-symbols}\label{sec:6j}
Given two relative groupoid algebras $\grAlgL$, $\grAlgLbis$ and a pair of representations defined by $(\mc{D}^{\rho_{AB}},V_{\rho_{AB}})$, $(\mc{D}^{\rho_{BC}},V_{\rho_{BC}})$, we constructed earlier the tensor product representation $ ((\mc{D}^{\rho_{AB}}\otimes \mc{D}^{\rho_{BC}}) \circ \Delta_B, V_{\rho_{AB}} \otimes V_{\rho_{BC}})$ of $\grAlgLter$. Let us now consider the \emph{quasi-invertible} algebra element $\Phi_{ABCD} \in  \grAlgL \otimes  \grAlgLbis \otimes  \grAlgLqua$ defined as
\begin{equation}
	\Phi_{ABCD} := \sum_{\substack{\fr g_1 \in {\rm Ob}(\Lambda(G_{AB})) \\ \fr g_2 \in {\rm Ob}(\Lambda(G_{BC})) \\ \fr g_3 \in {\rm Ob}(\Lambda(G_{CD}))}}
	\alpha^{-1}(\fr g_1, \fr g_2, \fr g_3)
	\big|\biGr{\fr g_1}{\mathbbm 1_A}{\mathbbm 1_B}\big\ra 
	\otimes
	\big|\biGr{\fr g_2}{\mathbbm 1_B}{\mathbbm 1_C}\big\ra
	\otimes 
	\big|\biGr{\fr g_3}{\mathbbm 1_C}{\mathbbm 1_D}\big\ra \; ,
\end{equation}
such that $\fr g_1$, $\fr g_2$ and $\fr g_3$ are composable morphisms in $\Lambda G$.
The cocycle conditions
\begin{align*}
	d^{(3)}\alpha(\fr a,\fr a^{-1}\fr g_{1}\fr b,\fr b^{-1}\fr g_{2}\fr c,\fr c^{-1}\fr g_{3}\fr d) & =1
	&
	d^{(3)}\alpha(\fr g_{1},\fr g_{2},\fr c,\fr c^{-1}\fr g_{3}\fr d)&=1
	\\
	d^{(3)}\alpha(\fr g_{1},\fr b,\fr b^{-1}\fr g_{2}\fr c,\fr c^{-1}\fr g_{3}\fr d) &= 1
	&
	d^{(3)}\alpha(\fr g_{1},\fr g_{2},\fr g_{3},\fr d) & =1
\end{align*}
imply the identity
\begin{align}
	\label{eq:quasiCoCoc}
	\frac{\zeta^{\Lambda(BCD)}_{\fr b,\fr c,\fr d}(\fr g_2,\fr g_3)  \, \zeta^{\Lambda(ABD)}_{\fr a,\fr b,\fr d}(\fr g_1,\fr g_2 \fr g_3) }{\zeta^{\Lambda(ACD)}_{\fr a,\fr c,\fr d}(\fr g_1 \fr g_2,\fr g_3) \, \zeta^{\Lambda(ABC)}_{\fr{\fr a,\fr b,\fr c}}(\fr g_1, \fr g_2)}
	=
	\frac{	\alpha(\fr g_1, \fr g_2, \fr g_3)}{\alpha(\fr a^{-1} \fr g_{1} \fr b,\fr b^{-1}\fr g_{2}\fr c,\fr c^{-1}\fr g_{3}\fr d)} \; ,
\end{align}
which in turn ensures that the comultiplication is \emph{quasi-coassociative}, i.e
\begin{equation}
	(\Delta_B \otimes {\rm id}) \Delta_{C}\big(\big|\biGr{\fr g}{\fr a}{\fr d}\big\ra \big) 
	= \Phi_{ABCD} \star \big[ ({\rm id} \otimes \Delta_{C})  \Delta_B\big(\big|\biGr{\fr g}{\fr a}{\fr d}\big\ra \big) \big] \star \Phi_{ABCD}^{-1} 
	\; , \q \forall \; \big|\biGr{\fr g}{\fr a}{\fr d}\big\ra\in \mathbb C[\Lambda(G_{AB})]^\alpha_{\phi \chi} \; .
\end{equation}
This signifies that the truncated tensor product of representations $(\mc{D}^{\rho_{AB}} \otimes_{B} \mc{D}^{\rho_{BC}}) \otimes_{C} \mc{D}^{\rho_{CD}}$ and $\mc{D}^{\rho_{AB}} \otimes_{B} (\mc{D}^{\rho_{BC}} \otimes_{C} \mc{D}^{\rho_{CD}})$ defined as 
\begin{align}
	(\mc{D}^{\rho_{AB}} \otimes_{B} \mc{D}^{\rho_{BC}}) \otimes_{C} \mc{D}^{\rho_{CD}}
	&:=
	(\mc{D}^{\rho_{AB}} \otimes \mc{D}^{\rho_{BC}} \otimes \mc{D}^{\rho_{CD}})
	\circ(\Delta_B \otimes {\rm id}) \Delta_{C}
	\\
	\mc{D}^{\rho_{AB}} \otimes_{B} (\mc{D}^{\rho_{BC}} \otimes_{C} \mc{D}^{\rho_{CD}})
	&:=
	(\mc{D}^{\rho_{AB}} \otimes \mc{D}^{\rho_{BC}} \otimes \mc{D}^{\rho_{CD}})
	\circ({\rm id} \otimes \Delta_{C})  \Delta_B
\end{align}
must be isomorphic as $\mathbb C[\Lambda(G_{AD})]^\alpha_{\phi \chi}$-modules. More specifically, it follows immediately from the quasi-coassociativity condition that the maps
\begin{equation}
	\Phi^{\rho_{AB},\rho_{BC},\rho_{CD}} := (\mc{D}^{\rho_{AB}} \otimes \mc{D}^{\rho_{AB}} \otimes \mc{D}^{\rho_{AB}})(\Phi_{ABCD}) \in {\rm End}(V_{\rho_{AB}} \otimes V_{\rho_{BC}} \otimes V_{\rho_{CD}})
\end{equation}
define intertwiners between the tensor product of representations above such that
\begin{equation}
	\Phi^{\rho_{AB},\rho_{BC},\rho_{CD}}
	[\mc{D}^{\rho_{AB}} \otimes_{B} (\mc{D}^{\rho_{BC}} \otimes_{C} \mc{D}^{\rho_{CD}})]
	=
	[(\mc{D}^{\rho_{AB}} \otimes_{B} \mc{D}^{\rho_{BC}}) \otimes_{C} \mc{D}^{\rho_{CD}}]
	\Phi^{\rho_{AB},\rho_{BC},\rho_{CD}}
	\; .
\end{equation}
Let us consider two vector spaces $V_{\rho_{AB}}$ and $V_{\rho_{BC}}$. These are spanned by vectors $|\rho_{AB}I_{AB} \ra$ and $| \rho_{BC}I_{BC} \ra$, respectively, such that the corresponding groupoid algebras act on these basis vectors from the right. We define the truncated tensor product of two such vectors as
\begin{align}
	|\rho_{AB}I_{AB} \ra \otimes_{B} |\rho_{BC}I_{BC} \ra &:=  \big(|\rho_{AB}I_{AB} \ra \otimes |\rho_{BC}I_{BC} \ra \big) \triangleright \Delta_B(\mathbbm{1}_{AC}) \;  ,
\end{align}
which span the vector space $V_{\rho_{AB}} \otimes_{B} V_{\rho_{BC}} \subset V_{\rho_{AB}} \otimes V_{\rho_{BC}}$. More specifically. we have
\begin{align}
	&|\rho_{AB}I_{AB} \ra \otimes_B |\rho_{BC}I_{BC} \ra=
	\sum_{\substack{ \rho_{AC}\\ I_{AC}}}
	\ket{\rho_{AB}\otimes_{B}\rho_{BC}; \rho_{AC}, I_{AC}}
	\CC{\rho_{AB}}{\rho_{BC}}{\rho_{AC}}{I_{AB}}{I_{BC}}{I_{AC}} \; ,
\end{align}
where we define
\begin{align}
	\ket{\rho_{AB}\otimes_{B}\rho_{BC}, \rho_{AC} I_{AC}}
	:=
	\sum_{I_{AB},I_{BC}}
	\overline{	\CC{\rho_{AB}}{\rho_{BC}}{\rho_{AC}}{I_{AB}}{I_{BC}}{I_{AC}}}
	\big(|\rho_{AB}I_{AB} \ra \otimes |\rho_{BC}I_{BC} \ra \big) \; .
\end{align}
Noting that
\begin{align}
	\ket{\rho_{AB}\otimes_{B} \rho_{BC},\rho_{AC}I_{AC}}
	(\mc{D}^{\rho_{AB}}\otimes_{B} \mc{D}^{\rho_{BC}})\big( \big| \biGr{\fr g}{\fr a}{\fr c} \big \ra \big)
	=
	\ket{\rho_{AB}\otimes_{B} \rho_{BC},\rho_{AC}I_{AC}}\mc{D}^{\rho_{AC}}\big( \big| \biGr{\fr g}{\fr a}{\fr c} \big\ra \big) \; ,
\end{align}
we realize that ${\rm Span}_{\mathbb{C}}\{ \ket{\rho_{AB}\otimes_{B}\rho_{BC},\rho_{AC}I_{AC} } \}_{\forall \, I_{AC}}\cong V_{\rho_{AC}}$ as $\grAlgLter$ representations through the map $\ket{\rho_{AB}\otimes_{B} \rho_{BC},\rho_{AC}I_{AC}}\mapsto \ket{\rho_{AC}I_{AC}}$.
Similarly, we can define the following truncated tensor product of vectors
\begin{align*}
	&\big( |\rho_{AB}I_{AB} \ra \otimes_B |\rho_{BC}I_{BC} \ra \big) \otimes_C |\rho_{CD}I_{CD}\ra
	\\
	& \q := 
	\big( |\rho_{AB}I_{AB} \ra \otimes |\rho_{BC}I_{BC} \ra  \otimes |\rho_{CD}I_{CD}\ra \big)
	\triangleright
	[(\Delta_B \otimes {\rm id})  \Delta_{C}](\mathbbm 1_{AD}) 
	\\[0.5em]
	& |\rho_{AB}I_{AB} \ra \otimes_{B} \big( |\rho_{BC}I_{BC} \ra  \otimes_C |\rho_{CD}I_{CD}\ra \big)
	\\
	& \q := 
	\big( |\rho_{AB}I_{AB} \ra \otimes |\rho_{BC}I_{BC} \ra  \otimes |\rho_{CD}I_{CD}\ra \big)
	\triangleright[({\rm id} \otimes \Delta_{C})  \Delta_B](\mathbbm 1_{AD})  \; ,
\end{align*}
which define basis vectors in $(V_{\rho_{AB}} \otimes _{B} V_{\rho_{BC}}) \otimes_{C} V_{\rho_{BC}}$ and $V_{\rho_{AB}} \otimes _{B} ( V_{\rho_{BC}} \otimes_{C} V_{\rho_{BC}})$, respectively. We then find that $	\Phi_{ABCD}$ induces the following isomorphism:
\begin{equation}
	\label{eq:isoVec}
	(V_{\rho_{AB}} \otimes _{B} V_{\rho_{BC}}) \otimes_{C} V_{\rho_{BC}} \cong
	V_{\rho_{AB}} \otimes _{B} ( V_{\rho_{BC}} \otimes_{C} V_{\rho_{BC}}) \; .
\end{equation}
Vectors $\big( |\rho_{AB}I_{AB} \ra \otimes_{B} |\rho_{BC}I_{BC} \ra \big) \otimes_{C} |\rho_{CD}I_{CD}\ra$ are typically not linearly independent, however a basis for the  vector space $(V_{\rho_{AB}} \otimes_{B} V_{\rho_{BC}}) \otimes_{C} V_{\rho_{BC}}$ is provided by the vectors
\begin{equation}
	\sum_{\{I\}}
	\overline{\CC{\rho_{AB}}{\rho_{BC}}{\rho_{AC}}{I_{AB}}{I_{BC}}{I_{AC}}}
	\overline{\CC{\rho_{AC}}{\rho_{CD}}{\rho_{AD}}{I_{AC}}{I_{CD}}{K_{AD}}}
	|\rho_{AB}I_{AB} \ra \otimes |\rho_{BC}I_{BC} \ra  \otimes |\rho_{CD}I_{CD}\ra \; .
\end{equation}
We obtain that $\Phi_{ABCD}$ acts on such basis vectors as 
\begin{align}
	\nn
	& \sum_{\rho_{AC}}\sum_{\{I\}}
	\SixJ{\rho_{AB}}{\rho_{BC}}{\rho_{CD}}{\rho_{AD}}{\rho_{AC}}{\rho_{BD}}
	\overline{\CC{\rho_{AB}}{\rho_{BC}}{\rho_{AC}}{I_{AB}}{I_{BC}}{I_{AC}}}
	\overline{\CC{\rho_{AC}}{\rho_{CD}}{\rho_{AD}}{I_{AC}}{I_{CD}}{K_{AD}}}
	|\rho_{AB}I_{AB} \ra \otimes |\rho_{BC}I_{BC} \ra  \otimes |\rho_{CD}I_{CD}\ra
	\triangleright \Phi_{ABCD}
	\\
	\label{eq:defSixJ}
	& \q =  
	\sum_{\{I\}}
	\overline{\CC{\rho_{AB}}{\rho_{BD}}{\rho_{AD}}{I_{AB}}{I_{BD}}{K_{AD}}}
	\overline{\CC{\rho_{BC}}{\rho_{CD}}{\rho_{BD}}{I_{BC}}{I_{CD}}{I_{BD}}}
	|\rho_{AB}I_{AB} \ra \otimes |\rho_{BC}I_{BC} \ra  \otimes |\rho_{CD}I_{CD}\ra
\end{align}
such that the so-called $6j$-symbols are defined as
\begin{equation*}
	\SixJ{\rho_{AB}}{\rho_{BC}}{\rho_{CD}}{\rho_{AD}}{\rho_{AC}}{\rho_{BD}}
	\!\! :=
	\frac{1}{d_{\rho_{AD}}} \!
	\sum_{\{I\}} \!
	\alpha(\mathfrak{o}_{i_{AB}},\mathfrak{o}_{i_{BC}},\mathfrak{o}_{i_{CD}})
	\CC{\rho_{AB}}{\rho_{BC}}{\rho_{AC}}{I_{AB}}{I_{BC}}{I_{AC}}
	\CC{\rho_{AC}}{\rho_{CD}}{\rho_{AD}}{I_{AC}}{I_{CD}}{I_{AC}}
	\overline{\CC{\rho_{AB}}{\rho_{BD}}{\rho_{AD}}{I_{AB}}{I_{BD}}{I_{AD}}}
	\overline{\CC{\rho_{BC}}{\rho_{CD}}{\rho_{BD}}{I_{BC}}{I_{CD}}{I_{BD}}} ,
\end{equation*}
where the notation is the one of definition \eqref{eq:defD} of the representation matrices. This establishes the isomorphism \eqref{eq:isoVec}. A detailed proof of the defining relation \eqref{eq:defSixJ} can be found in app.~\ref{sec:app_SixJ}.

Furthermore, given the vector space $((V_{\rho_{AB}} \otimes_{B} V_{\rho_{BC}}) \otimes_{C} V_{\rho_{CD}}) \otimes_{D} V_{\rho_{DE}}$, we find that
\begin{align}
	\nn
	&[({\rm id} \otimes {\rm id } \otimes \Delta_{D_\chi})(\Phi_{ABCE})] \star
	[(\Delta_B  \otimes {\rm id} \otimes {\rm id })(\Phi_{ACDE})]
\end{align}
and
\begin{align}
	(\mathbbm 1_{AB} \otimes \Phi_{BCDE}) \star
	[({\rm id} \otimes \Delta_{C} \otimes {\rm id })(\Phi_{ABDE})] \star 
	(\Phi_{ABCD} \otimes \mathbbm 1_{DE})
\end{align}
induce the same isomorphism. This is referred to as the so-called \emph{pentagon identity} and ensures the self-consistency of the quasi-coassociativity. A proof of the pentagon identity can be found in app.~\ref{sec:app_pent}. 

In a similar vein, it can be shown that the regular $\mathbb{C}[\Lambda(G_{BB})]^{\alpha}_{\psi\psi}$-module satisfies the so-called \emph{triangle identity} such that the following diagram commutes
\begin{align}
	\begin{tikzcd}[ampersand replacement=\&, column sep=1.4em, row sep=1.3em]
	|[alias=A]|(\rho_{AB}\otimes_{B}\mathbb{C}[\Lambda(G_{BB})]^{\alpha}_{\psi\psi})\otimes_{B}\rho_{BC}
	\&\&\&\&
	|[alias=C]|\rho_{AB}\otimes_{B}(\mathbb{C}[\Lambda(G_{BB})]^{\alpha}_{\psi\psi}\otimes_{B}\rho_{BC})
	\\\\
	\&\&
	|[alias=BB]|\rho_{AB}\otimes_{B}\rho_{BC}
	\arrow[from=A,to=C,"\Phi_{ABBC}"]
	\arrow[from=C,to=BB,"\cong"]
	\arrow[from=A,to=BB,"\cong"']
	\end{tikzcd}
\end{align}
as $\mathbb{C}[\Lambda(G_{AB})]^{\alpha}_{\phi\varphi}$-modules for all $\mathbb{C}[\Lambda(G_{AB})]^{\alpha}_{\phi\psi}$-modules $\rho_{AB}$ and $\mathbb{C}[\Lambda(G_{BC})]^{\alpha}_{\psi\varphi}$-modules $\rho_{BC}$.

\subsection{Canonical basis for (2+1)d boundary excited states\label{sec:canonicalBasis}}

So far we have been dealing with the groupoid algebra $\grAlgL$, which is isomorphic to the (3+1)d tube algebra derived in \eqref{sec:tube3D}. We have defined its simple modules, which classify elementary string-like excitations terminating at gapped boundaries, and introduced a comultiplication map that defines a notion of concatenation for these string-like excitations. Furthermore, we constructed the Clebsch-Gordan series and $6j$-symbols associated with this comultiplication map. As mentioned earlier, we have been using the language of relative groupoid algebras, since it unifies both the tube algebras in (2+1)d and in (3+1)d. More specifically, we explained earlier how to obtain the (2+1)d algebra from  the (3+1)d one by restricting the object in $\Lambda(G_{AB})$ to group variables in $G$ and by replacing the loop groupoid 3-cocycle $\alpha \equiv \sfT(\pi) \in Z^3(\Lambda  G, \rU(1))$, where $\pi \in Z^4(G,\rU(1))$, by a group 3-cocycle $\alpha \in Z^3(G,\rU(1))$. We shall now use this mechanism to adapt all the notions derived so far to the study of elementary point-like excitations at the interface between two gapped boundaries in (2+1)d. Thanks to our formulation, the notations remain almost identical. Concretely, it simply amounts to replacing $\fr g \in {\rm Ob}(\Lambda(G_{AB}))$ by $g \in G$, and $(\fr a, \fr b) \in \Lambda A \times \Lambda B$ by $(a,b) \in A \times B$, and to picking $\alpha$ in $H^3(G,\rU(1))$, the other cocycle data descending from it. Note that replacing $(\fr a, \fr b)$ by $(a,b)$ is merely formal as we have often identified the morphisms $\fr a$ and $\fr b$ with the group variables they are characterized by for notational convenience.

Using the definition of the representation matrices together with the Clebsch-Gordan series, we shall now illustrate the mathematical structures introduced earlier by defining a \emph{complete} and \emph{orthonormal} basis of excited states for any pattern of elementary point-like excitations in (2+1)d. The same basis can also be used to define ground state subspaces in the absence of excitations. Naturally, the same construction could be carried out in (3+1)d since we have derived all the relevant notions in this case, which encompasses the (2+1)d one. However, we choose to focus in (2+1)d where it is easier to visualise the construction.

First, let us derive the canonical basis for a pair of dual elementary point-like excitations living at the interfaces of two gapped boundaries labelled by the data $(A,\phi)$ and $(B,\psi)$. This corresponds to the situation depicted in \eqref{eq:cyl} so that we are merely looking for a canonical basis for the vector space $\grAlg$. For each simple module labelled by $\rho_{AB}$, this basis is defined by the set of elements $|\rho_{AB} IJ \ra \in  \grAlg$, with $I,J \in \{1,\ldots,d_{\rho_{AB}} \}$, such that
\begin{equation}
\label{eq:FtransCyl}
|\rho_{AB} IJ \ra = \Big(\frac{d_{\rho_{AB}}}{|A||B|}\Big)^{\frac{1}{2}}
\!\!
\sum_{\substack{g \in G \\ (a,b) \in A \times B}}
\!\!
\overline{\mc{D}^{\rho_{AB}}_{IJ}\big( \big|\biGr{g}{a}{b}\big\ra \big)}
\,
\big|\biGr{g}{a}{b}\big\ra \; .
\end{equation}
This transformation defines an isomorphism such that the inverse is provided by the formula
\begin{equation}
\label{eq:FtransInvCyl}
\big|\biGr{g}{a}{b}\big\ra= \Big(\frac{1}{|A||B|}\Big)^{\frac{1}{2}}
\sum_{\rho} d_{\rho_{AB}}^\frac{1}{2}
\sum_{I,J}
{\mc{D}^{\rho_{AB}}_{IJ}\big( \big|\biGr{g}{a}{b}\big\ra \big)}\,
|\rho_{AB} IJ \ra \; .
\end{equation}
The latter formula expresses the fact that a given state describing such point-like boundary excitations can be written as a sum of states describing elementary excitations.
It follows immediately from the orthonormality \eqref{eq:ortho} of the representation matrices that this basis is orthonormal:
\begin{align}
\nn
\big\la \rho_{AB}' I' J' \big| \rho_{AB} I J \big\ra &=
\frac{d_{\rho_{AB}}^{\frac{1}{2}} d_{\rho'_{AB}}^{\frac{1}{2}}}{|A||B|}
\!\!\!\!\!\!
\sum_{\substack{g,g' \in G \\ (a,b),(a',b') \in A \times B}}
\!\!\!\!\!\!
{\mc{D}^{\rho'_{AB}}_{I'J'}\big( \big|\biGr{g'}{a'}{b'}\big\ra}
\overline{\mc{D}^{\rho_{AB}}_{IJ}\big( \big|\biGr{g}{a}{b}\big\ra \big)}
\,
\big\la \biGr{g'}{a'}{b'}\big| \biGr{g}{a}{b} \big\ra
\\
& = \delta_{\rho_{AB}',\rho_{AB}} \, \delta_{I',I} \, \delta_{J',J}
\end{align}
and complete:
\begin{align}
	\nn
	\sum_{\rho_{AB},I,J} \big\la \rho_{AB} I J \big| \rho_{AB} I J \big\ra &=
	\sum_{\rho_{AB},I,J}\frac{d_{\rho_{AB}}}{|A||B|}
	\sum_{\substack{g \in G \\ (a,b) \in A \times B}}
	{\mc{D}^{\rho_{AB}}_{IJ}\big( \big|\biGr{g}{a}{b}\big\ra}
	\overline{\mc{D}^{\rho_{AB}}_{IJ}\big( \big|\biGr{g}{a}{b}\big\ra \big)}
	\\
	& = \sum_{\substack{g \in G \\ (a,b) \in A \times B}} 1 = |G| \cdot |A| \cdot |B| = \big| \grAlg \big| \; .
\end{align}
Crucially, the canonical basis diagonalizes the $\star$-product (see proof in app.~\ref{sec:app_diag}):
\begin{equation}
	\label{eq:diagProd}
	|\rho_{AB} IJ \ra \star | \rho_{AB}' I' J' \ra 
	=
	|A|^\frac{1}{2} |B|^\frac{1}{2}
	\frac{\delta_{\rho_{AB},\rho_{AB}'} \, \delta_{J,I'}}{d_{\rho_{AB}}^\frac{1}{2}}| \rho IJ' \ra \; .
\end{equation}
As a useful corollary, we have that
\begin{align}
	\label{eq:cor1}
	\big|\biGr{g}{a}{b}\big\ra \star
	| \rho_{AB} IJ \ra 
	&= 
	\sum_{I'}
	\mc{D}^{\rho_{AB}}_{II'}\big( \big|\biGr{g}{a}{b}\big\ra \big)
	|\rho I'J \ra
	\\
	\label{eq:cor2}
	| \rho_{AB} IJ \ra
	\star
	\big|\biGr{g}{a}{b}\big\ra 
	&= 
	\sum_{I'}
	\mc{D}^{\rho_{AB}}_{J'J}\big( \big|\biGr{g}{a}{b}\big\ra \big) 
	|\rho IJ' \ra \; .
\end{align}
Let $Z_{\grAlg}$ be the centre of $\grAlg$ consisting of all elements $| \uppsi \ra \in \grAlg$ that satisfy
\begin{equation}
|\uppsi \ra \star \big|\biGr{g}{a}{b}\big\ra 
=
	\big|\biGr{g}{a}{b}\big\ra \star | \uppsi \ra \; , \q \forall \, \big|\biGr{g}{a}{b}\big\ra \in \grAlg \; .
\end{equation}
Let us consider the states 
\begin{equation}
| \rho_{AB} \ra := \frac{1}{d^\frac{1}{2}_{\rho_{AB}}}\sum_{I}| \rho_{AB}II \ra \; .
\end{equation}
It follows immediately from corollaries \eqref{eq:cor1} and \eqref{eq:cor2} that these states are central, i.e.
\begin{equation}
	| \rho_{AB} \ra \star \big|\biGr{g}{a}{b}\big\ra  = \big|\biGr{g}{a}{b}\big\ra  \star | \rho_{AB} \ra \; , \q \forall \, \big|\biGr{g}{a}{b}\big\ra \in \grAlg \; ,
\end{equation}
from which we can easily deduce that $| \rho_{AB} \ra$ form a complete and orthonormal basis for the centre:
\begin{equation}
	Z_{\grAlg} = {\rm Span}_{\mathbb C}\big\{ | \rho_{AB} \ra \big\}_{\forall \, \rho_{AB}} \; .
\end{equation}
We now would like to show that this centre describes the ground state subspace of our model for the \emph{annulus} $\mathbb O$ depicted below:
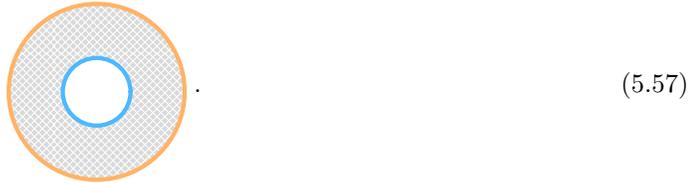
\begin{equation}
	\annulus{0.9} \; .
\end{equation}
A triangulation $\mathbb O_\triangle$ for $\mathbb O$ can be inferred from $\mathfrak{T}[\mathbb I]$ defined in \eqref{eq:basisTube} by imposing the identifications $\snum{(0)} \equiv \snum{(1)}$, $\snum{(0')} \equiv \snum{(1')}$ and $\snum{(00')} \equiv \snum{(11')}$. It further follows that we can identify the space of coloured graph-states on $\mathbb O_\triangle$ as the subspace of coloured graph-states on $\mathfrak{T}[\mathbb I]$ that satisfy $g = a^{-1}gb$. The ground state subspace can be finally obtained by enforcing the twisted gauge invariance at the two vertices via the Hamiltonian projector $\mathbb P_{\mathbb O_\triangle}$. This operator can be easily deduced from the one appearing in the definition of the (2+1)d open tube algebra:
\begin{equation}
	\label{eq:annProj}
	\mathbb P_{\mathbb O_\triangle} =
	\frac{1}{|A||B|}
	\sum_{\substack{g \in G \\ (a,b) \in A \times B}}
	\sum_{(\tilde a,\tilde b) \in A \times B} 
	\delta_{g,a^{-1}gb} \,
	\frac{\vartheta^{AB}_g(a, \tilde a|b , \tilde b)}{\vartheta^{AB}_g(\tilde a , \tilde a ^{-1}a \tilde a| \tilde b, \tilde b^{-1}b \tilde b)} \big|
	\, \biGr{\tilde a^{-1}g\tilde b}{\tilde a^{-1}a \tilde a}{\tilde b ^{-1} b \tilde b} \big\ra \big \la \biGr{g}{a}{b} \big| \; .
\end{equation}
Crucially, this operator can be identically expressed in terms of algebra elements in $\grAlg$ as follows (cf. proof in app.~\ref{sec:app_gsProjAnn})
\begin{equation}
	\mathbb P_{\mathbb O_\triangle}	 
	=
	\frac{1}{|A||B|}
	\sum_{\substack{g \in G \\ (a,b) \in A \times B}}
	\sum_{\substack{\tilde g \in G \\ (\tilde a, \tilde b) \in A \times B}}
	\Big( \big| \biGr{\tilde g}{\tilde a}{\tilde b} \big\ra^{-1} \star \big| \biGr{g}{a}{b} \big\ra \star \big| \biGr{\tilde g}{\tilde a}{\tilde b} \big\ra \Big) \big\la \biGr{g}{a}{b} \big| \; ,
\end{equation}
where
\begin{equation}
	\big| \biGr{\tilde g}{\tilde a}{\tilde b} \big\ra^{-1} = 	\frac{1}{\vartheta^{AB}_{\tilde g}(\tilde a,\tilde a^{-1}|\tilde b,\tilde b^{-1})} 	\big| \biGr{\tilde a^{-1}\tilde g \tilde b}{\tilde a^{-1}}{\tilde b^{-1}} \big\ra \; .
\end{equation}
Note furthermore that we can express the identity algebra element in $\grAlg$ as
\begin{equation}
	\big| \mathbbm 1_{AB} \big\ra = \sum_{\substack{\tilde g \in G \\ (\tilde a, \tilde b) \in A \times B}} \big| \biGr{\tilde g}{\tilde a}{\tilde b} \big\ra^{-1}  \star \big| \biGr{\tilde g}{\tilde a}{\tilde b} \big\ra
\end{equation}
such that
\begin{equation}
	\big| \mathbbm 1_{AB} \big\ra \star \big| \biGr{g}{a}{b} \big\ra
	= \big| \biGr{g}{a}{b} \big\ra
	= \big| \biGr{g}{a}{b} \big\ra \star \big| \mathbbm 1_{AB} \big\ra \; , \q \forall \, \big| \biGr{g}{a}{b} \big\ra \in \grAlg \; .
\end{equation}
It implies that the image of the Hamiltonian projector $\mathbb P_{\mathbb O_\triangle}$ is spanned by states $| \uppsi \ra \in \grAlg$ satisfying
\begin{equation}
	| \uppsi \ra \star \big|\biGr{g}{a}{b}\big\ra  = \big|\biGr{g}{a}{b}\big\ra  \star | \uppsi \ra \; , \q \forall \, \big|\biGr{g}{a}{b}\big\ra \in \grAlg \; ,
\end{equation}
which is precisely the definition of the centre of $| \uppsi \ra \in \grAlg$. We deduce that the ground state subspace on $\mathbb O_\triangle$ is spanned by the states $| \rho_{AB} \ra$:
\begin{equation}
	\mc{V}^{G,A,B}_{\alpha, \phi, \psi}[\mathbb O_\triangle] = {\rm Im} \, \mathbb P_{\mathbb O_\triangle} = Z_{\grAlg} = {\rm Span}_{\mathbb C}\big\{ | \rho_{AB} \ra\big\}_{\forall \, \rho_{AB}} \; .	
\end{equation}
As an immediate consequence of this statement is the fact that the ground state degeneracy of the annulus equals the number of elementary boundary point-like excitations at the interface of two gapped boundaries. This mimics the well-know result that the number of bulk point-like excitations equals the ground state degeneracy on the torus.

\bigskip \noindent 
Let us pursue our construction by defining the canonical basis associated with the following configuration:
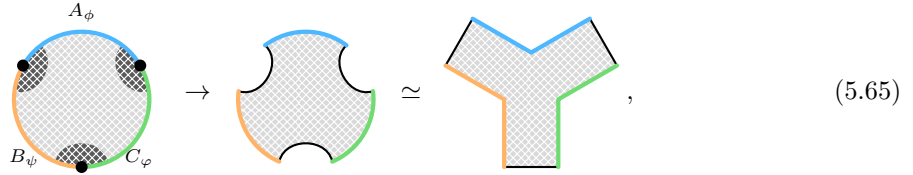
\begin{equation}
\trinion{0.9}{1} \;\; \to \;\; \trinion{0.9}{2} \;\; \simeq \;\; \trinionTr{0.9}{1} \; ,
\end{equation}
i.e. the two-disk $\mathbb D^2$ from which local neighbourhoods at the interface of the three gapped boundaries have been removed. This manifold is referred to as the \emph{thrice-punctured two-disk} and is denoted by $\mathbb Y$. We choose a triangulation $\mathbb Y_\triangle$ for this manifold and consider the following space of coloured graph-states:
\begin{align*}
	{\rm Span}_{\mathbb C} \Bigg\{\Bigg| \mathfrak{g} \Bigg[\!\! \trinionTr{1.2}{3} \!\! \Bigg] \Bigg\ra \!\! \Bigg\}_{\forall \mathfrak{g} \in {\rm Col}(\mathbb Y_\triangle,G,A,B,C)} \!\!\!\!\!\!\!\!\!\!
	&\equiv 
	{\rm Span}_{\mathbb C} \Bigg\{\Bigg| \!\!\!\!  \trinionTr{1.2}{2} \!\!\!\! \Bigg\ra \!\! \Bigg\}_{\substack{\forall \,  g_1,g_2 \in G \\ \hspace{-0.5em}\forall \, a,a' \in A \\ \hspace{-0.6em} \forall \, b,b' \in B \\ \hspace{-0.6em}\forall \, c,c' \in C }} 
	\\
	& \equiv
	| g_1,a,b,g_2,b',c, a',c' \ra_{\mathbb Y_\triangle}\; .
\end{align*}
We are interested in the ground state subspace $\mc{V}^{G,A,B,C}_{\alpha,\phi,\psi,\varphi}[\mathbb Y_\triangle]$ on this manifold. In order to obtain this Hilbert space, we need to apply the Hamiltonian projector $\mathbb P_{\mathbb Y_\triangle}$ simultaneously at all three physical boundary vertices. This operator is obtained by evaluating the partition function \eqref{eq:relativepartitionfunction} on the relative pinched interval cobordism
\begin{equation}
	\vspace{-2em}
	\trinionPinched{0.8}
\end{equation}
and its action explicitly reads
\begin{align}
	\label{eq:projTrinion}
	&\mathbb P_{\mathbb Y_\triangle}\big(| g_1,a,b,g_2,b',c,g_3,a',c' \ra_{{\mathbb Y}_\triangle} \big)
	\\
	\nn
	& \q =
	\frac{1}{|A||B||C|}
	\sum_{\substack{\tilde a \in A \\ \tilde b \in B \\ \tilde c \in C}} \;\;
	\frac{\vartheta^{AC}_{a'g_1g_2c'^{-1}}(a',\tilde a|c', \tilde c)}{\vartheta^{AB}_{g_1}(\tilde a, \tilde a^{-1}a| \tilde b, \tilde b^{-1}b) \, \vartheta^{BC}_{g_2}(\tilde b, \tilde b^{-1}b'| \tilde c, \tilde c^{-1}c) \, \zeta^{ABC}_{\tilde a, \tilde b, \tilde c}(g_1, g_2)} 
	\\[-0.5em]
	& \hspace{7.7em} \times
	| \tilde a^{-1}g_1\tilde b,\tilde a^{-1}a,\tilde b^{-1}b,\tilde b^{-1}g_2 \tilde c,\tilde b^{-1} b',\tilde c^{-1}c,a'\tilde a , c' \tilde c \ra_{{\mathbb Y}_\triangle} \; .
\end{align}
Let us now define the following basis states
\begin{align*}
	&|\rho_{AB}I_{AB},\rho_{BC}I_{BC},\rho_{AC}I_{AC} \ra_{{\mathbb Y}_\triangle}
	\\
	& \q :=
	\sum_{\{g \in G\}} \sum_{\substack{a,a' \in A \\ b,b' \in B \\ c,c' \in C}}\sum_{\{J\}}
	\; \; 
	\overline{\mc{D}^{\rho_{AB}}_{J_{AB}I_{AB}}\big( \big|\biGr{g_1}{a}{b}\big\ra \big)} \, 
	\overline{\mc{D}^{\rho_{BC}}_{J_{BC}I_{BC}} \big( \big|\biGr{g_2}{ b'}{ c}\big\ra\big)}
	\CC{\rho_{AB}}{\rho_{BC}}{\rho_{AC}}{J_{AB}}{J_{BC}}{J_{AC}}
	\overline{\mc{D}^{\rho_{AC}}_{I_{AC}J_{AC}} \big( \big|\biGr{a'g_1g_2c'^{-1}}{a'}{c'}\big\ra\big)}
	\\[-2em]
	& \hspace{8.8em}\times 
	| g_1,a,b,g_2,b',c,a',c' \ra_{{\mathbb Y}_\triangle} \; .
\end{align*}
We can show using the invariance property \eqref{eq:gaugeConj} of the Clebsch-Gordan coefficients that these basis states diagonalise the action of the Hamiltonian projector, i.e. for every $\{\rho_xI_x\}_{x=AB,BC,AC}$ we have
\begin{equation}
	\label{eq:diagTrinion}
	\mathbb P_{\mathbb Y_\triangle}\big(|\rho_{AB}I_{AB},\rho_{BC}I_{BC},\rho_{AC}I_{AC} \ra_{{\mathbb Y}_\triangle}\big) = |\rho_{AB}I_{AB},\rho_{BC}I_{BC},\rho_{AC}I_{AC} \ra_{{\mathbb Y}_\triangle} \; .
	\end{equation}
A proof of this crucial relation can be found in app.~\ref{sec:app_diagProj}. We refer to these states as the canonical basis states for $\mathbb Y_\triangle$. It follows from the orthogonality and the completeness of the representation matrices as well as the Clebsch-Gordan series, that this basis is orthogonal and complete.

It is now possible to use the canonical basis states we have derived so far in order to define excited states associated with more complicated boundary patterns. For instance, the case of $\mathbb D^2$ with four different gapped boundaries can be treated easily by noticing that the manifold resulting from removing local neighbourhoods at every interface can be realised as the gluing of two copies of $\mathbb Y_\triangle$. Similarly, canonical basis states for this manifold are obtained via the $\star$-product by contracting two states of $\mathbb Y_\triangle$ along one magnetic index. Interestingly, two different bases can be defined following this scheme, but they are equivalent. This is ensured by the quasi-coassociativity, and more specifically the isomorphism \eqref{eq:isoVec}. As a matter of fact, the two bases can be explicitly related to each other via the $6j$-symbols as defined in \eqref{eq:defSixJ}, which was the motivation for introducing them. More generally, any number of gapped boundaries can be treated in a similar fashion by gluing several copies $\mathbb Y_\triangle$ according to a \emph{fusion binary tree}. Thanks to the quasi-coassociativity, the choice of tree is not relevant as the corresponding bases are all equivalent.

\tikzset{
	every picture/.append style={
		line cap=butt,
		line width=0.4pt
	}
}

\section{Gapped boundaries and higher algebras}\label{sec:Highercat}

\emph{In this section, we describe a higher categorical construction capturing the salient features of the gapped boundary excitations considered in the previous sections. We begin by reviewing the definitions of monoidal categories and bicategories before introducing the theory of module categories. For more details on such constructions, see for example \cite{etingof2016tensor,schommerpries2011classification,bartlett2009unitary,douglas2018fusion}. Building upon such notions, we then demonstrate the relation between gapped boundary excitations and bicategories of module categories. In particular we review that the bicategory $\mathsf{MOD}(\mathsf{Vec}^{\alpha}_{G})$ provides a convenient description of gapped  boundary excitations in (2+1)d Dijkgraaf-Witten theory \cite{kitaev2012models}, and show that} $\mathsf{MOD}(\mathsf{Vec}^{{\ssfT}(\pi)}_{\Lambda G})$ \emph{describes string-like bulk excitations terminating at the boundary in (3+1)d Dijkgraaf-Witten theory. 
}

\subsection{Higher category theory}

We begin this section by first introducing \emph{higher category theory}. In order to motivate the ethos of higher category theory, it is illuminating to first consider the notion of \emph{categorification}. Generally, categorification refers to a collection of techniques in which statements about sets are translated into statements about categories. Let us consider a simple example. Given a pair of sets $X,Y$ and a triple of functions $f,g,h:X\rightarrow Y$, it is natural to pose relations between such functions in terms of equations. For instance, we may have $f=g$ and $g=h$ as functions from $X$ to $Y$, from which we can infer the relation $f=h$ by transitivity. In this setting, categorification is the process whereby each set $X$ is replaced by a category $\mc{C}_X$, and each function $f:X\rightarrow Y$ is sent to a functor $F_{f}:\mc{C}_{X}\rightarrow \mc{C}_{Y}$. Using the additional structure proper to categories, we have a choice about the way we lift the equations $f=g$ and $g=h$. We could either require the corresponding functors to be equal, i.e. $F_{f}=F_{g}$ and $F_{g}=F_{h}$ implying $F_{f}=F_{h}$, or alternatively, we could instead require only the existence of natural isomorphisms, i.e. $\eta_{fg}:F_{f}\xrightarrow{\sim} F_{g}$ and $\eta_{gh}:F_{g}\xrightarrow{\sim} F_{h}$. In the latter case, we use equations on the natural transformations in order to prescribe a natural transformation $\eta_{fh}=\eta_{fg}\circ\eta_{gh}:F_{f}\rightarrow F_{h}$ replacing transitivity.

Building upon the idea of categorification, let us now introduce \emph{bicategories}, which will form the model of higher category theory utilised in the following discussion. Given a (small) category $\mc{C}$, recall that we denote by ${\rm Hom}_\mc{C}(X,Y)$ the set of (1-)morphisms (hom-set) between the objects $X,Y\in {\rm Ob}(\mc{C})$. Roughly speaking, a bicategory is obtained by applying the categorification mechanism spelt out above to such sets of morphisms. More specifically, we replace ${\rm Hom}_{\mc{C}}(X,Y)$ with a category that we denote by $\mathsf{Hom}_\mc{C}(X,Y)$. The composition function $\circ:{\rm Hom}_\mc{C}(X,Y)\times {\rm Hom}_\mc{C}(Y,X)\rightarrow {\rm Hom}_\mc{C}(X,Z)$ is then replaced with a \emph{composition bifunctor} $\otimes:\mathsf{Hom}_\mc{C}(X,Y)\times \mathsf{Hom}_\mc{C}(Y,Z)\rightarrow \mathsf{Hom}_\mc{C}(X,Z)$. Moreover, equations between morphisms are replaced with natural transformations between functors together with equations defined for such natural transformations. With this idea in mind, we now define our notion of bicategory:
\begin{definition}[\emph{Bicategory}\label{def:bicategory}]
	A bicategory $\mc{B}\mathpzc{i}$ consists of:
	\begin{enumerate}[itemsep=0.4em,parsep=1pt,leftmargin=3em]
		\item[${\ssss \bullet}$] A set of objects ${\rm Ob}(\mc{B}\mathpzc{i})$.
		\item[${\ssss \bullet}$] For each pair of objects $X,Y\in {\rm Ob}(\mc{B}\mathpzc{i})$, a category $\mathsf{Hom}_{\mc{B}\mathpzc{i}}(X,Y)$, whose objects and morphisms are referred to as 1- and 2-morphisms, respectively. Given a 1-morphism $f\in \mathsf{Hom}_{\mc{B}\mathpzc{i}}(X,Y)$, $X =: {\rm s}(f)$ and $Y =: {\rm t}(f)$ are referred to as the `source' and the `target' objects of $f$, respectively. The composition of 2-morphisms in $\mathsf{Hom}_{\mc{B}\mathpzc{i}}(X,Y)$ is designated as the `vertical' composition. 
		\item[${\ssss \bullet}$] For each triple of objects $X,Y,Z\in{\rm Ob}(\mc{B}\mathpzc{i})$, a binary functor $\otimes: \mathsf{Hom}_{\mc{B}\mathpzc{i}}(X,Y) \times \mathsf{Hom}_{\mc{B}\mathpzc{i}}(Y,Z) \to \mathsf{Hom}_{\mc{B}\mathpzc{i}}(X,Z)$ designated as the `horizontal' composition.
		\item[${\ssss \bullet}$] For each object $X\in{\rm Ob}(\mc{B}\mathpzc{i})$, a 1-morphism $\mathbbm 1_{X}\in {\rm Ob}(\mathsf{Hom}_{\mc{B}\mathpzc{i}}(X,X))$, and for each
		morphism $f: X \to Y$, a pair of natural isomorphism $\ell_{f}:\mathbbm  1_{X}\otimes f\rightarrow f$ and $r_{Y}:f\otimes \mathbbm  1_{Y}\rightarrow f$ called the `left' and `right' unitors, respectively.
		\item[${\ssss \bullet}$] For each triple of composable 1-morphisms $f,g,h$, a natural isomorphism $\alpha_{f,g,h}:(f\otimes g)\otimes h\rightarrow f\otimes (g\otimes h)$ called the 1-associator.
	\end{enumerate}

	This data is subject to coherence relations encoded in the commutativity of the diagrams
	\begin{align*}
		\label{eq:pentagonbicat}
		\begin{tikzcd}[ampersand replacement=\&, column sep=1.6em, row sep=1.3em]
			\&\&|[alias=B]|((f\otimes g)\otimes h)\otimes k
			\\\\
			|[alias=AA]|(f\otimes(g\otimes h))\otimes k
			\&\&\&\&
			|[alias=CC]|(f\otimes g)\otimes (h\otimes k)
			\\\\
			|[alias=AAA]|f\otimes((g\otimes h)\otimes k)
			\&\&\&\&
			|[alias=CCC]|f\otimes (g\otimes (h\otimes k))
			\arrow[from=B,to=CC,"\alpha_{f\otimes g,h,k}", sloped]
			\arrow[from=CC,to=CCC,"\alpha_{f,g,h\otimes k}"]
			\arrow[from=B,to=AA,"\alpha_{f,g,h}\otimes{\rm id}_{k} ", sloped]
			\arrow[from=AA,to=AAA,"\alpha_{f,g\otimes h,k}"']
			\arrow[from=AAA,to=CCC,"{\rm id}_{f}\otimes \alpha_{g,h,k}"']
		\end{tikzcd}
	\end{align*}
	and
	\begin{align}
		\begin{tikzcd}[ampersand replacement=\&, column sep=1.6em, row sep=1.3em]
			|[alias=A]|(f\otimes \mathbbm 1_{{\rm t}(f)})\otimes g
			\&\&\&\&
			|[alias=C]|f\otimes (\mathbbm 1_{{\rm t}(f)} \otimes g)
			\\\\
			\&\&
			|[alias=BB]|f\otimes g
			\arrow[from=A,to=C,"\alpha_{f,\mathbbm 1_{{\rm t}(f)},g}"]
			\arrow[from=C,to=BB,"{\rm id}_{f}\otimes \ell_{g}"]
			\arrow[from=A,to=BB,"r_{f}\otimes {\rm id}_{g}"']
		\end{tikzcd}
	\end{align}
	for all composable 1-morphisms $f,g,h,k$, referred to as the pentagon and the triangle relations, respectively.
\end{definition}
\noindent
As in conventional category theory, it is customary to depict relations in a bicategory using diagrammatic calculus. Unlike the directed graph structure utilised in category theory, the diagrammatic presentation of bicategories is given in terms of so-called \emph{pasting diagrams} of the form
\begin{align}
	\begin{tikzcd}[ampersand replacement=\&]
		|[alias=A]|X  \!\!
		\&\&
		|[alias=B]| \phantom{X} \!\!\!\!\!\!\! Y
		%%%
		\arrow[from=A,to=B,"f",bend left=40,""{name=X,below}]
		\arrow[from=A,to=B,"g"',bend right=40,""{name=Y,above}]
		\arrow[from=X,to=Y,Rightarrow,"F"]
	\end{tikzcd}
\end{align}
where $X,Y\in{\rm Ob}(\mc{B}\mathpzc{i})$ are objects, $f,g\in {\rm Ob}(\mathsf{Hom}_{\mc{B}\mathpzc{i}}(X,Y))$ are 1-morphisms and $F\in{\rm Hom}_{\mathsf{Hom}_{\mc{B}\mathpzc{i}}(X,Y)}(f,g)$ is a 2-morphism.
In this notation, horizontal and vertical compositions are depicted as
\begin{align*}
	\begin{tikzcd}[ampersand replacement=\&]
		|[alias=A]|X \!\!
		\&\&
		|[alias=B]| \phantom{X} \!\!\!\!\!\!\! Y
		\&\&
		|[alias=C]| \phantom{X} \!\!\!\!\!\!\! Z
		%%%
		\arrow[from=A,to=B,"f",bend left=40,""{name=X,below}]
		\arrow[from=A,to=B,"g"',bend right=40,""{name=Y,above}]
		\arrow[from=X,to=Y,Rightarrow,"F"]
		\arrow[from=B,to=C,"f'",bend left=40,""{name=XX,below}]
		\arrow[from=B,to=C,"g'"',bend right=40,""{name=YY,above}]
		\arrow[from=XX,to=YY,Rightarrow,"F'"]
	\end{tikzcd}
	=
	\begin{tikzcd}[ampersand replacement=\&]
		|[alias=A]|X  \!\!
		\&\&
		|[alias=C]| \phantom{X} \!\!\!\!\!\!\! Z
		%%%
		\arrow[from=A,to=C,"f\otimes f'",bend left=40,""{name=X,below}]
		\arrow[from=A,to=C,"g\otimes g'"',bend right=40,""{name=Y,above}]
		\arrow[from=X,to=Y,Rightarrow,"F\otimes F'"]
	\end{tikzcd}
	\q {\rm and} \q 
	\begin{tikzcd}[ampersand replacement=\&]
		|[alias=A]|X \hspace{-0.36em} \&\& |[alias=B]| \phantom{X} \!\!\!\!\!\!\! Y
		\arrow[from=A,to=B,bend left=50,"f",""{name=U,below}]
		\arrow[from=A,to=B,""{name=V,above},""{name=W,below}]
		\arrow[from=A,to=B,bend right=50,"f''"',""{name=X,above}]
		\arrow[from=U,to=V,Rightarrow,"F"]
		\arrow[from=W,to=X,Rightarrow,"G"]
	\end{tikzcd}
	=
	\begin{tikzcd}[ampersand replacement=\&]
		|[alias=A]|X \!\!  \&\& |[alias=B]|  \phantom{X} \!\!\!\!\!\!\! Y
		\arrow[from=A,to=B,bend left=40,"f",""{name=U,below}]
		\arrow[from=A,to=B,bend right=40,"f''"',""{name=X,above}]
		\arrow[from=U,to=X,Rightarrow,"FG"]
	\end{tikzcd} ,
\end{align*}
respectively. Explicit examples of bicategories will be provided in sec.~\ref{sec:sAlg} and \ref{sec:MOD}

\subsection{Higher groupoid algebra $\mathsf{Vec}_{\mc G}^\alpha$}

We shall now apply the idea of categorification to groupoid algebras, yielding a notion of `higher groupoid algebra'. First, let us review the relation between monoids and categories. A \emph{monoid} is defined by a set $X$ equipped with a function $\cdot:X\times X\rightarrow X$ called the product, and a distinguished element $\mathbbm 1\in X$ called the unit, satisfying the relations $\mathbbm 1\cdot x=x=x\cdot \mathbbm 1$, $\forall \, x\in X$. Alternatively, a monoid can be defined as a (small) category $\mc{C}$ with a single object $\bul$ and ${\rm Hom}(\mc C) = {\rm Hom}_{\mc{C}}(\bul,\bul)$ such that the composition function $\circ:{\rm Hom}_{\mc{C}}(\bul,\bul)\times {\rm Hom}_{\mc{C}}(\bul,\bul)\rightarrow {\rm Hom}_{\mc{C}}(\bul,\bul)$ provides the monoid product on ${\rm Hom}_{\mc{C}}(\bul,\bul)$, and the identity morphism ${\rm id}_{\bul}$ provides the corresponding monoid unit. Using this presentation of a monoid as a one-object category, we recover upon categorification the notion of \emph{monoidal} category as a one-object bicategory: given a bicategory $\mc{B}\mathpzc{i}$ with a single object ${\rm Ob}(\mc{B}\mathpzc{i})=\{\bul\}$, the category of homorphisms $\mathsf{Hom}_{\mc{B}\mathpzc{i}}(\bul,\bul)$ defines a monoidal category equipped with a tensor product structure provided by the bifunctor $\otimes:\mathsf{Hom}_{\mc{B}\mathpzc{i}}(\bul,\bul)\times \mathsf{Hom}_{\mc{B}\mathpzc{i}}(\bul,\bul)\rightarrow \mathsf{Hom}_{\mc{B}}(\bul,\bul)$. In particular, the 1-associator in $\mc{B}\mathpzc{i}$ induces the (0-)associator in the monoidal category $\mathsf{Hom}_{\mc{B}\mathpzc{i}}(\bul,\bul)$.

Akin to the categorification of a monoid to a monoidal category, one can consider a categorification of an algebra over a field. Instead of presenting the general case, we shall restrict ourselves to the categorification of groupoid algebras. Recall that given a finite groupoid $\mc{G}$, the (complex) groupoid algebra $\mathbb{C}[\mc{G}]$ is the algebra defined over the vector space ${\rm Span}_{\mathbb{C}}\{ \ket{\fr g}\,|\, \forall \,  \fr g\in {\rm Hom}(\mc{G}) \}$ with algebra product $\ket{\fr g}\star \ket{\fr g'} := \delta_{{\rm t}(\fr g),{\rm s}(\fr g')}  \, \ket{\fr g \fr g'} $. One natural categorification of $\mathbb{C}[\mc{G}]$ is given by  replacing the complex field with the (symmetric) monoidal category $\mathsf{Vec}$ of finite dimensional complex-vector spaces, which yields the monoidal category of groupoid-graded vector spaces:
\begin{definition}[\emph{Category of $\mc{G}$-graded vector spaces}]\label{exmp:groupoidgradedvec}
	Let $\mc{G}$ be a finite groupoid. A $\mc{G}$-graded vector space is a vector space of the form $V = \bigoplus_{\fr g \in {{\rm Hom}(\mathcal G)}}V_{\fr g}$. We call a $\mc{G}$-graded vector space $V$ `homogeneous' of degree ${\fr g} \in {\rm Hom}(\mc{G})$ if $V_{\fr g'}$ is the zero vector space $\underline{0}$ for all $\fr g'\neq \fr g$.
	The monoidal category $\mathsf{Vec}_{\mc{G}}$ is then defined as the category whose objects are $\mathcal{G}$-graded complex-vector spaces, and morphisms are grading preserving linear maps. The tensor product is defined on homogeneous components $V_{\fr g}$ and $W_{\fr g'}$ according to
	\begin{align}
		V_{\fr g}\otimes W_{\fr g'}=
		\begin{cases} 
			(V\otimes W)_{\fr g \fr g'}  &\text{if  ${\rm t}(\fr g)={\rm s}(\fr g')$} 
			\\ \underline{0} &\text{otherwise}
		\end{cases} 
	\end{align}
	with unit object $\mathbbm{1}=\bigoplus_{\fr g\in {\rm Hom}(\mc{G})}\delta_{{\rm id}_{\fr g}}$. There are $|{\rm Hom}(\mathcal G)|$ simple objects denoted by $\mathbb{C}_{\fr g}, \forall \, \fr g\in {\rm Hom}(\mathcal G)$.
	Every object is isomorphic to a direct sum of simple objects, making $\mathsf{Vec}_{\mc{G}}$ semi-simple. 
	Finally, the associator is given by the canonical map
	\begin{align}
		{\rm id}_{\mathbb C_{\fr g \fr g'  \fr g''}  }:
		(U_{\fr g}\otimes V_{\fr g'})\otimes W_{\fr g''}
		\xrightarrow{\sim}
		U_{\fr g}\otimes (V_{\fr g'}\otimes W_{\fr g''})\; .
	\end{align}
\end{definition}
\noindent
Note that by choosing the groupoid to be the delooping of a finite group, we recover the more familiar fact that the category of $G$-graded vector spaces is a categorification of the notion of group algebra. 
Analogously to the twisting of a groupoid algebra by a groupoid 2-cocycle, we can twist the associator of $\mathsf{Vec}_{\mc{G}}$ by a normalised groupoid $3$-cocycle $\alpha\in Z^{3}(\mc{G},\rU(1))$ so as to define the monoidal category $\mathsf{Vec}^{\alpha}_{\mc{G}}$, whereby the associator on simple objects is provided by
\begin{align}
	\alpha_{\mathbb{C}_{\fr g},\mathbb{C}_{\fr g'},\mathbb{C}_{\fr g''}}=\alpha(\fr g,\fr g',\fr g'')\cdot {\rm id}_{\mathbb{C}_{\fr g\fr g'\fr g''}}:
	(\mathbb{C}_{\fr g}\otimes\mathbb{C}_{\fr g'})\otimes\mathbb{C}_{\fr g''}
	\xrightarrow{\sim}
	\mathbb{C}_{\fr g}\otimes(\mathbb{C}_{\fr g'}\otimes\mathbb{C}_{\fr g''})\; .
\end{align}
The monoidal category $\mathsf{Vec}^{\alpha}_{\mc{G}}$ has the additional property of being a \emph{multi-fusion category}:
\begin{definition}[\emph{Multifusion category}]
	A category $\mc{C}$ is called multi-fusion if $\mc{C}$ is a finite semi-simple, $\mathbb C$-linear, abelian, rigid monoidal category such that tensor product $\otimes:\mc{C}\times\mc{C}\rightarrow\mc{C}$ is bilinear on morphisms. If additionally ${\rm Hom}_{\mc{C}}(\mathbbm{1},\mathbbm{1})\cong \mathbb C$ then we call $\mc{C}$ a fusion category.
\end{definition}
\noindent
We shall not expand on this definition here, but instead  refer the reader to the chapter 4 of \cite{etingof2016tensor}. Conceptually, the observation that $\mathsf{Vec}^{\alpha}_{\mc{G}}$ is a multi-fusion category plays a similar role to semi-simplicity in the theory of algebras. Recall that given a semi-simple algebra $A$, every module is isomorphic to a direct sum of simple modules. These simple modules can be found via the notion of \emph{primitive orthogonal idempotents}. An idempotent in an algebra $A$ is an element $e\in A$ such that $e\cdot e=e$, and a pair of idempotents $e,e'\in A$ are orthogonal if $e\cdot e'= \delta_{e,e'} \, e$. Such an idempotent is called primitive if it cannot be written as sum of non-trivial idempotents. Specifying a complete set of primitive orthogonal idempotents $\{e_{1},\ldots,e_{n}\}$ for $A$, we can define a simple right $A$-module $M_{i}=e_{i}\cdot A$, for each $i\in 1,\ldots,n$. In the following, we will review the notion of \emph{module category} over a multi-fusion category, categorifying the notion of module over a semi-simple algebra. In this setting the analogue of idempotent will be given by so called \emph{separable algebra objects}.

\subsection{Module categories}

In this part, we introduce the notions of \emph{module category} over multi-fusion category $\mc{C}$, and module category functors following closely \cite{etingof2016tensor}. These happen to be relevant notions to describe gapped boudaries and their excitations \cite{kitaev2012models}. However, as we explain below, we use in practice an equivalent description in terms of separable algebra objects. First, let us define a module category:

\begin{definition}[$\mc{C}$-Module category]
	Given a multi-fusion category $\mc{C}\equiv (\mc C, \otimes , \mathbbm 1, \ell, r, \alpha)$, a (left) $\mc{C}$-module category is defined by a triple $(\mc M , \odot, \dot{\alpha})$ consisting of a category $\mc M$, an action bifunctor $\odot:\mc{C}\times \mc{M}\rightarrow \mc{M}$ and a natural isomorphism
	\begin{align}
		\dot{\alpha}_{X,Y,M}:(X\otimes Y)\odot M\xrightarrow{\sim}X\odot (Y\odot M) \, ,
		\q \forall \,  X,Y\in {\rm Ob}(\mc{C})\;  {\rm and}\; M\in {\rm Ob}(\mc{M}) \, ,
	\end{align}
	referred to as the module associator, such that the diagram
	\begin{align}\label{eq:modcatpentagon}
		\begin{tikzcd}[ampersand replacement=\&, column sep=1.6em, row sep=1.3em]
			\&\&|[alias=B]|((X\otimes Y)\otimes Z)\odot M
			\\\\
			|[alias=AA]|(X\otimes(Y\otimes Z))\odot M
			\&\&\&\&
			|[alias=CC]|(X\otimes Y)\odot (Z\odot M)
			\\\\
			|[alias=AAA]|X\odot((Y\otimes Z)\odot M)
			\&\&\&\&
			|[alias=CCC]|X\odot (Y\odot (Z\odot M))
			\arrow[from=B,to=CC,"\!\! \dot{\alpha}_{X\otimes Y,Z,M}", sloped]
			\arrow[from=CC,to=CCC,"\dot{\alpha}_{X,Y,Z\odot M}"]
			\arrow[from=B,to=AA,"\alpha_{X,Y,Z}\otimes{\rm id}_{M} \; ", sloped]
			\arrow[from=AA,to=AAA,"\dot{\alpha}_{X,Y\otimes Z,M}"']
			\arrow[from=AAA,to=CCC,"{\rm id}_{X}\otimes \dot{\alpha}_{Y,Z,M}"']
		\end{tikzcd}
	\end{align}
	commutes for every $X,Y,Z\in {\rm Ob}(\mc{C})$ and $M\in {\rm Ob}(\mc{M})$. Additionally there is a \emph{unit isomorphism}
	$\ell_{M}:\mathbbm{1}\odot M\xrightarrow{\sim} M$, 
	where $\mathbbm{1}$ is the tensor unit of $\mc{C}$,
	such that the following diagram commutes:
	\begin{align}\label{eq:modcattriangle}
		\begin{tikzcd}[ampersand replacement=\&, column sep=1.6em, row sep=1.3em]
			|[alias=A]|(X\otimes \mathbbm 1)\odot M
			\&\&\&\&
			|[alias=C]|X\otimes (\mathbbm 1\odot M)
			\\\\
			\&\&
			|[alias=BB]|X\odot M
			\arrow[from=A,to=C,"\dot{\alpha}_{X,\mathbbm 1,M}"]
			\arrow[from=C,to=BB,"{\rm id}_{X}\otimes \ell_{M}"]
			\arrow[from=A,to=BB,"r_{X}\otimes {\rm id}_{M}"']
		\end{tikzcd} \; ,
	\end{align}
	for all $X\in {\rm Ob}(\mc{C})$, $M\in {\rm Ob}(\mc{M})$.
\end{definition}
\noindent 
Every module category can be decomposed into so-called \emph{indecomposable} module categories \cite{2001math.....11139O}:
\begin{definition}[Indecomposable module category]
	A $\mc{C}$-module category $\mc{M}$ is said to be `indecomposable' when $\mc{M}$ is not equivalent to a direct sum of non-zero $\mc{C}$-module categories.
\end{definition}

\noindent
Indecomposable module categories will turn out to be the relevant data to label gapped boundaries. To describe excitations, we further require the notion module category functors:

\begin{definition}[Module category functor]
	Given a multi-fusion category $\mc{C}$ and a pair $(\mc M_1, \mc M_2)$ of $\mc C$-module categories with module associators $\dot{\alpha}$ and $\ddot{\alpha}$, respectively, a $\mc C$-module functor is a pair $(F,s)$ where $F: \mc M_1 \to \mc M_2$ is a functor, and $s$ is natural isomorphism given by
	\begin{align}
		s_{X,M}:F(X\odot M)\rightarrow X\odot F(M) \, , \q \forall \, X \in {\rm Ob}(\mc C) \; {\rm and} \; M \in {\rm Ob}(\mc M_1) \; ,
	\end{align}
	such that the diagram
	\begin{align}
		\begin{tikzcd}[ampersand replacement=\&, column sep=1.8em, row sep=1.3em]
			|[alias=A]|F(X\odot (Y\odot M))
			\&\&
			|[alias=B]|F((X\otimes Y)\odot M)
			\&\&
			|[alias=C]|(X\otimes Y)\odot F(M)
			\\
			\\
			|[alias=AA]|X\odot F(Y\odot M)
			\&\&
			\&\&
			|[alias=CC]|X\odot(Y\odot F(M))
			\arrow[from=B,to=A,"{F(\dot{\alpha}_{X,Y,M})}"']
			\arrow[from=A,to=AA,"{s_{X,Y\otimes M}}"']
			\arrow[from=AA,to=CC,"{{\rm id}_{X}\odot s_{Y,M}}"']
			\arrow[from=B,to=C,"{s_{X\otimes Y,M}}"]
			\arrow[from=C,to=CC,"{\ddot{\alpha}_{X,Y,F(M)}}"]
		\end{tikzcd}	
	\end{align}
	commutes for every $X,Y \in {\rm Ob}(\mc{C})$ and $M\in {\rm Ob}(\mc{M})$.
\end{definition}
\noindent
We are almost ready to define a bicategory, the remaining ingredient is a notion of morphism for module functors:
\begin{definition}[Morphism of module functors]
	Given a multi-fusion category $\mc C$ and two $\mc C$-module functors $(F,s)$ and $(F',s')$, a morphism of module functors between $F$ and $F$' is a natural transformation  $\eta:F \to  F'$ such that the diagram
	\begin{align}
		\begin{tikzcd}[ampersand replacement=\&, column sep=1.8em, row sep=1.3em]
			|[alias=A]|F(X\odot M)
			\&\&
			|[alias=B]|X\odot F(M)
			\\\\
			|[alias=AA]|F'(X\odot M)
			\&\&
			|[alias=BB]|X\odot F'(M)
			\arrow[from=A,to=B,"s_{X,M}"]
			\arrow[from=B,to=BB,"{\rm id}_{X}\odot\eta_{M}"]
			\arrow[from=A,to=AA,"\eta_{X\odot M}"']
			\arrow[from=AA,to=BB,"s'_{X,M}"]
		\end{tikzcd}
	\end{align}
commutes for every $X \in {\rm Ob}(\mc C)$ and $M \in {\rm Ob}(\mc M)$.
\end{definition}
\noindent
Putting everything together, we obtain the following definition of a bicategory of module categories

\begin{definition}[Bicategory of module categories]
	Given a multi-fusion category $\mc{C}$, we denote by $\mathsf{MOD}(\mc{C})$ the bicategory with objects, $\mc{C}$-module categories, 1-morphisms, $\mc{C}$-module functors, and 2-morphisms, $\mc{C}$-module natural transformations.
\end{definition}
\noindent
The remainder of this section is dedicated to providing a more practical formulation of this bicategory using the fact that for a multi-fusion category $\mc C$, every indecomposable $\mc{C}$-module category is equivalent to the \emph{category of module objects} for a separable algebra object in $\mc{C}$ \cite{etingof2016tensor}. Using this latter formulation, we shall then explain how the bicategory of module categories is indeed the relevant notion to describe gapped boundaries and their excitations in gauge models of topological phases. 

\subsection{Algebra objects in $\mathsf{Vec}^{\alpha}_{\mc{G}}$\label{sec:algObj}}

Let us now present the notion of \emph{algebra objects} in the multi-fusion category $\mathsf{Vec}^{\alpha}_{\mc{G}}$ thought as a categorification of the groupoid algebra over $\mc{G}$. In the subsequent discussion, we will build upon this notion in order to define module categories over higher groupoid algebras as a categorification of modules over semi-simple algebras.

\begin{definition}[Algebra object]
	Given a multi-fusion category $\mc{C}\equiv (\mc C, \otimes , \mathbbm 1, \ell, r, \alpha)$, an (associative) algebra object in $\mc{C}$ is defined by a triple $(A,m,u)$ consisting of an object $A$ as well as morphisms $m:A\otimes A\rightarrow A$ and $u:\mathbbm{1}\rightarrow A$ in $\mc{C}$ referred to as multiplication and unit, respectively, such that the diagrams below commute:
	\begin{enumerate}[itemsep=0.4em,parsep=1pt,leftmargin=3em]
		\item[${\ssss \bullet}$] {\emph Associativity}: 
		\begin{align}
			\begin{tikzcd}[ampersand replacement=\&, column sep=1.8em, row sep=1.3em]
				|[alias=A]|(A\otimes A)\otimes A \&\& |[alias=B]|A\otimes (A\otimes A) \&\& |[alias=C]| A\otimes A
				\\
				\\
				|[alias=AA]|A\otimes A \&\&\&\& |[alias=CC]|A
				\arrow[from=A,to=B,"\alpha"]
				\arrow[from=B,to=C,"{\rm id}_{A}\otimes m"]
				\arrow[from=AA,to=CC,"m"']
				\arrow[from=A,to=AA,"m\otimes {\rm id}_{A}"']
				\arrow[from=C,to=CC,"m"]
			\end{tikzcd},
		\end{align}
		\item[${\ssss \bullet}$] {\emph Unit}:
		\begin{align}
			\begin{tikzcd}[ampersand replacement=\&, column sep=1.8em, row sep=1.3em]
				|[alias=A]|A
				\&\&
				|[alias=B]|A\otimes \mathbbm{1}
				\&\&
				|[alias=C]| A\otimes A
				%%%%%
				\\
				\\
				%%%%%
				|[alias=AA]|\mathbbm{1}\otimes A
				\&\&
				|[alias=BB]|A\otimes A
				\&\&
				|[alias=CC]| A
				%%%%%
				%%%%%
				\arrow[from=A,to=B,"r^{-1}"]
				\arrow[from=B,to=C,"{\rm id}_{A}\otimes u"]
				\arrow[from=C,to=CC,"m"]
				\arrow[from=A,to=AA,"\ell ^{-1}"']
				\arrow[from=AA,to=BB,"u\otimes{\rm id}_{A}"']
				\arrow[from=BB,to=CC,"m"']
				\arrow[sha, from=A,to=CC,"{\rm id}_{A}",sloped]
			\end{tikzcd},
		\end{align}
	\end{enumerate}

	where $\alpha, \ell ,r $ refer to the associator, left unitor and right unitor for the monoidal structure of $\mc{C}$, respectively.
\end{definition}
\noindent
Given the above definition, an important observation is that algebra objects in the fusion category $\mathsf{Vec}$ correspond to associative, unital, finite-dimensional, complex algebras. Let us now consider algebra objects in $\mathsf{Vec}^{\alpha}_{\mc{G}}$. For each $(\mc{G},\alpha)$-subgroupoid $(\mc{A},\phi)$, as defined in sec.~\ref{sec:relatGr}, we construct an algebra object $\mc{A}_{\phi} \equiv (\bigoplus_{\fr{a}\in{\rm Hom}(\mc{A})}\mathbb{C}_{\fr{a}},m,u)$ with multiplication and unit defined according to
\begin{equation*}
	\begin{array}{ccccl}
		m & : & \mc{A}_\phi \otimes \mc{A}_\phi & \to & \mc{A}_\phi
		\\
		& : & {\fr a} \otimes {\fr a'} &\mapsto & \delta_{{\rm t}(\fr a),{\rm s}(\fr a')} \, \phi(\fr a, \fr a')\,  {\fr a \fr a '}
	\end{array} 
	\q {\rm and} \q 
	u(\mathbbm{1}_{\mathsf{Vec}^{\alpha}_{\mc{G}}}):=\sum_{X\in {\rm Ob}(\mc{A}_\phi)}{\rm id}_{X} \; , 
\end{equation*}
respectively. In particular, we remark that the algebra object $\mc{A}_{\phi}$ in $\mathsf{Vec}^{\alpha}_{\mc{G}}$ corresponds to a generalisation of a twisted groupoid algebra over
$\mc{A}$, where the twisting by a 2-cocycle is instead given by the 2-cochain $\phi$. Since $\phi$ is not a groupoid 2-cocycle, algebra objects are not associative as conventional algebras, but instead are only associative within $\mathsf{Vec}^{\alpha}_{\mc{G}}$ due to the condition $d^{(2)}\phi=\alpha^{-1}{\sss |}_{\mc{A}}$. We leave it to the reader to check that every algebra object in $\mathsf{Vec}^{\alpha}_{\mc{G}}$ is in one-to-one correspondence with a $(\mc{G},\alpha)$-subgroupoid and $\mathsf{Vec}^{\alpha}_{\mc{G}}$ algebra objects.

Given an algebra object $A$ in a multi-fusion category $\mc{C}$, we are interested in modules over $A$ referred to as \emph{$A$-module objects}:
\begin{definition}[Right module object]
	Let $\mc{C}$ be a multi-fusion category and $A \equiv (A,m,u)$ an algebra object  in $\mc{C}$. A right module object over $A$ (or right $A$-module) consists of a pair $(M,p)$, with $M \in {\rm Ob}(\mc C)$  and $p:M\otimes A\rightarrow M \in {\rm Hom}(\mc C)$ such that the diagrams below commute:
	\begin{enumerate}[itemsep=0.4em,parsep=1pt,leftmargin=3em]
		\item[${\ssss \bullet}$] {\emph Compatibility}:
		\begin{align}
			\label{eq:compatibilityModuleObject}
			\begin{tikzcd}[ampersand replacement=\&, column sep=1.8em, row sep=1.3em]
				|[alias=A]|(M\otimes A)\otimes A
				\&\&
				\&\&
				|[alias=C]| M\otimes A
				%%%%%
				\\
				\\
				%%%%%
				|[alias=AA]| M\otimes(A\otimes A)
				\&\&
				|[alias=BB]| M\otimes A
				\&\&
				|[alias=CC]| M
				%%%%
				%%%%
				\arrow[from=A,to=C,"p\otimes {\rm id}_{A}"]
				\arrow[from=C,to=CC,"p"]
				\arrow[from=A,to=AA,"\alpha"']
				\arrow[from=AA,to=BB,"{\rm id}_{M}\otimes m"']
				\arrow[from=BB,to=CC,"p"']
			\end{tikzcd} , 
		\end{align} 
		\item[${\ssss \bullet}$] {\emph Unit}:
		\begin{align}
			\begin{tikzcd}[ampersand replacement=\&, column sep=1.8em, row sep=1.3em]
				|[alias=A]|M
				\&\&
				|[alias=B]|M
				%%%%%
				\\
				\\
				%%%%%
				|[alias=AA]|M\otimes \mathbbm 1
				\&\&
				|[alias=BB]|M\otimes A
				\arrow[from=A,to=B,"{\rm id}_{M}"]
				\arrow[from=BB,to=B,"p"']
				\arrow[from=A,to=AA,"r^{-1}"']
				\arrow[from=AA,to=BB,"u"']
			\end{tikzcd} .
		\end{align} 
	\end{enumerate}
\end{definition}
\noindent
Homorphisms between modules over a given algebra object are then defined in an obvious way:
\begin{definition}[Module object homomorphism] Given an algebra object $A$ in a multifusion category $\mc C$, let $(M_{1},p_{1})$ and  $(M_{2},p_{2})$ be two right $A$-modules. An $A$-homomorphism between  these $A$-modules is a morphism $f\in {\rm Hom}_{\mc{C}}(M_{1},M_{2})$ such that the diagram
	\begin{align}
		\begin{tikzcd}[ampersand replacement=\&, column sep=1.8em, row sep=1.3em]
			|[alias=A]|M_{1}\otimes A
			\&\&
			|[alias=B]|M_{2}\otimes A
			%%%
			\\
			\\
			%%%
			|[alias=AA]|M_{1}
			\&\&
			|[alias=BB]|M_{2}
			\arrow[from=A,to=B,"f\otimes{\rm id}_{A}"]
			\arrow[from=B,to=BB,"p_{2}"]
			\arrow[from=A,to=AA,"p_{1}"']
			\arrow[from=AA,to=BB,"f"']
		\end{tikzcd}
	\end{align}
commutes.
\end{definition}
\noindent
It follows from the definition above that $A$-homomorphisms between a pair of $A$-module objects $(M_{1},p_{1})$ and $(M_{1},p_{1})$ in $\mc{C}$ define a subspace of ${\rm Hom}_{\mc{C}}(M_{1},M_{2})$, which is notated via ${\rm Hom}_{A}(M_{1},M_{2})$ in the following. Moreover, composing $A$-homomorphisms yields another $A$-homomorphism so that we can define a category of $A$-modules as follows:
\begin{definition}[Category of module objects]
	Given a multi-fusion category $\mc{C}$ and an algebra object $A=(A,m,u)$, we define the category $\mathsf{Mod}_{\mc{C}}(A)$ as the category with objects $A$-module objects in $\mc{C}$ and morphisms $A$-module homomorphisms.
\end{definition}
\noindent
In a similar vein, we can define a left $A$-module objects and left $A$-module homomorphisms. We leave it to the reader to derive the corresponding axioms. Combining both left and right modules over an algebra object yields the notion \emph{bimodule object}:
\begin{definition}[Bimodule object]
	Let $\mc C$ be a multi-fusion category and $(A,B)$ a pair of algebra objects in $\mc C$. We define an $(A,B)$-bimodule object in $\mc C$ as a triple $(M,p,q)$ such that $(M,p)$ is a right $B$-module object, $(M,q)$ is a left $A$-module object and the diagram	\begin{align}
		\label{eq:compatibilityBimoduleObject}
		\begin{tikzcd}[ampersand replacement=\&, column sep=1.8em, row sep=1.3em]
			|[alias=A]|(A\otimes M)\otimes B
			\&\&
			\&\&
			|[alias=C]|M\otimes B
			\\
			\\
			|[alias=AA]|A\otimes (M\otimes B)
			\&\&
			|[alias=BB]|A\otimes M
			\&\&
			|[alias=CC]|M
			\arrow[from=A,to=C,"q\otimes{\rm id}_{B}"]
			\arrow[from=C,to=CC,"p"]
			\arrow[from=A,to=AA,"\alpha"']
			\arrow[from=AA,to=BB,"{\rm id}_{A}\otimes p"']
			\arrow[from=BB,to=CC,"q"']
		\end{tikzcd}.
	\end{align}
	commutes.
\end{definition}
\noindent
Noticing that the monoidal identity of any multi-fusion category $\mc{C}$ naturally defines an algebra object,  we can identify the $(\mathbbm{1},A)$-bimodule $(M,\ell_{M},p)$, for a given algebra object $A$, with the right $A$-module $(M,p)$, and similarly the $(A,\mathbbm{1})$-bimodule $(M,r_{M},q)$ with the left $A$-module $(M,q)$.

\begin{definition}[Bimodule object homomorphism]
	Let $(M_{1},p_{1},q_{1})$ and $(M_{2},p_{2},q_{2})$ be a pair of $(A,B)$-bimodule objects in a multi-fusion category $\mc{C}$. An $(A,B)$-homomorphism between these $(A,B)$-bimodules is a morphism $f\in{\rm Hom}_{\mc{C}}(M_{1},M_{2})$ such that $f:(M_{1},p_{1})\rightarrow (M_{2},p_{2})$ is a right $B$-module homomorphism, $f:(M_{1},q_{1})\rightarrow (M_{2},q_{2})$ is a left $A$-module homomorphism, and the following diagram commutes:
	\begin{align*}
		\begin{tikzcd}[ampersand replacement=\&, column sep=1.7em, row sep=1.3em]
			|[alias=X]|(A\otimes M_{1})\otimes B
			\&\&\&\&\&\&\&\&
			|[alias=Z]|M_{1}\otimes B
			\\
			\\
			%%%%%%%%%%%%%%%%%%%%
			\&\&|[alias=A]|(A\otimes M_{2})\otimes B
			\&\&
			\&\&
			|[alias=C]|M_{2}\otimes B
			\\[-0.5em]
			\\
			\&\&|[alias=AA]|A\otimes (M_{2}\otimes B)
			\&\&
			|[alias=BB]|A\otimes M_{2}
			\&\&
			|[alias=CC]|M_{2}
			%%%%%%%%%%%%%%%%%%%
			\\
			\\
			|[alias=XX]|A\otimes (M_{1}\otimes B)
			\&\&\&\&
			|[alias=YY]|A\otimes M_{1}
			\&\&\&\&
			|[alias=ZZ]|M_{1}
			%%%%%%%%%
			\arrow[from=A,to=C,"q_{2}\otimes{\rm id}_{B}"]
			\arrow[from=C,to=CC,"p_{2}"]
			\arrow[from=A,to=AA,"\alpha"']
			\arrow[from=AA,to=BB,"{\rm id}_{A}\otimes p_{2}"']
			\arrow[from=BB,to=CC,"q_{2}"']
			%%%%%%%%%%%%%%%%%%%%%%%%%
			\arrow[from=X,to=Z,"q_{1}\otimes{\rm id}_{B}"]
			\arrow[from=Z,to=ZZ,"p_{1}"]
			\arrow[from=X,to=XX,"\alpha"']
			\arrow[from=XX,to=YY,"{\rm id}_{A}\otimes p_{1}"']
			\arrow[from=YY,to=ZZ,"q_{1}"']
			%%%%%%%%%%%%%%%%%%%%%%%%%
			\arrow[from=X,to=A,"\;\; ({\rm id}_{A}\otimes f)\otimes{\rm id}_{B}", sloped]
			\arrow[from=XX,to=AA,"\;\; {\rm id}_{A}\otimes (f\otimes{\rm id}_{B})"',sloped]
			\arrow[from=YY,to=BB,"{\rm id}_{A}\otimes f",]
			\arrow[from=Z,to=C,"f\otimes {\rm id}_{B}"',sloped]
			\arrow[from=ZZ,to=CC,"f",sloped]
			%%%%%%%%%%%%%%%%%%%%%%%%%
		\end{tikzcd}.
	\end{align*}
\end{definition}
\noindent
It follows from the definition that $(A,B)$-homomorphisms between a pair of $(A,B)$-bimodule object $(M_1,p_1,q_1)$ and $(M_2,p_2,q_2)$ in $\mc C$ define a subspace of ${\rm Hom}_{\mc C}(M_1,M_2)$, which will be denoted by ${\rm Hom}_{A,B}(M_1,M_2)$ in the following. Moreover, composing two $(A,B)$-homomorphisms yields another $(A,B)$-homomorphism so that we can define the following category of $(A,B)$-bimodules:
\begin{definition}[Category of bimodule objects]
	Given a multi-fusion category $\mc{C}$ and a pair of algebra objects $A$ and $B$, we define the category $\mathsf{Bimod}_{\mc{C}}(A,B)$ as the category with objects $(A,B)$-bimodules and morphisms $(A,B)$-bimodule homomorphisms.
\end{definition}
\noindent
Let us now go back to our example of interest, namely the higher groupoid algebras $\mathsf{Vec}^{\alpha}_{\mc{G}}$, and describe the corresponding bimodule objects. We consider a pair $(\mc{A},\phi)$, $(\mc{B},\psi)$ of $(\mc{G},\alpha)$-subgroupoids, and the corresponding algebra objects $\mc{A}_{\phi} \equiv (\bigoplus_{\fr{a}\in{\rm Hom}(\mc{A})}\mathbb{C}_{\fr{a}},m_\mc{A},u_\mc{A})$, and $\mc{B}_{\psi} \equiv (\bigoplus_{\fr{b}\in{\rm Hom}(\mc{B})}\mathbb{C}_{\fr{b}},m_\mc{B},u_\mc{B})$. Let $(M,p,q)$ be an $(\mc A_\phi, \mc B_\psi)$-bimodule in $\mathsf{Vec}^\alpha_\mc{G}$ such that $M = \bigoplus_{\fr g \in {\rm Hom}(\mc{G})}M_{\fr g}$, $p: M \otimes \mc{B}_\psi \to M$ and $q: \mc{A}_\phi \otimes M \to M$. Let us consider the $\mathsf{Vec}^{\alpha}_{\mc{G}}$ morphism $\qp \equiv q \circ ({\rm id}_\mc{A} \otimes p)$ such that
\begin{equation*}
	\begin{array}{ccccl}
		\qp & : & \mc{A}_\phi \otimes (M \otimes \mc{B}_\psi) & \to & M
		\\
		& : & \mathbb C_{\fr a} \otimes (M_\fr{g} \otimes \mathbb C_{\fr b}) &\mapsto & 	
		\delta_{{\rm t}(\fr a), {\rm s}(\fr g)} \, \delta_{{\rm s}(\fr b), {\rm t}(\fr g)} \, [M_\fr{g} \triangleright \qp \, (M_{\fr g}, \mathbb C_{\fr a}, \mathbb C_{\fr b})] \in M_{\fr a \fr g \fr b}
	\end{array} 
\end{equation*}
where $\qp \, (M_{\fr g}, \mathbb C_{\fr a}, \mathbb C_{\fr b}) :M_{\fr{g}}\rightarrow M_{\fr{a}\fr{g}\fr{b}}$ is a linear map which includes a $\mc{G}$-grading shift. In virtue of the compatibility conditions satisfied by $p$ and $q$, the diagram
\begin{align}
	\label{eq:bimoduleVecG}
	\begin{tikzcd}[ampersand replacement=\&, column sep=2.2em, row sep=1.3em]
		|[alias=A]|\mc{A}_{\phi} \otimes((\mc{A}_{\phi} \otimes (M \otimes \mc{B}_{\psi}))\otimes \mc B_\psi)
		\&\&\&
		|[alias=B]|\mc{A}_{\phi} \otimes (M \otimes \mc{B}_\psi)
		%%%%%
		\\
		\\
		%%%%%
		|[alias=AA]|\mc{A}_{\phi} \otimes (M \otimes \mc{B}_\psi)
		\&\&\&
		|[alias=BB]|M
		\arrow[from=A,to=B,"{\rm id}_{\mc A}\otimes (\mathrlap{\hspace{1.5pt} p}q \; \otimes {\rm id}_{\mc B})"]
		\arrow[from=B,to=BB,"\mathrlap{\hspace{1.5pt} p}q \;"]
		\arrow[from=A,to=AA,"{}_{\mc A}m_{\mc B}"']
		\arrow[from=AA,to=BB,"\mathrlap{\hspace{1.5pt} p}q \;"']
	\end{tikzcd} ,
\end{align}
commutes, where ${}_{\mc A}m_{\mc B}$ decomposes as 
\begin{align}
	{}_{\mc A}m_{\mc B} :
	\mc{A}_{\phi} \otimes((\mc{A}_{\phi} \otimes (M \otimes \mc{B}_\psi))\otimes \mc B_\psi)
	&\xrightarrow{{\rm id}_{\mc A}\otimes \alpha_{\mc A,M \otimes \mc{B},\mc B}}
	\mc{A}_{\phi} \otimes(\mc{A}_{\phi} \otimes ((M \otimes \mc{B}_\psi)\otimes \mc B_\psi))
	%%%%%%%%%%%%%
	\nonumber\\
	&\xrightarrow{{\rm id_{A}}\otimes({\rm id_{A}}\otimes \alpha_{M,\mc B, \mc B})}
	\mc{A}_{\phi} \otimes(\mc{A}_{\phi} \otimes (M\otimes (\mc B_\psi \otimes \mc B_\psi)))
	%%%%%%%%%%%%
	\nonumber\\
	&\xrightarrow{{\rm id}_{\mc A}\otimes ({\rm id}_{\mc A}\otimes ({\rm id}_{M}\otimes m_{\mc B}))}
	\mc{A}_{\phi} \otimes(\mc{A}_{\phi} \otimes (M \otimes \mc{B}_\psi))
	%%%%%%%%%%%%
	\nonumber\\
	&\xrightarrow{\alpha^{-1}_{\mc A,A,M \otimes \mc{B}}}
	(\mc{A}_{\phi} \otimes \mc A_\phi)\otimes (M \otimes \mc{B}_\psi)
	%%%%%%%%%%%%%
	\nonumber\\
	&\xrightarrow{m_{\mc A}\otimes {\rm id}_{M \otimes \mc{B}}}
	\mc{A}_{\phi} \otimes(M \otimes \mc{B}_\psi) \; .
\end{align}
Furthermore, it acts on non-zero basis vectors $(\fr a , \fr a', \fr b , \fr b') \in \mathbb{C}_{\mathfrak{a}} \times \mathbb C_{\fr a'} \times \mathbb C_{\fr b} \times \mathbb C_{\fr b'}$ and $v_{\fr g}\in M_{\fr g}$ as
\begin{align}\label{eq:bimodactioncompositon}
	{}_{\mc A}m_{\mc B} :
	\fr{a}' \otimes((\fr{a}\otimes (v_{\fr{g}}\otimes \fr{b}))\otimes \fr{b'})
	\mapsto
	\delta_{{\rm t}(\fr{a}'),{\rm s}(\fr{a})} \,
	\delta_{{\rm t}(\fr{b}),{\rm s}(\fr{b}')} \,
	\varpi_{\fr g}^{\mc A \mc B}(\fr a, \fr a' | \fr b , \fr b') \, 
	[\fr{a'a}\otimes (v_{\fr{g}}\otimes \fr{bb'})]
\end{align}
for any set of $\fr{a}',\fr{a},\fr{g},\fr{b},\fr{b}'$ composable morphisms in $\mc{G}$,
where we introduced the cocycle data
\begin{equation}
	\varpi_{\fr g}^{\mc A \mc B}(\fr a, \fr a' | \fr b , \fr b')
	 :=
	 \frac{\alpha(\fr{a},\fr{gb},\fr{b}')\, \alpha(\fr{g},\fr{b},\fr{b}')}{\alpha(\fr{a}',\fr{a},\fr{gbb}')}\phi(\fr{a}',\fr{a})\psi(\fr{b},\fr{b}') \; .
\end{equation}
Writing
\begin{equation*}
	\begin{array}{ccccl}
		\qp & : & \mc{A}_\phi \otimes (M \otimes \mc{B}_\psi) & \to & M
		\\
		& : & \fr a  \otimes (v_\fr{g} \otimes  \fr b) &\mapsto & 	
		 v_\fr{g} \triangleright \qp \, (v_{\fr g},\fr a , \fr b)\in M_{\fr a \fr g \fr b}
	\end{array} \; , 
\end{equation*}
it follows from equation \eqref{eq:bimodactioncompositon} that $\qp \, (v_{\fr g},\fr a , \fr b) \in{\rm End}(M)$ satisfies the algebra
\begin{align}\label{eq:weakcompositionbimodaction}
	\qp \, (v_{\fr g},\fr a , \fr b)
	\triangleright
	\qp \, (v_{\fr g'},\fr a' , \fr b')
	=
	\delta_{{\rm t}(\fr{a}'),{\rm s}(\fr{a})} \,
	\delta_{{\rm t}(\fr{b}),{\rm s}(\fr{b}')} \,
	\delta_{\fr g',\fr a \fr g \fr b} \,
	\varpi_{\fr g}^{\mc A \mc B}(\fr a, \fr a' | \fr b , \fr b')
	\, \qp \, (v_{\fr g},\fr a '\fr a , \fr b \fr b')
\end{align}
for all $\fr{g},\fr{g'}\in {\rm Hom}_{\mc{G}}({\rm Ob}(\mc{A}),{\rm Ob}(\mc{B})) )$, ${\fr{a}}\in {\rm Hom}_{\mc{A}}(-,{\rm s}(\fr{g}))$, ${\fr{a}'}\in {\rm Hom}_{\mc{A}}(-,{\rm s}(\fr{g'}))$, ${\fr{b}}\in {\rm Hom}_{\mc{B}}({\rm t}(\fr{g}),-)$ and ${\fr{b'}}\in {\rm Hom}_{\mc{B}}({\rm t}(\fr{g'}),-)$. Such data can be concisely described by introducing the groupoid  $\widetilde{\mc{G}}_{{\mc A} \mc{B}}$ with object set ${\rm Hom}_{\mc{G}}({\rm Ob}(\mc A),{\rm Ob}(\mc B))$ and morphism set given by
\begin{align}
	\biGr{\fr g}{}{\fr a, \fr b}
	\fr{a}\fr{g}\fr{b}
	\equiv 
	\biGr{\fr g}{}{\fr a, \fr b} \; ,
\end{align}
for all $\fr{g}\in {\rm Ob}(\widetilde{\mc{G}}_{\mc A \mc{B}})$, $\fr{a}\in {\rm Hom}_{\mc A}(-,{\rm s}(\fr{g}))$ and $\fr{b}\in {\rm Hom}_{\mc B}({\rm t}(\fr{g}),-)$.
Composition is defined by
\begin{equation}
	\biGr{\fr g}{}{\fr a, \fr b}
	\biGr{\fr{a}\fr{g}\fr{b}}{}{\fr a', \fr b'}
	\fr{a}'\fr{a}\fr{g}\fr{b}\fr{b}'
	=
	\biGr{\fr g}{}{\fr a' \fr a, \fr b \fr b'}
	\fr{a}'\fr{a}\fr{g}\fr{b}\fr{b}' \; ,	
\end{equation}  
for all composable pairs $(\fr{a}',\fr{a}) \in \mc A^2_{\rm comp.}$ and $(\fr{b},\fr{b}') \in \mc B^2_{\rm comp.}$.
Noting that $[\varpi^{\mc{A}\mc{B}}] \in H^{2}(\widetilde{\mc{G}}_{\mc{A} \mc{B}} ,\rU(1))$ defines a $\widetilde{\mc{G}}_{\mc{A} \mc{B}}$ 2-cocycle, $\qp$ can then be described via a weak functor
\begin{equation*}
	\begin{array}{ccccl}
		F_{\mathrlap{\hspace{1.5pt} p}q_{\, M}} & : & \widetilde{\mc{G}}_{\mc{A} \mc{B}} & \to & \mathsf{Vec}
		\\
		& : & \fr  g \in {\rm Ob}(\widetilde{\mc{G}}_{\mc{A} \mc{B}}) &\mapsto & 	
		M_{\fr g} \subset M
		\\
		& : & \biGr{\fr g}{}{\fr a, \fr b} \;  \in {\rm Hom}(\widetilde{\mc{G}}_{\mc{A} \mc{B}}) & \mapsto & \qp \, (v_{\fr g}, \fr a , \fr b):M_{\fr g} \to M_{\fr a \fr g \fr b}
	\end{array} \; , 
\end{equation*}
such that every isomorphism $\qp \, (v_{\fr g}, \fr a , \fr b)$ satisfies the composition relation \eqref{eq:weakcompositionbimodaction}. Using the equivalence between representations and modules of algebraic structures, we can thus view the pair $(M, F_{\mathrlap{\hspace{1.5pt} p}q_{\, M}})$ as a module over the twisted groupoid algebra $\mathbb{C}[\widetilde{\mc{G}}_{\mc{A} \mc{B}}]^{\varpi^\mc{AB}}$.
Considering the diagram
\begin{align}
	\begin{tikzcd}[ampersand replacement=\&, column sep=1.8em, row sep=1.3em]
		|[alias=A]|\mc A_\phi \otimes(M_{1}\otimes \mc B_\psi)
		\&\&\&
		|[alias=B]|\mc A_\phi \otimes (M_{2}\otimes \mc B_\psi)
		%%%%%
		\\
		\\
		%%%%%
		|[alias=AA]|M_{1}
		\&\&\&
		|[alias=BB]|M_{2}
		\arrow[from=A,to=B,"{\rm id}_{\mc A}\otimes (f\otimes {\rm id}_{\mc B})"]
		\arrow[from=B,to=BB,"\mathrlap{\hspace{1.5pt} p}q_{\;2}"]
		\arrow[from=A,to=AA,"\mathrlap{\hspace{1.5pt} p}q_{\; 1}"']
		\arrow[from=AA,to=BB,"f"']
	\end{tikzcd} , 
\end{align}
for a pair of $(\mc A_\phi, \mc B_\psi)$-bimodules $(M_{1},\qp_{\, 1}(-))$ and $(M_{2},\qp_{\, 2}(-))$,
we conclude that an $(\mc A_\phi, \mc B_\psi)$-bimodule homomorphism is defined via a natural transformation $f:\qp_1 \to \qp_2$, or equivalently, as an \emph{intertwiner} for representations of $\mathbb{C}[\widetilde{\mc{G}}_{\mc{A} \mc{B}}]^{\varpi^\mc{AB}}$. Putting everything together, we obtain the equivalence $\mathsf{Bimod}_{\mathsf{Vec}^\alpha_\mc{G}}(\mc{A}_\phi,\mc{B}_\psi) \simeq \mathsf{Mod}(\mathbb{C}[\widetilde{\mc{G}}_{\mc{A} \mc{B}}]^{\varpi^\mc{AB}})$.

\subsection{Bicategory of separable algebra objects in $\mathsf{Vec}^{\alpha}_{\mc{G}}$}\label{sec:sAlg}

Pursuing our construction, we shall now introduce a special class of algebra objects known as \emph{separable algebra objects}. We will then construct a bicategory whose objects are separable objects, and morphisms are bimodule objects between them. First, let us define what it means for an algebra object to be separable:

\begin{definition}[Separable algebra object] 
	Let $\mc C$ be a multi-fusion category and $A \equiv (A,m,u)$ an algebra object in $\mc C$. The algebra object $A$ is said to be `separable' if the multiplication map $m: A\otimes A \to A$ admits a `section' map $\Delta: A \to A\otimes A$ such that
	\begin{align*}
		A\xrightarrow{\Delta}A\otimes A\xrightarrow{m}A=A\xrightarrow{{\rm id}_{A}}A \; ,
	\end{align*}
	as an $(A,A)$-bimodule homomorphism.
\end{definition}
\noindent
Let us now define a binary functor. Let $A,B,C$ be three separable algebra objects in a multi-fusion category $\mc{C}$, $M_{AB} \equiv (M_{AB},q_{A},p_{B}) \equiv (M_{AB} , \qp_{M_{AB}})$ an $(A,B)$-bimodule, and $M_{BC} \equiv (M_{BC},q_{B},p_{C}) \equiv (M_{BC}, \qp_{M_{BC}})$ a $(B,C)$-bimodule. Using this data, we want to construct an $(A,C)$-bimodule, which we shall notate via $(M_{AB} \otimes_B M_{BC} , \qp_{M_{AB}} \otimes_B \qp_{M_{BC}})$. First, let us define the morphism $\qp_{M_{AB}\otimes M_{BC}}: A\otimes ((M_{AB}\otimes M_{BC})\otimes C ) \to M_{AB}\otimes M_{BC}$ that decomposes as
\begin{align}
	\label{eq:bimodcompbimod}
	&A\otimes ((M_{AB}\otimes M_{BC})\otimes C )
	\xrightarrow{{\rm id}_{A}\otimes \alpha_{M_{AB},M_{BC},C}}
	A\otimes (M_{AB}\otimes (M_{BC}\otimes C ))
	\\
	\nn
	& \q \xrightarrow{\alpha^{-1}_{A,M_{AB},M_{BC}\otimes C}}
	(A\otimes M_{AB})\otimes (M_{BC}\otimes C )
	\xrightarrow{  ( \ell_{A \otimes M_{AB}}^{-1})\otimes {\rm id}_{M_{BC}\otimes C}}
	((A\otimes M_{AB})\otimes \mathbbm{1})\otimes (M_{BC}\otimes C )
	\\
	\nn
	& \q \xrightarrow{  ({\rm id}_{A\otimes M_{AB}}\otimes u_B)\otimes {\rm id}_{M_{BC}\otimes C}} 
	((A\otimes M_{AB})\otimes B)\otimes (M_{BC}\otimes C )
	\\
	\nn
	& \q \xrightarrow{  ({\rm id}_{A\otimes M_{AB}}\otimes \Delta_B)\otimes {\rm id}_{M_{BC}\otimes C}} 
	((A\otimes M_{AB})\otimes (B\otimes B))\otimes (M_{BC}\otimes C )
	\\
	\nn
	& \q \xrightarrow{\alpha^{-1}_{A\otimes M_{AB},B,B}\otimes {\rm id}_{M_{BC}\otimes C}}
	(((A\otimes M_{AB})\otimes B)\otimes B)\otimes (M_{BC}\otimes C )
	\\
	\nn
	& \q \xrightarrow{(\alpha_{A,M_{AB},B}\otimes {\rm id}_{B})\otimes {\rm id}_{M_{BC}\otimes C}}
	((A\otimes (M_{AB}\otimes B))\otimes B)\otimes (M_{BC}\otimes C )
	\\
	\nn
	& \q \xrightarrow{\alpha_{A\otimes (M_{AB}\otimes B),B,M_{BC}\otimes C}}
	(A\otimes (M_{AB}\otimes B))\otimes (B\otimes (M_{BC}\otimes C ))
	\xrightarrow{\mathrlap{\hspace{1.5pt} p}q_{\; M_{AB}} \otimes \mathrlap{\hspace{1.5pt} p}q_{\; M_{BC}}}
	M_{AB}\otimes M_{BC} \; .
\end{align}
Using this morphism, let us further define the endomorphism $e_{M_{AB} \otimes M_{BC}} : M_{AB} \otimes M_{BC} \to M_{AB} \otimes M_{BC}$ that decomposes as 
\begin{align}
	\label{eq:seperabilityidempotent}
	M_{AB}\otimes M_{BC}
	&\xrightarrow{r^{-1}_{M_{AB}\otimes M_{BC}}}
	(M_{AB}\otimes M_{BC})\otimes \mathbbm{1}
	\xrightarrow{\ell^{-1}_{(M_{AB}\otimes M_{BC})\otimes \mathbbm{1}}}
	\mathbbm{1}\otimes((M_{AB}\otimes M_{BC})\otimes \mathbbm{1})
	\\
	\nn
	& \xrightarrow{u_{A}\otimes({\rm id}_{M_{AB}\otimes M_{BC}}\otimes u_{C})}
	A\otimes((M_{AB}\otimes M_{BC})\otimes C)
	\xrightarrow{\mathrlap{\hspace{1.5pt} p}q_{\; M_{AB} \otimes M_{BC}} }
	M_{AB}\otimes M_{BC} \; .
\end{align}
By the requirement that $\Delta : B \to B\otimes B$ is a $(B,B)$-bimodule section to the $(B,B)$-bimodule homomorphism $m : B\otimes B \to B$, together with the compatibility conditions spelt out above and the naturalness of the associator $\alpha$, we can show that $e_{M_{AB}\otimes M_{BC}}$ is an \emph{idempotent endomorphism} in $\mc{C}$, i.e. $e_{M_{AB} \otimes M_{BC}} \circ e_{M_{AB} \otimes M_{BC}} = e_{M_{AB} \otimes M_{BC}}$.
The requirement that the multi-fusion category $\mc C$ is abelian ensures that every idempotent is a \emph{split idempotent}: 
\begin{definition}[Split idempotent]
	An idempotent $a \xrightarrow{e} a$ is called split when there exists an object $b$ and morphisms $a\xrightarrow{\mathsf s}b$, $b\xrightarrow{\mathsf r} a$ such that $b\xrightarrow{\mathsf r\circ \mathsf s}b=b\xrightarrow{{\rm id}_{b}}b$ and $a\xrightarrow{\mathsf s\circ \mathsf r}a=a\xrightarrow{e}a$.
\end{definition}
\noindent
We define the object $M_{AB}\otimes_{B}M_{BC}\in\mc{C}$ as a choice of splitting object for the idempotent $e_{M_{AB}\otimes M_{BC}}$ such that 
$M_{AB}\otimes M_{BC}\xrightarrow{\mathsf s_{M_{AB} ,M_{BC}}}M_{AB}\otimes_{B} M_{BC}$ and
$M_{AB}\otimes_{B} M_{BC}\xrightarrow{\mathsf r_{M_{AB} , M_{BC}}}M_{AB}\otimes M_{BC}$, 
where $\mathsf s_{M_{AB} , M_{BC}}\circ \mathsf r_{M_{AB} , M_{BC}}=e_{M_{AB}\otimes M_{BC}}$ and $\mathsf r_{M_{AB} , M_{BC}}\circ \mathsf s_{M_{AB} , M_{BC}}={\rm id}_{M_{AB}\otimes M_{BC}}$.
Crucially, a choice of splitting object is unique up to isomorphism, and independent of a choice of section up to isomorphism. Using this data, let us further define the following morphism:
\begin{align}
	 \qp_{M_{AB}} \otimes_B \qp_{M_{BC}}:=
	\mathsf r_{M_{AB}, M_{BC}}\circ 
	( \qp_{M_{AB} \otimes M_{BC}})
	\circ
	\mathsf s_{M_{AB} ,M_{BC}} \; .
\end{align}
Putting everything together, we obtain that $(M_{AB} \otimes_B M_{BC} , \qp_{M_{AB}} \otimes_B \qp_{M_{BC}})$ defines an $(A,C)$-bimodule in $\mc{C}$. So we have obtained a way to define an $(A,C)$-bimodule out of an $(A,B)$- and a $(B,C)$-bimodule given three separable algebra objects $A,B,C$. This can expressed in terms of the bifunctor
\begin{align}
	\otimes_{B}:\mathsf{Bimod}_{\mc{C}}(A,B)
	\times
	\mathsf{Bimod}_{\mc{C}}(B,C)
	\rightarrow
	\mathsf{Bimod}_{\mc{C}}(A,C) \; ,
\end{align}
where objects $M_{AB}\in {\rm Ob}(\mathsf{Bimod}_{\mc{C}}(A,B))$ and $M_{BC}\in {\rm Ob}(\mathsf{Bimod}_{\mc{C}}(B,C))$ are mapped via
\begin{align}
	\otimes_{B}:M_{AB}\times M_{BC}\mapsto M_{AB}\otimes_{B} M_{BC} \; ,
\end{align}
and bimodule homomorphisms $f_{AB}\in {\rm Hom}(\mathsf{Bimod}_{\mc{C}}(A,B))$, $f_{BC}\in {\rm Hom}(\mathsf{Bimod}_{\mc{C}}(B,C))$ are sent to
\begin{align}
	\otimes_{B}:f_{AB}\times f_{BC}\mapsto f_{AB}\otimes_{B} f_{BC} :=\mathsf s_{M_{AB} , M_{BC}}\circ  (f_{AB}\otimes f_{BC}) \circ \mathsf r_{M_{AB} , M_{BC}} \; .
\end{align}
In order to obtain a bicategory, we are left to define a left unitor, a right unitor and an associator. Considering $A$ and $B$ as $(A,A)$- and $(B,B)$-bimodules, respectively, one can verify that for any $(A,B)$-bimodule $M_{AB}$ 
\begin{align}
	A\otimes_{A} M_{AB}\cong M_{AB}\cong M_{AB}\otimes_{B}B \; ,
\end{align}
as $(A,B)$-bimodule in $\mc{C}$. This property demonstrates that an algebra $A$ seen as an $(A,A)$-bimodule defines a notion a unit morphism for an algebra object $A$. The corresponding left unitor isomorphism, which is an $(A,B)$-bimodule, is defined via the maps
\begin{align*}
	A\otimes_{A}M_{AB}&\xrightarrow{\mathsf r_{A , M_{AB}}}A\otimes M_{AB} \xrightarrow{q_{A}}M_{AB}
\end{align*}
and
\begin{align*}
	M_{AB} & \xrightarrow{\ell^{-1}_{M_{AB}}}\mathbbm{1}\otimes M_{AB}
	\xrightarrow{\Delta\otimes{\rm id}_{M_{AB}}}(A\otimes A)\otimes M_{AB}
	\xrightarrow{\alpha_{A,A,M_{AB}}}A\otimes (A\otimes M_{AB})
	\\
	& \xrightarrow{{\rm id}_{A}\otimes q_{A}}A\otimes M_{AB}
	\xrightarrow{\mathsf s_{A , M_{AB}}}A\otimes_{A} M_{AB} \; ,
\end{align*}
which can be shown to satisfy the triangle relations. The right unitor can be defined in a similar fashion. Finally, for any quadruple of algebra objects $A,B,C,D$ and $(A,B)$-bimodule $M_{AB}$, $(B,C)$-bimodule $M_{BC}$ and $(C,D)$-bimodule $M_{CD}$, the morphism
\begin{align*}
	&(M_{AB}\otimes_{B} M_{BC})\otimes_{C}M_{CD}
	\xrightarrow{\mathsf r_{M_{AB}\otimes M_{BC} , M_{CD}}}
	(M_{AB}\otimes_{B} M_{BC})\otimes M_{CD}
	\\
	& \q \xrightarrow{\mathsf r_{M_{AB} , M_{BC}}\otimes {\rm id}_{M_{CD}} }
	(M_{AB}\otimes M_{BC})\otimes M_{CD}
	\xrightarrow{\alpha_{M_{AB},M_{BC},M_{BC}} }
	M_{AB}\otimes (M_{BC}\otimes M_{CD})
	\\
	& \q \xrightarrow{{\rm id}_{M_{AB}}\otimes \mathsf s_{M_{BC},M_{CD}} }
	M_{AB}\otimes (M_{BC}\otimes_{C}M_{CD})
	\xrightarrow{{\rm id}_{M_{AB}}\otimes \mathsf s_{M_{BC},M_{CD}} }
	M_{AB}\otimes_{B} (M_{BC}\otimes_{C}M_{CD})
\end{align*}
defines an isomorphism of $(A,D)$-bimodules in $\mc{C}$ satisfying the pentagon relation. Putting everything together, we obtain the following bicategory:

\begin{definition}[Bicategory of separable algebra objects]
	Given a multi-fusion $\mc{C}$, we notate via $\mathsf{sAlg}(\mc{C})$ the bicategory with objects, separable algebras objects in $\mc{C}$, and hom-category $\mathsf{Hom}_{\mathsf{sAlg}(\mc{C})}(A,B):=\mathsf{Bimod}_{\mc{C}}(A,B)$ for all separable algebra objects $A,B$ in $\mc{C}$. The composition bifunctor is provided by $\otimes_B:\mathsf{Bimod}_{\mc{C}}(A,B)\times \mathsf{Bimod}_{\mc{C}}(B,C)\rightarrow \mathsf{Bimod}_{\mc{C}}(A,C)$ as defined in this section.
\end{definition}

\bigskip \noindent
Let us now apply the definition above to the multi-fusion category $\mathsf{Vec}^{\alpha}_{\mc{G}}$. First of all, every algebra object in $\mathsf{Vec}^{\alpha}_{\mc{G}}$ can be shown to be separable. Indeed, given an algebra object $\mc A_\phi$ in $\mathsf{Vec}^{\alpha}_{\mc{G}}$, a choice of section $\Delta: \mc A_\phi \to \mc A_\phi\otimes \mc A_\phi$ is provided by the following map on basis elements:
\begin{align}
	\label{eq:section}
	\Delta: \fr{a} \mapsto
	\frac{1}{|{\rm Hom}_{\mc{A}}({\rm s}(\fr{a}),-)|}
	\sum_{\substack{\fr{a}_{1},\fr{a}_{2}\in {\rm Hom}(\mc A)\\ \fr{a}_{1}\fr{a}_{2}=\fr{a}}}
	\frac{1}{\phi(\fr{a}_{1},\fr{a}_{2})}
	\fr{a}_{1}\otimes \fr{a}_{2}\; .
\end{align}
Algebra objects equipped with the section defined above form  the objects of the bicategory $\mathsf{sAlg}(\mathsf{Vec}^\alpha_{\mc G})$. Let $A_\phi$, $B_\psi$, $C_\varphi$ be three objects in  $\mathsf{sAlg}(\mathsf{Vec}^\alpha_{\mc G})$, we consider the 1-morphisms   $M_{\mc A \mc B}\equiv(M_{\mc A \mc B},\qp_{M_{\mc{A}\mc{B}}})\in {\rm Ob}(\mathsf{Bimod}_{ \mathsf{Vec}^{\alpha}_{\mc{G}} }(\mc A_\phi, \mc B_\psi))$ and $M_{\mc B \mc C}\equiv(M_\mc{BC},\qp_{M_{\mc{B}\mc{C}}})\in {\rm Ob}(\mathsf{Bimod}_{ \mathsf{Vec}^{\alpha}_{\mc{G}} }(\mc B_\psi,\mc C_{\varphi}))$. Following \eqref{eq:bimodcompbimod}, the map $\qp_{M_{\mc{A}\mc{B}} \otimes M_\mc{BC}}$ acts on basis elements of $\mathbb C_{\fr a}\otimes (([M_\mc{AB}]_{\fr g_1} \otimes [M_\mc{BC}]_{\fr g_2}) \otimes \mathbb C_{\fr c})$ as

\begin{align*}
	\qp_{M_{\mc{A}\mc{B}} \otimes M_\mc{BC}} : 
	\fr{a}\otimes ((v^{\mc A \mc B}_{\fr{g}_{1}}\otimes v^\mc{BC}_{\fr{g}_{2}})\otimes \fr{c})
	\mapsto \frac{1}{|{\rm Hom}(\mc{B})|}
	\! \sum_{\fr{b}\in {\rm Hom}(\mc{B})} \! 
	& \frac{
		\alpha(\fr{g}_{1},\fr{g}_{2},\fr c) \, \alpha(\fr{a},\fr{g}_{1},\fr{b}) \, \alpha(\fr{ag}_{1}\fr{b},\fr{b}^{-1},\fr{g}_{2}\fr{c})
	}
	{
		\psi(\fr{b},\fr{b}^{-1}) \, \alpha(\fr{a},\fr{g}_{1},\fr{g}_{2}\fr{c}) \, \alpha(\fr{ag}_{1},\fr{b},\fr{b}^{-1})
	}
	\\
	&v^\mc{AB}_{\fr{g}_{1}}\triangleright \qp \, (v^\mc{AB}_{\fr g_1},\fr a , \fr b)
	\otimes 
	v^\mc{BC}_{\fr{g}_{2}}\triangleright \qp \, (v^\mc{BC}_{\fr g_2},\fr b^{-1} , \fr c) \, .
\end{align*}
Applying the formula above to $\fr a = {\rm id}_{\rm s(\fr g_1)}$ and $\fr c = {\rm id}_{\rm t(\fr g_2)}$, we obtain that the map $\mathsf s_{M_\mc{AB} , M_\mc{BC}}:M_\mc{AB}\otimes M_\mc{BC}\rightarrow M_\mc{AB}\otimes_\mc{B} M_\mc{BC}$ acts on basis elements as
\begin{align*}
	\mathsf s_{M_\mc{AB}, M_\mc{BC}} : 
	v^\mc{AB}_{\fr{g}_{1}} \otimes v^\mc{BC}_{\fr{g}_{2}} 
	\mapsto
	\frac{1}{|{\rm Hom}(\mc{B})|}
	\sum_{\fr{b}\in {\rm Hom}(\mc{B})}
	&\frac{1}{\psi(\fr{b},\fr{b}^{-1})}
	\frac{
		\alpha(\fr{g}_{1}\fr{b},\fr{b}^{-1},\fr{g}_{2})
	}
	{
		\alpha(\fr{g}_{1},\fr{b},\fr{b}^{-1})
	}
	\\
	&v^\mc{AB}_{\fr{g}_{1}}\triangleright \qp \, (v^\mc{AB}_{\fr g_1},{\rm id}_{\rm s(\fr g_1)} , \fr b)
	\otimes 
	v^\mc{BC}_{\fr{g}_{2}}\triangleright \qp \, (v^\mc{BC}_{\fr g_2},\fr b^{-1} , {\rm id}_{\rm t(\fr g_2)}) \; ,
\end{align*}
whereas $\mathsf r_{M_\mc{AB} , M_\mc{BC}}:M_\mc{AB}\otimes_{B} M_\mc{BC}\to M_\mc{AB} \otimes M_\mc{BC}$ is given by the inclusion. We can finally check that the binary functor simplifies such that
\begin{align}
	\qp_{M_{\mc{A}\mc{B}}}\otimes_{B} \qp_{M_{\mc{B}\mc{C}}}= \qp_{M_{\mc{A}\mc{B}} \otimes M_\mc{BC}} \; .
\end{align}
Left unitor, right unitor and associator can now be readily obtained. Finally, let us remark that the above bifunctor can be conveniently rephrased as a comultiplication map $\widetilde{\Delta}_B:\mathbb C[\widetilde{\mc{G}}_\mc{AC}]^{\varpi^\mc{AC}} \to  \mathbb C[\widetilde{\mc{G}}_\mc{AB}]^{\varpi^\mc{AB}} \otimes \mathbb C[\widetilde{\mc{G}}_\mc{BC}]^{\varpi^\mc{BC}} $
defined by
\begin{align}
	\widetilde{\Delta}_B\big( \big|\biGr{\fr g}{}{\fr a , \fr c} \big\ra \big)
	:= \frac{1}{|{\rm Hom}(\mc B)|}
	\!\!\!\!\!\!\! \sum_{\substack{
			\fr{g}_{1} \in {\rm Ob}(\widetilde{\mc G}_\mc{AB}) \\ \fr{g}_{2} \in {\rm Ob}(\widetilde{\mc G}_\mc{BC})
			\\
			\fr{g}_{1}\fr{g}_{2}=\fr{g}\\
			\fr{b} \in {\rm Hom}_\mc{B}(\rm t(\fr g_1), \rm t(\fr g_2))}
		} \!\!\!\!\!\!\!\!\!
	&\frac{
		\alpha(\fr{g}_{1},\fr{g}_{2},\fr{c}) \, \alpha(\fr{a},\fr{g}_{1},\fr{b}) \, \alpha(\fr{ag}_{1}\fr{b},\fr{b}^{-1},\fr{g}_{2}\fr{c})
	}
	{
		\psi(\fr b , \fr b^{-1}) \, \alpha(\fr{g},\fr{g}_{1},\fr{g}_{2}\fr{g}) \, \alpha(\fr{g}\fr{g}_{1},\fr{g},\fr{g}^{-1})
	}	
	\big| \biGr{\fr g_1}{}{\fr a , \fr b} \big \ra \otimes | \biGr{\fr g_2}{}{\fr b^{-1} ,\fr c}\big \ra .
\end{align}

\subsection{Bicategory of $\mathsf{Vec}^{\alpha}_{\mc{G}}$-module categories}\label{sec:MOD}

We are now ready to describe the bicategory $\mathsf{MOD}(\mathsf{Vec}^\alpha_\mc{G})$ by spelling out equivalence with the bicategory $\mathsf{sAlg}(\mc{C})$ described above. In the following, we will describe how this is the relevant structure to describe boundary excitations in gauge models of topological phases. 

Letting $\mc A_\phi$ be a (separable) algebra object in $\mathsf{Vec}^{\alpha}_{\mc{G}}$, the category $\mathsf{Mod}_{\mathsf{Vec}^\alpha_\mc{G}}(\mc A_\phi)$ of right $\mc A_\phi$-modules is a left module category for $\mathsf{Vec}^{\alpha}_{\mc{G}}$. Let us spell out this correspondence. The module functor
\begin{align}
	\odot:&
	\mathsf{Vec}^{\alpha}_{\mc{G}}
	\times\mathsf{Mod}_{\mathsf{Vec}^\alpha_\mc{G}}(\mc A_\phi)
	\rightarrow
	\mathsf{Mod}_{\mathsf{Vec}^\alpha_\mc{G}}(\mc A_\phi)
\end{align}
is defined on objects $V\in {\rm Ob}(\mathsf{Vec}^{\alpha}_{\mc{G}})$ and $(M_{\mc A},p_{\mc A})\in {\rm Ob}(\mathsf{Mod}_{\mathsf{Vec}^\alpha_\mc{G}}(\mc A_\phi)$ by
\begin{align}
	\odot: V\times M_{\mc A}\mapsto V\otimes M_{\mc A} \; ,
\end{align}
where $V\otimes M_{\mc A}\in {\rm Ob}(\mathsf{Mod}_{\mathsf{Vec}^\alpha_\mc{G}}(\mc A_\phi))$ is the $\mc A_\phi$-module with action defined by the following composition of morphisms in $\mathsf{Vec}^{\alpha}_{\mc{G}}$:
\begin{align}
	(V\otimes M_{\mc A})\otimes \mc A_\phi \xrightarrow{\alpha_{V,M_{\mc A},\mc A}}V\otimes (M_{\mc A}\otimes \mc A_\phi)\xrightarrow{{\rm id}_{V}\otimes p_{\mc A}}V\otimes M_{\mc A}\;.
\end{align}
The functor takes morphisms to their tensor product over the field $\mathbb{C}$. The module associator $\dot{\alpha}_{V,W,M}$ reduces to the associator in $\mathsf{Hom}^\alpha_\mc{G}$ such that for $V,W\in {\rm Ob}(\mathsf{Vec}^{\alpha}_{\mc{G}})$ one has
\begin{align}
	(V\otimes W)\otimes M_\mc{A} \xrightarrow{\alpha_{V,W,M_{\mc A}}} V\otimes (W\otimes M_{\mc A}) \; .
\end{align}
A $\mathsf{Vec}^{\alpha}_{\mc{G}}$-module category $\mathsf{Mod}_{\mathsf{Vec}^\alpha_\mc{G}}(\mc A_\phi)$ is then indecomposable if and only if the algebra object $\mc A_\phi$ is not isomorphic a direct sum of two non-trivial algebra objects \cite{etingof2016tensor}.

Let us now describe $\mathsf{Vec}^{\alpha}_{\mc{G}}$-module functors. Let $\mc A_\phi$ and $\mc B_\psi$ be any pair of algebra objects in $\mathsf{Vec}^{\alpha}_{\mc{G}}$, with $\mathsf{Mod}_{\mathsf{Vec}^{\alpha}_{\mc{G}}}(\mc A_\phi)$ and $\mathsf{Mod}_{\mathsf{Vec}^{\alpha}_{\mc{G}}}(\mc B_\psi)$ the corresponding category of module objects. To each $(\mc A_\phi , \mc B_\psi)$-bimodule object $M_{\mc A \mc B}$, we can define a  $\mathsf{Vec}^{\alpha}_{\mc{G}}$-module functor
\begin{align}
	-\otimes_{\mc A}M_{\mc A \mc B}:\mathsf{Mod}_{\mathsf{Vec}^{\alpha}_{\mc{G}}}(\mc A_\phi)\rightarrow \mathsf{Mod}_{\mathsf{Vec}^{\alpha}_{\mc{G}}}(\mc B_\psi) \; ,
\end{align}
which acts on objects $M_{\mc A}\in {\rm Ob}(\mathsf{Mod}_{\mathsf{Vec}^{\alpha}_{\mc{G}}}(\mc A_\phi )) $ via the map
\begin{align}
	-\otimes_{\mc A}M_{\mc A \mc B}:M_{\mc A}\mapsto M_{\mc A}\otimes_{\mc A}M_{\mc A \mc B} \; ,
\end{align}
and sends morphisms $f \in {\rm Hom}(\mathsf{Mod}_{\mathsf{Vec}^{\alpha}_{\mc{G}}}(\mc A_\phi))$ to $f \otimes {\rm id}_{M_{\mc A \mc B}}$. The natural isomorphism
\begin{align}
	s:
	(\mathsf{Vec}^{\alpha}_{\mc{G}}\otimes \mathsf{Mod}_{\mathsf{Vec}^{\alpha}_{\mc{G}}}(\mc A_\phi))
	\otimes_{\mc A}
	\mathsf{Bimod}(\mc A_\phi,\mc B_\psi)
	\rightarrow
	\mathsf{Vec}^{\alpha}_{\mc{G}}\otimes (\mathsf{Mod}_{\mathsf{Vec}^{\alpha}_{\mc{G}}}(\mc A_\phi)\otimes_{\mc A}\mathsf{Bimod}(\mc A_\phi,\mc B_\psi))
\end{align}
is given on objects $V \in {\rm Ob}(\mathsf{Vec}^{\alpha}_{\mc{G}})$ and $M_{\mc A}\in\mathsf{Mod}_{\mathsf{Vec}^{\alpha}_{\mc{G}}}(\mc A_\phi)$ via the associator $\alpha$ in $\mathsf{Vec}^{\alpha}_{\mc{G}}$ such that:
\begin{align}
	\nn
	s_{V,M_{\mc A}}:(V\otimes M_{\mc A})\otimes_{\mc A}M_{\mc A \mc B}&\xrightarrow{\mathsf{r}_{V\otimes M_{\mc A},M_{\mc A \mc B}}}(V\otimes M_{\mc A})\otimes M_{\mc A \mc B}
	\\
	\nn
	&\xrightarrow{\alpha_{V,M_{\mc A},M_{\mc A \mc B}}}V\otimes (M_{\mc A}\otimes M_{\mc A \mc B})
	\\&\xrightarrow{{\rm id}_{V}\otimes \mathsf{s}_{M_{\mc A},M_{\mc A \mc B}}}V\otimes (M_{\mc A}\otimes_{\mc A}M_{\mc A \mc B})\;.
\end{align}
In a similar vein, morphisms of $\mathsf{Vec}^{\alpha}_{\mc{G}}$-module functors are induced by natural transformations between bimodules. Together, this yields the desired equivalence:

\begin{proposition}\label{prop:MOD-sAlg}
	There exists an equivalence of bicategories between $\mathsf{sAlg}(\mathsf{Vec}^\alpha_\mc{G})$ and $\mathsf{MOD}(\mathsf{Vec}^\alpha_\mc{G})$ by sending separable algebra objects in $\mathsf{Vec}^\alpha_\mc{G}$ to their category of (right) modules in $\mathsf{Vec}^\alpha_\mc{G}$, bimodule objects $M_{\mc A \mc B}\in \mathsf{Hom}_{\mathsf{sAlg}(\mathsf{Vec}^\alpha_\mc{G})}(\mc A_\phi,\mc B_\psi)$ are sent to the $\mathsf{Vec}^\alpha_\mc{G}$-module functor $-\otimes_{\mc A} M_{\mc A \mc B}:\mathsf{Mod}_{\mathsf{Vec}^\alpha_\mc{G}}(\mc A)\rightarrow \mathsf{Mod}_{\mathsf{Vec}^\alpha_\mc{G}}(\mc B)$ and bimodule natural transformations are sent to morphisms of $\mathsf{Vec}^\alpha_\mc{G}$-module functors.
\end{proposition}

\subsection{Bicategory of boundary excitations in (2+1)d gauge models}\label{sec:bicatmodel2D}

Using the technology developed in this section, we are now ready to describe gapped boundaries and their excitations in (2+1)d gauge models of topological phases within the language of bicategories. More specifically, we shall define a bicategory $\mathsf{Bdry}_G^\alpha$ whose objects are given by gapped boundary conditions, 1-morphisms provide gapped boundary excitations, and 2-morphisms define fusion processes of gapped boundary excitations. We shall then demonstrate that $\mathsf{Bdry}_G^\alpha$ is equivalent, as a bicategory, to $\mathsf{MOD}(\mathsf{Vec}^{\alpha}_{G})$.

Let us begin with a brief review of the  results obtained in the first part of this manuscript within the tube algebra approach. Hamiltonian realisations of (2+1)d Dijkgraf-Witten theory are defined in terms of pairs $(G,\alpha)$, where $G$ is a finite group and $\alpha$ is a normalised 3-cocycle in $H^{3}(G,\rU(1))$. In sec.~\ref{sec:DW}, it was argued that gapped boundaries can be indexed by pairs $(A,\phi)$, where $A\subset G$ is a subgroup of $G$ and $\phi\in C^{2}(A,\rU(1))$ is a 2-cochain satisfying the condition $d^{(2)}\phi=\alpha^{-1}{\sss |}_{A}$. In sec.~\ref{sec:tube2D}, we showed that boundary excitations at the interface of two one-dimensional gapped boundaries labelled by $(A,\phi)$ and $(B,\psi)$, respectively, were classified via representations of the boundary tube algebra that is isomorphic to the  twisted groupoid algebra $\mathbb{C}[G_{AB}]^{\alpha}_{\phi \psi}$.

We now collect the previous results into a bicategory $\mathsf{Bdry}_G^\alpha$. The objects of $\mathsf{Bdry}_G^\alpha$ are given by the set of all gapped boundary conditions $\{(A,\phi)\}$. For each pair $(A,\phi)$, $(B,\psi)$ of gapped boundary conditions, we assign the hom-category
\begin{align}
	\mathsf{Hom}_{\mathsf{Bdry}_G^\alpha}((A,\phi),(B,\psi)):=\mathsf{Mod}(\mathbb{C}[G_{AB}]^{\alpha}_{\phi \psi})\; ,
\end{align}
where $\mathsf{Mod}(\mathbb{C}[G_{AB}]^{\alpha}_{\phi \psi})$ denotes the category of $\mathbb{C}[G_{AB}]^{\alpha}_{\phi \psi}$-modules and intertwiners. In this way, the 1-morphisms $\rho_{AB}\in {\rm Ob}(\mathsf{Hom}_{\mathsf{Bdry}_G^\alpha}((A,\phi),(B,\psi)))$ correspond to boundary excitations incident at the interface between gapped boudaries labelled by $(A,\phi)$ and $(B,\psi)$. The composition bifunctor
\begin{align}
	\otimes:
	\mathsf{Mod}(\mathbb{C}[G_{AB}]^{\alpha}_{\phi \psi})
	\times
	\mathsf{Mod}(\mathbb{C}[G_{BC}]^{\alpha}_{\psi \varphi})
	\rightarrow
	\mathsf{Mod}(\mathbb{C}[G_{AC}]^{\alpha}_{\phi \varphi})
\end{align}
is defined on 1-morphisms $\rho_{AB}\in {\rm Ob}(\mathsf{Mod}(\mathbb{C}[G_{AB}]^{\alpha}_{\phi \psi}))$ and $\rho_{BC}\in {\rm Ob}(\mathsf{Mod}(\mathbb{C}[G_{BC}]^{\alpha}_{\psi \varphi}))$ via
\begin{align}
	\otimes:\rho_{AB}\times\rho_{BC}\mapsto
	\rho_{AB}\otimes_{B} \rho_{BC}:=
	(\rho_{AB}\otimes \rho_{BC})\triangleright \Delta_{B}(\mathbbm{1}_{AC}) \; ,
\end{align}
as described in sec.~\ref{sec:comult}, and on 2-morphisms $f_{AB}:\rho_{AB}\rightarrow \rho_{AB}' \in {\rm Hom}(\mathsf{Mod}(\mathbb{C}[G_{AB}]^{\alpha}_{\phi \psi}))$, $f_{BC}:\rho_{BC}\rightarrow \rho_{BC}'\in {\rm Hom}(\mathsf{Mod}(\mathbb{C}[G_{BC}]^{\alpha}_{\psi \varphi}))$ via
\begin{align}
	\otimes: f_{AB} \times f_{BC} \mapsto ( f_{AB} \otimes_B f_{BC} 
	:  \rho_{AB}\otimes_{B}\rho_{BC}\rightarrow \rho_{AB}'\otimes_{B}\rho_{BC}' ) \; ,
\end{align}
where the morphism on the r.h.s decomposes as
\begin{align}
	f_{AB}\otimes_{B}f_{BC}:\rho_{AB}\otimes_{B}\rho_{BC}
	\xhookrightarrow{}
	\rho_{AB}\otimes \rho_{BC}\xrightarrow{f_{AB}\otimes f_{BC}}\rho_{AB}'\otimes\rho_{BC}'\to \rho_{AB}'\otimes_{B}\rho_{BC}' \; .
\end{align}
In the sequence of linear maps above, the first arrow notates the injection of $\rho_{AB}\otimes_{B}\rho_{BC}$ into $\rho_{AB}\otimes \rho_{BC}$, and the last arrow notates the projection map
\begin{align}
	\rho_{AB}'\otimes \rho_{BC}'\mapsto (\rho_{AB}' \otimes \rho_{BC}') \triangleright \Delta_{B}(\mathbbm{1}_{AC})=\rho_{AB}' \otimes_{B}\rho_{BC}' \, .
\end{align}
Furthermore, a 2-morphism of the form $\zeta:\rho_{AB}\otimes_{B}\rho_{BC}\rightarrow \rho_{AC}\in {\rm Hom}(\mathsf{Mod}(\mathbb{C}[G_{AC}]^{\alpha}_{\phi \varphi}))$ is an intertwiner interpreted as describing the process of fusing a pair of boundary excitations at the interfaces of gapped boundaries labelled by $(A,\phi)$, $(B,\psi)$ and $(B,\psi)$, $(C,\varphi)$, respectively:
\begin{equation}
	\bdryInterfaceBis{0.9} \; .
\end{equation}
The identity morphism associated with the object $(A,\phi)$ is given by the regular module $\mathbb{C}[G_{AA}]^{\alpha}_{\phi \phi}\in {\rm Ob}(\mathsf{Mod}(\mathbb{C}[G_{AA}]^{\alpha}_{\phi \phi}))$ with left and right unitors the intertwiner isomorphisms
\begin{align}
	\ell:\mathbb{C}[G_{AA}]^{\alpha}_{\phi \phi}\otimes_{A} \rho_{AB}\xrightarrow{\sim}\rho_{AB}
	\; , \q
	r:\rho_{AB}\otimes_{B}\mathbb{C}[G_{BB}]^{\alpha}_{\psi \psi}\xrightarrow{\sim}\rho_{AB} \; ,
\end{align}
as described in sec.~\ref{sec:comult}. Finally, the 1-associator for a triple of 1-morphisms $\rho_{AB}\in {\rm Ob}(\mathsf{Mod}(\mathbb{C}[G_{AB}]^{\alpha}_{\phi \psi}))$, $\rho_{BC}\in {\rm Ob}(\mathsf{Mod}(\mathbb{C}[G_{BC}]^{\alpha}_{\psi \varphi}))$, $\rho_{CD}\in {\rm Ob}(\mathsf{Mod}(\mathbb{C}[G_{CD}]^{\alpha}_{\varphi\chi}))$  is given by the intertwiner isomorphism in ${\rm Hom}(\mathsf{Mod}(\mathbb{C}[G_{AD}]^{\alpha}_{\phi \chi}))$
\begin{align}
	\Phi^{\rho_{AB}\rho_{BC} \rho_{CD}}:(\rho_{AB}\otimes_{B}\rho_{BC})\otimes_{C} \rho_{CD}\rightarrow \rho_{AB}\otimes_{B}(\rho_{BC}\otimes_{C} \rho_{CD}) \; ,
\end{align}
as described explicitly in sec.~\ref{sec:comult}.\footnote{Recall that the derivations in sec.~\ref{sec:comult}, and more generally in sec.~\ref{sec:rep}, were carried out explicitly for the boundary tube algebra in (3+1)d. However, we explained that the (2+1)d boundary tube algebra, which is the one relevant here, is obtained as a limiting case.}  It follows from the results of the first part of this manuscript that such data satisfy the pentagon and triangle relations ensuring we do obtain a bicategory.

\bigskip \noindent
So we have recast our results obtained in the first part of this manuscript in terms of the boundary tube algebra and its representation theory as the bicategory $\mathsf{Bdry}_G^\alpha$. We shall now establish the following equivalence of bicategories:
\begin{align}
	\label{eq:bicatequivboundaryMOD}
	\mathsf{Bdry}_G^\alpha
	\simeq \mathsf{MOD}(\mathsf{Vec}^{\alpha}_{G}) \; .
\end{align}
More precisely, we shall establish the equivalence of the bicategories $\mathsf{Bdry}_G^\alpha\simeq \mathsf{sAlg}(\mathsf{Vec}^{\alpha}_{G})$, from which we can induce the equivalence above through prop.~\ref{prop:MOD-sAlg}, by noting equivalence of bicategories is \emph{transitive}. First, we need to introduce a notion of homomorphism between bicategories:
\begin{definition}[Strict homomorphism of bicategories\label{def:stricthomomoprphism}]
	Given a pair of bicategories $\mc{B}\mathpzc{i}$ and ${\mc{B}\mathpzc{i}}'$, a \emph{strict homomorphism} $\mc{F}:\mc{B}\mathpzc{i} \to {\mc{B}\mathpzc{i}}'$ of bicategories consists of 
	\begin{enumerate}[itemsep=0.4em,parsep=1pt,leftmargin=3em]
		\item[${\ssss \bullet}$] a function $\mc{F}:{\rm Ob}(\mc{B}i)\rightarrow {\rm Ob}(\mc{B}i')$,
		\item[${\ssss \bullet}$] a family of functors $\mc{F}_{XY}:\mathsf{Hom}_{\mc{B}\mathpzc{i}}(X,Y) \to \mathsf{Hom}_{{\mc{B}\mathpzc{i}}'} (\mc{F}(X), \mc{F}(Y))$ referred to as \emph{hom-functors}, for each pair of objects $X,Y \in{\rm Ob}(\mc{B}\mathpzc{i})$, 
	\end{enumerate}

	such that
	\begin{align*}
		\mc{F}_{X,Y}(f)\otimes \mc{F}_{Y,Z}(g)&=\mc{F}_{X,Z}(f\otimes g) & 
		\mathbbm{1}^{{\mc{B}\mathpzc{i}}'}_{\mc{F}(X)}&=\mc{F}_{X,X}(\mathbbm{1}^{\mc{B}\mathpzc{i}}_{X})
		\\
		\mc{F}(\alpha^{\mc{B}\mathpzc{i}}_{f,g,h})&=\alpha^{{\mc{B}\mathpzc{i}}'}_{\mc{F}_{X,Y}(f),\mc{F}_{Y,Z}(g),\mc{F}_{Z,W}(h)}
		&
		\mc{F}(r^{\mc{B}\mathpzc{i}}_{X})=r^{{\mc{B}\mathpzc{i}}'}_{\mc{F}(X)} & \;\; , \;\;  \mc{F}(\ell^{\mc{B}\mathpzc{i}}_{X})=\ell^{{\mc{B}\mathpzc{i}}'}_{\mc{F}(X)}  \; ,
	\end{align*}
	for all objects $W,X,Y,Z\in {\rm Ob}(\mc{B}\mathpzc{i})$ and morphisms $f\in {\rm Ob}(\mathsf{Hom}_{\mc{B}\mathpzc{i}}(X,Y))$, $g\in {\rm Ob}(\mathsf{Hom}_{\mc{B}\mathpzc{i}}(Y,Z))$, $h\in {\rm Ob}(\mathsf{Hom}_{\mc{B}\mathpzc{i}}(Z,W))$.
\end{definition}
\noindent
Recall that a functor between categories defines an equivalence if and only if it  is \emph{full}, \emph{faithful} and essentially \emph{surjective}. In a similar vein, a sufficient condition for a strict homomorphism of bicategories $\mc{F}$ to define an equivalence of bicategories is that the map is surjective on objects, and the functors $\mc{F}_{X,Y}$ for all $X,Y\in {\rm Ob}(\mc{B}\mathpzc{i})$ define equivalences of the categories $\mathsf{Hom}_{\mc{B}\mathpzc{i}}(X,Y)\simeq \mathsf{Hom}_{{\mc{B}\mathpzc{i}}'}(\mc{F}(X),\mc{F}(Y))$.

Using this sufficient condition, let us now establish the equivalence of bicategories $\mc{F}:\mathsf{Bdry}_G^\alpha\xrightarrow{\simeq} \mathsf{sAlg}(\mathsf{Vec}^{\alpha}_{G})$. We begin by defining the function $\mc{F}:{\rm Ob}(\mathsf{Bdry}_G^\alpha)\rightarrow {\rm Ob}(\mathsf{sAlg}(\mathsf{Vec}^{\alpha}_{G}))$. It is given by sending each boundary condition $(A,\phi)$ to the corresponding separable algebra object $A_{\phi}$ in $\mathsf{Vec}^{\alpha}_{G}$. From the previous discussion, we know that both boundary conditions and separable algebra objects are indexed by subgroups of $G$ and 2-cochains satisfying the compatibility conditions with $\alpha$. It follows that the function $\mc{F}$ is a bijection, and thus surjective. The hom-functors are required to define the following equivalence of categories: 
\begin{align*}
	\mathsf{Hom}_{\mathsf{Bdry}_G^\alpha}((A,\phi),(B,\psi))
	:=\mathsf{Mod}(\mathbb{C}[G_{AB}]^{\alpha}_{\phi \psi})
	\simeq
	\mathsf{Mod}(\mathbb{C}[\widetilde{G}_{AB}]^{\varpi^{AB}})
	=:
	\mathsf{Hom}_{\mathsf{sAlg}(\mathsf{Vec}^{\alpha}_{G})}(A_{\phi},B_{\psi}) \; ,
\end{align*}
where the groupoid $\widetilde G_{AB}$ and its 2-cocycle $\varpi^{AB}$ is obtained by applying the definition at the end of sec.~\ref{sec:algObj} to the delooping of $G$.
In order to establish this equivalence, it suffices to demonstrate the isomorphism of twisted groupoid algebras $\mathbb{C}[\widetilde{G}_{AB}]^{\varpi^{AB}} \simeq \mathbb{C}[G_{AB}]^{\alpha}_{\phi \psi} \equiv \mathbb C[G_{AB}]^{\vartheta^{AB}}$, for all boundary conditions $(A,\phi)$ and $(B,\psi)$. The equivalence $\mathsf{Mod}(\mathbb{C}[G_{AB}]^{\alpha}_{\phi \psi}) \simeq \mathsf{Mod}(\mathbb{C}[\tilde{G}_{AB}]^{\varpi^{AB}})
$ of their module categories then follows by pre-composition. 
Noting from the definition that both groupoids have the same dimension, the isomorphism is provided by the following map on basis elements:
\begin{align}
	\label{eq:isoGroupoid}
	\big| \biGr{ g}{}{ a,  b} \big\ra
	\mapsto
	\frac{\phi({a}^{-1},{a})}{\alpha(a^{-1},{a},{g}  b)}
	\big| \biGr{g}{a^{-1}}{b} \big\ra
	 \;, \q \forall \,
	\big| \biGr{ g}{}{ a,  b} \big\ra
	\in \mathbb{C}[\widetilde{G}_{AB}]^{\varpi^{AB}} \; . 
\end{align}
Furthermore, one can check that such an isomorphism is compatible with the respective comultiplication maps through the following commuting diagram
\begin{align}
	\begin{tikzcd}[ampersand replacement=\&, column sep=1.8em, row sep=1.3em]
		|[alias=A]|\mathbb{C}[G_{AC}]^{\alpha}_{\phi \varphi}
		\&\&
		|[alias=B]|\mathbb{C}[G_{AB}]^{\alpha}_{\phi \psi}\otimes \mathbb{C}[G_{BC}]^{\alpha}_{\psi \varphi}
		\\
		\\
		|[alias=AA]|\mathbb{C}[\widetilde{G}_{AC}]^{\varpi^{AC}}
		\&\&
		|[alias=BB]|\mathbb{C}[\widetilde{G}_{AB}]^{\varpi^{AB}}\otimes \mathbb{C}[\widetilde{G}_{BC}]^{\varpi^{BC}}
		\arrow[from=A,to=B,"\Delta_{B}"]
		\arrow[<->,from=B,to=BB,"\simeq"]
		\arrow[<->,from=A,to=AA,"\simeq"']
		\arrow[from=AA,to=BB,"\widetilde{\Delta}_{B}"']
	\end{tikzcd} \; .
\end{align} 
Commutativity is ensured by the relation
\begin{align*}
	&\frac{\phi(a^{-1},a)}{\alpha(a^{-1},a,g_{1}b)}
	\frac{\psi(b,b^{-1})}{\alpha(b,b^{-1},g_{2}c)}
	\frac{
		\alpha(g_{1},g_{2},c) \, \alpha(a,g_{1},b)\, \alpha(ag_{1}b,b^{-1},g_{2}c)
	}
	{
		\psi(b,b^{-1}) \, \alpha(a,g_{1},g_{2}c) \, \alpha(ag_{1},b,b^{-1})
	}
	\\
	& \q =
	\frac{\phi(a^{-1},a)}{\alpha(a^{-1},a,g_{1}g_{2}c)}
	\frac{
		\alpha(g_{1},g_{2},c) \, \alpha(a^{-1},ag_{1}b,b^{-1}g_{2}c)
	}{\alpha(g_{1},b,b^{-1}g_{2}c)} \; ,
\end{align*}
which follows from the cocycle relation
\begin{equation*}
	d^{(3)}\alpha(a^{-1},a,g_{1}b,b^{-1}g_{2}c) =1 \; , \q 
	d^{(3)}\alpha(a^{-1},a,g_{1}g_{2}c)  =1 \; , \q 
	d^{(3)}\alpha(ag_{1},b,b^{-1},g_{2}c)  =1 \; .
\end{equation*}
Since the composition functors in both bicategories are induced from the respective comultiplication maps, it can be verified that  such hom-functors satisfy the conditions of a strict homomorphism of bicategories, hence establishing the required equivalence of bicategories.

\subsection{Pseudo-algebra objects and gapped boundaries in (3+1)d gauge models}\label{sec:pseudoalgob}

In the previous discussion, we argued that, given a lattice Hamiltonian realisation of (2+1)d Dijkgraaf-Witten theory with input data $(G,\alpha)$, gapped boundary conditions are in bijection with algebra objects in the fusion category $\mathsf{Vec}^{\alpha}_{G}$. We shall now outline the analogue of this statement for lattice Hamiltonian realisations of (3+1)d Dijkgraaf-Witten theory. 

Given a fixed input data $(G,\pi)$, where $G$ is a finite group and $\pi$ is normalized group 4-cocycle in $H^{4}(G,\rU(1))$, it has been argued that the relevant category theoretical structure is provided by the \emph{monoidal} bicategory $\mathsf{2Vec}^{\pi}_{G}$ of $G$-graded 2-vector spaces \cite{PhysRevB.96.045136, lan2017classification, delcamp2018gauge, Kong:2019brm, Bullivant:2019fmk, douglas2018fusion}. Let us begin by describing the salient features of the monoidal bicategory $\mathsf{2Vec}$ as a categorification of $\mathsf{Vec}$. There exist several definitions of this bicategory, see e.g. \cite{kapranov19942, Baez:2003fs, Lurie:2009keu}, in the following we shall consider $\mathsf{2Vec}$ as the bicategory of finite dimensional, semi-simple $\mathsf{Vec}$-module categories, $\mathsf{Vec}$-module functors and $\mathsf{Vec}$-module functor homomorphisms. As customary, objects of $\mathsf{2Vec}$ will be referred to as \emph{2-vector spaces}. There is a single \emph{simple} object provided by the $\mathsf{Vec}$-module category $\mathsf{Vec}$, which implies that for all objects $X\in {\rm Ob}(\mathsf{2Vec})$, there exists a $\mathsf{Vec}$-module equivalence $X\simeq \bigbplus_i  \mathsf{Vec}$. The monoidal structure of $\mathsf{2Vec}$ is defined on objects via the weak 2-functor
\begin{equation*}
	\begin{array}{ccccl}
	\btimes \! & : & \mathsf{2Vec}\times \mathsf{2Vec} & \to & \mathsf{2Vec}
	\\
	& : & X\times Y &\mapsto & 	X \btimes Y
	\end{array} \; ,
\end{equation*}
for all $X, Y \in {\rm Ob}(\mathsf{2Vec})$, where $\! \btimes \! $ denotes the \emph{Deligne} tensor product of abelian categories \cite{deligne2007categories}.
In particular, for a pair of 2-vector spaces $X$ and $Y$, the Deligne tensor product yields the category $X\btimes Y$, whose set of objects is ${\rm Ob}(X\btimes Y):={\rm Ob}(X)\times {\rm Ob}(Y)$ and set of morphisms given by ${\rm Hom}(X\btimes Y):={\rm Hom}(X)\otimes_{\mathbb{C}} {\rm Hom}(Y)$. The composition in ${\rm Hom}(X\btimes Y)$ is induced from the ones in ${\rm Hom}(X)$ and ${\rm Hom}(Y)$, accordingly. This monoidal structure is equipped with a \emph{pseudo-natural} adjoint equivalence of $\mathsf{Vec}$-module categories\footnote{Although we use a similar notation, the associator of the monoidal structure is not to be confused with the 1-associator natural isomorphism of the underlying bicategory.}
\begin{align}
	(X \btimes Y)\btimes Z \xrightarrow{\alpha_{X,Y,Z}} X\btimes (Y\btimes Z) \; ,
\end{align}
together with a  $\mathsf{Vec}$-module functor isomorphism $\pi$ known as the \emph{pentagonator}:
\begin{align}
	\begin{tikzcd}[ampersand replacement=\&, column sep=1.6em, row sep=1.3em]
		\&\&|[alias=B]|((X\btimes Y)\btimes Z)\btimes W
		\\\\
		|[alias=AA]|(X\btimes (Y \btimes Z)) \btimes W
		\&\&\&\&
		|[alias=CC]|(X\btimes Y) \btimes (Z \btimes W)
		\\\\
		|[alias=AAA]|X\btimes ((Y \btimes Z) \btimes W)
		\&\&\&\&|[alias=CCC]|X\btimes (Y \btimes ((Z \btimes W))
		\arrow[from=B,to=CC,"\;\; \alpha_{X\btimesFFt Y,Z,W}", sloped]
		\arrow[from=CC,to=CCC,"\alpha_{\alpha_{X,Y,Z\btimesFFt W}}"]
		\arrow[from=B,to=AA,"\alpha_{X,Y,Z}\btimesFt {\rm id}_{W} \;\; ", sloped]
		\arrow[from=AA,to=AAA,"\alpha_{X,Y\btimesFFt Z,W}"',""{name=X,right}]
		\arrow[from=AAA,to=CCC,"{\rm id}_{X}\btimesFt \alpha_{Y,Z,W}"']
		\arrow[Rightarrow,from=X,to=CC,"\pi_{X,Y,Z,W}", shorten <= 2ex, shorten >= 2ex, sloped]
	\end{tikzcd} \; .
\end{align}
Both $\alpha$ and $\pi$ can be shown to evaluate to the identity 1- and 2-morphisms, respectively. Note that the \emph{pseudo-naturality} of $\alpha$ specifies that for any triple of 2-vector spaces $X,Y,Z$ and $\mathsf{Vec}$-module functors $f_{X}:X\rightarrow X'$, $f_{Y}:Y\rightarrow Y'$ and $f_{Z}:Z\rightarrow Z'$ there exists a 2-isomorphism
\begin{align}
	\begin{tikzcd}[ampersand replacement=\&, column sep=1.8em, row sep=1.3em]
		|[alias=A]|(X\btimes Y)\btimes Z
		\&\&
		|[alias=B]|X\btimes (Y\btimes Z)
		\\\\
		|[alias=AA]|(X'\btimes Y')\btimes Z'
		\&\&
		|[alias=BB]|X'\btimes (Y'\btimes Z')
		\arrow[from=A,to=B,"\alpha_{X,Y,Z}"]
		\arrow[from=B,to=BB,"f_{X}\btimesFt (f_{Y}\btimesFt f_{Z})"]
		\arrow[from=A,to=AA,"(f_{X}\btimesFt f_{Y})\btimesFt f_{Z}"']
		\arrow[from=AA,to=BB,"\alpha_{X,Y,Z}"']
		\arrow[from=B,to=AA,Rightarrow,"\simeq", shorten <= 2ex, shorten >= 2ex, sloped]
	\end{tikzcd} \; .
\end{align}
Henceforth, we shall not draw arrows for such 2-isomorphisms but instead notate the 2-cell with the $\simeq$ symbol. 

Akin to a monoidal category, the monoidal bicategory $\mathsf{2Vec}$ admits a monoidal unit $\mathbbm 1 \in{\rm Ob}(\mathsf{2Vec})$, which is  equipped with the $\mathsf{Vec}$-module category pseudo-natural adjoint equivalences 
\begin{align}
	X\btimes \mathbbm 1 \xrightarrow{r_{X}} X \q {\rm and} \q  \mathbbm 1 \btimes X\xrightarrow{\ell_{X}}X \; , 
\end{align}
for all  $X \in {\rm Ob}(\mathsf{2Vec})$, together with $\mathsf{Vec}$-module functor isomorphisms $\tau_1, \tau_2, \tau_3$ referred to as \emph{triangulators}:
\begin{gather}
	\begin{tikzcd}[ampersand replacement=\&, column sep=1.8em, row sep=1.3em]
		|[alias=A]|(\mathbbm 1\btimes X)\btimes Y
		\&\&
		|[alias=B]|\mathbbm 1 \btimes (X\btimes Y)
		\\\\
		\&\&|[alias=BB]|X\btimes Y
		\arrow[from=A,to=B,"\alpha_{\mathbbm 1,X,Y}"]
		\arrow[from=B,to=BB,"\ell_{X\btimesFFt Y}"]
		\arrow[from=A,to=BB,"\ell_{X}\btimesFt {\rm id}_{Y}"',""{name=X,above}]
		\arrow[Rightarrow,from=B,to=X,"\tau_1"]
	\end{tikzcd}
	, \q
	\begin{tikzcd}[ampersand replacement=\&, column sep=1.8em, row sep=1.3em]
		|[alias=A]|(X\btimes \mathbbm 1)\btimes Y
		\&\&
		|[alias=B]|X\btimes (\mathbbm 1\btimes Y)
		\\\\
		\&\&|[alias=BB]|X\btimes Y
		\arrow[from=A,to=B,"\alpha_{X,\mathbbm 11,Y}"]
		\arrow[from=B,to=BB,"{\rm id}_X \btimesFt \ell_{Y}"]
		\arrow[from=A,to=BB,"r_{X}\btimesFt {\rm id}_{Y}"',""{name=X,above}]
		\arrow[Rightarrow,from=B,to=X,"\tau_2"]
	\end{tikzcd}
	, 
	\\
	\begin{tikzcd}[ampersand replacement=\&, column sep=1.8em, row sep=1.3em]
		|[alias=A]|(X\btimes Y)\btimes \mathbbm 1
		\&\&
		|[alias=B]|X\btimes (Y\btimes \mathbbm 1)
		\\\\
		\&\&|[alias=BB]|X\btimes Y
		\arrow[from=A,to=B,"\alpha_{X,Y,\mathbbm 1}"]
		\arrow[from=B,to=BB,"{\rm id}_X \btimesFt r_{Y}"]
		\arrow[from=A,to=BB,"r_{X\btimesFFt Y}"',""{name=X,above}]
		\arrow[Rightarrow,from=B,to=X,"\tau_3"]
	\end{tikzcd} . 
\end{gather}
These isomorphisms can be all be shown to evaluate to the identity 1- and 2-morphisms, respectively.
More generally, for an arbitrary monoidal bicategory, such data is subject to a series of coherence data which we shall not provide here, instead pointing the reader to e.g. \cite{kapranov19942, gurski2011loop,schommerpries2011classification}.

Having described the most notable features of $\mathsf{2Vec}$, we now describe the monoidal bicategory $\mathsf{2Vec}^{\pi}_{G}$, which is obtained following a process analogous to the lift of $\mathsf{Vec}$ to $\mathsf{Vec}^{\alpha}_{G}$. Let $G$ be a finite group and $\pi$ a normalised group 4-cocycle in $H^{4}(G,\rU(1))$. A $G$-graded 2-vector space is a 2-vector space of the form
$X=\bigbplus_{g\in G} X_{g}$. We call a $G$-graded 2-vector space \emph{homogeneous} of degree $g\in G$ if $X=X_{g}$. The monoidal bicategory $\mathsf{2Vec}^{\pi}_{G}$ is then defined as the bicategory whose objects  are given by $G$-graded 2-vector spaces, 1-morphisms are $G$-grading preserving $\mathsf{Vec}$-module functors, and 2-morphisms are $\mathsf{Vec}$-module functor homomorphisms. The simple objects of $\mathsf{2Vec}^{\pi}_{G}$ are given by the categories $\mathsf{Vec}_{g}$, for all $g\in G$, and every object is equivalent to a direct sum of simple objects. The monoidal structure of $\mathsf{2Vec}^{\pi}_{G}$ is given on homogeneous components via the weak 2-functor
\begin{align}
	\btimes \! : \mathsf{Vec}_{g}\times \mathsf{Vec}_{g'}\rightarrow \mathsf{Vec}_{gg'} \; ,
\end{align}
for all $g,g'\in G$. Since $\pi$ is a normalised representative of $[\pi]\in H^{4}(G,\rU(1))$, the adjoint equivalences
\begin{gather}
	(\mathsf{Vec}_{g}\btimes\mathsf{Vec}_{g'})\btimes\mathsf{Vec}_{g''}
	\xrightarrow{\alpha_{\mathsf{Vec}_{g},\mathsf{Vec}_{g'},\mathsf{Vec}_{g''}}}
	\mathsf{Vec}_{g}\btimes(\mathsf{Vec}_{g'}\btimes\mathsf{Vec}_{g''}) \; ,
	\\
	\mathsf{Vec}_{g}\btimes \mathsf{Vec}_{\mathbbm 1_{G}}\xrightarrow{r_{\mathsf{Vec}_{g}}} \mathsf{Vec}_{g},
	\; , \q
	\mathsf{Vec}_{\mathbbm 1_{G}}\btimes \mathsf{Vec}_{g}\xrightarrow{\ell_{\mathsf{Vec}_{g}}}\mathsf{Vec}_{g}
\end{gather}
are the identity 1-morphisms, the triangulators  $\tau_1,\tau_2,\tau_3$ are the identity 2-morphisms, whereas the pentagonator 2-isomorphism is given by $\pi_{\mathsf{Vec}_{g},\mathsf{Vec}_{g'},\mathsf{Vec}_{g''},\mathsf{Vec}_{g'''}}:=\pi(g,g',g'',g''') \cdot {\rm id}_{\mathsf{Vec}_{gg'g''g'''}}$ for all $g,g',g'',g'''\in G$. It is straightforward to verify that the requirement that $\pi$ is a 4-cocycle ensures the coherence relations for the pentagonator are satisfied.

Having defined the monoidal bicategory $\mathsf{2Vec}^{\pi}_{G}$, we shall now argue that gapped boundary conditions in (3+1)d gauge models of topological phases correspond to \emph{pseudo-algebra objects} \cite{LACK2000179} in $\mathsf{2Vec}^{\pi}_{G}$, categorifying the relation between algebra objects in $\mathsf{Vec}^{\alpha}_{G}$ and gapped boundaries in (2+1)d gauge models:
\begin{definition}[Pseudo-algebra object]
	Let $\mc{B}\mathpzc{i} \equiv (\mc{B}\mathpzc{i}, \btimes, \mathbbm 1,\alpha, r , \ell , \pi , \tau_1 , \tau_2 , \tau_3)$ be a monoidal bicategory. A pseudo-algebra object in $\mc{B}\mathpzc{i}$ is a sextuple $(A,m,u, \varsigma_m, \varsigma_r, \varsigma_\ell)$ consisting of an object $A\in {\rm Ob}(\mc{B}\mathpzc{i})$, a pair of 1-morphisms $m:A\btimes A \to A$, $u:\mathbbm{1} \to A$, and a triple of 2-isomorphisms $\varsigma_m, \varsigma_r,\varsigma_\ell$ defined according to
	\begin{gather*}
		\begin{tikzcd}[ampersand replacement=\&, column sep=1.8em, row sep=1.3em]
			|[alias=A]|(A\btimes A)\btimes A \&\& |[alias=B]|A\btimes (A\btimes A) \&\& |[alias=C]| A\btimes A
			\\
			\\
			|[alias=AA]|A\btimes A \&\&\&\& |[alias=CC]|A
			\arrow[from=A,to=B,"\alpha"]
			\arrow[from=B,to=C,"{\rm id}_{A}\btimesFt m"]
			\arrow[from=AA,to=CC,"m"']
			\arrow[from=A,to=AA,"m \btimesFt {\rm id}_{A}"']
			\arrow[from=C,to=CC,"m"]
			%%%
			\arrow[from=C,to=AA,"\varsigma_m",Rightarrow, shorten <= 2ex, shorten >= 2ex]
		\end{tikzcd} \; ,
		\\
		\begin{tikzcd}[ampersand replacement=\&, column sep=1.8em, row sep=1.3em]
			|[alias=A]|A
			\&[-1em]\&
			|[alias=B]|A\btimes \mathbbm 1
			\&\&
			|[alias=C]|A\btimes A
			\&[-1em]\&
			|[alias=D]|A
			\arrow[from=A,to=B,"r^{-1}"]
			\arrow[from=B,to=C,"{\rm id}_A \btimesFt u",""{name=Y,below,xshift=0.1em}]
			\arrow[from=C,to=D,"m"]
			\arrow[from=A,to=D,bend right,"{\rm id}_A"',""{name=X,above,xshift=0.025em}]
			\arrow[from=Y,to=X,"\varsigma_r",Rightarrow, shorten <= 0.5ex, shorten >= 0.5ex]
		\end{tikzcd} \; , \q
		\begin{tikzcd}[ampersand replacement=\&, column sep=1.8em, row sep=1.3em]
			|[alias=A]|A
			\&[-1em]\&
			|[alias=B]|\mathbbm 1 \btimes A
			\&\&
			|[alias=C]|A\btimes A
			\&[-1em]\&
			|[alias=D]|A
			\arrow[from=A,to=B,"\ell^{-1}"]
			\arrow[from=B,to=C,"u \btimesFt {\rm id}_A",""{name=Y,below,xshift=0.1em}]
			\arrow[from=C,to=D,"m"]
			\arrow[from=A,to=D,bend right,"{\rm id}_A"',""{name=X,above, xshift=0.025em}]
			\arrow[from=Y,to=X,"\varsigma_\ell",Rightarrow,shorten <= 0.5ex, shorten >= 0.5ex]
		\end{tikzcd} \; ,
	\end{gather*}
	
	and subject to the following coherence relations:
	\begin{align*}
		\begin{tikzcd}[ampersand replacement=\&, column sep=2.7em, row sep=1.3em]
			|[alias=A]|((A\btimes A)\btimes A)\btimes A
			\&[-2.7em]\phantom{(A \btimes A)\btimes(A \btimes A)}\&[2.5em]
			|[alias=B]|(A\btimes A)\btimes A
			\&[0.4em] \phantom{A \btimes A}\&[-4em]
			|[alias=C]|A\btimes A
			\\\\
			|[alias=AA]|(A\btimes (A\btimes A))\btimes A
			\&\&
			|[alias=BB]|(A\btimes A)\btimes A
			\\[-0.65em]
			\&
			\phantom{(A\btimes A)\btimes(A\btimes A)}
			\&\&
			\phantom{A\btimes A}
			\&\&[-2em]
			|[alias=DDD]|A
			\\[-0.65em]
			|[alias=AAAA]|A\btimes ((A\btimes A)\btimes A)
			\&\&
			|[alias=BBBB]|A\btimes (A\btimes A)
			\\\\
			|[alias=AAAAA]|A\btimes (A\btimes (A\btimes A))
			\&\&
			|[alias=BBBBB]|A\btimes (A\btimes A)
			\&\&
			|[alias=CCCCC]|A\btimes A
			\arrow[from=A,to=B,"(m\btimesFt {\rm id}_A)\btimesFt {\rm id}_A"]
			\arrow[from=B,to=C,"m\btimesFt {\rm id}_A"]
			\arrow[from=C,to=DDD,"m"]
			\arrow[from=A,to=AA,"\alpha_{A,A,A}\btimesFt {\rm id}_A"']
			\arrow[from=AA,to=AAAA,"\alpha_{A,A\btimesFFt A,A}"']
			\arrow[from=AAAA,to=AAAAA,"{\rm id}_A\btimesFt \alpha_{A,A,A}"']
			\arrow[from=AAAA,to=BBBB,"({\rm id}_A\btimesFt {\rm id}_A)\btimesFt m"]
			\arrow[from=BBBBB,to=CCCCC,"{\rm id}_A\btimesFt m"']
			\arrow[from=CCCCC,to=DDD,"m"']
			\arrow[from=AA,to=BB,"(A\btimesFt m)\btimesFt {\rm id}_A"',""{name=X,above}]
			\arrow[from=AAAAA,to=BBBBB,"{\rm id}_A\btimesFt(m\btimesFt {\rm id}_A)"',""{name=Y,above}]
			\arrow[from=BB,to=BBBB,"\alpha_{A,A,A}"]
			\arrow[from=BB,to=C,"m\btimesFt {\rm id}_A", sloped]
			\arrow[from=BBBB,to=CCCCC,"{\rm id}_A\btimesFt m",sloped,""{name=W}]
			%%%
			\arrow[from=AA,to=BBBB,Rightarrow,white,"{\color{black}\simeq}" description]
			\arrow[from=W,to=C,Rightarrow,"\varsigma_m",shorten <= 2ex, shorten >= 2ex]
			\arrow[from=X,to=B,Rightarrow,"\varsigma_m \btimesFt {\rm id}_A",sloped, shorten <= 2ex, shorten >= 2ex]
			\arrow[from=Y,to=BBBB,Rightarrow,"{\rm id}_A\btimesFt \varsigma_m",sloped,shorten <= 2ex, shorten >= 2ex]
		\end{tikzcd} 
	\end{align*}
	is equal to
	\begin{align*}
		\begin{tikzcd}[ampersand replacement=\&, column sep=2.7em, row sep=1.3em]
			|[alias=A]|((A\btimes A)\btimes A)\btimes A
			\&[-2.7em]\&[2.5em]
			|[alias=B]|(A\btimes A)\btimes A
			\&[0.4em]\&[-4em]
			|[alias=C]|A\btimes A
			\\\\
			|[alias=AA]|(A\btimes (A\btimes A))\btimes A
			\&\&
			|[alias=CC]|A\btimes (A\btimes A)
			\\[-0.65em]
			\&
			|[alias=BBB]|(A\btimes A)\btimes(A\btimes A)
			\&\&
			|[alias=CCC]|A\btimes A
			\&\&[-2em]
			|[alias=DDD]|A
			\\[-0.65em]
			|[alias=AAAA]|A\btimes ((A\btimes A)\btimes A)
			\&\&
			|[alias=CCCC]|(A\btimes A)\btimes A
			\\\\
			|[alias=AAAAA]|A\btimes (A\btimes (A\btimes A))
			\&\&
			|[alias=BBBBB]|A\btimes (A\btimes A)
			\&\&
			|[alias=CCCCC]|A\btimes A
			\arrow[from=A,to=B,"(m\btimesFt {\rm id}_A)\btimesFt {\rm id}_A",""{name=Z,above}]
			\arrow[from=B,to=C,"m\btimesFt {\rm id}_A"]
			\arrow[from=C,to=DDD,"m"]
			\arrow[from=A,to=AA,"\alpha_{A,A,A}\btimesFt {\rm id}_A"']
			\arrow[from=AA,to=AAAA,"\alpha_{A,A\btimesFFt A,A}"',""{name=X}]
			\arrow[from=AAAA,to=AAAAA,"{\rm id}_A\btimesFt \alpha_{A,A,A}"']
			\arrow[from=BBBBB,to=CCCCC,"{\rm id}_A\btimesFt m"']
			\arrow[from=CCCCC,to=DDD,"m"']
			\arrow[from=AAAAA,to=BBBBB,"{\rm id}_A\btimesFt(m\btimesFt {\rm id}_A)"',""{name=Y,above}]
			\arrow[from=A,to=BBB,"\alpha_{A\btimesFt A,A,A}",sloped]
			\arrow[from=BBB,to=AAAAA,"\alpha_{A,A,A\btimesFt A}"',sloped]
			\arrow[from=BBB,to=CC,"\!\! m\btimesFt ({\rm id}_A\btimesFt {\rm id}_A)",sloped]
			\arrow[from=CC,to=CCC,"{\rm id}_A\btimesFt m",sloped,""{name=P}]
			\arrow[from=BBB,to=CCCC,"\!\! ({\rm id}_A\btimesFt {\rm id}_A)\btimesFt m"',sloped]
			\arrow[from=CCCC,to=CCC,"m\btimesFt {\rm id}_A"',sloped,""{name=Q}]
			\arrow[from=CCC,to=DDD,"m"]
			\arrow[from=B,to=CC,"\alpha_{A,A,A}"]
			\arrow[from=CCCC,to=BBBBB,"\alpha_{A,A,A}"]
			%%%%
			\arrow[from=CC,to=CCCC,Rightarrow,white,"{\color{black}\simeq}" description]
			\arrow[from=X,to=BBB,Rightarrow,"\pi", shorten <= 1ex, shorten >= 1ex]
			\arrow[from=Y,to=BBB,Rightarrow,white,"{\color{black}\simeq}" description]
			\arrow[from=Z,to=BBB,Rightarrow,white,"{\color{black}\simeq}" description]
			\arrow[from=P,to=C,Rightarrow,"\varsigma_m",shorten <= 2ex, shorten >= 2ex]
			\arrow[from=Q,to=CCCCC,Rightarrow,"\varsigma_m",shorten <= 3.2ex, shorten >= 2ex]
		\end{tikzcd}
	\end{align*}
	and
	\begin{align*}
		\begin{tikzcd}[ampersand replacement=\&, column sep=2.5em, row sep=1.6em]
			\&[-2em]\&|[alias=B]|(A\btimes \mathbbm 1)\btimes A
			\&\&|[alias=C]|(A\btimes A)\btimes A
			\&\&|[alias=D]|A\btimes A
			\\\\
			|[alias=AA]|A\btimes A
			\&\&\&\&\&\&\&\&[-2em]
			|[alias=EE]|A
			\\\\
			\&\&|[alias=BBB]|A\btimes (\mathbbm 1\btimes A)
			\&\&|[alias=CCC]|A\btimes (A\btimes A)
			\&\&|[alias=DDD]|A\btimes A
			\arrow[from=AA,to=B,"r^{-1}\btimesFt {\rm id}_A"]
			\arrow[from=B,to=C,"{(A\btimesFt u)\btimesFt {\rm id}_A}"]
			\arrow[from=C,to=D,"m\btimesFt {\rm id}_A"]
			\arrow[from=D,to=EE,"m"]
			\arrow[from=AA,to=BBB,"{\rm id}_A\btimesFt \ell^{-1}"']
			\arrow[from=BBB,to=CCC,"{{\rm id}_A\btimesFt (u\btimesFt {\rm id}_A)}"']
			\arrow[from=CCC,to=DDD,"{\rm id}_A\btimesFt m"']
			\arrow[from=DDD,to=EE,"m"']
			%%%%%
			\arrow[from=AA,to=D,"{\rm id}_A\btimesFt {\rm id}_A"',sloped,shorten >= 1ex,""{name=XX,above}]
			\arrow[from=AA,to=EE,"m",""{name=X,above},""{name=Y,below}]
			\arrow[from=AA,to=DDD,"{\rm id}_A\btimesFt {\rm id}_A",sloped,shorten >= 1ex, ""{name=YY,below}]
			\arrow[Rightarrow,from=B,to=XX,"\hspace{-1.7em} \varsigma_r \btimesFt {\rm id}_A"', sloped]
			\arrow[Rightarrow,from=D,to=X,"{\color{black}\simeq}" description,white]
			\arrow[Rightarrow,from=Y,to=DDD,"{\color{black}\simeq}" description,white]
			\arrow[Rightarrow,from=BBB,to=YY,"\hspace{-2.2em} {\rm id}_A\btimesFt \varsigma_\ell^{-1}",sloped]
		\end{tikzcd}
	\end{align*}
	is equal to
	\begin{align*}
		\phantom{.}\begin{tikzcd}[ampersand replacement=\&, column sep=2.5em, row sep=1.6em]
			\&[-2em]\&|[alias=B]|(A\btimes \mathbbm 1)\btimes A
			\&\&|[alias=C]|(A\btimes A)\btimes A
			\&\&|[alias=D]|A\btimes A
			\\\\
			|[alias=AA]|A\btimes A
			\&\&\&\&\&\&\&\&[-2em]
			|[alias=EE]|A
			\\\\
			\&\&|[alias=BBB]|A\btimes (\mathbbm 1\btimes A)
			\&\&|[alias=CCC]|A\btimes (A\btimes A)
			\&\&|[alias=DDD]|A\btimes A
			\arrow[from=AA,to=B,"r^{-1}\btimesFt {\rm id}_A",sloped,""{name=X,below}]
			\arrow[from=B,to=C,"{({\rm id}_A\btimesFt u)\btimesFt {\rm id}_A}"]
			\arrow[from=C,to=D,"m\btimesFt {\rm id}_A"]
			\arrow[from=D,to=EE,"m",sloped]
			\arrow[from=AA,to=BBB,"{\rm id}_A\btimesFt \ell^{-1}"',sloped]
			\arrow[from=BBB,to=CCC,"{{\rm id}_A\btimesFt (u\btimesFt {\rm id}_A)}"']
			\arrow[from=CCC,to=DDD,"{\rm id}_A\btimesFt m"',""{name=Y,above}]
			\arrow[from=DDD,to=EE,"m"',sloped]
			\arrow[from=B,to=BBB,"\alpha_{A,\mathbbm 1,A}"]
			\arrow[from=C,to=CCC,"\alpha_{A,A,A}"']
			\arrow[from=BBB,to=X,"\tau_2",Rightarrow, shorten <= 2ex, shorten >= 2ex]
			\arrow[from=B,to=CCC,Rightarrow,white,"{\color{black}\simeq}" description]
			\arrow[from=Y,to=D,Rightarrow,"\varsigma_m", shorten <= 2ex, shorten >= 2ex]
		\end{tikzcd} .
	\end{align*}
\end{definition}

\noindent
Given the above definition, a first observation is that a pseudo-algebra object in $\mathsf{2Vec}$ corresponds to a finite-dimensional, semi-simple monoidal category. This relies in particular on the fact that semi-simple abelian categories always have a unique structure of semi-simple $\mathsf{Vec}$-module category \cite{neuchl1997representation}. Let us now apply this definition to $\mathsf{2Vec}^{\pi}_{G}$. For each pair $(A,\lambda)$, where $A\subset G$ is a subgroup and $\lambda\in C^{3}(A,\rU(1))$ is a 3-cochain satisfying the condition $d^{(3)}\lambda=\pi^{-1}{\sss |}_{A}$, we construct a pseudo-algebra object $\mathsf{Vec}_{A,\lambda} \equiv (\bigbplus_{a \in A}\mathsf{Vec}_a,m,u,\varsigma_m,\varsigma_r,\varsigma_\ell)$ such that: the multiplication
$m:\mathsf{Vec}_{A,\lambda}\btimes \mathsf{Vec}_{A,\lambda}\to \mathsf{Vec}_{A,\lambda}$ is given on homogeneous components via the functor $m_{\mathsf{Vec}_{a},\mathsf{Vec}_{a'}}:\mathsf{Vec}_{a}\btimes \mathsf{Vec}_{a'}\mapsto \mathsf{Vec}_{aa'}$ for all $a,a'\in A$, the unit map $u$ is defined in an obvious way, the 2-isomorphisms $\varsigma_r$ and $\varsigma_\ell$ are trivial, and the 2-isomorphism
\begin{align}
	\varsigma_m:\alpha_{\mathsf{Vec}_{A,\lambda},\mathsf{Vec}_{A,\lambda},\mathsf{Vec}_{A,\lambda}}\circ({\rm id}_{\mathsf{Vec}_{A,\lambda}}\circ m\btimes m)\Rightarrow (m\btimes{\rm id}_{\mathsf{Vec}_{A,\lambda}})\circ m
\end{align}
defines an associator for the product map $m$ that is determined by $\lambda$. This associator acts on homogenous components labelled by $a,a',a''\in A$ as
\begin{align*}
	\lambda_{a,a',a''}:\alpha_{\mathsf{Vec}_{a},\mathsf{Vec}_{a'},\mathsf{Vec}_{a''}}
	\!\! \circ ({\rm id}_{\mathsf{Vec}_{a}} \! \btimes m_{\mathsf{Vec}_{a'},\mathsf{Vec}_{a''}}) \! \circ m_{\mathsf{Vec}_{a},\mathsf{Vec}_{a'a''}}
	\! \Rightarrow
	(m_{\mathsf{Vec}_{a},\mathsf{Vec}_{a'}} \! \btimes {\rm id}_{\mathsf{Vec}_{a''}}) \! \circ  m_{\mathsf{Vec}_{aa'},\mathsf{Vec}_{a''}} .
\end{align*}
The condition $d^{(3)}\lambda=\pi^{-1}{\sss |}_{A}$ demonstrates that $\mathsf{Vec}_{A,\lambda}$ is not a monoidal category in the conventional sense since the associator $\lambda$ fails to satisfy the pentagon equation \eqref{eq:pentagonbicat}. Instead, the associator satisfies the following equation on homogeneous components labelled by $a,a',a'',a''' \in A$:
\begin{align}
	(\lambda_{a,a',a''}\btimes{\rm id}_{\mathsf{Vec}_{a'''}})
	\circ
	\lambda_{a,a'a'',a'''}
	\circ
	({\rm id}_{\mathsf{Vec}_{a}}\btimes \lambda_{a',a'',a'''})\circ\pi_{a,a',a'',a'''}
	=
	\lambda_{aa',a'',a'''}\circ \lambda_{a,a',a''a'''} \; .
\end{align}
In this way, we see that $\mathsf{Vec}_{A,\lambda}$ defines a monoidal category which is associative inside 2$\mathsf{Vec}^{\pi}_{G}$ but not as a conventional monoidal category. This result provides a categorification of the observation that an algebra object $A_{\phi}$ in $\mathsf{Vec}^{\alpha}_{G}$ defines a twisted groupoid algebra, which is associative inside $\mathsf{Vec}^{\alpha}_{G}$ but not as a conventional algebra.

\subsection{Bicategory of gapped boundary excitations in (3+1)d gauge models}\label{sec:gappedexcitationsbicat}

Mimicking the analysis carried out in sec.~\ref{sec:bicatmodel2D}, we shall now introduce a category theoretical formulation of gapped boundaries in (3+1)d gauge models and string-like excitations terminating at gapped boundaries, which we studied from a tube algebra point of view in sec.~\ref{sec:tube3D}. In particular, we shall define a bicategory  $\mathsf{2Bdry}_G^\pi$ that is analogous to $\mathsf{Bdry}_G^\alpha$. We shall then relate this construction to the work of Kong et al. in \cite{Kong:2019brm} arguing that $\mathsf{2Bdry}_G^\pi$ forms a full sub-bicategory of $\mc{Z}(\mathsf{2Vec}^{\pi}_{G})$, i.e. the \emph{centre} of $\mathsf{2Vec}^{\pi}_{G}$.

Let us begin with a brief review of the results obtained in the first part of this manuscript within the tube algebra approach. Hamiltonian realisations of (3+1)d Dijkgraaf-Witten theory are defined in terms of pairs $(G,\pi)$, where $G$ is a finite group and $\pi$ a normalised 4-cocycle in $H^4(G,\rU(1))$. In sec.~\ref{sec:bdryHam}, it was argued that gapped boundaries can be indexed by pairs $(A,\lambda)$, where $A \subset G$ is a subgroup of $G$ and $\lambda \in C^3(A,\rU(1))$ is a 3-cochain satisfying the condition $d^{(3)}\lambda = \pi^{-1}{\sss |}_A$. In the previous section, we explained that such data is in bijection with pseudo-algebra objects $\mathsf{Vec}_{A,\lambda}$ in $\mathsf{2Vec}^{\pi}_{G}$. Moreover, we showed in sec.~\ref{sec:tube3D} within the tube algebra approach that given a pair of two-dimensional gapped boundaries labelled by $(A,\lambda)$ and $(B,\mu)$, respectively, string-like excitations threading through the bulk from the former boundary to the latter were defined as modules of the twisted relative groupoid algebra $\mathbb{C}[\Lambda(G_{AB})]^{{\textsf{\scriptsize T}}(\pi)}_{{\textsf{\scriptsize T}}(\lambda){\textsf{\scriptsize T}}(\mu)}$, where $\Lambda(G_{AB}) \equiv \Lambda G_{\Lambda A \Lambda B}$ and ${\sfT}: Z^{\bullet}(G,\rU(1)) \to Z^{\bullet -1}(\Lambda  G, \rU(1))$.\footnote{Recall that $\Lambda G$ refers to the loop groupoid of the group $G$ treated as a one-object groupoid (see sec.~\ref{sec:tube3D}).} Via the introduction of a comultiplication map, we further described the concatenation of such string-like excitations in sec.~\ref{sec:rep}.

Let us now collect these results into a bicategory $\mathsf{2Bdry}_G^\pi$, in a way akin to the definition of $\mathsf{Bdry}_G^\alpha$. The objects of $\mathsf{2Bdry}_G^\pi$ are given by pairs $(\Lambda A,{\sfT}(\lambda))$ for every gapped boundary condition labelled by $(A,\lambda)$.
Given a pair of objects $(\Lambda A,{\sfT}(\lambda))$, $(\Lambda B,{\sfT}(\mu))$, we define the hom-category
\begin{align}
	\mathsf{Hom}_{\mathsf{2Bdry}_G^\pi}\big((\Lambda A,{\sfT}(\lambda)),(\Lambda B,{\sfT}(\mu))\big):=
	\mathsf{Mod}\big(\mathbb{C}[\Lambda(G_{AB})]^{{\ssfT}(\pi)}_{{\ssfT}(\lambda){\ssfT}(\mu)}\big) \; ,
\end{align}
where $\mathsf{Mod}(\mathbb{C}[\Lambda(G_{AB})]^{{\ssfT}(\pi)}_{{\ssfT}(\lambda){\ssfT}(\mu)})$ denotes the category of $\mathbb{C}[\Lambda(G_{AB})]^{{\ssfT}(\pi)}_{{\ssfT}(\lambda){\ssfT}(\mu)}$-modules and intertwiners. The composition functors, associator and unitors are given analogously to the construction of $\mathsf{Bdry}_G^\alpha$.

From this definition, we interpret the objects $(\Lambda A,{\sfT}(\lambda))$ of $\mathsf{2Bdry}_G^\pi$ as defining boundary conditions for the endpoints of a string-like excitation that terminates on a gapped boundary labelled by $(A,\lambda)$. An isomorphism class of objects in $\Lambda A$ specifies possible fluxes for a string-like excitation terminating on the boundary $(A,\lambda)$. This flux corresponds to the closed holonomy going along the non-contractible cycle perpendicular to the length of the string. Given a pair of objects $(\Lambda A,{\sfT}(\lambda))$, $(\Lambda B,{\sfT}(\mu))$ a 1-morphism $\rho_{AB}\in {\rm Ob}(\mathsf{Hom}_{\mathsf{2Bdry^\pi_G}}((A,\lambda),(B,\mu)))$ specifies a magnetic quantum number describing the gauge orbit of parallel transports along the length of the string---generically, such a parallel transport must be compatible with the possible boundary conditions for the endpoints of the string---as well as a charge quantum number decomposing the symmetries of the gauge action on the string. In this way, we view such strings as dyonic excitations. The bifunctor on 1-morphisms provides a notion of concatenation for a pair of string-like excitations that share a boundary endpoint, as described in sec.~\ref{sec:comult}. The 2-morphisms correspond to intertwiners, so that a 2-morphism of the form $\zeta: \rho_{AB}\otimes_B \rho_{BC} \to \rho_{AC}$ can be interpreted as implementing the renormalization of a pair of concatenated string-like excitations. Identity 1-morphisms and unitors are defined analoguously to $\mathsf{Bdry}_G^\alpha$. Similarly, the 1-associator for a triple of 1-morphisms $\rho_{AB}$, $\rho_{BC}$, $\rho_{CD}$ in the appropriate hom-categories is given by the intertwiner isomorphism $\Phi^{\rho_{AB}\rho_{BC} \rho_{CD}}:(\rho_{AB}\otimes_{B}\rho_{BC})\otimes_{C} \rho_{CD}\to \rho_{AB}\otimes_{B}(\rho_{BC}\otimes_{C} \rho_{CD})$, as described explicitly in sec.~\ref{sec:comult}.  

\bigskip \noindent
It is well-known that, given a lattice Hamiltonian realisation of (2+1)d Dijkgraaf-Witten theory with input data $(G,\alpha)$, algebraic properties of the (bulk) anyonic excitations can be encoded into the \emph{centre} $\mc{Z}(\mathsf{Vec}^{\alpha}_{G})$ of the fusion category $\mathsf{Vec}^{\alpha}_{G}$, this centre being in particular a \emph{braided} monoidal category. The objects of $\mc{Z}(\mathsf{Vec}^{\alpha}_{G})$ are interpreted as the elementary excitations of the model, or anyons, and the morphisms implement space-time processes of such anyons. The monoidal structure describes the fusion and splitting processes of the excitations, whereas the braiding structure encodes their exchange statistics. Recently, Kong et al. studied in \cite{Kong:2019brm} the analogue of this result in (3+1)d. The relevant category theoretical structure in (3+1)d being the monoidal bicategory $\mathsf{2Vec}^{\pi}_{G}$, they computed the braided monoidal bicategory ${\mc{Z}}(\mathsf{2Vec}^{\pi}_{G})$ obtained as the categorified centre of $\mathsf{2Vec}^{\pi}_{G}$, arguing that such a bicategory should describe string-like excitations and their statistics in (3+1)d gauge models. More specifically,  they demonstrated that as a bicategory ${\mc{Z}}(\mathsf{2Vec}^{\pi}_{G})$ is equivalent to the bicategory $\mathsf{MOD}(\mathsf{Vec}^{{\ssfT}(\pi)}_{\Lambda G})$. Using this equivalence, they suggested that objects of ${\mc{Z}}(\mathsf{2Vec}^{\pi}_{G})$ could be interpreted as string-like topological excitations, 1-morphisms as particle-like topological excitations, and 2-morphisms as instantons. Relating this bicategory to the boundary tube algebra in (3+1)d, we shall argue that objects of ${\mc{Z}}(\mathsf{2Vec}^{\pi}_{G})$ should be interpreted as boundary conditions for the endpoints of a string-like excitation---such a boundary condition specifying in particular allowed fluxes for the excitation---the 1-morphisms as quantum numbers associated with string-like topological excitations that are constrained by a choice of endpoints boundary conditions, and 2-morphisms as implementing the renormalisation of concatenated string-like excitations.

In order to establish the interpretation spelt out above, we begin by showing that $\mathsf{2Bdry}_G^\pi$ is equivalent as a bicategory to a full sub-bicategory $\sfpartial \mathsf{MOD}(\mathsf{Vec}^{{\ssfT}(\pi)}_{\Lambda G})$ of $\mathsf{MOD}(\mathsf{Vec}^{{\ssfT}(\pi)}_{\Lambda G})$.
Our argument mirrors the equivalence of bicategories $\mathsf{Bdry}_G^\alpha\simeq \mathsf{MOD}(\mathsf{Vec}^{\alpha}_{G})$ established in sec.~\ref{sec:bicatmodel2D}. Utilising prop.~\ref{prop:MOD-sAlg}, we know that, up to equivalence, all $\mathsf{Vec}^{{\ssfT}(\pi)}_{\Lambda G}$-module categories can be expressed as the category of module objects for an algebra object in $\mathsf{Vec}^{{\ssfT}(\pi)}_{\Lambda G}$. Moreover, we established in sec.~\ref{sec:algObj} that all such algebra objects were indexed by $(\Lambda G, {\sfT}(\pi))$-subgroupoids, as defined in sec.~\ref{sec:relatGr}. Given the data $(A,\lambda)$ of gapped boundary condition in (3+1)d, we explained in sec.~\ref{sec:tube3D} that the loop groupoid $\Lambda A$ together with the groupoid 2-cochain ${\sfT}(\lambda)$ defines such a $(\Lambda G, {\sfT}(\pi))$-subgroupoid. Henceforth, we shall refer to groupoids of this form as $\sfpartial (\Lambda G, {\sfT}(\pi))$-subgroupoids. In this vein, we define the bicategory $\sfpartial\mathsf{MOD}(\mathsf{Vec}^{{\ssfT}(\pi)}_{\Lambda G})$ as the full sub-bicategory of $\mathsf{MOD}(\mathsf{Vec}^{{\ssfT}(\pi)}_{\Lambda G})$ whose objects are  $\mathsf{Vec}^{{\ssfT}(\pi)}_{\Lambda G}$-module categories induced from $\sfpartial(\Lambda G,{\sfT}(\pi))$-subgroupoids, and hom-categories are the corresponding ones in $\mathsf{MOD}(\mathsf{Vec}^{{\ssfT}(\pi)}_{\Lambda G})$. Similarly, we define $\sfpartial\mathsf{sAlg}(\mathsf{Vec}^{{\ssfT}(\pi)}_{\Lambda G})$ as the full sub-bicategory of $\mathsf{sAlg}(\mathsf{Vec}^{{\ssfT}(\pi)}_{\Lambda G})$, whose objects are  algebra objects in $\mathsf{Vec}^{{\ssfT}(\pi)}_{\Lambda G}$ of the form $\Lambda A_{{\ssfT}(\lambda)}$, and hom-categories are the corresponding categories of bimodule objects in $\mathsf{Vec}^{{\ssfT}(\pi)}_{\Lambda G}$. Mimicking our proof of the equivalence $\mathsf{Bdry}_G^\alpha\simeq \mathsf{sAlg}(\mathsf{Vec}^{\alpha}_{G})$, we can show the equivalence between $\mathsf{2Bdry}_G^\pi$ and $\sfpartial\mathsf{sAlg}(\mathsf{Vec}^{{\ssfT}(\pi)}_{\Lambda G})$. This equivalence relies in particular on the isomorphism $\mathbb C[\Lambda(G_{AB})]^{{\ssfT}(\pi)}_{{\ssfT}(\lambda){\ssfT}(\mu)} \equiv \mathbb{C}[\Lambda G_{\Lambda A\Lambda B}]^{\vartheta^{\Lambda A \Lambda B}} \simeq \mathbb C[\widetilde{\Lambda G}_{\Lambda A \Lambda B}]^{\varpi^{\Lambda A \Lambda B}}$ of twisted relative groupoid algebras, which is realised by an obvious generalisation of \eqref{eq:isoGroupoid}. Utilising the proof of prop.~\ref{prop:MOD-sAlg}, it follows that $\partial\mathsf{sAlg}(\mathsf{Vec}^{\alpha}_{G})\simeq \partial\mathsf{MOD}(\mathsf{Vec}^{{\ssfT}(\pi)}_{\Lambda G})$, hence establishing the equivalence
\begin{align}
	\mathsf{2Bdry}_G^\pi\simeq \sfpartial \mathsf{MOD}(\mathsf{Vec}^{{\ssfT}(\pi)}_{\Lambda G}) \; . 
\end{align}
Let us now explain how we can generalise our approach so as to obtain the bicategory $\mathsf{MOD}(\mathsf{Vec}^{{\ssfT}(\pi)}_{\Lambda G})$, which we recall was shown to be equivalent to $\mathcal{Z}(\mathsf{2Vec}^\pi_G)$. When considering the boundary tube algebra for the (3+1)d gauge models in sec.~\ref{sec:tube3D}, we could have allowed for a larger spectrum of boundary colourings beyond the ones inherited from the gapped boundary conditions. More specifically, we could have considered $G$-colourings that are provided by morphisms in any $(\Lambda G,{\sfT}(\pi))$-subgroupoid $(\mathcal{X},\phi)$ such that $d^{(2)}\phi = {\sfT}(\pi){\sss |}_\mc X^{-1}$. Given a pair of  $(\Lambda G,{\sfT}(\pi))$-subgroupoids $(\mathcal{X},\phi)$ and $(\mathcal{Y},\psi)$, we could have then considered $G$-coloured graph-states of the form
\begin{align}
	\big| \biGr{\fr g}{\fr x}{\fr y} \big\ra \equiv 
	\Bigg| \! \tubeThreeD{0.9}{0}{1}{g}{x}{y}{3} \Bigg\ra
	\equiv 
	\Bigg| \tubeThreeDBeauty{1}{2} \Bigg\ra
\end{align}
where we borrowed the notation from sec.~\ref{sec:tube3D} and
\begin{equation*}
	\fr x = x_2 \xrightarrow{x_1}x_2^{x_1} \in {\rm Hom}(\mc X) \; , \q 
	\fr y = y_2 \xrightarrow{y_1}y_2^{y_1} \in {\rm Hom}(\mc Y) \; , \q
	\biGr{\fr g}{\fr x}{\fr y}\; , \in {\rm Hom}(\Lambda G_{\mc X \mc Y}) \; , 
\end{equation*}
such that $\Lambda G_{\mc{X}\mc{Y}}$ denotes the relative groupoid over $\Lambda G$ defined by $\mc X$ and $\mc Y$. 
In this setting, there exists a natural multiplication of such boundary tubes defining an algebra isomorphic to the twisted groupoid algebra $\mathbb{C}[\Lambda G_{\mc X \mc Y}]^{{\ssfT}(\pi)}_{\phi\psi}$. Letting $\mathsf{String}_G^\pi$ denote the bicategory defined in the same manner as $\mathsf{2Bdry}_G^\pi$ with objects all $(\Lambda G,{\sfT}(\pi))$-subgroupoids and hom-categories
\begin{align}
	\mathsf{Hom}_{\mathsf{String}_G^\pi}((\mc X,\phi),(\mc Y,\psi)):=\mathsf{Mod}(\mathbb{C}[\Lambda G_{\mc X \mc Y}]^{{\ssfT}(\pi)}_{\phi\psi}) \; ,
\end{align}
we obtain the following equivalence of bicategories:
\begin{align}
	\mathsf{String}_G^\pi\simeq \mathsf{MOD}(\mathsf{Vec}^{{\ssfT}(\pi)}_{\Lambda G}) \; .
\end{align}
Utilising this equivalence of bicategories, together with the physical interpretation inherited from the tube algebra approach, we interpret the  $\mathsf{Vec}^{{\ssfT}(\pi)}_{\Lambda G}$-module category $\mathsf{Mod}_{\mathsf{Vec}^{{\textsf{\tiny T}}(\pi)}_{\Lambda G}}(\mc X_{\phi})$ for a  $(\Lambda G,{\sfT}(\pi))$-subgroupoid $(\mc X,\phi)$ as the \emph{2-Hilbert space} \cite{baez1997higher} of boundary conditions that appear at the endpoint of a string-like (bulk) excitation. As before, 1-morphisms are naturally interpreted as the quantum numbers of string-like excitations.

The motivation for calling $\mathsf{Vec}^{{\ssfT}(\pi)}_{\Lambda G}$-module categories $\mathsf{Mod}_{\mathsf{Vec}^{{\textsf{\tiny T}}(\pi)}_{\Lambda G}}(\mc X_{\phi})$ 2-Hilbert spaces is as follows. In finite-dimensional quantum mechanics, given a finite set $X$ of classical field configurations, the corresponding Hilbert space $\mc{H}[X]$ is given by the free vector space of functions $f:X\rightarrow\mathbb{C}$. Categorifying the set of classical field configurations to a groupoid $\mc{G}$, whose objects correspond to classical field configurations and morphisms, the symmetries of the field configurations. The category $[\mc{G},\mathsf{Vec}]^{\beta}$ of (weak) functors $F:\mc{G}\rightarrow \mathsf{Vec}$ for $[\beta]\in H^{2}(\mc{G},\rU(1))$ provides a natural categorification of $\mc{H}[X]$ which defines a finite 2-vector space (see sec.~\ref{sec:pseudoalgob}). The category $[\mc{G},\mathsf{Vec}]^{\beta}$ can then be shown to admit a categorification of the inner-product of finite Hilbert spaces given by the hom-functor
\begin{align}
	\langle -,-\rangle:([\mc{G},\mathsf{Vec}]^{\beta})^{\rm op}\boxtimes [\mc{G},\mathsf{Vec}]^{\beta}\rightarrow \mathsf{Vec} \; .
\end{align}
Recalling that $\mathsf{Mod}_{\mathsf{Vec}^{{\textsf{\tiny T}}(\pi)}_{\Lambda G}}(\mc X_{\phi})$ is defined by a category of weak functors from a groupoid to $\mathsf{Vec}$, the term 2-Hilbert space seems most appropriate.

\bigskip \noindent
We conclude this section by showing that, in general, objects in $\sfpartial\mathsf{MOD}(\mathsf{Vec}^{{\ssfT}(\pi)}_{\Lambda G})$ are not \emph{indecomposable} as $\mathsf{Vec}^{{\ssfT}(\pi)}_{\Lambda G}$-module categories. For convenience, we shall focus on the limiting case where the group $G$ is \emph{abelian}, but our analysis can be extended to the non-abelian scenario.  Analogously to indecomposable modules over an algebra, an indecomposable module category is a module category which is not equivalent to the direct sum of non-zero module categories. Using the equivalence between $\mathsf{Vec}^{{\ssfT}(\pi)}_{\Lambda G}$-module categories and the categories of module objects for a separable algebra object in $\mathsf{Vec}^{{\ssfT}(\pi)}_{\Lambda G}$, we have that a   $\mathsf{Vec}^{{\ssfT}(\pi)}_{\Lambda G}$-module category is indecomposable if only if the corresponding algebra object is not Morita equivalent to a direct sum of non-zero algebra objects. Given a (3+1)d gauge model with input data $(G,\pi)$, and a choice of gapped boundary condition $(A,\lambda)$, an algebra object $\Lambda A_{{\ssfT}(\lambda)}$ in  $\mathsf{Vec}^{{\ssfT}(\pi)}_{\Lambda G}$ naturally decomposes as a direct sum via
\begin{align}
	\Lambda A_{{\ssfT}(\lambda)}=\bigoplus_{a\in A}(\Lambda A_a)_{{\ssfT}_a(\lambda)} \; ,
\end{align}
where $\Lambda A_{a}$ denotes the groupoid with unique object $a\in A$ and set of morphisms $\{a\xrightarrow{a'}a\}_{\forall \, a'\in A}$. The 2-cochain ${\sfT}_a(\lambda)\in C^{2}(\Lambda A^{a},\rU(1))$ is then given by the restriction of ${\sfT}(\lambda)\in C^{2}(\Lambda A,\rU(1))$ to $\Lambda A_{a}$. This decomposition yields
\begin{align}
	\mathsf{Mod}_{\mathsf{Vec}^{{\textsf{\tiny T}}(\pi)}_{\Lambda G}}(\Lambda A_{{\ssfT}(\lambda)})
	\simeq
	\bigoplus_{a\in A}
	\mathsf{Mod}_{\mathsf{Vec}^{{{\textsf{\tiny T}}}(\pi)}_{\Lambda G}}((\Lambda A_a)_{{\ssfT}_a(\lambda)})
\end{align}
as $\mathsf{Vec}^{{\ssfT}(\pi)}_{\Lambda G}$-module categories, so that the category of module objects  is not indecomposable as a module category unless $A=\mathbbm 1_{G}$ is the trivial subgroup of $G$. Generically, for possibly non-abelian $G$ an indecomposable $\mathsf{Vec}^{{\ssfT}(\pi)}_{\Lambda G}$-module category can be specified by a triple $(\mc O,H,\phi)$ ,  where $\mc O$ denotes a conjugacy class of $G$, $H$ is a subgroup of the centralizer $Z_{o_1}\subseteq G$ for a representative $o_1 \in \mc O$, and $\phi\in C^{2}(H,\rU(1))$ is 2-cochain satisfying $d^{(2)}\phi={\sfT}(\pi){\sss |}_{H}$ \cite{Kong:2019brm}. The corresponding algebra object is then given by $(H_{o_1})_{\phi_{o_1}}$, where $H_{o_1}$ denotes the groupoid with unique object $o_1 \in \mc O$ and  hom-set $\{h:o_1 \to o_1 \}_{\forall \, h\in H}$ with composition given by multiplication in $H$, and the 2-cochain $\phi_{o_1}\in C^{2}(H_{o_1},\rU(1))$ is defined by the relation $\phi_{o_1}(h:o_1 \to o_1,h':o_1 \to o_1):=\phi(h,h')$ for all $h,h'\in H$.

\section{Discussion}

Gapped boundaries of topological models have been under scrutiny in the past years. Focusing on lattice Hamiltonian realisations of Dijkgraaf-Witten theory, a.k.a gauge models of topological phases, we studied gapped boundaries and their excitations in (2+1)d and (3+1)d. More specifically, the goal of this paper was two-fold: Apply the tube algebra approach to classify gapped boundary excitations and, using these results, elucidate the higher-category theoretical formalism relevant to describe gapped boundaries in (3+1)d.

As explained in detail in \cite{Bullivant:2019fmk}, local operators of lattice Hamiltonian realisations of Dijkgraaf-Witten theory can be conveniently expressed in terms of the partition function of the theory applied to so-called pinched interval cobordisms. We introduced a generalisation of the Dijkgraaf-Witten partition function, from which the gapped boundary Hamiltonian operators could be defined in analogy with the bulk Hamiltonian operators using the language of relative pinched interval cobordisms.  Given gapped boundaries labelled by subgroups of the input group and cochains compatible with the input cocycle, we applied the tube algebra approach in order to reveal the algebraic structure underlying two types of excitations: $(i)$ Point-like excitations at the interface of two gapped boundaries in (2+1)d, where the `tube' has the topology of $\mathbb I \times \mathbb I$, and $(ii)$ string-like (bulk) excitations terminating at point-like gapped boundary excitations, where the `tube' has the topology of $(\mathbb S^1 \times \mathbb I)\times \mathbb I$. Crucially, both tube algebras can be related via a lifting (or dimensional reduction) argument, and as such can be studied in parallel. This statement was formalised using the notion of relative groupoid algebra. When applied to the input group treated as a one-object groupoid, this notion yields the (2+1)d tube algebra, whereas it yields the (3+1)d tube algebra when applied to the loop groupoid of the group. We subsequently studied the representation theory of the (3+1)d tube algebra in full detail, which encompasses the (2+1)d one as a limiting case, deriving the irreducible representations as well as the corresponding recoupling theory.

In the second part of this manuscript, we reformulated the previous statements in category theoretical terms. In (2+1)d, the relevant notion to describe gapped boundaries and their excitations is the bicategory $\mathsf{MOD}(\mathsf{Vec}^{\alpha}_{G})$ of module categories over the category $\mathsf{Vec}^{\alpha}_{G}$ of group-graded vector spaces. In practice, a module category can be obtained as a category of modules over an algebra object in the input category. The bicategory of module categories above can then be shown to be equivalent to a bicategory of separable algebra objects, such that objects correspond to the gapped boundary conditions and morphisms to representations of a groupoid algebra isomorphic to the (2+1)d tube algebra. The identification with the tube algebra allowed us to elucidate the physical interpretation of the category theoretical notions at play. Mimicking this (2+1)d construction, we further defined a bicategory that encodes the string-like excitations terminating at point-like excitations on gapped boundaries and found that is was equivalent to a \emph{sub-bicategory} of the bicategory $\mathsf{MOD}(\mathsf{Vec}^{{\ssfT}(\pi)}_{\Lambda G})$ of modules categories over the category $\mathsf{Vec}^{{\ssfT}(\pi)}_{\Lambda G}$ of loop-groupoid-graded vector spaces. Comparing with the work of Kong et al. \cite{Kong:2019brm}, $\mathsf{MOD}(\mathsf{Vec}^{{\ssfT}(\pi)}_{\Lambda G})$ is equivalent to the higher categorical centre $\mathcal{Z}(\mathsf{2Vec}_G^\pi)$ of the category $\mathsf{2Vec}_G^\pi$ of $G$-graded 2-vector spaces, which is the input category of (3+1)d gauge models. In virtue of the physical interpretation inherited from the tube algebra approach, we thus suggested that $\mathcal{Z}(\mathsf{2Vec}_G^\pi)$ describes dyonic bulk string-like excitations whose end-points are pinned to the boundary of the spatial manifold. This is the higher-dimensional analogue of the well-known statement that bulk point-like excitations in (2+1)d are described by the centre $\mathcal Z(\mathsf{Vec}^\alpha_G)$ of  the input category.

The distinction between the gapped boundary string-like excitations we focused on, and the more general ones encoded in the centre $\mathcal Z(\mathsf{2Vec}^\pi_G)$ can be appreciated from an extended TQFT point of view. We should think of $\mathcal Z(\mathsf{2Vec}^\pi_G)$ as describing the object the extended 4-3-2-1 Dijkgraaf-Witten TQFT assigns to the circle. It follows from our analysis that such extended TQFT is more general than what gapped boundary conditions provide. Working out the details of this more general scenario will be the purpose of another paper.

The study carried out in this manuscript can be generalized in several ways. First of all, we could study gapped domains walls instead of gapped boundaries and consider string-like excitations that terminate at gapped domains walls point-like excitations. In (2+1)d, the so-called \emph{folding trick} can be used in order to map a gapped domain wall configuration to a gapped boundary one. It would certainly be interesting to consider how this generalizes in higher dimensions. Once this more general scenario is well-understood, we could then apply our results to so-called \emph{fracton} models, which were recently suggested in \cite{wen2020systematic,aasen2020topological, Wang:2020yma} to have an interpretation in terms of defect TQFTs. A related question would be to study invertible domain walls such as duality defects and derive the underlying mathematical structure in category theoretical terms.

Another follow-up work pertains to the relation between the string-like excitations as described by $\mathcal {Z}(\mathsf{2Vec}^\pi_G)$ and the loop-like excitations of the model. In a recent paper \cite{Bullivant:2019fmk}, the authors  showed that loop-like excitations and their statistics were captured by the category of modules over the so-called \emph{twisted quantum triple algebra}. This algebra can be expressed as the twisted groupoid algebra $\mathbb C[\Lambda^2G]^{\ssfT^2(\pi)}$ of the loop groupoid of the loop groupoid of $G$. In comparison, recall that the twisted quantum double is isomorphic to $\mathbb C[\Lambda G]^{\ssfT(\alpha)}$ in this language. This groupoid algebra was shown by the authors to be isomorphic to the tube algebra associated with the manifold $\mathbb T^2 \times \mathbb I$, a local neighbourhood of a loop-like object being a solid torus. Intuitively, we may expect loop-like excitations to descend from the string-like ones via a tracing mechanism. This can be formalized using the notion of \emph{categorical trace}, building upon the fact that it maps a module category over $\mathsf{Vec}_G$ to a module over $\mathbb C[\Lambda G]$ \cite{bartlett2009unitary, ganter2006representation}. Another way to establish the connection between string-like and loop-like excitations consists in first realising that, as braided monoidal categories, we have the equivalences $\mathcal Z(\mathsf{Vec}_{\Lambda G}^{\ssfT(\pi)}) \simeq \mathsf{Mod}(\mathbb C[\Lambda^2 G]^{\ssfT^2(\pi)})$ and $\mathcal Z(\mathsf{Vec}_{\Lambda G}^{\ssfT(\pi)}) \simeq \mathsf{Dim}(\mathsf{MOD}(\mathsf{Vec}^{\ssfT(\pi)}_{\Lambda G}))$, where $\mathsf{Dim}$ denotes the \emph{dimension of a bicategory} \cite{bartlett2009unitary,Baez:1995xq} obtained via an appropriate categorification of the dimension of a vector space. The details of this correspondence will be presented in a forthcoming paper \cite{Bullivant:2021pkd}.

\newpage
\begin{center}
	\textbf{Acknowledgments}
\end{center}

\noindent
CD would like to thank David Aasen and Dominic Williamson for very useful discussions on closely related topics.
CD is funded by the European Research Council (ERC) under the European Union’s Horizon 2020 research and innovation programme through the ERC Starting Grant WASCOSYS (No. 636201) and the Deutsche Forschungsgemeinschaft (DFG, German Research Foundation) under Germany’s Excellence Strategy – EXC-2111 – 390814868. AB is funded by the EPSRC doctoral prize fellowship scheme.

\newpage
\titleformat{name=\section}[display]
{\normalfont}
{\footnotesize\centering {APPENDIX \thesection}}
{0pt}
{\large\bfseries\centering}
\appendix

\section{Representation theory of the relative groupoid algebra}
\emph{In this appendix, we collect the proofs of several important results of the representation theory of the relative groupoid algebra $\grAlgL$.}

\subsection{Proof of the orthogonality relations \eqref{eq:ortho}\label{sec:app_ortho}}

Let us confirm that the representation matrices as defined in \eqref{eq:defD} satisfy the orthogonality relation \eqref{eq:ortho}:
\begin{align*}
	&\frac{1}{|A||B|}
	\sum_{\biGrFoot{\fr g}{\fr a}{\fr b} \in \Lambda(G_{AB})}
	\overline{\mc{D}^{\rho_{AB}}_{IJ}\big( \big| \biGr{\fr g}{\fr a}{\fr b} \big\ra \big)} 
	\mc{D}^{\rho_{AB}'}_{I'J'}\big( \big| \biGr{\fr g}{\fr a}{\fr b} \big\ra \big) 
	\\
	& \q =
	\frac{1}{|A||B|}
	\sum_{\biGrFoot{\fr g}{\fr a}{\fr b} \in \Lambda(G_{AB})}
	\delta_{\fr g,\fr o_i} \, \delta_{\fr a^{-1}\fr g \fr b,\fr o_{j}} \, 
	\frac{\vartheta^{\Lambda(AB)}_{\mathfrak{o}_1}(\fr p_i^{-1}\fr a\fr p_j,\fr p_j^{-1}|\fr q_i^{-1}\fr b \fr q_j, \fr q_j^{-1})}{\vartheta^{\Lambda(AB)}_{\mathfrak{o}_1}(\fr p_i^{-1},\fr a|\fr q_i^{-1},\fr b)} \, 
	\overline{\mc{D}^R_{mn}\big (\big| \biGrZ{\fr p_i^{-1}\fr a\fr p_j}{\fr q_i^{-1}\fr b \fr q_j}\big\ra \big)}
	\\[-0.8em]
	& \hspace{8.4em} \times 	\delta_{\fr g,\fr o_i'} \, \delta_{\fr a^{-1}\fr g \fr b,\fr o_{j}'} \, 
	\frac{\vartheta^{\Lambda(AB)}_{\mathfrak{o}_1}(\fr p_i'^{-1},\fr a|\fr q_i'^{-1},\fr b)}{\vartheta^{\Lambda(AB)}_{\mathfrak{o}_1}(\fr p_i'^{-1}\fr a\fr p_j',\fr p_j'^{-1}|\fr q_i'^{-1}\fr b \fr q_j',\fr q_j'^{-1})} \, 
	\mc{D}^{R'}_{m'n'}\big (\big| \biGrZ{\fr p_i'^{-1}\fr a\fr p_j'}{\fr q_i'^{-1}\fr b \fr q_j'}\big\ra \big)
	\\
	& \q =
	\frac{1}{|A||B|}
	\sum_{\biGrFoot{\fr o_i}{\fr a}{\fr b} \in {\rm Hom}(\Lambda(G_{AB}))}
	\delta_{\mc{O}_{AB},\mc{O}_{AB}'} \,
	\delta_{i,i'} \,
	\delta_{j,j'} \,
	\delta_{\fr a^{-1}\fr o_i \fr b,\fr o_j}  \, 
	\overline{\mc{D}^R_{mn}\big (\big| \biGrZ{\fr p_i^{-1}\fr a\fr p_j}{\fr q_i^{-1}\fr b \fr q_j}\big\ra \big)}
	\mc{D}^{R'}_{m'n'}\big (\big| \biGrZ{\fr p_i^{-1}\fr a\fr p_j}{\fr q_i^{-1}\fr b\fr q_j}\big\ra \big)
	\\
	& \q =
	\frac{1}{|Z_{\mathcal{\mc O}_{AB}}|}
	\sum_{\substack{ (\fr a,\fr b) \in Z_{\mc{O}_{AB}}}}
	\delta_{\mc{O}_{AB},\mc{O}_{AB}'} \,
	\delta_{i,i'} \,
	\delta_{j,j'} \, 
	\overline{\mc{D}^R_{mn}\big (\big| \biGrZ{\fr a}{\fr b}\big\ra \big)}
	\mc{D}^{R'}_{m'n'}\big (\big| \biGrZ{\fr a}{\fr b}\big\ra \big) 
	= \frac{\delta_{\rho_{AB},\rho_{AB}'} \, \delta_{I,I'} \, \delta_{J,J'}}{d_{\rho_{AB}}} \; ,
\end{align*}
where we first expanded the representation matrices according to definition \eqref{eq:defD} and then used the orthogonality of the irreducible representation in $Z_{\mc{O}_{AB}}$ together with the relation $|Z_{\mc{O}_{AB}}| \cdot |\mc{O}_{AB}| = |A||B|$.

\subsection{Proof of the invariance property \eqref{eq:gaugeBIS}\label{sec:app_gaugeBIS}}
Let us prove the invariance property \eqref{eq:gaugeBIS}, which we reproduce below for convenience
\begin{align}
	\label{eq:gaugeTER}
	&\sum_{\{J\}}
	\mc{D}^{\rho_{AB}}_{I_{AB}J_{AB}}\big( \big|\biGr{\fr g_1}{\fr a}{\fr b}\big\ra \big)
	\mc{D}^{\rho_{BC}}_{I_{BC}J_{BC}} \big( \big|\biGr{\fr g_2}{\fr b'}{\fr c}\big\ra\big)
	\CC{\rho_{AB}}{\rho_{BC}}{\rho_{AC}}{J_{AB}}{J_{BC}}{J_{AC}}
	\mc{D}^{\rho_{AC}}_{J_{AC}I_{AC}} \big( \big|\biGr{\fr g_3}{\fr a'}{\fr c'}\big\ra\big)
	\\
	\nn
	& \q = 
	\frac{1}{|B|}\sum_{\fr{\tilde b} \in B}
	\frac{\vartheta^{\Lambda(AB)}_{\fr g_1}(\fr{a,\tilde a|b,\tilde b}) \, \vartheta^{\Lambda(BC)}_{\fr g_2}(\fr{b',\tilde b|c,\tilde c})  \, \zeta^{\Lambda(ABC)}_{\fr{\tilde a,\tilde b,\tilde c}}(\fr a^{-1} \fr g_1 \fr b,\fr b'^{-1}\fr g_2\fr c)}{\vartheta^{\Lambda(AC)}_{\fr g_3}(\fr{\tilde a, \tilde a^{-1}a'|\tilde c, \tilde c^{-1}c'})} 
	\delta_{\fr g_3,\fr a^{-1}\fr g_1 \fr b \fr b'^{-1}\fr g_2 \fr c}
	\\[-0.5em]
	\nn
	& \hspace{2.9em} \times 	
	\sum_{\{K\}}
	\mc{D}^{\rho_{AB}}_{I_{AB}K_{AB}}\big(\big|\biGr{\fr g_1}{\fr {a\tilde a}}{\fr {b\tilde b}}\big\ra \big)
	\mc{D}^{\rho_{BC}}_{I_{BC}K_{BC}}\big(\big|\biGr{\fr g_2}{\fr {b'\tilde b}}{\fr {c\tilde c}}\big\ra \big)
	\CC{\rho_{AB}}{\rho_{BC}}{\rho_{AC}}{K_{AB}}{K_{BC}}{K_{AC}}
	\mc{D}^{\rho_{AC}}_{K_{AC}I_{AC}}\big( \big|\biGr{\fr{\tilde a}^{-1}\fr g_3\fr{\tilde c}}{\fr{\tilde a^{-1}a'}}{\fr{\tilde c^{-1}c'}}\big\ra \big) \, ,
\end{align}
Let us consider the left-hand side of \eqref{eq:gaugeTER}. In virtue of the gauge invariance \eqref{eq:gauge} of the Clebsch-Gordan coefficients, this is equal to
\begin{align*}
	{\rm l.h.s}\eqref{eq:gaugeTER}
	&= \!\!\!
	\sum_{\fr g \in {\rm Hom}({\rm s}(\fr{\tilde a}), {\rm s}(\fr{\tilde c}))}
	\sum_{\{J,K\}}
	\mc{D}^{\rho_{AB}}_{I_{AB}J_{AB}}\big( \big|\biGr{\fr g_1}{\fr a}{\fr b}\big\ra \big)
	\mc{D}^{\rho_{BC}}_{I_{BC}J_{BC}} \big( \big|\biGr{\fr g_2}{\fr b'}{\fr c}\big\ra\big)
	\mc{D}^{\rho_{AC}}_{J_{AC}I_{AC}} \big( \big|\biGr{\fr g_3}{\fr a'}{\fr c'}\big\ra\big)
	\\[-0.6em]
	& \hspace{7.9em} \times 		
	(\mc{D}^{\rho_{AB}}_{J_{AB}K_{AB}} \otimes_B \mc{D}^{\rho_{BC}}_{J_{BC}K_{BC}}) \big(\big|\biGr{\fr g}{\fr{\tilde a}}{\fr{\tilde c}}\big\ra\big)
	\overline{	\mc{D}^{\rho_{AC}}_{J_{AC}K_{AC}}\big( \big|\biGr{\fr g}{\fr{\tilde a}}{\fr{\tilde c}}\big\ra \big)}
	\CC{\rho_{AB}}{\rho_{BC}}{\rho_{AC}}{K_{AB}}{K_{BC}}{K_{AC}}
	\\
	& = 
	\frac{1}{|B|}
	\sum_{\substack{\fr g_1' \in {\rm Ob}(\Lambda(G_{AB})) \\ \fr g_2' \in {\rm Ob}(\Lambda(G_{BC}))  \\ \fr{\tilde b} \in B}}\sum_{\{J,K\}}
	\mc{D}^{\rho_{AB}}_{I_{AB}J_{AB}}\big( \big|\biGr{\fr g_1}{\fr a}{\fr b}\big\ra \big)
	\mc{D}^{\rho_{BC}}_{I_{BC}J_{BC}} \big( \big|\biGr{\fr g_2}{\fr b'}{\fr c}\big\ra\big)
	\mc{D}^{\rho_{AC}}_{J_{AC}I_{AC}} \big( \big|\biGr{\fr g_3}{\fr a'}{\fr c'}\big\ra\big)
	\\[-2em]
	& \hspace{11.4em} \times 	
	\mc{D}^{\rho_{AB}}_{J_{AB}K_{AB}}\big(\big|\biGr{\fr g_1'}{\fr{\tilde a}}{\fr{\tilde b}}\big\ra \big)
	\mc{D}^{\rho_{BC}}_{J_{BC}K_{BC}}\big(\big|\biGr{\fr g_2'}{\fr{\tilde b}}{\fr{\tilde c}}\big\ra \big)
	\overline{	\mc{D}^{\rho_{AC}}_{J_{AC}K_{AC}}\big( \big|\biGr{\fr g_1'\fr g_2'}{\fr{\tilde a}}{\fr{\tilde c}}\big\ra \big)}
	\\
	&\hspace{11.4em} \times
	\zeta^{\Lambda(ABC)}_{\fr{\tilde a,\tilde b,\tilde c}}(\fr g_1',\fr g_2')
	\CC{\rho_{AB}}{\rho_{BC}}{\rho_{AC}}{K_{AB}}{K_{BC}}{K_{AC}} \, ,
\end{align*}
where we applied the definitions of the truncated tensor product $\otimes_B$ and the comultiplication map $\Delta_B$. Using
\begin{equation*}
		\overline{	\mc{D}^{\rho_{AC}}_{J_{AC}K_{AC}}\big( \big|\biGr{\fr g_1'\fr g_2'}{\fr{\tilde a}}{\fr{\tilde c}}\big\ra \big)}
		= \frac{1}{\vartheta_{\fr g_1'\fr g_2'}^{\Lambda(AC)}(\fr{\tilde a , \tilde a^{-1}| \tilde c , \tilde c^{-1}})}
		\mc{D}^{\rho_{AC}}_{K_{AC}J_{AC}}\big( \big|\biGr{\fr{\tilde a}^{-1}\fr g_1'\fr g_2'\fr{\tilde c}}{\fr{\tilde a}^{-1}}{\fr{\tilde c}^{-1}}\big\ra \big) \; ,
\end{equation*}
together with the fact that the representation matrices define algebra homomorphisms yields
\begin{align*}
	{\rm l.h.s}\eqref{eq:gaugeTER} &=
	\frac{1}{|B|}
	\sum_{\substack{\fr g_1', \fr g_2' \\ \fr{\tilde b} \in B}}
	\sum_{\{K\}} 
	\frac{	\vartheta^{\Lambda(AB)}_{\fr g_1}(\fr{a,\tilde a|b,\tilde b}) \, \vartheta^{\Lambda(BC)}_{\fr g_2}(\fr{b',\tilde b|c,\tilde c}) \, 
	\vartheta^{\Lambda(AC)}_{\fr{\tilde a}^{-1}\fr g_1'\fr g_2' \fr{\tilde c}}(\fr{\tilde a}^{-1},\fr a'|\fr{\tilde c}^{-1},\fr c')}{\vartheta^{\Lambda(AC)}_{\fr g_1'\fr g_2'}(\fr{\tilde a,\tilde a^{-1}|\tilde c,\tilde c^{-1}})}
	\\[-1.5em]
	& \hspace{6.7em} \times 	
	\mc{D}^{\rho_{AB}}_{I_{AB}K_{AB}}\big(\big|\biGr{\fr g_1}{\fr{a\tilde a}}{\fr{b\tilde b}}\big\ra \big)
	\mc{D}^{\rho_{BC}}_{I_{BC}K_{BC}}\big(\big|\biGr{\fr g_2}{\fr{b'\tilde b}}{\fr{c\tilde c}}\big\ra \big)
	\mc{D}^{\rho_{AC}}_{K_{AC}I_{AC}}\big( \big|\biGr{\fr{\tilde a}^{-1}\fr g_1'\fr g_2'\fr{\tilde c}}{\fr{\tilde a^{-1}a'}}{\fr{\tilde c^{-1}c'}}\big\ra \big)
	\\
	& \hspace{6.7em} \times
	\delta_{\fr g_1',\fr a^{-1}\fr g_1 \fr b} \,
	\delta_{\fr g_2',\fr b'^{-1}\fr g_2 \fr c} \,
	\delta_{\fr g_3,\fr g_1' \fr g_2'} \,
	\zeta^{\Lambda(ABC)}_{\fr{\tilde a,\tilde b,\tilde c}}(\fr g_1',\fr g_2') 
		\,
	\CC{\rho_{AB}}{\rho_{BC}}{\rho_{AC}}{K_{AB}}{K_{BC}}{K_{AC}}
	\\[0.5em]
	&  =
	\frac{1}{|B|} 
	\sum_{\fr{\tilde b} \in B}\sum_{\{K\}}
	\frac{	\vartheta^{\Lambda(AB)}_{\fr g_1}(\fr{a,\tilde a|b,\tilde b}) \, \vartheta^{\Lambda(BC)}_{\fr g_2}(\fr{b',\tilde b|c,\tilde c}) \, 
	\vartheta^{\Lambda(AC)}_{\fr{\tilde a}^{-1}\fr g_3 \fr{\tilde c}}(\fr{\tilde a^{-1},a'|\tilde c^{-1},c'})}{\vartheta^{\Lambda(AC)}_{\fr g_3}(\fr{\tilde a,\tilde a^{-1}|\tilde c,\tilde c^{-1}})}
	\\[-0.5em]
	& \hspace{5.4em} \times 	
	\mc{D}^{\rho_{AB}}_{I_{AB}K_{AB}}\big(\big|\biGr{\fr g_1}{\fr{a\tilde a}}{\fr{b\tilde b}}\big\ra \big)
	\mc{D}^{\rho_{BC}}_{I_{BC}K_{BC}}\big(\big|\biGr{\fr g_2}{\fr{b'\tilde b}}{\fr{c\tilde c}}\big\ra \big)
	\mc{D}^{\rho_{AC}}_{K_{AC}I_{AC}}\big( \big|\biGr{\fr{\tilde a}^{-1}\fr g_3\fr{\tilde c}}{\fr{\tilde a^{-1}a'}}{\fr{\tilde c^{-1}c'}}\big\ra \big)
	\\
	&\hspace{5.4em} \times
	\delta_{\fr g_3,\fr a^{-1} \fr g_1 \fr b \fr b'^{-1}\fr g_2 \fr c} \,
	\zeta^{\Lambda(ABC)}_{\fr{\tilde a,\tilde b,\tilde c}}(\fr a^{-1} \fr g_1 \fr b,\fr b'^{-1}\fr g_2 \fr c) \, 
	\CC{\rho_{AB}}{\rho_{BC}}{\rho_{AC}}{K_{AB}}{K_{BC}}{K_{AC}} \; .
\end{align*}
Finally, using $d^{(2)}\vartheta_{\fr g_3}^{\Lambda(AC)}(\fr{\tilde a, \tilde a^{-1}, a'|\tilde c ,\tilde c^{-1}, c'}) = 1$, we obtain
\begin{align*}
	{\rm l.h.s}\eqref{eq:gaugeTER} &=
	\frac{1}{|B|}\sum_{\fr{\tilde b} \in B} 	\sum_{\{K\}}
	\frac{	\vartheta^{\Lambda(AB)}_{\fr g_1}(\fr{a,\tilde a|b,\tilde b}) \, \vartheta^{\Lambda(BC)}_{\fr g_2}(\fr{b',\tilde b|c,\tilde c})  \, 
	\zeta^{\Lambda(ABC)}_{\fr{\tilde a,\tilde b,\tilde c}}(\fr a^{-1}\fr g_1 \fr b, \fr b'^{-1}\fr g_2 \fr c)}{\vartheta^{\Lambda(AC)}_{\fr g_3}(\fr{\tilde a, \tilde a^{-1}a'|\tilde c, \tilde c^{-1}c'})}
	\\[-0.5em]
	& \hspace{5.4em} \times 	
	\mc{D}^{\rho_{AB}}_{I_{AB}K_{AB}}\big(\big|\biGr{\fr g_1}{\fr{a\tilde a}}{\fr{b\tilde b}}\big\ra \big)
	\mc{D}^{\rho_{BC}}_{I_{BC}K_{BC}}\big(\big|\biGr{\fr g_2}{\fr{b'\tilde b}}{\fr{c\tilde c}}\big\ra \big)
	\mc{D}^{\rho_{AC}}_{K_{AC}I_{AC}}\big( \big|\biGr{\fr{\tilde a}^{-1}\fr g_3\fr{\tilde c}}{\fr{\tilde a^{-1}a'}}{\fr{\tilde c^{-1}c'}}\big\ra \big) 
	\\
	& \hspace{5.4em} \times
	\delta_{\fr g_3,\fr a^{-1}\fr g_1 \fr b \fr b'^{-1}\fr g_2 \fr c}
	\CC{\rho_{AB}}{\rho_{BC}}{\rho_{AC}}{K_{AB}}{K_{BC}}{K_{AC}} \; ,
\end{align*}
which is the right-hand side of \eqref{eq:gaugeTER}, as expected. Note that the above is true for every morphism $\fr{\tilde a}, \fr{\tilde c}$.

\newpage
\subsection{Proof of the defining relation of the $6j$-symbols\label{sec:app_SixJ}}

In this appendix, we confirm the definition of the $6j$-symbols
\begin{equation*}
	\SixJ{\rho_{AB}}{\rho_{BC}}{\rho_{CD}}{\rho_{AD}}{\rho_{AC}}{\rho_{BD}}
	\!\! :=
	\frac{1}{d_{\rho_{AD}}} \!
	\sum_{\{I\}} \!
	\alpha(\mathfrak{o}_{i_{AB}},\mathfrak{o}_{i_{BC}},\mathfrak{o}_{i_{CD}})
	\CC{\rho_{AB}}{\rho_{BC}}{\rho_{AC}}{I_{AB}}{I_{BC}}{I_{AC}}
	\CC{\rho_{AC}}{\rho_{CD}}{\rho_{AD}}{I_{AC}}{I_{CD}}{I_{AC}}
	\overline{\CC{\rho_{AB}}{\rho_{BD}}{\rho_{AD}}{I_{AB}}{I_{BD}}{I_{AD}}}
	\overline{\CC{\rho_{BC}}{\rho_{CD}}{\rho_{BD}}{I_{BC}}{I_{CD}}{I_{BD}}} ,
\end{equation*}
such that they satisfy the relation
\begin{align}
	\nn
	& \sum_{\rho_{AC}}\sum_{\{I\}}
	\SixJ{\rho_{AB}}{\rho_{BC}}{\rho_{CD}}{\rho_{AD}}{\rho_{AC}}{\rho_{BD}}
	\overline{\CC{\rho_{AB}}{\rho_{BC}}{\rho_{AC}}{I_{AB}}{I_{BC}}{I_{AC}}}
	\overline{\CC{\rho_{AC}}{\rho_{CD}}{\rho_{AD}}{I_{AC}}{I_{CD}}{K_{AD}}}
	|\rho_{AB}I_{AB} \ra \otimes |\rho_{BC}I_{BC} \ra  \otimes |\rho_{CD}I_{CD}\ra
	\triangleright \Phi_{ABCD}
	\\
	\label{eq:defSixJApp}
	& \q =  
	\sum_{\{I\}}
	\overline{\CC{\rho_{AB}}{\rho_{BD}}{\rho_{AD}}{I_{AB}}{I_{BD}}{K_{AD}}}
	\overline{\CC{\rho_{BC}}{\rho_{CD}}{\rho_{BD}}{I_{BC}}{I_{CD}}{I_{BD}}}
	|\rho_{AB}I_{AB} \ra \otimes |\rho_{BC}I_{BC} \ra  \otimes |\rho_{CD}I_{CD}\ra \; .
\end{align}
Inserting the definition of the $6j$-symbols into equation \eqref{eq:defSixJApp} and writing down explicitly the action of $\Phi_{ABCD}$ using \eqref{eq:defD}, we find that the left-hand side is equal to
\begin{align*}
	{\rm l.h.s}\eqref{eq:defSixJApp} &= 
	\frac{1}{d_{\rho_{AD}}}
	\sum_{\rho_{AC}}\sum_{\{I,J\}} \frac{\alpha(\mathfrak{o}_{j_{AB}},\mathfrak{o}_{j_{BC}},\mathfrak{o}_{j_{CD}})}{\alpha(\mathfrak{o}_{i_{AB}},\mathfrak{o}_{i_{BC}},\mathfrak{o}_{i_{CD}})}
	\\[-0.1em]
	& \hspace{6.5em} \times 
	\CC{\rho_{AB}}{\rho_{BC}}{\rho_{AC}}{J_{AB}}{J_{BC}}{J_{AC}}
	\CC{\rho_{AC}}{\rho_{CD}}{\rho_{AD}}{J_{AC}}{J_{CD}}{J_{AD}}
	\overline{\CC{\rho_{AB}}{\rho_{BD}}{\rho_{AD}}{J_{AB}}{J_{BD}}{J_{AD}}}
	\overline{\CC{\rho_{BC}}{\rho_{CD}}{\rho_{BD}}{J_{BC}}{J_{CD}}{J_{BD}}}
	\\
	& \hspace{6.5em} \times
	\overline{\CC{\rho_{AB}}{\rho_{BC}}{\rho_{AC}}{I_{AB}}{I_{BC}}{I_{AC}}}
	\overline{\CC{\rho_{AC}}{\rho_{CD}}{\rho_{AD}}{I_{AC}}{I_{CD}}{K_{AD}}}
	|\rho_{AB}I_{AB} \ra \otimes |\rho_{BC}I_{BC} \ra  \otimes |\rho_{CD}I_{CD}\ra
\end{align*}
The defining relation of the Clebsch-Gordan coefficients yields
\begin{align*}
	&	\CC{\rho_{AB}}{\rho_{BC}}{\rho_{AC}}{J_{AB}}{J_{BC}}{J_{AC}}
	\CC{\rho_{AC}}{\rho_{CD}}{\rho_{AD}}{J_{AC}}{J_{CD}}{J_{AD}}
	\overline{\CC{\rho_{AB}}{\rho_{BC}}{\rho_{AC}}{I_{AB}}{I_{BC}}{I_{AC}}}
	\overline{\CC{\rho_{AC}}{\rho_{CD}}{\rho_{AD}}{I_{AC}}{I_{CD}}{K_{AD}}}
	\\ 
	& \q =
	\frac{d_{\rho_{AC}}d_{\rho_{AD}}}{|A|^2|C||D|}
	\!\! \sum_{\substack{ \biGrFoot{\fr g}{\fr a}{\fr c} \in \Lambda(G_{AC}) \\  \biGrFoot{\fr g'}{\fr a'}{\fr d} \in \Lambda(G_{AD})}} \;\;
	(\mc{D}^{\rho_{AB}}_{J_{AB}I_{AB}} \otimes_B \mc{D}^{\rho_{BC}}_{J_{BC}I_{BC}}) \big(\big|\biGr{\fr g}{\fr a}{\fr c}\big\ra \big)
	\overline{\mc{D}^{\rho_{AC}}_{J_{AC}I_{AC}}\big( \big|\biGr{\fr g}{\fr a}{\fr c}\big\ra \big)}
	\\[-3em]
	& \hspace{13em} \times 
	(\mc{D}^{\rho_{AC}}_{J_{AC}I_{AC}} \otimes_C \mc{D}^{\rho_{CD}}_{J_{CD}I_{CD}}) \big(\big|\biGr{\fr g'}{\fr a'}{\fr d}\big\ra \big)
	\overline{\mc{D}^{\rho_{AD}}_{J_{AD}K_{AD}}\big( \big|\biGr{\fr g'}{\fr a'}{\fr d}\big\ra \big)}
	\\ 
	& \q =
	\frac{d_{\rho_{AC}}d_{\rho_{AD}}}{|A|^2|B||C|^2|D|}
	\!\!\!\! \sum_{\substack{\fr g_1, \fr g_2, \fr g_1', \fr g_2' \\ (\fr a, \fr c)  \in A \times C \\ (\fr a',\fr d)  \in A \times D \\ (\fr b,\fr c') \in B \times C}} \;\;
	\mc{D}^{\rho_{AB}}_{J_{AB}I_{AB}}\big( \big|\biGr{\fr g_1}{\fr a}{\fr b}\big\ra \big)  \mc{D}^{\rho_{BC}}_{J_{BC}I_{BC}} \big(\big|\biGr{\fr g_2}{\fr b}{\fr c}\big\ra \big)
	\overline{\mc{D}^{\rho_{AC}}_{J_{AC}I_{AC}}\big( \big|\biGr{\fr g_1 \fr g_2}{\fr a}{\fr c}\big\ra \big)}
	\\[-3.5em]
	& \hspace{13.3em} \times 
	\mc{D}^{\rho_{AC}}_{J_{AC}I_{AC}}\big(\big|\biGr{\fr g_1'}{\fr a'}{\fr c'}\big\ra \big)
	\mc{D}^{\rho_{CD}}_{J_{CD}I_{CD}}\big(\big|\biGr{\fr g_2'}{\fr c'}{\fr d}\big\ra \big)
	\overline{\mc{D}^{\rho_{AD}}_{J_{AD}K_{AD}}\big( \big|\biGr{\fr g_1'\fr g_2'}{\fr a'}{\fr d}\big\ra \big)}
	\\[0.2em]
	& \hspace{13.3em} \times 
	\zeta^{\Lambda(ABC)}_{\fr{a,b,c}}(\fr g_1,\fr g_2)	 \, \zeta^{\Lambda(ACD)}_{\fr{a',c',d}}(\fr g_1',\fr g_2') \; ,
\end{align*}
where the second sum is over $\fr g_1 \in {\rm Ob}(\Lambda(G_{AB}))$, $\fr g_2 \in {\rm Ob}(\Lambda(G_{BC}))$, $\fr g_1' \in {\rm Ob}(\Lambda(G_{AC}))$, $\fr g_2' \in {\rm Ob}(\Lambda(G_{CD}))$ and the corresponding morphisms, which we loosely identify with the group variables they are characterized by.
Furthermore, we have that
\begin{align}
	&\frac{1}{|A||C|}\sum_{\substack{\rho_{AC} \\ I_{AC},J_{AC}}} d_{\rho_{AC}}
	\mc{D}^{\rho_{AC}}_{J_{AC}I_{AC}}\big(|\biGr{\fr g_1'}{\fr a'}{\fr c'}\big\ra \big)
	\overline{\mc{D}^{\rho_{AC}}_{J_{AC}I_{AC}}\big( \big|\biGr{\fr g_1 \fr g_2}{\fr a}{\fr c}\big\ra \big)	}
	\\
	& \q =
	\delta_{\fr a ^{-1}\fr g_{1}\fr g_{2}\fr c, \fr a'^{-1}\fr g_1' \fr c'} \frac{1}{|A||C|}\sum_{\substack{\rho_{AC}}} d_{\rho_{AC}}
	{\rm tr}\big[ \mc{D}^{\rho_{AC}}\big(|\biGr{\fr g_1'}{\fr{a'a^{-1}}}{\fr{c'c^{-1}}}\big\ra \big) \big]
	= \delta_{\fr g_{1} \fr g_{2},\fr g_1'} \, \delta_{\fr a,\fr a'} \, \delta_{\fr c,\fr c'} \; ,
\end{align}
where we made use of the orthogonality relation \eqref{eq:ortho} so that
\begin{align*}
	&\CC{\rho_{AB}}{\rho_{BC}}{\rho_{AC}}{J_{AB}}{J_{BC}}{J_{AC}}
	\CC{\rho_{AC}}{\rho_{CD}}{\rho_{AD}}{J_{AC}}{J_{CD}}{J_{AD}}
	\overline{\CC{\rho_{AB}}{\rho_{BC}}{\rho_{AC}}{I_{AB}}{I_{BC}}{I_{AC}}}
	\overline{\CC{\rho_{AC}}{\rho_{CD}}{\rho_{AD}}{I_{AC}}{I_{CD}}{K_{AD}}}
	\\ 
	& \q =
	\frac{d_{\rho_{AD}}}{|A||B||C||D|}
	\!\sum_{\substack{\fr g_1, \fr g_2,\fr g_2' \\ (\fr a, \fr c)  \in A \times C \\ \fr d  \in D  \\ \fr b \in B}} \;\;
	\mc{D}^{\rho_{AB}}_{J_{AB}I_{AB}}\big( \big|\biGr{\fr g_1}{\fr a}{\fr b}\big\ra \big)  \mc{D}^{\rho_{BC}}_{J_{BC}I_{BC}} \big(\big|\biGr{\fr g_2}{\fr b}{\fr c}\big\ra \big)
	\\[-3.1em]
	& \hspace{12.7em} \times 
	\mc{D}^{\rho_{CD}}_{J_{CD}I_{CD}}\big(\big|\biGr{\fr g_2'}{\fr c}{\fr d}\big\ra \big)
	\overline{\mc{D}^{\rho_{AD}}_{J_{AD}K_{AD}}\big( \big|\biGr{\fr g_1 \fr g_2 \fr g_2'}{\fr a}{\fr d}\big\ra \big)}
	\\
	& \hspace{12.7em} \times 
	\zeta^{\Lambda(ABC)}_{\fr{a,b,c}}(\fr g_1,\fr g_2)	 \, \zeta^{\Lambda(ACD)}_{\fr{a,c,d}}(\fr g_1 \fr g_2,\fr g_2')  \; .
\end{align*}
Putting everything together so far, we obtain
\begin{align*}
	{\rm l.h.s}\eqref{eq:defSixJApp} &= 
	\frac{1}{|A||B||C||D|}
	\sum_{\substack{\fr g_1,\fr g_2,\fr g_2'  \\ (\fr a,\fr c)  \in A \times C \\ (\fr b,\fr d) \in B \times D}} 
	\sum_{\{I,J\}} \frac{\alpha(\mathfrak{o}_{j_{AB}},\mathfrak{o}_{j_{BC}},\mathfrak{o}_{j_{CD}})}{\alpha(\mathfrak{o}_{i_{AB}},\mathfrak{o}_{i_{BC}},\mathfrak{o}_{i_{CD}})}\,
	\zeta^{\Lambda(ABC)}_{\fr{a,b,c}}(\fr g_1,\fr g_2)	 \, \zeta^{\Lambda(ACD)}_{\fr{a,c,d}}(\fr g_1 \fr g_2,\fr g_2') 
	\\[-2em]
	& \hspace{12.5em} \times
	\mc{D}^{\rho_{BC}}_{J_{BC}I_{BC}}\big(\big|\biGr{\fr g_2}{\fr b}{\fr c}\big\ra \big)
	\mc{D}^{\rho_{CD}}_{J_{CD}I_{CD}}\big( \big|\biGr{\fr g_2'}{\fr c}{\fr d}\big\ra \big) 
	\overline{\CC{\rho_{BC}}{\rho_{CD}}{\rho_{BD}}{J_{BC}}{J_{CD}}{J_{BD}}}
	\\
	& \hspace{12.5em} \times 
	\mc{D}^{\rho_{AB}}_{J_{AB}I_{AB}} \big(\big|\biGr{\fr g_1}{\fr a}{\fr b}\big\ra \big)
	\overline{\mc{D}^{\rho_{AD}}_{J_{AD}K_{AD}}\big( \big|\biGr{\fr g_1\fr g_2 \fr g_2'}{\fr a}{\fr d}\big\ra \big)}
	\overline{\CC{\rho_{AB}}{\rho_{BD}}{\rho_{AD}}{J_{AB}}{J_{BD}}{J_{AD}}}
	\\
	& \hspace{12.5em} \times	
	|\rho_{AB}I_{AB} \ra \otimes |\rho_{BC}I_{BC} \ra  \otimes |\rho_{CD}I_{CD}\ra \; .
\end{align*}
In virtue of the definition of the representation matrices, we observe that we must have $\mathfrak{o}_{i_{AB}}=\fr a^{-1} \fr g_{1} \fr b$, $\mathfrak{o}_{i_{BC}}=\fr b^{-1} \fr g_2 \fr c$, $\mathfrak{o}_{i_{CD}}=\fr c^{-1}\fr g_2' \fr d$, $\mathfrak{o}_{j_{AB}}= \fr g_1$, $\mathfrak{o}_{j_{BC}}= \fr g_2$ and $\mathfrak{o}_{j_{CD}}= \fr g_2'$ in order for the whole expression not to vanish. Applying the quasi-coassociativity condition 
\begin{align}
	\frac{\zeta^{\Lambda(BCD)}_{\fr{b,c,d}}(\fr g_2, \fr g_2')  \, \zeta^{\Lambda(ABD)}_{\fr{a,b,d}}(\fr g_1, \fr g_2 \fr g_2') }{\zeta^{\Lambda(ACD)}_{\fr{a,c,d}}(\fr g_1 \fr g_2, \fr g_2') \, \zeta^{\Lambda(ABC)}_{\fr{a,b,c}}(\fr g_1, \fr g_2)}
	=
	\frac{	\alpha(\fr g_{1},\fr g_{2},\fr g_{2}')}{\alpha(\fr a^{-1}\fr g_{1}\fr b,\fr b^{-1}\fr g_{2} \fr c,\fr c^{-1}\fr g_{2}'\fr d)} \; ,
\end{align}
we obtain
\begin{align*}
	{\rm l.h.s}\eqref{eq:defSixJApp} &= 
	\frac{1}{|A||B||C||D|}
	\sum_{\substack{\fr g_1,\fr g_2,\fr g_2'  \\ (\fr a,\fr c)  \in A \times C \\ (\fr b,\fr d) \in B \times D}} 
	\sum_{\{I,J\}}
	\zeta^{\Lambda(BCD)}_{\fr{b,c,d}}(\fr g_2, \fr g_2')	 \, \zeta^{\Lambda(ABD)}_{\fr{a,b,d}}(\fr g_1, \fr g_2 \fr g_2')
	\\[-2.5em]
	& \hspace{12.5em} \times
	\mc{D}^{\rho_{BC}}_{J_{BC}I_{BC}}\big(\big|\biGr{\fr g_2}{\fr b}{\fr c}\big\ra \big)
	\mc{D}^{\rho_{CD}}_{J_{CD}I_{CD}}\big( \big|\biGr{\fr g_2'}{\fr c}{\fr d}\big\ra \big) 
	\overline{\CC{\rho_{BC}}{\rho_{CD}}{\rho_{BD}}{J_{BC}}{J_{CD}}{J_{BD}}}
	\\
	& \hspace{12.5em} \times 
	\mc{D}^{\rho_{AB}}_{J_{AB}I_{AB}} \big(\big|\biGr{\fr g_1}{\fr a}{\fr b}\big\ra \big)
	\overline{\mc{D}^{\rho_{AD}}_{J_{AD}K_{AD}}\big( \big|\biGr{\fr g_1\fr g_2 \fr g_2'}{\fr a}{\fr d}\big\ra \big)}
	\overline{\CC{\rho_{AB}}{\rho_{BD}}{\rho_{AD}}{J_{AB}}{J_{BD}}{J_{AD}}}
	\\
	& \hspace{12.5em} \times	
	|\rho_{AB}I_{AB} \ra \otimes |\rho_{BC}I_{BC} \ra  \otimes |\rho_{CD}I_{CD}\ra \; .
\end{align*}
Let us now insert the resolution of the identity
\begin{equation}
	\delta_{J_{BD},J_{BD}} = \sum_{\fr h, \fr h' \in {\rm Ob}(\Lambda(G_{BD}))}\sum_{I_{BD}}
		\mc{D}^{\rho_{BD}}_{J_{BD}I_{BD}} \big(\big|\biGr{\fr h'}{\fr b}{\fr d}\big\ra \big)
	\overline{\mc{D}^{\rho_{BD}}_{J_{BD}I_{BD}} \big(\big|\biGr{\fr h}{\fr b}{\fr d}\big\ra \big)}
	 \; ,
\end{equation}
where $\fr h$ and $\fr h'$ are implicitly identified via the algebra product.
As a special case of \eqref{eq:gaugeBIS}, we have
\begin{align*}
	&\sum_{\{J\}}\sum_{\substack{\fr c \in C}}
	\mc{D}^{\rho_{BC}}_{J_{BC}I_{BC}}\big(\big|\biGr{\fr g_2}{\fr b}{\fr c}\big\ra \big)
	\mc{D}^{\rho_{CD}}_{J_{CD}I_{CD}}\big( \big|\biGr{\fr g_2'}{\fr c}{\fr d}\big\ra \big) 
	\overline{\mc{D}^{\rho_{BD}}_{J_{BD}I_{BD}} \big(\big|\biGr{\fr h}{\fr b}{\fr d}\big\ra \big)}
	\overline{\CC{\rho_{BC}}{\rho_{CD}}{\rho_{BD}}{J_{BC}}{J_{CD}}{J_{BD}}}
	\\
	& \q = 	
	\sum_{\{J\}}\sum_{\substack{\fr c \in C}}
	\mc{D}^{\rho_{BC}}_{J_{BC}I_{BC}}\big(\big|\biGr{\fr g_2}{\fr b}{\fr c}\big\ra \big)
	\mc{D}^{\rho_{CD}}_{J_{CD}I_{CD}}\big( \big|\biGr{\fr g_2'}{\fr c}{\fr d}\big\ra \big) 
	\overline{\mc{D}^{\rho_{BD}}_{J_{BD}I_{BD}} \big(\big|\biGr{\fr g_2 \fr g_2'}{\fr b}{\fr d}\big\ra \big)}\,
	\overline{\CC{\rho_{BC}}{\rho_{CD}}{\rho_{BD}}{J_{BC}}{J_{CD}}{J_{BD}}}
	\delta_{\fr h,\fr g_2 \fr g_2'} \; .
\end{align*}
We can finally use the gauge invariance of the Clebsch-Gordan coefficients
\begin{align*}
	&\frac{1}{|B|}
	\sum_{\{J\}} \sum_{\substack{\fr h',\fr g_1 }} \!\!
	\zeta^{\Lambda(ABD)}_{\fr{a,b,d}}(\fr g_1,\fr h')
	\mc{D}^{\rho_{AB}}_{J_{AB}I_{AB}}\big( \big|\biGr{\fr g_1}{\fr a}{\fr b}\big\ra \big)
	\mc{D}^{\rho_{BD}}_{J_{BD}I_{BD}} \big(\big|\biGr{\fr h'}{\fr b}{\fr d}\big\ra \big)
	\overline{\mc{D}^{\rho_{AD}}_{J_{AD}K_{AD}}\big( \big|\biGr{\fr g_1 \fr h'}{\fr a}{\fr d}\big\ra \big)}\,  \overline{\CC{\rho_{AB}}{\rho_{BD}}{\rho_{AD}}{J_{AB}}{J_{BD}}{J_{AD}}}
	\\
	& \q = 
	\overline{\CC{\rho_{AB}}{\rho_{BD}}{\rho_{AD}}{I_{AB}}{I_{BD}}{K_{AD}}}
\end{align*}
and
\begin{align*}
	&\frac{1}{|C|}\sum_{\{J\}}  \sum_{\substack{\fr g_2,\fr g_2'}} \!\!
	\zeta^{\Lambda(BCD)}_{\fr{b,c,d}}(\fr g_2, \fr g_2')
	\mc{D}^{\rho_{BC}}_{J_{BC}I_{BC}}\big(\big|\biGr{\fr g_2}{\fr b}{\fr c}\big\ra \big)
	\mc{D}^{\rho_{CD}}_{J_{CD}I_{CD}}\big( \big|\biGr{\fr g_2'}{\fr c}{\fr d}\big\ra \big) 
	\overline{\mc{D}^{\rho_{BD}}_{J_{BD}I_{BD}} \big(\big|\biGr{\fr g_2 \fr g_2'}{\fr b}{\fr d}\big\ra \big)}\,
	\overline{\CC{\rho_{BC}}{\rho_{CD}}{\rho_{BD}}{J_{BC}}{J_{CD}}{J_{BD}}}
	\\
	& \q = \overline{\CC{\rho_{BC}}{\rho_{CD}}{\rho_{BD}}{I_{BC}}{I_{CD}}{I_{BD}}} \; ,
\end{align*}
so as to yield \eqref{eq:defSixJApp} as expected.

\subsection{Proof of the pentagon identity \label{sec:app_pent}}

As explained in the main text, the pentagon identity is the statement that the algebra elements
\begin{align}
	\nn
	&[({\rm id} \otimes {\rm id } \otimes \Delta_D)(\Phi_{ABCE})] \star 
	[(\Delta_{B}  \otimes {\rm id} \otimes {\rm id })(\Phi_{ACDE})]
\end{align}
and
\begin{align*}
	(\mathbbm 1_{AB} \otimes \Phi_{BCDE}) \star
	[({\rm id} \otimes \Delta_{C} \otimes {\rm id })(\Phi_{ABDE})] \star 
	(\Phi_{ABCD} \otimes \mathbbm 1_{DE})
\end{align*}
induce the same isomorphism on the four-particle vector space $((V_{\rho_{AB}} \otimes_B V_{\rho_{BC}}) \otimes_C V_{\rho_{CD}}) \otimes_D V_{\rho_{DE}}$. In light of the definition of the truncated tensor product of vector spaces, this can be demonstrated explicitly by showing the equality:
\begin{align}
	\nn
	&(\mathbbm 1_{AB} \otimes \Phi_{BCDE}) \star
	[({\rm id} \otimes \Delta_{C} \otimes {\rm id })(\Phi_{ABDE})] \star 
	(\Phi_{ABCD} \otimes \mathbbm 1_{DE}) \star
	\mathbbm 1_{(((AB)C)D)E}
	\\
	\label{eq:pentExp}
	& \q =
	[({\rm id} \otimes {\rm id } \otimes \Delta_D)(\Phi_{ABCE})] \star
	[(\Delta_{B}  \otimes {\rm id} \otimes {\rm id })(\Phi_{ACDE})] \star
	\mathbbm 1_{(((AB)C)D)E} \; ,
\end{align}
where we defined
\begin{equation*}
	\mathbbm 1_{(((AB)C)D)E} := 
	[(\Delta_B \otimes {\rm id}) \circ (\Delta_C \otimes {\rm id}) \circ \Delta_D](\mathbbm 1 _{AE}) \; .
\end{equation*}
Writing down explicitly the definition of the comultiplication maps, we have
\begin{align*}
	[({\rm id} \otimes \Delta_{C} \otimes {\rm id })(\Phi_{ABDE})]
	=
	\frac{1}{|C|} \!
	\sum_{\substack{\{\fr g\} \\ \fr c \in C}} \!
	\frac{\zeta^{\Lambda(BCD)}_{\mathbbm 1_B,c,\mathbbm 1_D}(\fr g_2, \fr g_3)}{\alpha(\fr g_1, \fr g_2 \fr g_3, \fr g_4)}
	\big|\biGr{\fr g_1}{\mathbbm 1_A}{\mathbbm 1_B}\big\ra
	\otimes
	\big|\biGr{\fr g_2}{\mathbbm 1_B}{\fr c}\big\ra
	\otimes
	\big|\biGr{\fr g_3}{\fr c}{\mathbbm 1_D}\big\ra
	\otimes
	\big|\biGr{\fr g_4}{\mathbbm 1_D}{\mathbbm 1_E}\big\ra ,
\end{align*}
\begin{equation*}
	[({\rm id} \otimes {\rm id } \otimes \Delta_D)(\Phi_{ABCE})]
	= 
	\frac{1}{|D|}
	\! \sum_{\substack{\{\fr g\} \\ \fr d \in D}} \!
	\frac{\zeta^{\Lambda(CDE)}_{\mathbbm 1_C,d,\mathbbm 1_E}(\fr g_3, \fr g_4)}{\alpha(\fr g_1, \fr g_2, \fr g_3 \fr g_4)}
	\big|\biGr{\fr g_1}{\mathbbm 1_A}{\mathbbm 1_B}\big\ra
	\otimes
	\big|\biGr{\fr g_2}{\mathbbm 1_B}{\mathbbm 1_C}\big\ra
	\otimes
	\big|\biGr{\fr g_3}{\mathbbm 1_C}{\fr d}\big\ra
	\otimes
	\big|\biGr{\fr g_4}{\fr d}{\mathbbm 1_E}\big\ra ,
\end{equation*}
\begin{equation*}
	[(\Delta_{B}  \otimes {\rm id} \otimes {\rm id })(\Phi_{ACDE})]
	=
	\frac{1}{|B|}
	\! \sum_{\substack{\{\fr g\} \\ \fr b \in B}} \!
	\frac{\zeta^{\Lambda(ABC)}_{\mathbbm 1_A,b,\mathbbm 1_C}(\fr g_1, \fr g_2)}{\alpha(\fr g_1 \fr g_2, \fr g_3, \fr g_4)}
	\big|\biGr{\fr g_1}{\mathbbm 1_A}{\fr b}\big\ra
	\otimes
	\big|\biGr{\fr g_2}{\fr b}{\mathbbm 1_C}\big\ra
	\otimes
	\big|\biGr{\fr g_3}{\mathbbm 1_C}{\mathbbm 1_D}\big\ra
	\otimes
	\big|\biGr{\fr g_4}{\mathbbm 1_D}{\mathbbm 1_E}\big\ra ,
\end{equation*}
and
\begin{align*}
	&\mathbbm 1_{(((AB)C)D)E}
	=
	\frac{1}{|B||C||D|}
	\sum_{\substack{\{\fr g\} \\ (\fr b,\fr c,\fr d) \in B \times C \times D}}
	\!\!\!\!
	\zeta^{\Lambda(ADE)}_{\mathbbm 1_A,\fr d,\mathbbm 1_E}(\fr g_1 \fr g_2 \fr g_3, \fr g_4) \,
	\zeta^{\Lambda(ACD)}_{\mathbbm 1_A,\fr c,\fr d}(\fr g_1 \fr g_2,\fr g_3) \,
	\zeta^{\Lambda(ABC)}_{\mathbbm 1_A,\fr b,\fr c}(\fr g_1, \fr g_2)
	\\[-1.8em]
	&\hspace{20em}
	\times
	\big|\biGr{\fr g_1}{\mathbbm 1_A}{\fr b}\big\ra
	\otimes
	\big|\biGr{\fr g_2}{\fr b}{\fr c}\big\ra
	\otimes
	\big|\biGr{\fr g_3}{\fr c}{\fr d}\big\ra
	\otimes
	\big|\biGr{\fr g_4}{\fr d}{\mathbbm 1_E}\big\ra \, .
\end{align*}
Applying the definition of the algebra product, we then obtain
\begin{align*}
	&[({\rm id} \otimes {\rm id } \otimes \Delta_D)(\Phi_{ABCE})] \star 
	[(\Delta_{B}  \otimes {\rm id} \otimes {\rm id })(\Phi_{ACDE})]
	\\
	& \q = 
	\frac{1}{|B||D|}\sum_{\substack{\{\fr g\} \\ (\fr b,\fr d) \in B \times D}}
	\frac{\zeta^{\Lambda(CDE)}_{\mathbbm 1_C,\fr d,\mathbbm 1_E}(\fr g_3, \fr g_4) \, \zeta^{\Lambda(ABC)}_{\mathbbm 1_A,\fr b,\mathbbm 1_C}(\fr g_1, \fr g_2)}{\alpha(\fr g_1, \fr g_2, \fr g_3 \fr g_4) \, \alpha(\fr g_1 \fr g_2, \fr g_3 \fr d,\fr d^{-1} \fr g_4)}
	\\[-1.2em]
	& \hspace{9.6em} \times \big|\biGr{\fr g_1}{\mathbbm 1_A}{\fr b}\big\ra
	\otimes
	\big|\biGr{\fr g_2}{\fr b}{\mathbbm 1_C}\big\ra
	\otimes
	\big|\biGr{\fr g_3}{\mathbbm 1_C}{\fr d}\big\ra
	\otimes
	\big|\biGr{\fr g_4}{\fr d}{\mathbbm 1_E}\big\ra
\end{align*}
and
\begin{align*}
	&(\mathbbm 1_{AB} \otimes \Phi_{BCDE}) \star
	[({\rm id} \otimes \Delta_{C} \otimes {\rm id })(\Phi_{ABDE})] \star 
	(\Phi_{ABCD} \otimes \mathbbm 1_{DE})
	\\
	& \q =
	\frac{1}{|C|}\sum_{\substack{\{ \fr g \} \\ \fr c \in C}}
	\frac{\zeta^{\Lambda(BCD)}_{\mathbbm 1_B,\fr c,\mathbbm 1_D}(\fr g_2, \fr g_3)}{\alpha(\fr g_2,\fr g_3,\fr g_4) \, \alpha(\fr g_1, \fr g_2 \fr g_3, \fr g_4) \, \alpha(\fr g_2,\fr g_2 \fr c,\fr c^{-1} \fr g_3)}
	\\[-0.9em]
	& \hspace{6.5em} \times
	\big|\biGr{\fr g_1}{\mathbbm 1_A}{\mathbbm 1_B}\big\ra
	\otimes
	\big|\biGr{\fr g_2}{\mathbbm 1_B}{\fr c}\big\ra
	\otimes
	\big|\biGr{\fr g_3}{\fr c}{\mathbbm 1_D}\big\ra
	\otimes
	\big|\biGr{\fr g_4}{\mathbbm 1_D}{\mathbbm 1_E}\big\ra \, .
\end{align*}
It remains to multiply both expression from the right by $\mathbbm 1_{(((AB)C)D)E}$. First, we compute the right-and side of \eqref{eq:pentExp}:
\begin{align*}
	{\rm r.h.s}\eqref{eq:pentExp} &=
	\frac{1}{|B|^2|C||D|^2} \!\!\!
	\sum_{\substack{\{ \fr g \} \\ \fr{b,b',c,d,d'}}} \!
	\frac{\zeta^{\Lambda(CDE)}_{\mathbbm 1_C,\fr d,\mathbbm 1_E}(\fr g_3, \fr g_4) \, \zeta^{\Lambda(ABC)}_{\mathbbm 1_A,b,\mathbbm 1_C}(\fr g_1, \fr g_2)}{\alpha(\fr g_1, \fr g_2, \fr g_3 \fr g_4) \, \alpha(\fr g_1 \fr g_2,\fr g_3 \fr d,\fr d^{-1} \fr g_4)}
	\\[-1em]
	& \hspace{10.1em}\!\! \times 
	\zeta^{\Lambda(ADE)}_{\mathbbm 1_A,\fr d',\mathbbm 1_E}(\fr g_1 \fr g_2 \fr g_3 \fr d,\fr d^{-1} \fr g_4) \,
	\zeta^{\Lambda(ACD)}_{\mathbbm 1_A,\fr c, \fr d'}(\fr g_1 \fr g_2,\fr g_3 \fr d) \,
	\zeta^{\Lambda(ABC)}_{\mathbbm 1_A,\fr b',\fr c}(\fr g_1 \fr b,\fr b^{-1}\fr g_2)
	\\[0.4em]
	& \hspace{10.1em}\!\! \times 
	\vartheta^{\Lambda(AB)}_{\fr g_1}(\mathbbm 1_A, \mathbbm 1_A|\fr b,\fr b') \,
	\vartheta^{\Lambda(BC)}_{\fr g_2}(\fr b, \fr b'| \mathbbm 1_C,\fr c) \,
	\vartheta^{\Lambda(CD)}_{\fr g_3}(\mathbbm 1_C, \fr c|\fr d,\fr d')
	\\
	& \hspace{10.1em} \!\! \times
	\vartheta^{\Lambda(DE)}_{\fr g_4}(\fr d, \fr d'|\mathbbm 1_E ,\mathbbm 1_E)  \big|\biGr{\fr g_1}{\mathbbm 1_A}{\fr b}\big\ra
	\otimes
	\big|\biGr{\fr g_2}{b}{\mathbbm 1_C}\big\ra
	\otimes
	\big|\biGr{\fr g_3}{\mathbbm 1_C}{\fr d}\big\ra
	\otimes
	\big|\biGr{\fr g_4}{\fr d}{\mathbbm 1_E}\big\ra  .
\end{align*}
Using the cocycle relations
\begin{equation*}
	\frac{\vartheta^{\Lambda(AB)}_{\fr g_1}(\mathbbm 1_A,\mathbbm 1_A|\fr b,\fr b') \, \vartheta^{\Lambda(BC)}_{\fr g_2}(\fr b, \fr b'|\mathbbm 1_C,\fr c)}{\vartheta^{\Lambda(AC)}_{\fr g_1 \fr g_2}(\mathbbm 1_A,\mathbbm 1_A|\mathbbm 1_C,\fr c)}
	=
	\frac{\zeta^{\Lambda(ABC)}_{\mathbbm 1_A,\fr b \fr b',\fr c}(\fr g_1, \fr g_2)}{\zeta^{\Lambda(ABC)}_{\mathbbm 1_A,\fr b,\mathbbm 1_C}(\fr g_1, \fr g_2) \, \zeta^{\Lambda(ABC)}_{\mathbbm 1_A,\fr b',\fr c}(\fr g_1 \fr b,\fr b^{-1}\fr g_2 \fr c)}
\end{equation*}
and 
\begin{align*}
	\frac{\vartheta^{\Lambda(CD)}_{\fr g_3}(\mathbbm 1_C,\fr c|\fr d,\fr d') \, \vartheta^{\Lambda(DE)}_{\fr g_4}(\fr d, \fr d'|\mathbbm 1_E,\mathbbm 1_E)}{\vartheta^{\Lambda(CE)}_{\fr g_3 \fr g_4}(\mathbbm 1_C,\fr c|\mathbbm 1_E,\mathbbm 1_E)}
	=
	\frac{\zeta^{\Lambda(CDE)}_{\fr c,\fr d \fr d',\mathbbm 1_E}(\fr g_3, \fr g_4)}{\zeta^{\Lambda(CDE)}_{\mathbbm 1_C,\fr d,\mathbbm 1_E}(\fr g_3, \fr g_4) \, \zeta^{\Lambda(CDE)}_{\fr c,\fr d',\mathbbm 1_E}(\fr g_3 \fr d,\fr d^{-1}\fr g_4)}
\end{align*}
as well as the quasi-coassociativity conditions
\begin{align*}
	\frac{\zeta^{\Lambda(CDE)}_{\fr c,\fr d',\mathbbm 1_E}(\fr g_3 \fr d,\fr d^{-1}\fr g_4)  \, \zeta^{\Lambda(BCE)}_{\mathbbm 1_B,\fr c,\mathbbm 1_E}(\fr g_1 \fr g_2, \fr g_3 \fr g_4) }{\zeta^{\Lambda(BDE)}_{\mathbbm 1_B,\fr d,\mathbbm 1_E}(\fr g_1 \fr g_2 \fr g_3 \fr d, \fr d^{-1}\fr g_4) \, \zeta^{\Lambda(BCD)}_{\mathbbm 1_B,\fr c,\fr d'}(\fr g_1 \fr g_2,\fr g_3 \fr d)}
	=
	\frac{	\alpha(\fr g_{1}\fr g_{2},\fr g_{3}\fr d,\fr d^{-1}\fr g_4)}{\alpha(\fr g_1 \fr g_2 \fr c,\fr c^{-1}\fr g_{3}\fr d \fr d',\fr d'^{-1}\fr d^{-1}\fr g_4)}
\end{align*}
and
\begin{align*}
	\frac{\zeta^{\Lambda(CDE)}_{\fr c,\fr d \fr d',\mathbbm 1_E}(\fr g_3, \fr g_4)  \, \zeta^{\Lambda(BCE)}_{\mathbbm 1_B,\fr c,\mathbbm 1_E}(\fr g_1 \fr g_2,\fr g_3\fr g_4) }{\zeta^{\Lambda(BDE)}_{\mathbbm 1_B,\fr d \fr d',\mathbbm 1_E}(\fr g_1 \fr g_2 \fr g_3, \fr g_4) \, \zeta^{\Lambda(BCD)}_{\mathbbm 1_B,\fr c,\fr d \fr d'}(\fr g_1 \fr g_2, \fr g_3)}
	=
	\frac{	\alpha(\fr g_{1}\fr g_{2},\fr g_{3},\fr g_4)}{\alpha(\fr g_1 \fr g_2\fr c,\fr c^{-1}\fr g_{3}\fr d \fr d',\fr d'^{-1}\fr d^{-1}\fr g_4)}
\end{align*}
yields
\begin{align*}
	{\rm r.h.s}\eqref{eq:pentExp} &= 
	\frac{1}{|B||C||D|}
	\sum_{\substack{ \{ \fr g \} \\ b,c,d}}
	\frac{\zeta^{\Lambda(CDE)}_{\mathbbm 1_C,\fr d,\mathbbm 1_E}(\fr g_1 \fr g_2 \fr g_3, \fr g_4) \,
		\zeta^{\Lambda(BCD)}_{\mathbbm 1_B,\fr c,\fr d}(\fr g_1 \fr g_2, \fr g_3) \,
		\zeta^{\Lambda(ABC)}_{\mathbbm 1_A,\fr b,\fr c}(\fr g_1, \fr g_2)}{	\alpha(\fr g_1, \fr g_2, \fr g_3 \fr g_4) \,
		\alpha(\fr g_1 \fr g_2, \fr g_3, \fr g_4)}
	\\[-1em]
	& \hspace{8.2em} \times
	\big|\biGr{\fr g_1}{\mathbbm 1_A}{\fr b}\big\ra
	\otimes
	\big|\biGr{\fr g_2}{\fr b}{\fr c}\big\ra
	\otimes
	\big|\biGr{\fr g_3}{\fr c}{\fr d}\big\ra
	\otimes
	\big|\biGr{\fr g_4}{\fr d}{\mathbbm 1_E}\big\ra \; .
\end{align*}

\bigskip \noindent
Let us repeat the same procedure in order to compute the left-hand side of \eqref{eq:pentExp}:
\begin{align*}
	{\rm l.h.s}\eqref{eq:pentExp} &=
	\frac{1}{|B||C|^2|D|}\sum_{\substack{\{ \fr g \} \\ \fr b,\fr c,\fr c',\fr d}}
	\frac{\zeta^{\Lambda(BCD)}_{\mathbbm 1_B,\fr c,\mathbbm 1_D}(\fr g_2, \fr g_3)}{\alpha(\fr g_2,\fr g_3,\fr g_4) \, \alpha(\fr g_1, \fr g_2 \fr g_3, \fr g_4) \, \alpha(\fr g_2,\fr g_2 \fr c,\fr c^{-1}\fr g_3)}
	\\[-1em]
	& \hspace{8.8em} \times 
	\zeta^{\Lambda(ADE)}_{\mathbbm 1_A,\fr d,\mathbbm 1_E}(\fr g_1 \fr g_2 \fr g_3, \fr g_4) \,
	\zeta^{\Lambda(ACD)}_{\mathbbm 1_A,\fr c',\fr d}(\fr g_1\fr g_2\fr c,\fr c^{-1}\fr g_3) \,
	\zeta^{\Lambda(ABC)}_{\mathbbm 1_A,\fr b,\fr c'}(\fr g_1,\fr g_2 \fr c)
	\\[0.4em]
	& \hspace{8.8em} \times 
	\vartheta^{\Lambda(AB)}_{\fr g_1}(\mathbbm 1_A, \mathbbm 1_A | \mathbbm 1_B,\fr b) \,
	\vartheta^{\Lambda(BC)}_{\fr g_2}(\mathbbm 1_B, \fr b | \fr c,\fr c') \, 
	\vartheta^{\Lambda(CD)}_{\fr g_3}(\fr c,\fr c' | \mathbbm 1_D,\fr d)
	\\
	& \hspace{8.8em} \times
		\vartheta^{\Lambda(DE)}_{\fr g_4}(\mathbbm 1_D, \fr d | \mathbbm 1_E,\mathbbm 1_E) 
	\big|\biGr{\fr g_1}{\mathbbm 1_A}{\mathbbm 1_B}\big\ra
	\otimes
	\big|\biGr{\fr g_2}{\mathbbm 1_B}{\fr c}\big\ra
	\otimes
	\big|\biGr{\fr g_3}{\fr c}{\mathbbm 1_D}\big\ra
	\otimes
	\big|\biGr{\fr g_4}{\mathbbm 1_D}{\mathbbm 1_E}\big\ra .
\end{align*}
Using the cocycle relation
\begin{equation*}
	\frac{\vartheta^{\Lambda(BC)}_{\fr g_2}(\mathbbm 1_B,\fr b|\fr c,\fr c') \, \vartheta^{\Lambda(CD)}_{\fr g_3}(\fr c,\fr c'|\mathbbm 1_D,\fr d)}{\vartheta^{\Lambda(BD)}_{\fr g_2 \fr g_3}(\mathbbm 1_B,\fr b|\mathbbm 1_D,\fr d)}
	=
	\frac{\zeta^{\Lambda(BCD)}_{\fr b, \fr c \fr c',d}(\fr g_2, \fr g_3)}{\zeta^{\Lambda(BCD)}_{\mathbbm 1_B,\fr c,\mathbbm 1_D}(\fr g_2, \fr g_3) \, \zeta^{\Lambda(BCD)}_{\fr b,\fr c',\fr d}(\fr g_2 \fr c,\fr c^{-1}\fr g_3)}
\end{equation*}
as well as the quasi-coassociativity conditions
\begin{align*}
	\frac{\zeta^{\Lambda(BCD)}_{\fr b,\fr c',\fr d}(\fr g_2 \fr c,\fr c^{-1}\fr g_3)  \, \zeta^{\Lambda(ABD)}_{\mathbbm 1_A,\fr b,\fr d}(\fr g_1 ,\fr g_2 \fr g_3) }{\zeta^{\Lambda(ACD)}_{\mathbbm 1_A,\fr c',\fr d}(\fr g_1 \fr g_2 \fr c,\fr c^{-1}\fr g_3) \, \zeta^{\Lambda(ABC)}_{\mathbbm 1_A,\fr b,\fr c'}(\fr g_1,\fr g_2 \fr c)}
	=
	\frac{	\alpha(\fr g_1,\fr g_2\fr c,\fr c^{-1}\fr g_3)}{\alpha(\fr g_1 \fr b,\fr b^{-1}\fr g_2\fr c\fr c',\fr c'^{-1}\fr c^{-1}\fr g_3\fr d)}
\end{align*}
and
\begin{align*}
	\frac{\zeta^{\Lambda(BCD)}_{\fr b, \fr c \fr c',\fr d}(\fr g_2, \fr g_3)  \, \zeta^{\Lambda(ABD)}_{\mathbbm 1_A,\fr b,\fr d}(\fr g_1 ,\fr g_2 \fr g_3) }{\zeta^{\Lambda(ACD)}_{\mathbbm 1_A, \fr c \fr c',\fr d}(\fr g_1 \fr g_2, \fr g_3) \, \zeta^{\Lambda(ABC)}_{\mathbbm 1_A,\fr b, \fr c \fr c'}(\fr g_1, \fr g_2)}
	=
	\frac{	\alpha(\fr g_1 ,\fr g_2, \fr g_3)}{\alpha(\fr g_1 \fr b,\fr b^{-1}\fr g_2 \fr c \fr c',\fr c'^{-1}\fr c^{-1}\fr g_3 \fr d)}
\end{align*}
yields 
\begin{align*}
	{\rm l.h.s}\eqref{eq:pentExp}
	&=	
	\frac{1}{|B||C||D|}
	\sum_{\substack{\{ \fr g \} \\ \fr b,\fr c,d}}
	\frac{\zeta^{\Lambda(ADE)}_{\mathbbm 1_A,\fr d,\mathbbm 1_E}(\fr g_1 \fr g_2 \fr g_3, \fr g_4) \,
		\zeta^{\Lambda(ACD)}_{\mathbbm 1_A,\fr c,\fr d}(\fr g_1 \fr g_2, \fr g_3) \,
		\zeta^{\Lambda(ABC)}_{\mathbbm 1_A,\fr b,\fr c}(\fr g_1, \fr g_2) }{\alpha(\fr g_1 ,\fr g_2, \fr g_3) \, \alpha(\fr g_1, \fr g_2 \fr g_3, \fr g_4) \, \alpha(\fr g_2, \fr g_3,\fr g_4)}
	\\[-1em]
	& \hspace{8.2em}
	\times
	\big|\biGr{\fr g_1}{\mathbbm 1_A}{\fr b}\big\ra
	\otimes
	\big|\biGr{\fr g_2}{\fr b}{\fr c}\big\ra
	\otimes
	\big|\biGr{\fr g_3}{\fr c}{\fr d}\big\ra
	\otimes
	\big|\biGr{\fr g_4}{\fr d}{\mathbbm 1_E}\big\ra \; .
\end{align*}
The equality between ${\rm l.h.s}\eqref{eq:pentExp}$ and ${\rm r.h.s}\eqref{eq:pentExp}$ finally follows from the groupoid 3-cocycle condition $d^{(3)}\alpha = 1$, hence the pentagon identity.

\newpage
\section{Canonical basis for boundary excitations in (2+1)d}
\emph{In this appendix, we collect the proofs of some properties crucial to the definition of the canonical basis presented in sec.~\ref{sec:canonicalBasis}.}

\subsection{Proof of the canonical algebra product \eqref{eq:diagProd}\label{sec:app_diag}}
Using transformations \eqref{eq:FtransCyl} and \eqref{eq:FtransInvCyl}, as well as the definition of the $\star$-product, we have
\begin{align*}
	&|\rho_{AB} I J \ra \star | \rho_{AB}' I' J' \ra 
	\\
	& \q = 
	\frac{(d_{\rho_{AB}} d_{\rho_{AB}'})^\frac{1}{2}}{|A||B|}
	\!\!\!\!\!
	\sum_{\substack{g,g' \in G \\ (a,b),(a',b') \in A \times B}}
	\!\!\!\!\!
	\overline{\mc{D}^{\rho_{AB}}_{IJ}\big( \big|\biGr{g}{a}{b}\big\ra \big)}
	\,
	\overline{\mc{D}^{\rho'_{AB}}_{I'J'}\big( \big|\biGr{g'}{a'}{b'}\big\ra \big)}
	\,
	\big|\biGr{g}{a}{b}\big\ra
	\star\big|\biGr{g'}{a'}{b'}\big\ra
	\\
	& \q =
	\frac{(d_{\rho_{AB}} d_{\rho_{AB}'})^\frac{1}{2}}{|A||B|}
	\!\!\!\!\!
	\sum_{\substack{g,g' \in G \\ (a,b),(a',b') \in A \times B}}
	\!\!\!\!\!
	\overline{\mc{D}^{\rho_{AB}}_{IJ}\big( \big|\biGr{g}{a}{b}\big\ra \big)}
	\,
	\overline{\mc{D}^{\rho'_{AB}}_{I'J'}\big( \big|\biGr{g'}{a'}{b'}\big\ra \big)}
	\, \delta_{g',a^{-1}gb} \, \vartheta^{AB}_{g}(a,a'|b,b') \,
	\big|\biGr{g}{aa'}{ab'}\big\ra	
	\\
	& \q =
	\frac{(d_{\rho_{AB}} d_{\rho_{AB}'})^\frac{1}{2}}{|A||B|}
	\!\!\!\!\!
	\sum_{\substack{g,g' \in G \\ (a,b),(a',b') \in A \times B}}
	\!\!\!\!\!
	\overline{\mc{D}^{\rho_{AB}}_{IJ}\big( \big|\biGr{g}{a}{b}\big\ra \big)}
	\,
	\overline{\mc{D}^{\rho'_{AB}}_{I'J'}\big( \big|\biGr{g'}{a'}{b'}\big\ra \big)}
	\delta_{g',a^{-1}xb} \, \vartheta_{g}(a,a'|b,b')
	\\[-0.5em]
	& \hspace{12em} \times \Big(\frac{1}{|A||B|}\Big)^{\frac{1}{2}}
	\sum_{\rho_{AB}''} d_{\rho_{AB}''}^\frac{1}{2}
	\sum_{I'',J''}
	\mc{D}^{\rho''_{AB}}_{I''J''}\big( \big|\biGr{g}{aa'}{bb'}\big\ra \big)
	|\rho_{AB}'' I''J'' \ra \; .
\end{align*}
But by linearity of the representation matrices, we have
\begin{equation}
	\delta_{g',a^{-1}gb} \, \vartheta^{AB}_{g}(a,a'|b,b') \, 
	\mc{D}^{\rho''_{AB}}_{I''J''}\big( \big| \biGr{g}{aa'}{bb'} \big\ra \big)
	=
	\sum_{K}	
	\mc{D}^{\rho''_{AB}}_{I''K}\big( \big| \biGr{g}{a}{b} \big\ra \big)
	\mc{D}^{\rho''_{AB}}_{KJ''}\big( \big| \biGr{g'}{a'}{b'} \big\ra \big) \; .
\end{equation}
Orthogonality of the representation matrices finally yields the desired expression
\begin{equation}
	|\rho_{AB} IJ \ra \star | \rho_{AB}' I' J' \ra 
	=
	|A|^\frac{1}{2} |B|^\frac{1}{2}
	\frac{\delta_{\rho_{AB},\rho_{AB}'} \, \delta_{J,I'}}{d_{\rho_{AB}}^\frac{1}{2}}| \rho_{AB} IJ' \ra \; .
\end{equation}

\subsection{Ground state projector on the annulus\label{sec:app_gsProjAnn}}
Let us evaluate the quantity
\begin{equation}
	\frac{1}{|A||B|}
	\sum_{\substack{g \in G \\ (a,b) \in A \times B}}
	\sum_{\substack{\tilde g \in G \\ (\tilde a, \tilde b) \in A \times B}}
	\Big( \big| \biGr{\tilde g}{\tilde a}{\tilde b} \big\ra^{-1} \star \big| \biGr{g}{a}{b} \big\ra \star \big| \biGr{\tilde g}{\tilde a}{\tilde b} \big\ra \Big) \big\la \biGr{g}{a}{b} \big| \; ,
\end{equation}
and confirm that it is equal to $\mathbb P_{\mathbb O_\triangle}$ as defined in \eqref{eq:annProj}. By direct computation, we have
\begin{align}
	\big| \biGr{g}{a}{b} \big\ra
	\star
	\big| \biGr{\tilde g}{\tilde a}{\tilde b} \big\ra
	=
	\delta_{\tilde g,a^{-1}gb} \,
	\vartheta^{AB}_g(a,\tilde a|b,\tilde b) \,
	\, \big| \biGr{g}{a\tilde a}{b \tilde b} \big\ra 
\end{align}
and
\begin{align}
	\big| \biGr{\tilde g}{\tilde a}{\tilde b} \big\ra^{-1}
	\star
	\big| \biGr{g}{a \tilde a}{b \tilde b} \big\ra
	=
	\delta_{\tilde g,g} \,
	\frac{\vartheta^{AB}_{\tilde a^{-1}g \tilde b}(\tilde a^{-1},a \tilde a|\tilde b^{-1},b \tilde b)}{\vartheta^{AB}_{\tilde g}(\tilde a, \tilde a^{-1}| \tilde b , \tilde b^{-1})} \,
	\, \big| \biGr{\tilde a^{-1}g \tilde b}{\tilde a^{-1}a\tilde a}{\tilde b^{-1}b \tilde b} \big\ra 
\end{align}
so that 
\begin{equation}
	\big| \biGr{\tilde g}{\tilde a}{\tilde b} \big\ra^{-1} \star \big| \biGr{g}{a}{b} \big\ra \star \big| \biGr{\tilde g}{\tilde a}{\tilde b} \big\ra
	=
	\delta_{\tilde g,g} \, \delta_{\tilde g, a^{-1}gb} \,
	\frac{\vartheta^{AB}_g(a,\tilde a| b \tilde b) \, \vartheta^{AB}_{\tilde a^{-1}g \tilde b}(\tilde a^{-1},a \tilde a|\tilde b^{-1},b \tilde b)}{\vartheta^{AB}_{\tilde g}(\tilde a, \tilde a^{-1}| \tilde b , \tilde b^{-1})} \,
	\, \big| \biGr{\tilde a^{-1}g \tilde b}{\tilde a^{-1}a\tilde a}{\tilde b^{-1}b \tilde b} \big\ra \; .
\end{equation}
Using the groupoid cocycle condition $d^{(2)}\vartheta^{AB}_g(\tilde a, \tilde a^{-1}, a \tilde a | \tilde b, \tilde b^{-1}, b \tilde b)$ and performing the summations finally yield the desired result. 
\subsection{Proof of the diagonalisation property \eqref{eq:diagTrinion}\label{sec:app_diagProj}}

Given the action of the Hamiltonian projector \eqref{eq:projTrinion} on $\mathbb Y_\triangle$, we show that the basis states defined as
\begin{align*}
	&|\rho_{AB}I_{AB},\rho_{BC}I_{BC},\rho_{AC}I_{AC} \ra_{{\mathbb Y}_\triangle}
	\\
	& \q :=
	\sum_{\substack{g_1,g_2 \in G \\ a,a' \in A \\ b,b' \in B \\ c,c' \in C}}\sum_{\{J\}}
	\; \; 
	\overline{\mc{D}^{\rho_{AB}}_{J_{AB}I_{AB}}\big( \big|\biGr{g_1}{a}{b}\big\ra \big)} \, 
	\overline{\mc{D}^{\rho_{BC}}_{J_{BC}I_{BC}} \big( \big|\biGr{g_2}{b'}{c}\big\ra\big)}
	\CC{\rho_{AB}}{\rho_{BC}}{\rho_{AC}}{J_{AB}}{J_{BC}}{J_{AC}}
	\overline{\mc{D}^{\rho_{AC}}_{I_{AC}J_{AC}} \big( \big|\biGr{a'g_1g_2c'^{-1}}{a'}{c'}\big\ra\big)}
	\\[-2em]
	& \hspace{6.7em}\times 
	| g_1,a,b,g_2,b',c,a',c' \ra_{{\mathbb Y}_\triangle}
\end{align*}
satisfy the relation
\begin{equation}
	\mathbb P_{\mathbb Y_\triangle}\big(|\rho_{AB}I_{AB},\rho_{BC}I_{BC},\rho_{AC}I_{AC} \ra_{{\mathbb Y}_\triangle}\big) = |\rho_{AB}I_{AB},\rho_{BC}I_{BC},\rho_{AC}I_{AC} \ra_{{\mathbb Y}_\triangle} \; .
\end{equation}
By direct computation, we have
\begin{align*}
	&\mathbb P_{\mathbb Y_\triangle}\big(|\rho_{AB}I_{AB},\rho_{BC}I_{BC},\rho_{AC}I_{AC} \ra_{{\mathbb Y}_\triangle}\big)
	\\
	& \q = 
	\sum_{\substack{\{g \in G\} \\ a,a' \in A \\ b,b' \in B \\ c,c' \in C}}\sum_{\{J\}}
	\; \; 
	\overline{\mc{D}^{\rho_{AB}}_{J_{AB}I_{AB}}\big( \big|\biGr{g_1}{a}{b}\big\ra \big)} \, 
	\overline{\mc{D}^{\rho_{BC}}_{J_{BC}I_{BC}} \big( \big|\biGr{g_2}{b'}{c}\big\ra\big)}
	\CC{\rho_{AB}}{\rho_{BC}}{\rho_{AC}}{J_{AB}}{J_{BC}}{J_{AC}}
	\overline{\mc{D}^{\rho_{AC}}_{I_{AC}J_{AC}} \big( \big|\biGr{a'g_1g_2c'^{-1}}{a'}{c'}\big\ra\big)}
	\\[-2.5em]
	& \hspace{6em} \times 
	\frac{1}{|A||B||C|} 
	\sum_{\substack{\tilde a \in A \\ \tilde b \in B \\ \tilde c \in C}} \;\;
	\frac{\vartheta^{AC}_{a'g_1g_2c'^{-1}}(a',\tilde a|c', \tilde c)}{\vartheta^{AB}_{g_1}(\tilde a, \tilde a^{-1}a| \tilde b, \tilde b^{-1}b) \, \vartheta^{BC}_{g_2}(\tilde b, \tilde b^{-1}b'| \tilde c, \tilde c^{-1}c) \, \zeta^{ABC}_{\tilde a, \tilde b, \tilde c}(g_1,g_2)}
	\\[-0.5em]
	& \hspace{12.9em} \times
	| \tilde a^{-1}g_1\tilde b,\tilde a^{-1}a,\tilde b^{-1}b,\tilde b^{-1}g_2 \tilde c,\tilde b^{-1} b',\tilde c^{-1}c,a'\tilde a , c' \tilde c \ra_{{\mathbb Y}_\triangle} .
\end{align*}
Using the invariance property \eqref{eq:gaugeConj} of the Clebsch-Gordan series, we can rewrite the previous quantity as
\begin{align*}
	&\mathbb P_{\mathbb Y_\triangle}\big(|\rho_{AB}I_{AB},\rho_{BC}I_{BC},\rho_{AC}I_{AC} \ra_{{\mathbb Y}_\triangle}\big)
	\\
	& \q = 
	\frac{1}{|A||B|^2|C|}  
	\sum_{\substack{\{g \in G\} \\ \tilde a ,a,a' \in A \\ \tilde b, \tilde b', b,b' \in B \\ \tilde c,c,c' \in C}}
	\sum_{\{J\}}
	\; \; 
	\overline{\mc{D}^{\rho_{AB}}_{J_{AB}I_{AB}}\big( \big|\biGr{\tilde a^{-1}g_1\tilde b'}{\tilde a^{-1}a}{\tilde b'^{-1}b}\big\ra \big)} \, 
	\overline{\mc{D}^{\rho_{BC}}_{J_{BC}I_{BC}} \big( \big|\biGr{\tilde b'^{-1}g_2\tilde c}{\tilde b'^{-1}b'}{\tilde c^{-1}c}\big\ra\big)}
	\\[-2.5em]
	& \hspace{12em} \times
	\CC{\rho_{AB}}{\rho_{BC}}{\rho_{AC}}{J_{AB}}{J_{BC}}{J_{AC}}
	\overline{\mc{D}^{\rho_{AC}}_{I_{AC}J_{AC}} \big( \big|\biGr{a'g_1g_2c'^{-1}}{a'\tilde a}{c' \tilde c}\big\ra\big)}
	\\
	& \hspace{12em} \times
	\frac{\vartheta^{AB}_{g_1}(\tilde a, \tilde a^{-1}a| \tilde b', \tilde b'^{-1}b) \, \vartheta^{BC}_{g_2}(\tilde b', \tilde b'^{-1}b'| \tilde c, \tilde c^{-1}c) \, \zeta^{ABC}_{\tilde a, \tilde b', \tilde c}(g_1,g_2)}{\vartheta^{AB}_{g_1}(\tilde a, \tilde a^{-1}a| \tilde b, \tilde b^{-1}b) \, \vartheta^{BC}_{g_2}(\tilde b, \tilde b^{-1}b'| \tilde c, \tilde c^{-1}c) \, \zeta^{ABC}_{\tilde a, \tilde b, \tilde c}(g_1,g_2)}
	\\
	& \hspace{12em} \times
	| \tilde a^{-1}g_1\tilde b,\tilde a^{-1}a,\tilde b^{-1}b,\tilde b^{-1}g_2 \tilde c,\tilde b^{-1} b',\tilde c^{-1}c,a'\tilde a , c' \tilde c \ra_{{\mathbb Y}_\triangle} \; .
\end{align*}
Let us now use the fact that
\begin{equation*}
	\frac{\zeta^{ABC}_{\tilde a, \tilde b', \tilde c}(g_1,g_2)}{\zeta^{ABC}_{\tilde a, \tilde b, \tilde c}(g_1,g_2)} 
	= 
	\zeta^{ABC}_{\mathbbm 1_A, \tilde b^{-1}\tilde b',\mathbbm 1_C}(\tilde a^{-1}g_1\tilde b, \tilde b^{-1}g_2 \tilde c) \, 
	\vartheta^{AB}_{g_1}(\tilde a , \mathbbm 1_A| \tilde b, \tilde b^{-1}\tilde b') \, 
	\vartheta^{BC}_{g_2}(\tilde b, \tilde b^{-1}\tilde b'|\tilde c, \mathbbm 1_C)
\end{equation*}
as well as the groupoid cocycle conditions
\begin{equation*}
	d^{(2)}\vartheta^{AB}_{g_1}(\tilde a, \mathbbm 1_A, \tilde a^{-1}a|\tilde b, \tilde b^{-1}\tilde b', \tilde b'^{-1}b) = 1 \q \text{and} \q d^{(2)}\vartheta^{BC}_{g_2}(\tilde b, \tilde b^{-1}\tilde b',\tilde b'^{-1}b|\tilde c, \mathbbm 1_C, \tilde c^{-1}c)
\end{equation*}
in order to rewrite
\begin{align*}
	&\frac{\vartheta^{AB}_{g_1}(\tilde a, \tilde a^{-1}a| \tilde b', \tilde b'^{-1}b) \, \vartheta^{BC}_{g_2}(\tilde b', \tilde b'^{-1}b'| \tilde c, \tilde c^{-1}c) \, \zeta^{ABC}_{\tilde a, \tilde b', \tilde c}(g_1,g_2)}{\vartheta^{AB}_{g_1}(\tilde a, \tilde a^{-1}a| \tilde b, \tilde b^{-1}b) \, \vartheta^{BC}_{g_2}(\tilde b, \tilde b^{-1}b'| \tilde c, \tilde c^{-1}c) \, \zeta^{ABC}_{\tilde a, \tilde b, \tilde c}(g_1,g_2)} 
	\\
	& \q = \vartheta^{AB}_{\tilde a^{-1}g_1\tilde b}(\mathbbm 1_A, \tilde a^{-1}a| \tilde b^{-1}\tilde b',\tilde b'^{-1}b) \, 
	\vartheta^{BC}_{\tilde b^{-1}g_2\tilde c}(\tilde b^{-1}\tilde b', \tilde b'^{-1}b'| \mathbbm 1_C, \tilde c^{-1}c)
	\, \zeta^{ABC}_{\mathbbm 1_A, \tilde b^{-1}\tilde b',\mathbbm 1_C}(\tilde a^{-1}g_1\tilde b, \tilde b^{-1}g_2 \tilde c) \; .
\end{align*}
Performing a simple relabelling of summation variables, we then obtain
\begin{align*}
	&\mathbb P_{\mathbb Y_\triangle}\big(|\rho_{AB}I_{AB},\rho_{BC}I_{BC},\rho_{AC}I_{AC} \ra_{{\mathbb Y}_\triangle}\big)
	\\
	& \q = 
	\frac{1}{|A||B|^2|C|} 
	\sum_{\substack{\{g \in G\} \\ \tilde a ,a,a' \in A \\ \tilde b, \tilde b', b,b' \in B \\ \tilde c,c,c' \in C}}
	\sum_{\{J\}}
	\; \; 
	\overline{\mc{D}^{\rho_{AB}}_{J_{AB}I_{AB}}\big( \big|\biGr{g_1\tilde b^{-1} \tilde b'}{a}{\tilde b'^{-1}b}\big\ra \big)} \, 
	\overline{\mc{D}^{\rho_{BC}}_{J_{BC}I_{BC}} \big( \big|\biGr{\tilde b'^{-1}\tilde bg_2}{\tilde b'^{-1}b'}{c}\big\ra\big)}
	\\[-2.5em]
	& \hspace{12em} \times
	\CC{\rho_{AB}}{\rho_{BC}}{\rho_{AC}}{J_{AB}}{J_{BC}}{J_{AC}}
	\overline{\mc{D}^{\rho_{AC}}_{I_{AC}J_{AC}} \big( \big|\biGr{g_3}{a'}{c'}\big\ra\big)}
	\\
	& \hspace{12em} \times
	\vartheta^{AB}_{g_1}(\mathbbm 1_A, a| \tilde b^{-1}\tilde b',\tilde b'^{-1}b) \, 
	\vartheta^{BC}_{g_2}(\tilde b^{-1}\tilde b', \tilde b'^{-1}b'| \mathbbm 1_C, c)
	\, \zeta^{ABC}_{\mathbbm 1_A, \tilde b^{-1}\tilde b',\mathbbm 1_C}( g_1, g_2 ) 
	\\
	& \hspace{12em} \times
	| g_1,a,\tilde b^{-1}b,g_2,\tilde b^{-1} b',c,a',c' \ra_{{\mathbb Y}_\triangle} \; .
\end{align*}
Moreover, let us notice that \eqref{eq:gaugeConj} induces
\begin{align*}
	&\frac{1}{|B|} 
	\sum_{\tilde b'}
	\sum_{\{J\}}
	\; \; 
	\overline{\mc{D}^{\rho_{AB}}_{J_{AB}I_{AB}}\big( \big|\biGr{g_1\tilde b^{-1} \tilde b'}{a}{\tilde b'^{-1}b}\big\ra \big)} \, 
	\overline{\mc{D}^{\rho_{BC}}_{J_{BC}I_{BC}} \big( \big|\biGr{\tilde b'^{-1}\tilde bg_2}{\tilde b'^{-1}b'}{c}\big\ra\big)}
	\\
	& \hspace{4.2em} \times
	\CC{\rho_{AB}}{\rho_{BC}}{\rho_{AC}}{J_{AB}}{J_{BC}}{J_{AC}}
	\overline{\mc{D}^{\rho_{AC}}_{I_{AC}J_{AC}} \big( \big|\biGr{a'g_1g_2c'^{-1}}{a'}{c'}\big\ra\big)}
	\\
	& \hspace{4.2em} \times
	\vartheta^{AB}_{g_1}(\mathbbm 1_A, a| \tilde b^{-1}\tilde b',\tilde b'^{-1}b) \, 
	\vartheta^{BC}_{g_2}(\tilde b^{-1}\tilde b', \tilde b'^{-1}b'| \mathbbm 1_C, c)
	\, \zeta^{ABC}_{\mathbbm 1_A, \tilde b^{-1}\tilde b',\mathbbm 1_C}( g_1, g_2 ) \,
	\\
	& \q =
	\sum_{\{J\}}
	\overline{\mc{D}^{\rho_{AB}}_{J_{AB}I_{AB}}\big( \big|\biGr{g_1}{a}{\tilde b^{-1}b}\big\ra \big)} \,
	\overline{\mc{D}^{\rho_{BC}}_{J_{BC}I_{BC}} \big( \big|\biGr{g_2}{\tilde b^{-1}b'}{c}\big\ra\big)}
	\CC{\rho_{AB}}{\rho_{BC}}{\rho_{AC}}{J_{AB}}{J_{BC}}{J_{AC}}
	\overline{\mc{D}^{\rho_{AC}}_{I_{AC}J_{AC}} \big( \big|\biGr{a'g_1g_2c'^{-1}}{a'}{c'}\big\ra\big)} \; .
\end{align*}
A final relabelling of summation variables yields the desired result.

\bibliographystyle{JHEP}
\bibliography{ref_cat}

\end{document}